 \newcommand{\mydriver}{pdflatex} 

\documentclass[12pt,\mydriver]{Files/thesis}  

\usepackage{titlesec}
   \titleformat{\chapter}
      {\normalfont\large}{Chapter \thechapter:}{1em}{}
\usepackage{tikz}
\usepackage{times}
\usepackage[english]{babel}
\usepackage{subcaption}
\usepackage{amsmath}
\usepackage{amsfonts}
\usepackage{amssymb}
\usepackage{braket}
\usepackage{hhline}
\usepackage{graphicx}
\usepackage{cite}
\usepackage{lscape}
\usepackage{indentfirst}
\usepackage{latexsym}
\usepackage{multirow}
\usepackage{epstopdf}
\usepackage{tabls}
\usepackage{amsthm}
\usepackage{thmtools}
\usepackage{wrapfig}
\usepackage{Files/slashbox}
\usepackage{float}
\usepackage{bm}
\usepackage{caption}
\usepackage{supertabular}
\usepackage{amsmath}
\usepackage{physics}
\usepackage{graphicx}
\usepackage{amssymb}
\usepackage{enumitem}
\usepackage{mathtools}
\usepackage{siunitx}
\usepackage{newtxmath}
\usepackage{titling}
\usepackage{authblk}
\newcommand{\SFF}{\textrm{SFF}}
\newcommand{\Zs}{Z_\text{enh}}
\newcommand{\q}{q}
\newcommand{\disAv}[1]{\mathbb{E}\left[  {#1} \right] }
\newcommand{\trp}{\textrm{TRP}}

\newcommand{\bbeta}{{\mathfrak b}}

\usepackage{xcolor}
\usepackage{physics}
\usepackage{graphicx} 
\usepackage[onehalfspacing]{setspace} 

\newtheoremstyle{lemma}
  {\topsep}
  {\topsep}
  {\itshape}
  {}
  {\bfseries}
  {}
  {.5em}
  {\thmname{#1}\thmnumber{ #2}\thmnote{ (#3)}}
\theoremstyle{lemma}

\usepackage[colorlinks=true,urlcolor=black,linkcolor=blue,citecolor=blue]{hyperref}

\usepackage{tocloft,calc}
\renewcommand{\cftlottitlefont}{\hspace*{\fill}\large}
\renewcommand{\cftafterlottitle}{\hspace*{\fill}}
\renewcommand{\cftloftitlefont}{\hspace*{\fill}\large}
\renewcommand{\cftafterloftitle}{\hspace*{\fill}}
\renewcommand{\cftchapaftersnum}{:\ }
\renewcommand{\cftchappresnum}{\chaptername\space}
\makeatletter
\g@addto@macro\appendix{%
  \addtocontents{toc}{%
    \protect%
  }%
}
\renewcommand{\baselinestretch}{2}
\begin{document}
\pagestyle{empty}

\hbox{\ } \vspace{.7in}
\renewcommand{\baselinestretch}{1}
\small \normalsize

\begin{center}
\large{{ABSTRACT}}

\vspace{3em}

\end{center}
\hspace{-.15in}
\begin{tabular}{ll}
Title of Dissertation:    & {\large  SPECTRAL STATISTICS, HYDRODYNAMICS}\\
&                     {\large  AND QUANTUM CHAOS}\\
\ \\
&                          {\large  Michael Winer} \\
&                           {\large Doctor of Philosophy, 2024} \\
\ \\
Dissertation Directed by: & {\large  Professor Brian Swingle} \\
&               {\large  Department of Physics } \\
\end{tabular}

\vspace{3em}

\renewcommand{\baselinestretch}{2}
\large \normalsize

One of the central problems in many-body physics, both classical and quantum, is the relations between different notions of chaos. Ergodicity, mixing, operator growth, the eigenstate thermalization hypothesis, and spectral chaos are defined in terms of completely different objects in different contexts, don't necessarily co-occur, but still seem to be manifestations of closely related phenomena.

In this dissertation, we study the relation between two notions of chaos: thermalization and spectral chaos. We define a quantity called the Total Return Probability (TRP) which measures how a system forgets its initial state after time $T$, and show that it is closely connected to the Spectral Form Factor (SFF), a measure of chaos deriving from the energy level spectrum of a quantum system.

The main thrust of this work concerns hydrodynamic systems- systems where locality prevents charge or energy from spreading quickly, this putting a throttle on thermalization. We show that the detailed spacings of energy levels closely capture the dynamics of these locally conserved charges.

We also study spin glasses, a phase of matter where the obstacle to thermalization comes not from locality but from the presence of too many neighbors. Changing one region requires changing nearby regions which requires changing nearby-to-nearby regions, until only catastrophic realignments of the whole system can fully explore phase space. In spin glasses we find our clearest analytic link between thermalization and spectral statistics. We analytically calculate the spectral form factor in the limit of large system size and show it is equal to the TRP.

Finally, in the conclusion, we discuss some ideas for the future of both the SFF and the TRP. 

\thispagestyle{empty} \hbox{\ } \vspace{1.5in}
\renewcommand{\baselinestretch}{1}
\small\normalsize
\begin{center}

\large{{SPECTRAL STATISTICS, HYDRODYNAMICS \\
AND QUANTUM CHAOS}}\\
\ \\
\ \\
\large{by} \\
\ \\
\large{Michael Winer}
\ \\
\ \\
\ \\
\ \\
\normalsize
Dissertation submitted to the Faculty of the Graduate School of the \\
University of Maryland, College Park in partial fulfillment \\
of the requirements for the degree of \\
Doctor of Philosophy \\
2024
\end{center}

\vspace{7.5em}

\noindent Advisory Committee: \\
\hbox{\ }\hspace{.5in}Professor Victor Galitski, Chair \\
\hbox{\ }\hspace{.5in}Professor Brian Swingle, Research Advisor \\
\hbox{\ }\hspace{.5in}Professor Maissam Barkeshli \\
\hbox{\ }\hspace{.5in}Professor Jonathan Rosenberg, Dean's Representative \\
\hbox{\ }\hspace{.5in}Professor Jay Sau\\


\thispagestyle{empty}
\hbox{\ }

\vfill
\renewcommand{\baselinestretch}{1}
\small\normalsize

\vspace{.5in}

\begin{center}
\large{\copyright \hbox{ }Copyright by\\
Michael Winer  
\\
2024}
\end{center}

\vfill

\newpage 

\pagestyle{plain} \pagenumbering{roman} \setcounter{page}{2}

\phantomsection
\addcontentsline{toc}{chapter}{Acknowledgements}

\renewcommand{\baselinestretch}{2}
\small\normalsize
\hbox{\ }
 
\vspace{-.5in}

\begin{center}
\large{Acknowledgements} 
\end{center} 

\vspace{1ex}

First, I need to thank my advisor, Brian Swingle. Brian, you are an incredible scientist, an excellent teacher, a skilled advocate, and a talented mentor. You helped me make the transition from physics student to physicist. You helped me learn about hundreds of scientific topics, and dozens of non-scientific ones, and have been the most important positive force in my graduate school experience.

I am also incredibly grateful to Victor Galitski, who has been a valuable ally for more years than I care to calculate. Your group helped show me how science can be a team sport, and gave me experience as a mentee, a mentor, and everything in between. Thank you for your support, scientific and otherwise.

I am grateful to the rest of my committee: Jay, Jonathan, and Maissam. Thank you for your work and help both on this committee and throughout my time at Maryland.

I am grateful to thank my family. Dad, who taught me that thinking can be fun, Jess who taught me that fun can be fun, and Mom, who has supported me through every twist and turn of my life. 

I have had many mentors throughout every stage of my life. These include exceptional teachers like Mrs. Manchester, Mr. Schafer, Mr. Rose, Mr. Stein, and Mr. Schwartz. It includes my professors at MIT, especially Jesse Thaler, Hong Liu, and Krishna Rajagopal. And it includes my scientific and personal mentors in grad school, especially Shao-kai Jian, Chris Baldwin, and Christopher White.

I owe a lot to the many members of the Swingle and Galitski groups I have worked with across time and space. Thank you to Subhayan, Yixu, Gong, Stefano, Shenglong, Brianna, Val, Greg, Divij, Nadie, Connor and especially Tiangang. Thank you also to Yunxiang, Andrey, Laura, Amit, Gautam, Musa, Alireza, Masoud, and especially Richard.

Perhaps most important in aggregate have been my friends. Ben, Jeremy, Nathan, Eric, Matthew, Victor, Bendeguz, MJ, Anna, Jackie, Rita, Kevin, Saranesh, John, Stuart, and others too numerous to list. You have been my teachers and my students, my support, my competition, and everything in between. And of course I need to thank my housemates, Saurabh, Deepak, Ed, Chung-chun, Yuxuan, Saketh, and Captain Billy. You guys have turned houses into homes and kitchens into pigsties. Thank you for making these past five years so wonderful.  
\phantomsection
 \addcontentsline{toc}{chapter}{Summary of Research Contributions}
\renewcommand{\baselinestretch}{2}
\small\normalsize
\hbox{\ }

\vspace{-.3in}

\begin{center}
\large{Summary of Research Contributions}
\end{center}
The research in this thesis has been conducted over the course of my graduate school career by myself, Brian Swingle, Richard Barney, and Chris Baldwin, and appears in papers \cite{winerprx,Winer_2022,Winer:2022gqz,Winer:2022ciz}.  In compliance
with the guidelines of the University of Maryland and of the Physics Department’s Graduate
Director, Chapters \ref{chapter:hydro}-\ref{chapter:glass} of this thesis reproduce the bodies of references \cite{winerprx,Winer_2022,Winer:2022gqz,Winer:2022ciz} verbatim. Appendices \ref{chapter:app1}-\ref{chapter:app4} reproduce the appendices of those papers exactly. The alterations are minor changes to formatting and an updating of some of the references for older papers. I will now report in detail the contributions I made to each chapter of this dissertation.

Chapter \ref{chapter:intro} consists of introductory material to this thesis, including background material and context for understanding quantum chaos. It is solely the work of the author.

Chapter \ref{chapter:hydro} and Appendix \ref{chapter:app1} were originally published as reference \cite{winerprx}. Both authors contributed to the conceptual development of the paper, with me having the initial idea and Brian providing valuable suggestions. All of the detailed calculations and numerical simulations are mine. The version that appears in this paper was not the final version published in PRX, but an earlier version submitted to arxiv as 2012.01436v2. This version contained more introductory and background material, and I judged it would be more accessible to potential readers of this dissertation.

Chapter \ref{chapter:ssb} and Appendix \ref{chapter:app2} were originally published as reference \cite{Winer_2022}, a follow-up generalizing the results of \cite{winerprx} to systems with spontaneous symmetry breaking. Again, I performed most of the concrete calculations and simulations.

Chapter \ref{chapter:soundPole} and Appendix \ref{chapter:app3} were originally published as reference \cite{Winer:2022gqz}, a follow-up to reference \cite{Winer_2022}. As with the rest of the series, it was a collaborative effort where I did most of the specific calculations and numerics while Brian and I shared the higher-level conceptual decisions.

Chapter \ref{chapter:glass} and Appendix \ref{chapter:app4} were originally published as reference \cite{Winer:2022ciz}. Chronologically, this paper actually started before any of the other projects. The inciting question (what do the spectral statistics of a spin glass look like) grew out of a discussion between Chris Baldwin, Brian Swingle and Victor Galitski. I contributed the answer (the spectral statistics count TAP states) as well as the initial calculations backing them up. Chris Baldwin in particular clarified the calculations, Richard checked them, and Brian and Victor provided guidance on directions and presentation.  

    \cleardoublepage
    \phantomsection
    \addcontentsline{toc}{chapter}{Table of Contents}
    \renewcommand{\contentsname}{\normalfont\large Table of Contents}
\renewcommand{\baselinestretch}{1}
\small\normalsize
\tableofcontents 

\newpage

\phantomsection
\addcontentsline{toc}{chapter}{List of Figures}
    \renewcommand{\contentsname}{List of Figures}
\listoffigures 
\newpage

\newpage
\setlength{\parskip}{0em}
\renewcommand{\baselinestretch}{2}
\small\normalsize
\setcounter{page}{1}
\pagenumbering{arabic}

\renewcommand{\thechapter}{1}

\chapter{Introduction}
\label{chapter:intro}
\section{Overview}
Quantum chaos is one of the most dynamic areas in contemporary physics, lying at the intersection of quantum information, many-body physics, and statistical mechanics. One of the most important diagnostics of quantum chaos is the Spectral Form Factor or SFF, a statistical quantity diagnosing repulsion between a Hamiltonian's energy eigenvalues.

In this thesis, I study the spectral form factor in four contexts: in diffusive hydrodynamic systems, systems with a spontaneously broken symmetry, systems with sound, and finally in a quantum spin glass.

The fundamental thread connecting these works is that in addition to carrying information about chaoticity, the SFF contains fine-grained information about the thermalization of the system or lack thereof. We find that in hydrodynamics, the SFF encodes information about the slowest modes and the diffusion of energy throughout the system. In systems with Goldstone modes, the Goldstone dynamics is also encoded in the SFF. For glassy systems, the transition at low temperatures from an ergodic liquid phase to a non-ergodic glassy phase is mirrored by changes in the SFF. 

\section{A Whirlwind Tour of Classical Chaos}
\label{sec:classChaos}
The history of chaos theory goes back to the 19th century, with giants like Maxwell and Poincare noticing that many of the most important systems in physics display unpredictable behavior. Canonical examples include chaotic billiards, three-body gravitational systems, fluids with low viscosity, and theme parks full of dinosaurs. 
\begin{figure}
    \centering
    \includegraphics[scale=0.5]{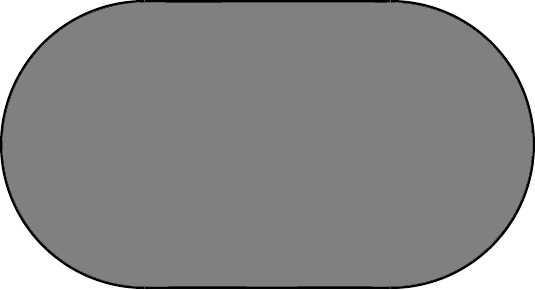}
    \includegraphics[scale=0.5]{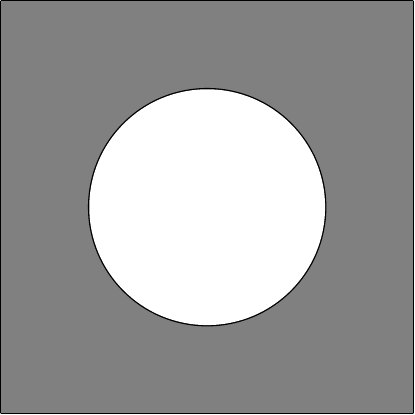}
    \caption{Two examples of chaotic billiards. The shape on the left in called a Bunimovich stadium, the shape on the right is a Sinai billiard. In either one of these systems, a particle bouncing around the walls will exhibit chaotic motion. In contrast, for rectangular, circular, or elliptical billiards, the motion is known not to be chaotic.}
    \label{fig:chaoticBilliards}
\end{figure}

The most basic concept is simple: there are many maps (such as the logistic map $x_{n+1}=rx_n(1-x_n)$ for most values of $r$ between 3.56... and 4) and many systems of differential equations (such as Newtons Laws for three equally massive bodies interacting through gravitation), where two nearby points will diverge after a small number of iterations. This effect is called the \textbf{butterfly effect}, named for the (untested) meteorological maxim that a butterfly flapping its wings in Brazil can set off a tornado in Texas.

\begin{figure}
    \centering
    \includegraphics[scale=0.4]{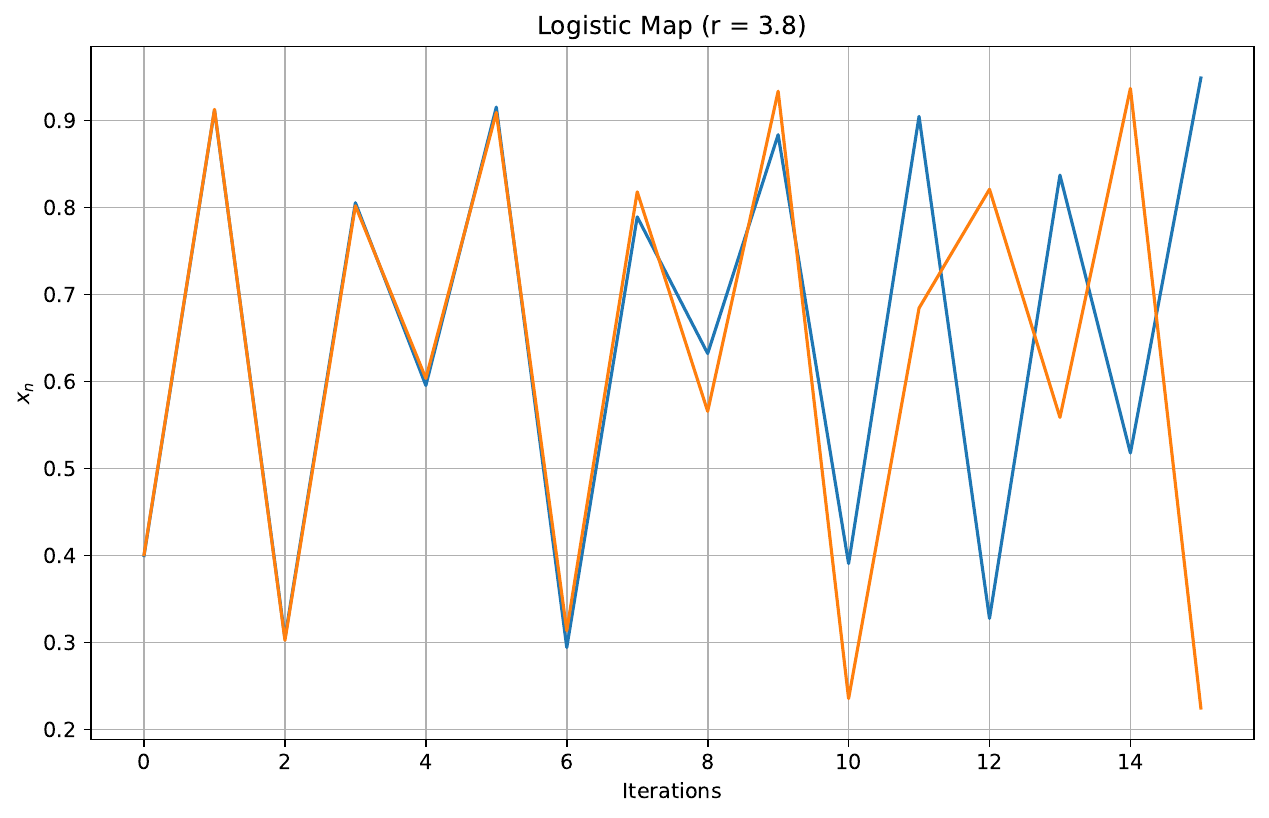}
    \caption{The iterated logistic map $x_{n+1}=rx_n(1-x_n)$ for $r=3.8$, with initial values $0.4$ (blue) and $0.401$ (orange). The two lines diverge exponentially, at a rate given by the Lyaponuv exponent.}
    \label{fig:logistic}
\end{figure}

More precisely, classical chaotic systems are characterized by a sensitive dependence on initial conditions, illustrated in figure \ref{fig:logistic}. Mathematically, if we start a chaotic system with two sets of initial conditions $x_i(0)$ and $x_i'(0)=x_i(0)+\delta x_i(0)$, we expect the insmall difference $\delta x_i (t)$ to grow exponentially as $e^{\lambda t}$ for some positive $\lambda$, possibly saturating when the difference $\delta x$ reaches some macroscopic value. 
The rate $\lambda$ at which such small differences increase is called the \textbf{Lyapunov exponent}. Mathematically we can write this in terms of the norm of the Jacobian:
\begin{equation}
    \lambda=\lim_{t\to \infty} \frac 1t \log \Bigg\lvert \frac{\partial x_j(t)}{\partial x_i(0)}\Bigg\rvert,
    \label{eq:lyapunovDef}
\end{equation}
with an analogous rule for discrete maps. For Hamiltonian systems, which take center stage in physics, we can replace the partial derivatives with Poisson brackets:
\begin{equation}
    \lambda=\lim_{t\to \infty} \frac 1t \log \Big\lvert \left\{x_j(t),x_i(0)\right\}\Big\rvert,
    \label{eq:lyapunovPoisson}.
\end{equation}
Because the largest eigenvalues of the matrix grow exponentially, we can replace the norm of the entire matrix with the size of any one element.

\begin{figure}
    \centering
    \includegraphics[scale=0.3]{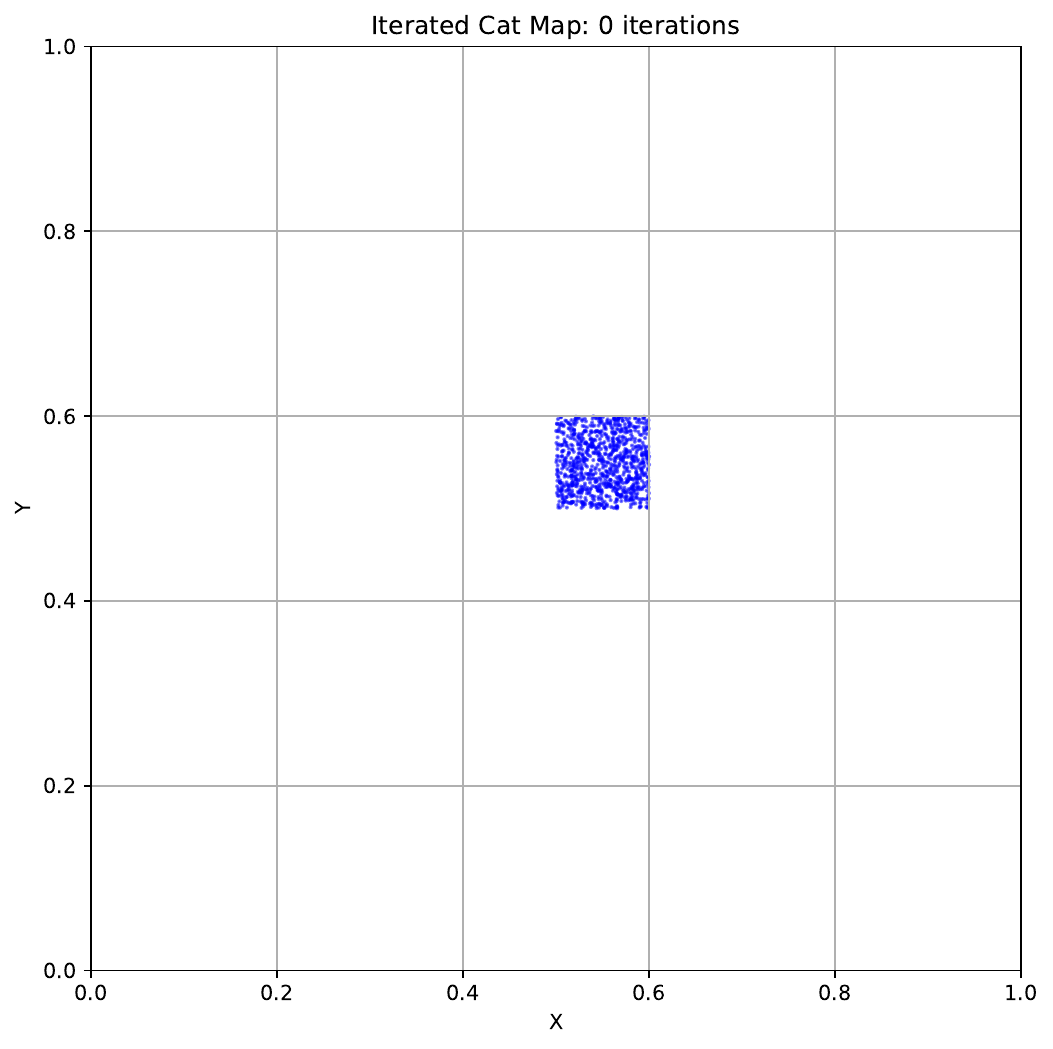}\includegraphics[scale=0.3]{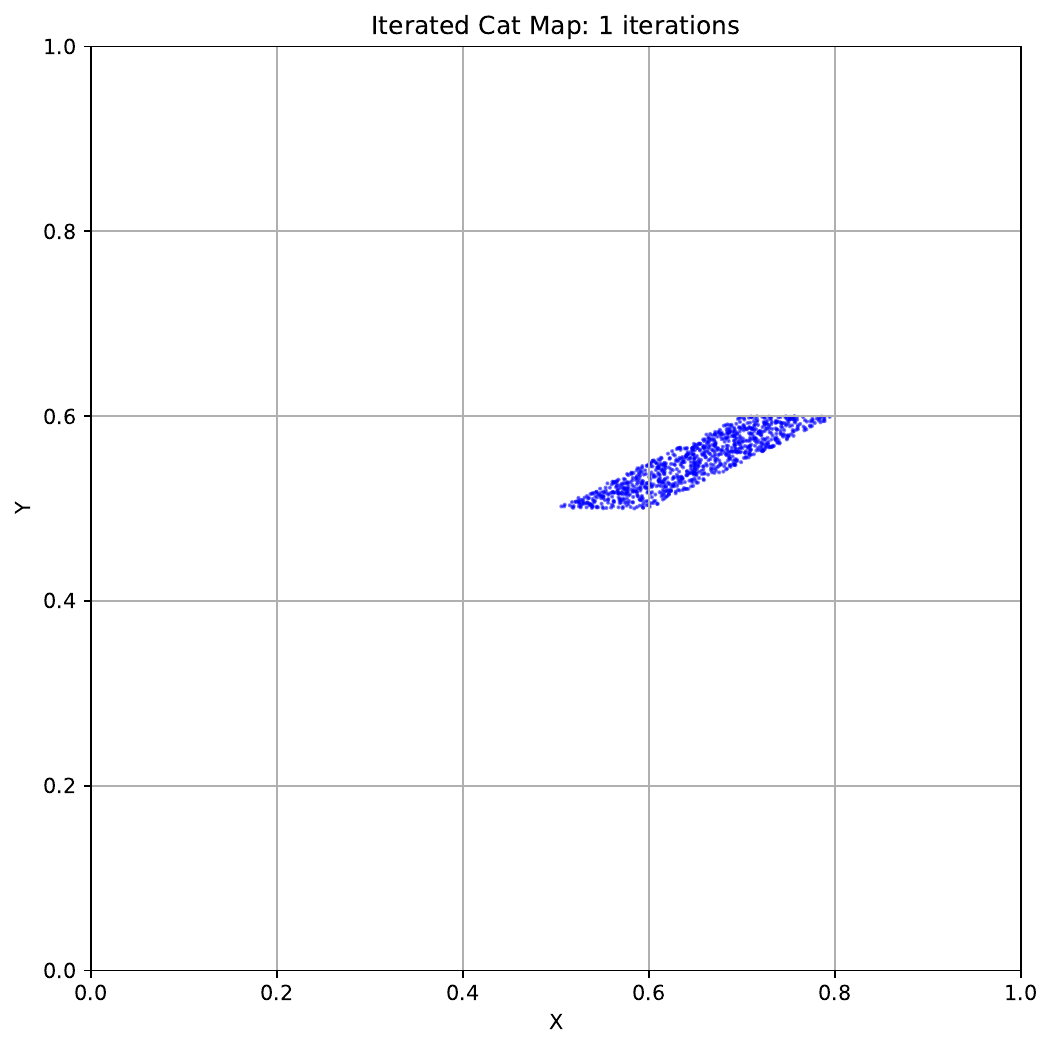}\\
    \includegraphics[scale=0.3]{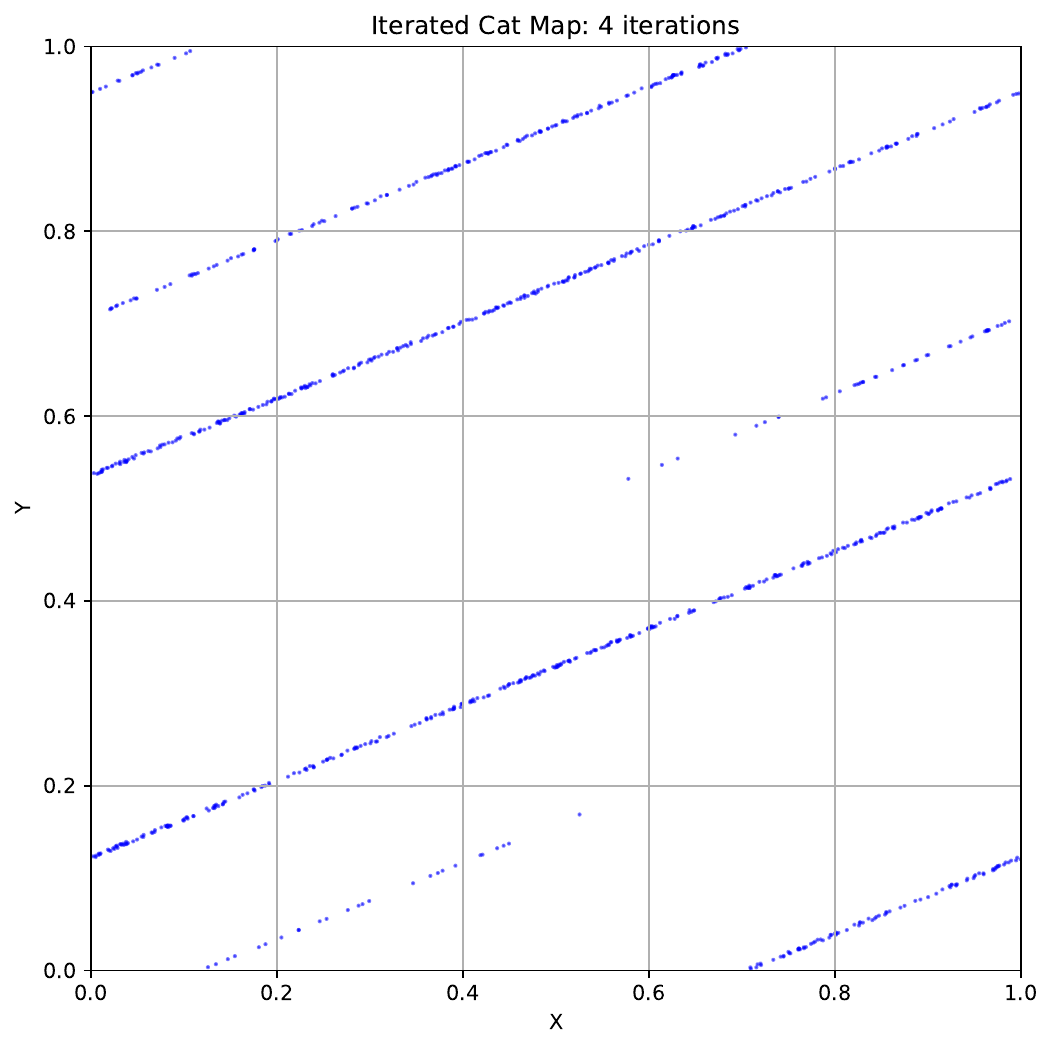}\includegraphics[scale=0.3]{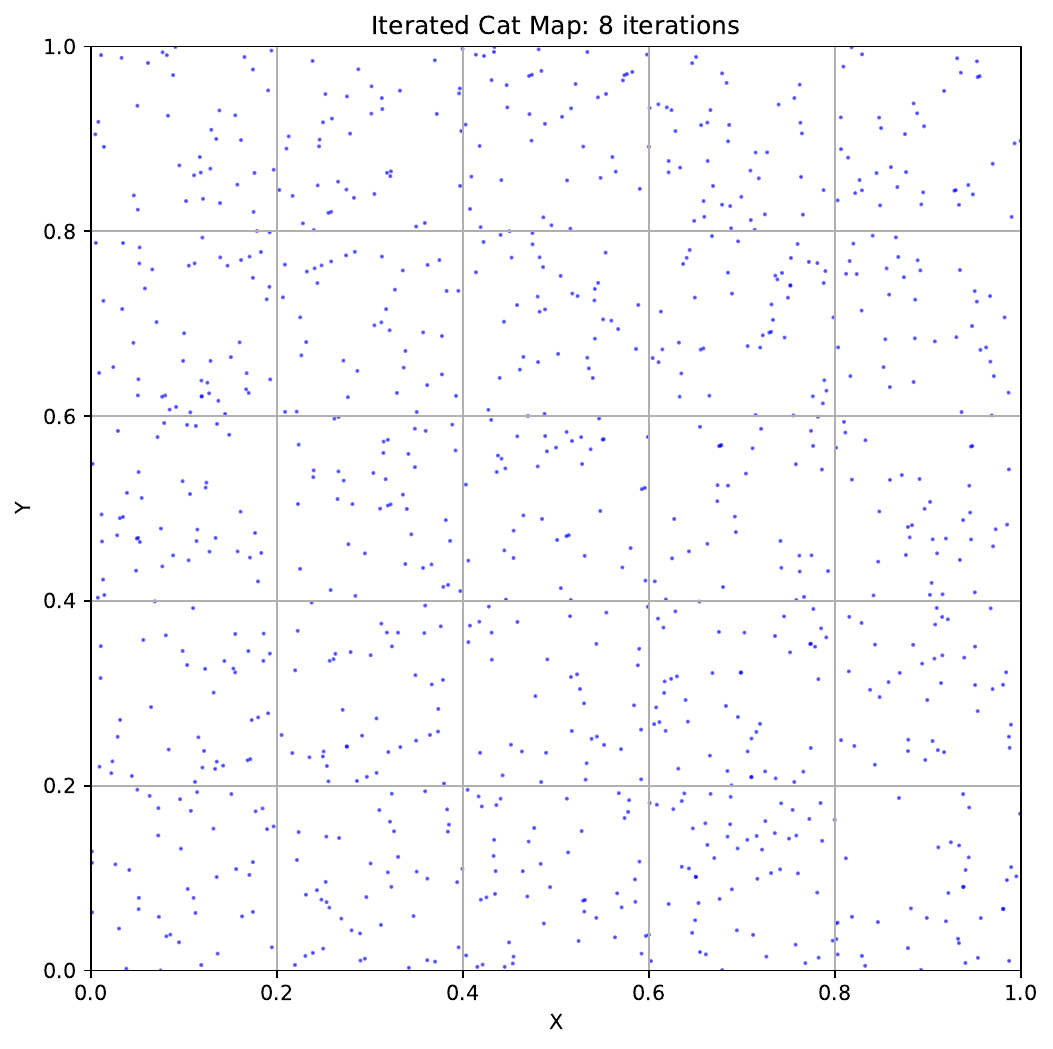}
    \caption{The so-called Arnold cat map, defined by $(x_{n+1},y_{n+1})=(2x+y \textrm{ mod }1,x+y \textrm{ mod }1)$. A cloud of points initially filling a small square soon swells up to fill a seemingly random subset of the full phase space.}
    \label{fig:catmap}
\end{figure}

This simplest notion of chaos is entirely local: do small changes in initial condition grow in a linearized regime? There are at least two other notions of classical chaos:
\begin{itemize}
    \item \textbf{Operator growth}: Do expressions for simple variables remain simple under time evolution? For instance the lack of closed-form solutions to the three-body problem in gravity was one of the first great discoveries in the prehistory of chaos theory. In contrast, in a harmonic oscillator the variable $x(0)$ evolves to $x(t)=x(0) \cos(\omega t)+p(0)\frac{1}{\sqrt{mk}}\sin (\omega t)$, a linear function of the phase space variables regardless of $t$. In chaotic systems, the operator $x(t)$ will depend on a very high-order polynomial in the initial phase space variables $x(0), p(0)$.
    
    As it turns out, operator growth- the growth in the length of expressions for operators- is intimately related to Lyapunov growth- the growth of small deviations in phase space. A good example of this is the repeated application of the map $x_{n+1}=F(x)\equiv \sin(K x_{n})$ to the region $x \in [-1,1]$, for some $K\gg 1$. The Lyapunov growth is simple: give or take some factors the derivative in equation \ref{eq:lyapunovDef} picks up a factor of $K$ each time we iterate, and our Lyapunov exponent is approximately $\log K$.  
    Conversely, Taylor expanding $F(x)$, we need to retain to at least the first $T$ terms, where $T$ is of order $K$.\footnote{This is because the $T$th term in the Taylor expansion of $sin(K)$ is $\frac{K^T}{T!}$ which only becomes small when $T>K$.} 
    This means that after $n$ maps we have an order $K^n$ polynomial. Thus the complexity of an operator grows exponentially with exactly the Lyapunov exponent. This connection can be made more precisely quantitative \cite{Qi_2019,PhysRevX.9.031048,Nahum_2018}.
    \item \textbf{Mixing}: Does a small region of phase space evolve to quickly spread over the entire system? For instance Arnold's cat map (see figure \ref{fig:catmap}) will quickly map a small cloud of points to fill the entire phase space, while a harmonic oscillator will merely rotate the cloud around. It is important to note that quickly filling phase space requires some Lyapunov character (how can the cloud fill phase space if it isn't being stretched?), it is a stronger notion. For instance, in some systems such as glasses, dynamical constraints prevent the cloud from filling up all of phase space, instead constraining it to a small subset of all points allowed by conservation laws. (There is a closely related concept called \textbf{ergodicity} which is the property that a single point traces out all of phase space as it evolves. In practice, these two properties of often found together and many physicists use the phrases interchangeably, but they are distinct concepts. For instance a particle on a torus obeying the dynamical equation $\frac{dx}{dt}=1, \frac{dy}{dt}=\sqrt 2$ is ergodic but not mixing.)
\end{itemize}
\begin{figure}
    \centering
    \includegraphics[scale=0.5]{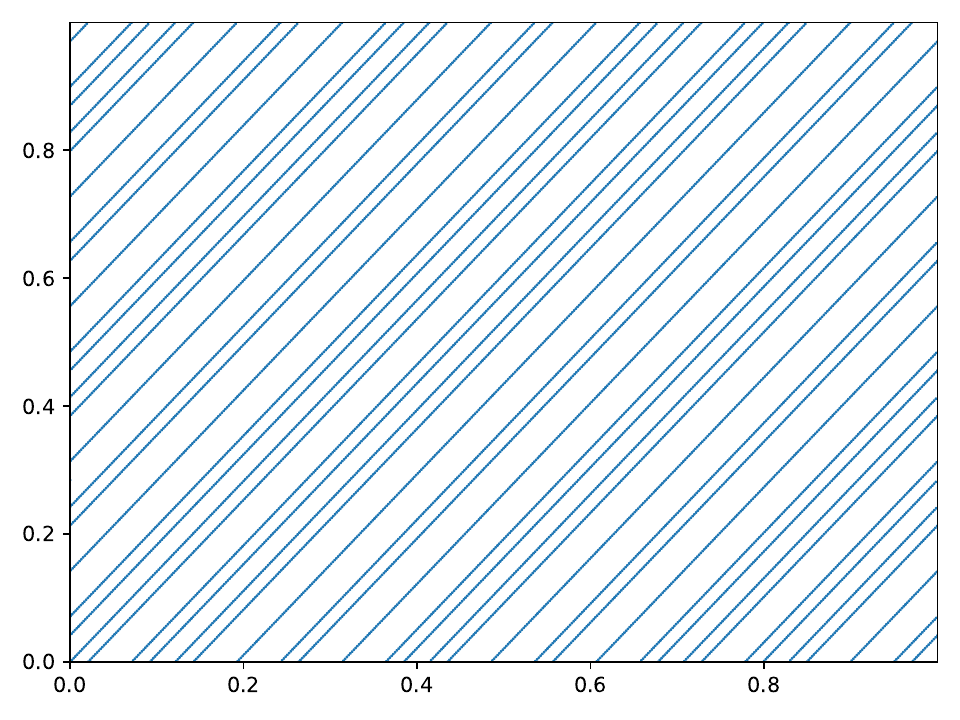}
    \caption{The dynamics of  $x'=1, y'=\sqrt 2$ on the unit torus up to $t=20$. This map will cover every part of phase space equally, but will never mix different regions, nor exhibit Lyapunov or operator growth. Despite this and other important counterexamples, for most scientific purposes the concepts of mixing and ergodic are used interchangeably.}
    \label{fig:root2}
\end{figure}
The last of these, mixing, underlies the all-important process of \textbf{thermalization}.

Very often, we want to predict observables of some many-body system: how much force will this gas exert on the wall of its container? How will this block of iron react to a magnetic field? Ideally, we may wish to calculate this from first principles: take the initial conditions of the system and evolve them according to the equations of motion. But for any system of a certain size, this simply isn't feasible: one can never measure the $10^{25}$ numbers characterizing a macroscopic object. A hint about how to bypass this problem comes from looking at figure \ref{fig:catmap}, where a small region of phase space spreads quickly to cover the entire region uniformly. This means the enormous $10^{25}$ variable calculation we might have performed is useless: no matter what the initial conditions, we know the final distribution of states is uniform over phase space (or perhaps the region of phase space allowed by conservation laws). But it also means that regardless of the distribution of initial conditions of our system, we are justified in assuming that after a short amount of time, it is a cloud covering all of phase space uniformly. Averaging quantities over phase space- or the region of phase space at a given energy and charge density- is the bread and butter of many-body physics, both classical and quantum.

\section{Quantum Chaos}
\label{sec:quanChaos}
At first blush, it seems like there should be no notion of chaos in quantum systems. After all, the Schrodinger equation is perfectly linear: $i \partial_t \Psi=\hat H \psi$. Not only that, it is unitary, perfectly preserving all distances in phase space. How can such an equation give rise to chaotic behavior?

A hint comes from considering the butterfly effect in a semiclassical system. We consider two wavefunctions $\Psi_1(t)$ and $\Psi_2(t)$ that are both localized at nearby points at time $t=0.$ As the system evolves, $\braket{\Psi_1}{\Psi_2}$ remains constant by unitarity, but we know from classical chaos that $\bra{\Psi_1}\hat x\ket{\Psi_1}$ and $\bra{\Psi_2}\hat x\ket{\Psi_2}$ diverge exponentially. 

This analogy makes it clear that quantum chaos is not to be found in the evolution of wavefunction coefficients, but in terms of operators. We can translate equation \ref{eq:lyapunovPoisson} into the language of quantum mechanics by replacing the Poisson bracket with a commutator:
\begin{equation}
    \lambda=\lim_{t\to \infty} \frac 1t \log \Big\lvert \left[\hat x_j(t),\hat x_i(0)\right]\Big\rvert.
    \label{eq:lyapunovQuantum}
\end{equation}
This can be rewritten as
\begin{equation}
    \lambda=\frac 12 \lim_{t\to \infty} \frac 1t \log \bra{\psi} \left[\hat x_j(t),\hat x_i(0)\right]^2\ket{\psi}.
\end{equation}
In addition to Lyapunov chaos, and operator growth, there are other notions of chaos analogous to the classical notions of mixing and ergodicity. The most important of these is the \textbf{eigenstate thermalization hypothesis.}

\subsection{The Eigenstate Thermalization Hypothesis}
The Eigenstate Thermalization Hypothesis (ETH) \cite{Deutsch_2018,PhysRevA.43.2046,PhysRevE.50.888,Rigol_2008} posits that for many isolated quantum systems, individual energy eigenstates can exhibit statistical properties of thermal equilibrium. Mathematically, this is often expressed by the behavior of the matrix elements of observables $\hat O$ in the energy eigenbasis to thermodynamic quantities at a temperature corresponding to that entropy. If we write $O_{ij}=\bra{i}\hat O\ket{j}$ for energy eigenstate $i,j$ far from the ground state, then ETH states that we can approximate
\begin{equation}
    O_{ij}\approx O_{\textrm{Thermal}}\left(\frac{E_i+E_j}2\right)\delta_{ij}+ r_{ij}f\left(\frac{E_i+E_j}{2},E_i-E_j\right).
\end{equation}
where $O_{\textrm{Thermal}}(E)$ is the thermal value of operator $O$ at energy $E$, $f$ is a smooth function of both of its inputs (whose value is of order $e^{-S/2}$), and $r_{ij}$ is a complex Gaussian random variable with variance $1$ satisfying $r_{ij}=r_{ji}^*.$

For systems obeying the ETH, the long-time expectation values of any operator (represented by, say, $\frac 1T \int_0^T\bra{\psi}\hat O(t)\ket{\psi} dt$) are necessarily thermal, with the $O_{\textrm{thermal}}$ diagonal elements dominating the oscillating off-diagonal elements. The ETH has a host of other interpretations, including treating the eigenstates as an approximate error-correcting code \cite{Bao_2019}, and viewing the operators as random matrices.
\section{Eigenvalue Statistics For Random Matrices}
This brings us to the main topic of this thesis, the eigenvalues of Hamiltonians in chaotic systems. The ETH says that for chaotic systems in the energy eigenbasis, local operators $\hat O$ are in some sense random matrices. On the other hand, the \textbf{Bohigas-Giannoni-Schmit conjecture} says that chaotic Hamiltonians are themselves random matrices. To understand this statement, we need a better understanding of random matrices.
\subsection{Random Matrices}
\textbf{Random Matrix Theory} (RMT) began as an offshoot of statistics. In fact, the earliest results are due to Wishart, who studied the covariance matrices of random data. The applications of RMT to physics, however, began in the 1950s with the work of Eugene Wigner. Wigner defined the three classic models of random matrix theory:
\begin{itemize}
    \item \textbf{Gaussian Orthogonal Ensemble} (GOE) matrices. These are random symmetric real $N\times N$ matrices whose diagonal elements are i.i.d.\footnote{Independent and Identically Distributed} Gaussians with variance $\frac{2}{N+1}$ and off diagonal elements are Gaussians with variance $\frac{1}{N+1}$. The `orthogonal' in their name comes from the fact that the ensemble has a statistical symmetry under conjugation by $N\times N$ orthogonal matrices, that is while the individual elements do not have the symmetry, the distribution is identical under orthogonal conjugation. 
    \item \textbf{Gaussian Unitary Ensemble} (GUE) matrices. These are random Hermitian $N\times N$ matrices whose diagonal elements are i.i.d. Gaussians with variance $\frac{1}{N}$ and off diagonal elements are complex Gaussians with variance $\frac{1}{2N}$ for their real and imaginary parts. The `unitary' in their name comes from the fact that the ensemble has a statistical symmetry under conjugation by $N\times N$ unitary matrices. 
    \item \textbf{Gaussian Symplectic Ensemble} (GSE) matrices, the strangest of the bunch. These are random self-adjoint quaternionic $N\times N$ matrices whose diagonal elements are i.i.d. Gaussians with variance $\frac{1}{2N-1}$ and off diagonal elements are quarternionic Gaussians with variance $\frac{2}{2N-1}$ for their four components (real $i$, $j$, and $k$). The `Symplectic' in their name comes from the fact that the ensemble has a statistical symmetry under conjugation by a group related to the symplectic group.
\end{itemize}
For obvious reasons, these three ensembles of random matrices (GOE, GUE, and GSE) are often called the three GXE ensembles.

Among Wigner's seminal results in random matrix theory was his celebrated \textbf{semicircle law}. For each of the three GXE ensembles, Wigner proved that the histogram of eigenvalues is a semicircle.
\begin{figure}
    \centering
    \includegraphics[scale=0.7]{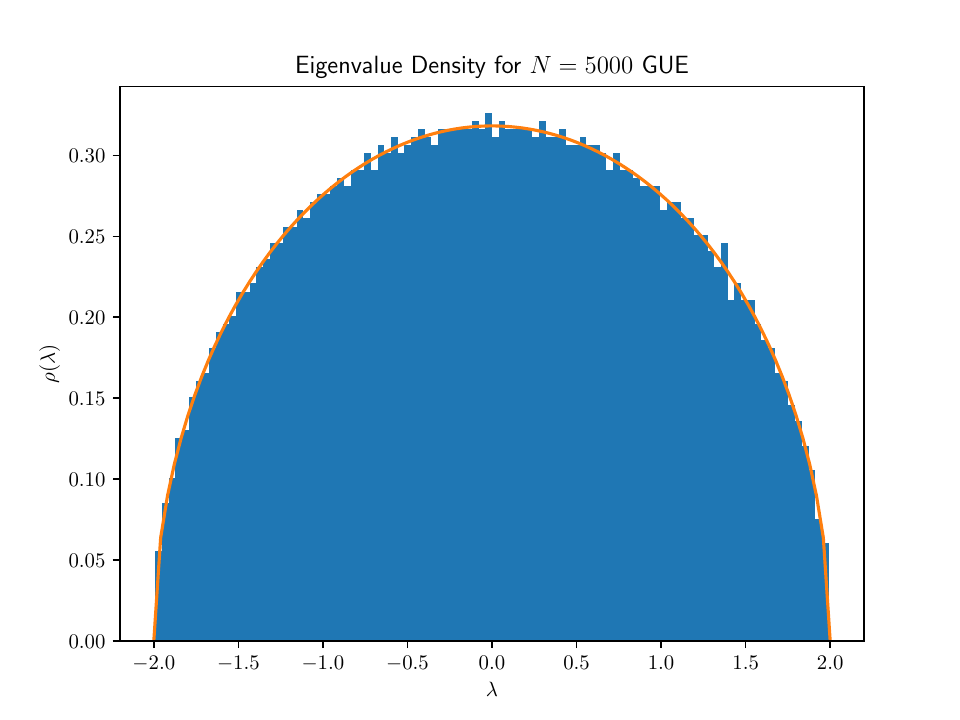}
    \caption{A histogram of the eigenvalues of a single $5000\times 5000$ GUE random matrix (blue), along with the perfect semicircular curve (orange). Notice that even though each bar in the histogram represents dozens of energy levels, the fluctuations- differences between who consecutive bars in the histogram) are often just one or two levels, far less than Poissonian statistics would suggest.}
    \label{fig:introHist}
\end{figure}
Even more consequentially, Wigner identified a phenomenon called \textbf{level repulsion}.
\subsection{Level Repulsion in Random Matrices}
The phenomenon of level repulsion is easiest to see in $2\times 2$ matrices. Consider a $2\times 2$ GUE matrix: 
\begin{equation}M=m_0I+m_1 \sigma_x+m_2 \sigma_y+m_3 \sigma_z,\end{equation} 
with $m_0,m_1,m_2,m_3$ drawn independently from Gaussian distributions with variance $\frac 1{2N}=\frac 14.$ The eigenvalues of $M$ are exactly
\begin{equation}
    \lambda_1,\lambda_2=m_0\pm\sqrt{m_1^2+m_2^2+m_3^2}.
\end{equation}
The gap between the two eigenvalues is thus $2\sqrt{m_1^2+m_2^2+m_3^2}$. We see then that for this gap to be smaller than $\epsilon$, we need each of $m_1, m_2$ and $m_3$ to be smaller than $\epsilon/2$. The probability of this is proportional to $\epsilon^3$, far smaller than the $O(\epsilon)$ probability we'd expect if the eigenvalues were picked independently from a distribution. The probability of the gap being between $\epsilon$ and $\epsilon+d\epsilon$ (for small $\epsilon$) is proportional to $\epsilon^\beta$, where $\beta=2$. We can perform the same analysis for a GOE or GSE matrix, and find that they have $\beta=1$ and $\beta=4$ respectively, as opposed to the $\beta=0$ behavior we would expect if the energy levels were independent of each other. 

\begin{figure}
    \centering
    \includegraphics[scale=0.5]{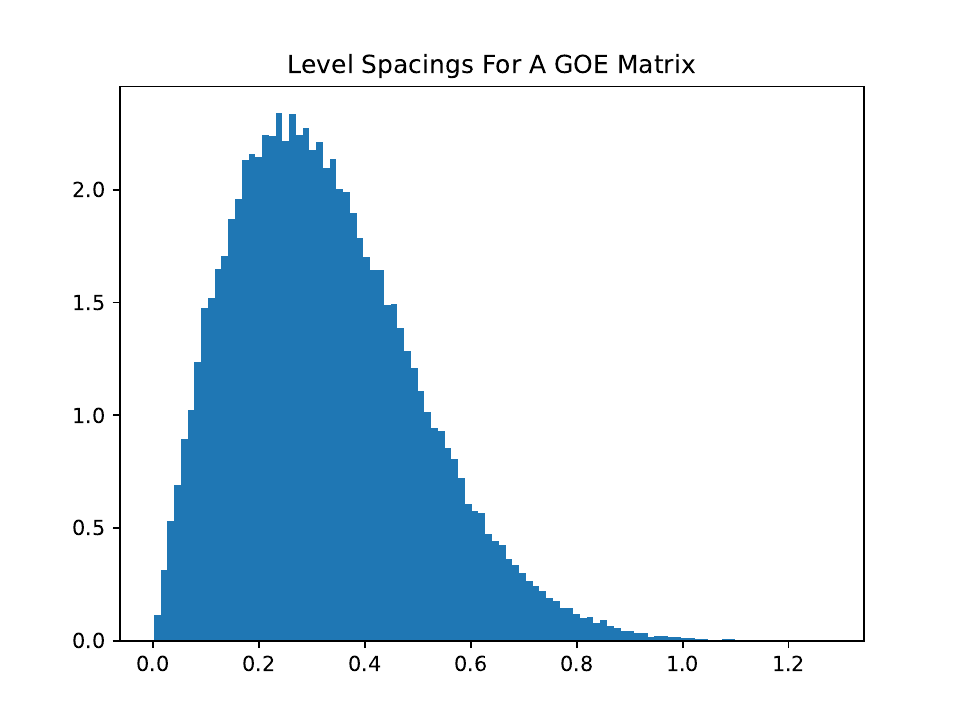}
    \includegraphics[scale=0.5]{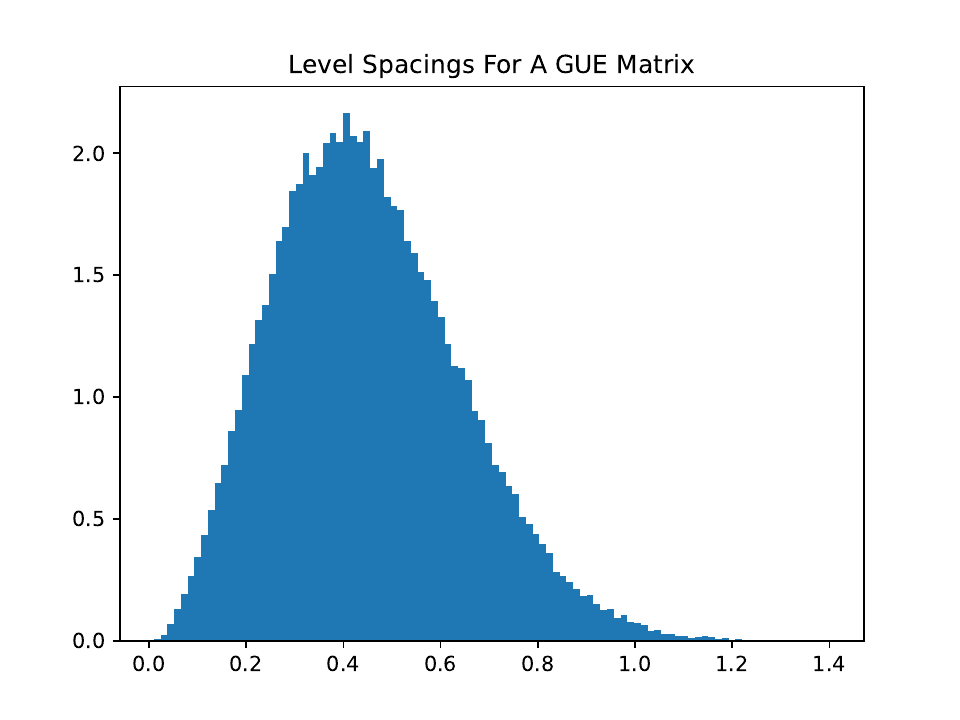}
    \includegraphics[scale=0.5]{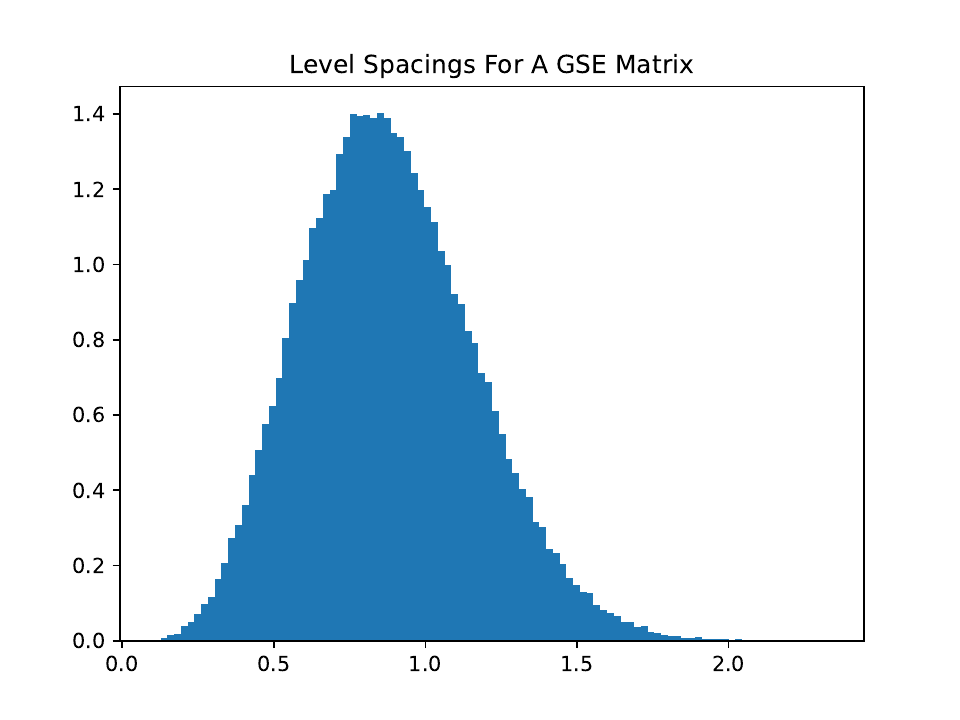}
    \caption{Spacing distributions at the middle of the spectrum for the three classic ensembles, averaged over many realizations. Note the linear increase for GOE, the quadratic for GUE, and the even slower quartic increase from zero for the GSE.}
    \label{fig:GXEspacings}
\end{figure}

One can show that this same repulsive behavior holds for large $N$. See figure \ref{fig:GXEspacings} for histograms of nearest-level spacings for the three GXE ensembles. Note that all of these densities go to zero for small gaps, meaning the chance of two levels near each other is vanishingly small.
\subsection{The Spectral Form Factor}
The most important diagnostic of level repulsion is the \textbf{Spectral Form Factor}, or SFF. The Spectral Form Factor for a given matrix $H$ is a function of time $T$. In words, it is the square of the Fourier Transform of the squared magnitude of the Fourier Transform of the density of states. Let's build this up.

We start with $H$'s density of states $\rho(E)$ defined as
\begin{equation}
    \rho(E)=\sum_{\lambda} \delta(E-\lambda),
\end{equation}
just a train of delta functions at the eigenvalues of $H$. The Fourier Transform of $\rho$ is
\begin{equation}
    Z(iT)\equiv \int_{-\infty}^\infty e^{-iET}\rho(E) dE=\sum_{\lambda}e^{-i\lambda T}=\tr e^{-iHT},
\end{equation}
where we choose the letter $Z$ to echo the partition function in thermodynamics: $Z(\beta)=\tr e^{-\beta H}$.

The SFF can be written in terms of $Z$ as
\begin{equation}
    \SFF(T)=Z(iT)Z(-iT)=\tr e^{-iHT}\tr e^{iHT}=\sum_{\lambda_1,\lambda_2}e^{-i(\lambda_1-\lambda_2) T}.
    \label{eq:orderedSFFdef}
\end{equation}
What is the interpretation of this mathematical object? In a word, it represents the wiggliness of the eigenvalues histogram at resolution $1/T$. When there are large fluctuations of size $~1/T$, the SFF is large, when the fluctuations are suppressed, the SFF is small.

\begin{figure}
    \centering
    \includegraphics[scale=0.7]{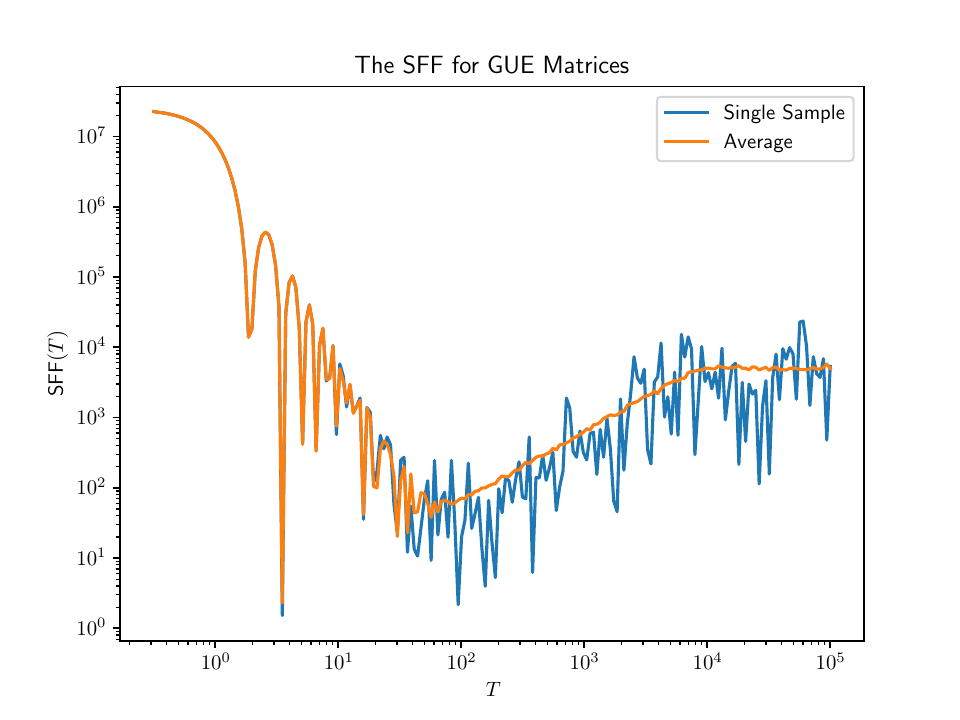}
    \includegraphics[scale=0.7]{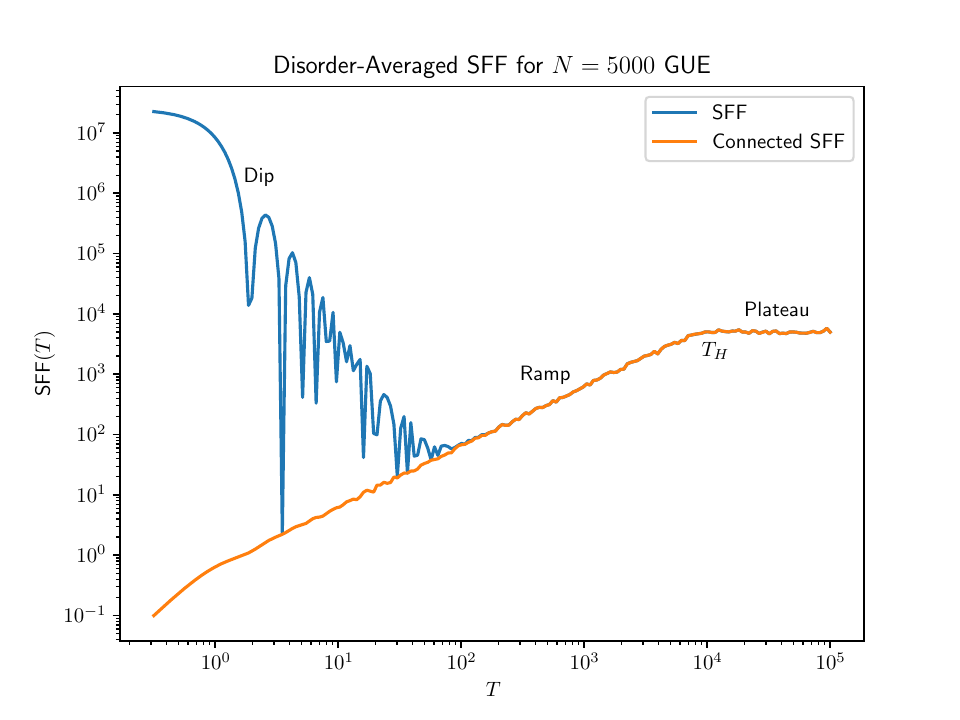}
    \caption{(Top) A graph of the equation \ref{eq:orderedSFFdef} for a single realization of disorder for a $5000\times5000$ GUE matrix (blue) versus the same quantity averaged over an ensemble of matrices (orange). Notice that this is a log plot, so the blue curve fluctuates by orders of magnitude over a very small range. (Bottom) The same averaged SFF (Blue) versus the connected SFF (orange) with important features pointed out.} 
    \label{fig:SFFgraphs}
\end{figure}

The top of figure \ref{fig:SFFgraphs} shows a graph of the SFF for $5000\times5000$ GUE random matrices. In particular, let's look at the blue line, which represents the SFF of a single matrix drawn from this ensemble. Note that the later part of the graph is incredibly jagged and random-looking, even though the average over the ensemble (in orange) is smooth. We say that the SFF is not \textbf{well-averaging}. This means that the SFF of a single matrix drawn from an ensemble is often quite different from the SFF averaged over the entire ensemble. Examples of well-averaging quantities include the largest eigenvalues, the average density of eigenvalues between $0$ and $1$, and the earlier part of the SFF. 
Due to the wild fluctuations of later part of the SFF from instance to instance of a given ensemble, it is often important to discuss the \textbf{Disorder-Averaged SFF}. This is just an average over the ensemble of interest of the SFF. Since the Disorder-Averaged SFF is the average of a square, it makes sense to break it into connected and disconnected parts. Denoting the disorder average of by $\disAv{\bullet}$, we have
\begin{equation}
\begin{split}
    \SFF_{\textrm{disc}}(T)=\Big | \disAv{Z(iT)} \Big|^2\\
    \SFF(T)=\disAv{\Big|Z(iT)\Big|^2}\\
    \SFF_{\textrm{con}}(T)=\SFF(T)-\SFF_{\textrm{disc}}(T)
\end{split}
\end{equation}
The bottom of figure \ref{fig:SFFgraphs} shows the full and connected SFF on a log-log plot. Note that at early times the SFF for the GUE matrices is dominated by the disconnected SFF, while nontrivial behavior at late time seems to be dominated by the connected part. The figure shows some of the most important features of the disorder-averaged SFF, discussed in more detail below:
\begin{itemize}
\item The \textbf{dip}, (sometimes called the \textbf{slope}) occurs at early times. The dip comes from the disconnected piece of the SFF (and thus its precise shape is non-universal, being connected to the thermodynamics of the system).
The dip reflects the loss of constructive interference --- the different terms of $\textrm{Tr} e^{-iHT}=\sum_{\lambda}e^{-i\lambda T}$ acquire different phase factors as $T$ increases from zero.
\item The \textbf{ramp}, occurs at intermediate times. It is a prolonged period of linear growth in the disorder-averaged SFF, particularly the connected part (see figure \ref{fig:SFFgraphs}). The ramp is arguably the most interesting of the three regimes, and it will be the focus of this thesis.
In the canonical GXE matrix ensembles, ramplike behavior is a consequence of the connected two-point function of level densities in a GXE matrix ensemble\cite{mehta2004random}
\begin{equation}
\mathbb{E} \left[ \rho \left( E + \frac{\omega}{2} \right) \rho \left( E - \frac{\omega}{2} \right) \right] - \mathbb{E} \left[ \rho \left( E + \frac{\omega}{2} \right) \right] \mathbb{E} \left[ \rho \left( E - \frac{\omega}{2} \right) \right] \sim -\frac{1}{\beta \pi^2 \omega^2},
\label{eq:twoPoint}
\end{equation}
where $\beta = 1$, $2$, $4$ in the orthogonal, unitary, and sympletic ensembles respectively \cite{mehta2004random}.
The right-hand side of equation \ref{eq:twoPoint} being negative is a mathematical realization of level repulsion.
Taking the Fourier transform of equation \ref{eq:twoPoint} with respect to $\omega$ shows that the SFF contains a term proportional time $T$.
Such a linear-in-$T$ ramp is often taken as a defining signature of level repulsion. The ramp continues until a timescale of order $L$, where $L$ is the dimension of the Hilbert space.

\item The \textbf{plateau}, occurs at late times. The plateau is a signature of the discrete nature of the energy spectrum.
At times much larger than the inverse level spacing, one expects that all off-diagonal terms in the double-trace of the SFF sum to effectively zero, meaning that
\begin{equation}
\textrm{SFF}(T) = \sum_{m,n=1}^L e^{-i(E_m - E_n)T}  \sim L.
\end{equation}
In other words, the plateau for the SFF is enforced by the fact that at long times, the spectral form factor has to (on average) equal the Hilbert space dimension. The time to reach the plateau value is the so-called \textbf{Heisenberg time} $T_H=\frac{\hat \rho}{2\pi}$, proportional to the inverse level spacing $\hat \rho$. For a large system with $N$ degrees of freedom, the density of states $\hat \rho$ is exponentially large in $N$, making the Heisenberg time far longer than any physically relevant timescale. For this and other reasons, the plateau is the most difficult of the three regions to access physically, though see \cite{saad2019late,saad2022convergent,winer2023reappearance,Altland_2021,Keating_2007}.
\end{itemize}

Importantly, note the different roles of the ramp and the plateau. The plateau is a consequence of the discrete, non-degenerate energy spectrum (though it can be modified to account for degeneracies). It appears at long enough times for any system, representing the fundamental discrete bumpiness of histograms. It is, in a way, analogous to shot noise in electrical circuits, and cannot be eliminated.

By contrast, the ramp is optional. Not every possible matrix will have one, only those with level repulsion. The ramp represents a suppression of fluctuations for small values of $T$, this suppression is a consequence of a level repulsion which is seen in GXE and related ensembles, but not in other ensembles such as certain choices of Rosenzweig-Porter ensembles \cite{RP,Kravtsov_2015,Guhr1996,Guhr1997} or, as we will see, the Hamiltonians of integrable quantum systems \cite{BT,Marklof}.

\section{Spectral Statistics and Quantum Chaos}
\label{sec:specQuantumChaos}
Since the 1980s, we have known that level repulsion isn't just a property of random matrices. It is also a property of the Hamiltonians of chaotic systems \cite{bgs1,bgs2}. This observation, that the energy levels of chaotic Hamiltonians behave like the energy levels of random matrices, is known as the \textbf{Bohigas-Giannoni-Schmit conjecture}. The intuition for this fact comes from the Eigenstate Thermalization Hypothesis: if generic operators like $\hat V(x)$ look like random matrices in the energy eigenbasis, then requiring two eigenvalues of $\hat H_0+\hat V(x)$ to be near each other requires the zeroing of $\beta+1$ different parts of the matrix (for instance in the GUE ensemble with $\beta=2$, degeneracy between two levels $i$ and $j$ means that $\hat V$ contributes no $\sigma_x$, $\sigma_y$, or $\sigma_z$ to the $ij$ subspace). When the matrix whose spectral form factor we calculate is a Hamiltonian, the SFF takes on additional physical meaning. Remembering that the unitary evolution operator is given by $\hat U(T)=e^{-i\hat H T}$, the spectral form factor can be written
\begin{equation}
    \SFF(T)=\tr \hat U(T) \tr \hat U^\dagger(T).
\end{equation}

This provides a link between the SFF evaluated at $T$ and the quantum dynamics after time $T.$

A number of works have used capitalized on this link to show the existence of the ramp in chaotic quantum systems. Two particularly important bodies of work are those using periodic orbit theory and those using large-$N$ methods and wormholes.

The \textbf{periodic orbit theory} of spectral statistics \cite{M_ller_2005,McD,Casati1980,Berry1981} calculates the trace of $\hat U$ semiclassically. It writes the trace as a path integral
\begin{equation}
    \tr \hat U(T)=\tr e^{-i\hat H T}=\int dx_0 \left(\int \mathcal D x(t) e^{iS[x]/\hbar} \bigg |_{x(0)=x(T)=x_0}\right).
    \label{eq:periodicInt}
\end{equation}
In other words, the path integral ranges over all functions $x(t)$ periodic with period $T$. To attempt to evaluate this integral, we take the semiclassical limit $\hbar \to 0$. In this case, the integral in equation \ref{eq:periodicInt} becomes, schematically
\begin{equation}
    I=\int \exp\left(Mf(x)\right)dx
\end{equation}
for some big number $M=\frac 1 \hbar$. Such integrals can be well-approximated using the \textbf{saddle point method}, closely related to Laplace's method for real integrands and steepest descent for complex integrands. The basic idea is that 
\begin{equation}
    I\approx \sum A \exp\left(Mf(x_{crit})\right),
\end{equation}
where $A^{-2}$ is the determinant of $\pi$ times the Hessian of $-f$ at $x_{crit}$ and the sum ranges over values of $x_{crit}$ satisfying
\begin{equation}\frac{\partial f}{\partial x}\big|_{x=x_{crit}}=0.\label{eq:saddlePointVague}\end{equation}
When the function $f$ is an action functional $S[x(t)]$, equation \ref{eq:saddlePointVague} tells us that the path integral is dominated by stationary points of the action. These are, of course, solutions to the classical equations of motion, allowing us to retrieve classical physics from the quantum path integral.
With this in mind, we can write equation \ref{eq:periodicInt} as 
\begin{equation}
    \tr \hat U(T)=\tr e^{-i\hat H T}\approx \sum_{\gamma} A_{\gamma}e^{iS_\gamma/\hbar}.
\end{equation}
The sum ranges over all periodic orbits $\gamma$ with period $T$. $S_\gamma$ is the action of the orbit, and $A_{\gamma}$ is an amplitude representing the Hessian of the action of the orbit. Physically, it is also connected to the stability of the orbit (to what extent do initial conditions near $x_0$ lead to final conditions near $x_0$?).

For chaotic systems at long times, there are exponentially many such periodic orbits, corresponding to the the Lyapunov exponent. Likewise, each periodic orbit has only an exponentially small amplitude $A_\gamma$. Finally, remember that there is an extra factor of $T$ from the fact that any periodic orbit $x(t)$ can be translated by a time $0\leq \Delta<T$ to a different periodic orbit $x(t)\to x(t+\Delta)$.

\begin{figure}
    \centering
    \includegraphics[scale=0.2]{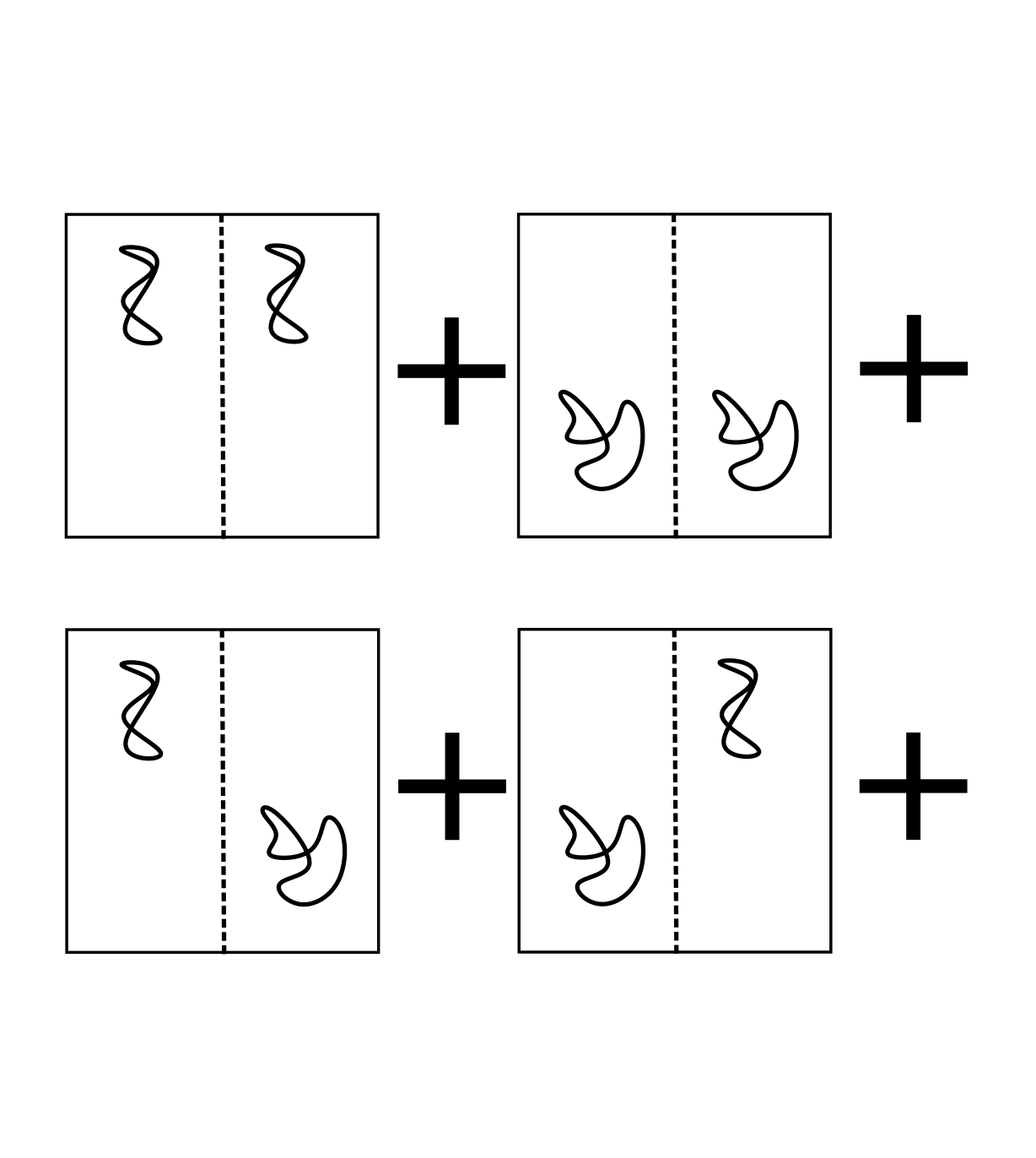}
    \caption{A sum over pairs of periodic orbits. At long times, there will be exponentially many choices for both $\gamma_1$ and $\gamma_2$. We show terms where $\gamma_1=\gamma_2$ (top) and where $\gamma_1\neq \gamma_2$ (bottom). The diagonal approximation says that the bottom terms average out to zero and can be safely ignored.}
    \label{fig:poTheoryPic}
\end{figure}

The SFF $\tr \hat U(T)\tr \hat U^\dagger(T)$ can now be written as a sum of pairs of orbits
\begin{equation}
    \SFF(T)\approx \sum_{\gamma_1\gamma_2} A_{\gamma_1}A^*_{\gamma_2}e^{i(S_{\gamma_1}-S_{\gamma_2})/\hbar}.
    \label{eq:SFFDoubleSum}
\end{equation}
This sum is well represented by figure \ref{fig:poTheoryPic}, which shows four possible pairs of orbits (out of exponentially many).
The final ingredient in the periodic orbit theory of the ramp is to treat the phases $S_\gamma$ as independent random variables. In this cases, terms in sum \ref{eq:SFFDoubleSum} average out to zero unless $\gamma_1=\gamma_2$, in which case the contribution is positive. Pictorially, the first two pictures in figure \ref{fig:poTheoryPic} still contribute positive amounts to the sum, but the bottom two terms are zeroed out on average. This approximation, treating the terms of \ref{eq:SFFDoubleSum} as zero unless $\gamma_1=\gamma_2$ is called the \textbf{Diagonal Approximation}. At sufficiently long times, other contributions arise, giving rise to the plateau.

In addition the the periodic orbit theory, another strategy for deriving the ramp is the \textbf{Wormhole Approach}. The canonical paper using this strategy is \cite{saad2019semiclassical}, which focuses on the \textbf{SYK model}. The SYK model is a useful toy model of quantum chaos. It is an ensemble rather than a single model, with Hamiltonians drawn from
\begin{equation}
    \hat H=i^{q/2}\sum_{1\leq i_1<i_2...<i_q\leq N} j_{i_1i_2...i_q}\hat \psi_{i_1}\hat \psi_{i_2}...\hat \psi_{i_q}.
\end{equation}
The $\hat \psi$s represent $N$ Majorana fermion operators and obey $\{\hat \psi_i,\hat\psi_j\}=\delta_{ij}$. The $j$s are drawn independently from a normal distribution with mean zero and variance $\frac{J^2(q-1)!}{N^{q-1}}$. In words, the Hamiltonain is $q$-body interactions aong $N$ Majorana fermions with random coefficients of strength proportional to $J$.

Using the fact that thermal partition functions are related to path integrals on the thermal circle \cite{Altland_Simons_2010,Abrikosov:107441}, the partition function of the SYK model can be written
\begin{equation}
    Z=\int \mathcal D \psi\exp\left(\int_0^\beta -\sum_i \psi_i\partial_t\psi_i-i^{q/2}\sum_{1\leq i_1<i_2...<i_q\leq N} j_{i_1i_2...i_q}\hat \psi_{i_1}\hat \psi_{i_2}...\hat \psi_{i_q}dt \right)
\end{equation}

Because the Hamiltonian for the SYK model is a random variable, quantities like the partition function, free energy, relaxation rate and energy eigenvalues are random variables. The disorder-averaged thermodynamics and dynamics can be calculated by a path integral in terms of \textbf{mean field} variables.

We define
\begin{equation}
    G(t_1,t_2)=\frac 1N \sum_{i}\hat \psi_i(t_1)\hat \psi_i(t_2)
\end{equation}

By adding in a fat unity
\begin{equation}
    1=\int \mathcal D \tilde G \mathcal D \tilde \Sigma \exp \left(\int dt_1dt_2 \tilde \Sigma(t_1,t_2)\left(\sum_i\psi_i(t_1)\psi_i(t_2)-N\tilde G(t_1,t_2)\right)\right)
\end{equation}
and integrating out the $j$s and $\psi$s, we can get a new partition function written only as a path integral over $\tilde G$ and $\tilde \Sigma$.
In terms of these new variables, we can write the disorder-average of the partition function precisely as
\begin{equation}
    \disAv{Z}=\int \mathcal D \tilde G\mathcal D \tilde \Sigma \exp \left[\frac 12 N\left(\log \det (\partial_t-\tilde \Sigma)-\int dt_1dt_2 \left(\tilde \Sigma(t_1,t_2)\tilde G(t_1,t_2)-\frac{J^2}{q}\tilde G(t_1,t_2)^q\right)\right)\right]
    \label{eq:SYKFullAction}
\end{equation}
When $N$ is large, we can again use the saddle-point method to evaluate this partition function, and derive all the thermodynamic quantities of the SYK model in terms of the bilocal (function of two times) correlation function $G$ and the self-energy $\Sigma$ at some saddle point of the action in equation \ref{eq:SYKFullAction}.

Physicists have long known that the techniques in equation \ref{eq:SYKFullAction} can apply beyond thermal partition functions. For instance the Schwinger Keldysh contour adds forward and backwards legs to to evolution, demanding we calculate $\tr e^{-\beta \hat H}e^{-iT \hat H}e^{iT \hat H}$.Calculations of the SYK's quantum Lyapunov exponent, which take up much of \cite{Maldacena_2016} require the more complicated four-legged contour which goes around the circle, then forwards, backwards, forwards and backwards. 

\begin{figure}
    \centering
    \includegraphics[scale=0.5]{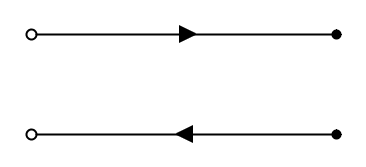}
    \caption{Evaluating path integrals on the SFF contour allows the calcuation of the Spectral Form Factor. The top of the two legs represents a time circle of real time period $T$, the bottom represents one with period $-T$. Given fields $\Psi$ (for instance the position coordinates of a particle, the fermion fields of an SYK model, the fields of a gauge theory, or anything else) the Spectral Form Factor can be calculated by integrating over all $\Psi_{top}(t),\Psi_{bottom}(t)$ with appropriate boundary conditions. For mean-feild models like the SYK, $\Psi_{top}$ and $\Psi_{bottom}$ can be replaced by a bilocal correlation field $G$, which has components connecting the forward and backwards legs of the contour.}
    \label{fig:sffContourImage}
\end{figure}

The key insight of \cite{saad2019semiclassical} was that the disorder-averaged SFF can be calculated by analyzing these seem correlations on the correct contour: in this case the so-called SFF contour (see figure \ref{fig:sffContourImage}).

The SFF contour has two legs, one with period $T$ and the other with period $-T$. Given a set of fields $\Psi$ which define our system, we have a copy of $\Psi$ on each of the two legs. For the case of a mechanical particle, $\Psi$ would represent the coordinates and would be periodic on each of the two legs. For the case of the SYK model, $\Psi$ represents the fermionic fields and is antiperiodic on each leg. For mean-field systems, we can integrate out the $\Psi$s on both contours to get a bilocal correlation field $G$ which connects not only different times, but also can connect the two contours. For large $N$, the SFF is dominated by special configurations of $G$ (and $\Sigma$) which are stationary points of the integrand of \ref{eq:SYKFullAction}. 

\cite{saad2019semiclassical} investigates such special $G$ configurations and finds two families of such solutions. One is contour diagonal, with $G$ taking on nonzero-values only for different points on the same contour. This solution starts off contributing an exponentially (in $N$) large amount to the SFF, but decays with time. In fact, one can show that this solution, in which the contours are literally disconnected and uncorrelated, is responsible for the disconnected SFF (that is to say, the slope seen in figure \ref{fig:SFFgraphs}).

The authors find another solution, related to the thermofield double, which reproduces the ramp, including both the overall factor of $T$ and the correct prefactors. Because of the gravity dual, this solution is called the wormhole solution, though it is perfectly well-defined even in systems that have nothing to do with gravity. 

This solution relies on two essential facts that are present in a wide array of models: a mean-field description at large $N$, and the exponential decay of $G$. As we shall see, this decay is very important. Indeed, signatures of $G$ appear in corrections to the linear ramp, and systems with non-decaying correlations (such as spin glasses) have fundamentally different behaviors in their spectral statistics.
\section{The Total Return Probability: Ergodicity and the Spectral Form Factor}
The primary contribution of this thesis is the study of how ergodicity, thermalization, and a lack thereof can modify the RMT ramp, even in systems that have Lyapunov chaos and operator growth. In particular, we show that in many contexts, the ramp is multiplied by a factor we call the \textbf{Total Return Probability} (TRP), a diagnostic of related to how thermalized a system is after time $T$.

A central result which will be discussed at length in chapter \ref{chapter:hydro} is that for systems which thermalize slowly, the connected SFF can be well approximated at times below the Heisenberg time as
\begin{equation}
    \SFF_{\textrm{con}}(T)=\int \trp(T;E) \frac{T}{\beta \pi}dE
    \label{eq:TRPSFF}
\end{equation}

It is useful to juxtapose this result with the idea of \textbf{Random Matrix Universality}. This notion states that one can understand the spectral properties of chaotic Hamiltonians, including the spectral form factor or level-spacing rations, purely by treating them as random matrices drawn from an appropriate ensemble. By contrast, equation \ref{eq:TRPSFF} shows that at short times, very physical facts about the system can affect the spectral statistics.

\subsection{Total Return Probability as a Measure of Thermalization}
Suppose we have a chaotic (in the Lyapunov sense) system whose configuration space can be partitioned into sectors $i$.
\begin{figure}
    \centering
    \includegraphics[scale=0.5]{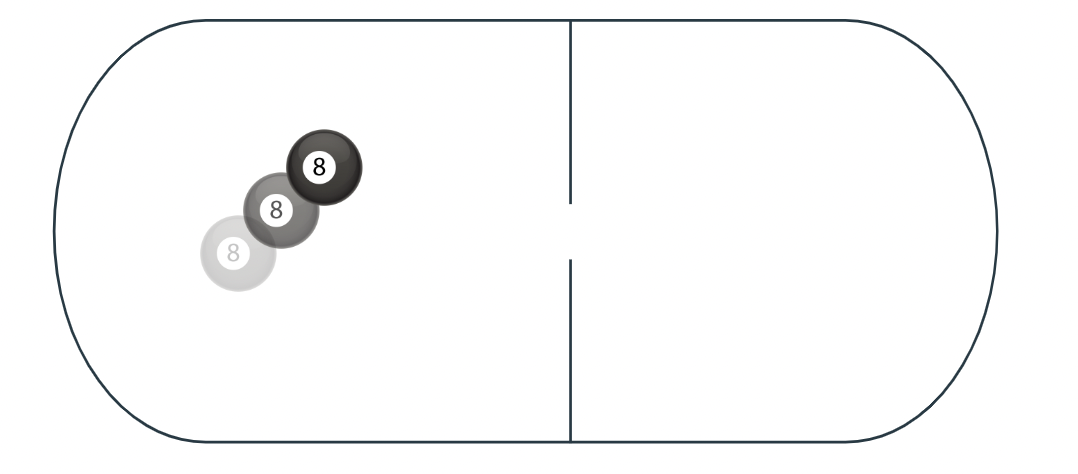}
    \caption{A (quantum) billiard divided into two chambers. Motion within each chamber is chaotic, and a state in a given chamber soon equilibrates to a Gibbs state spread over that chamber. Over a separate, possibly much longer timescale, the particle can seep through the small hole in the wall, allowing for transitions between two chambers. Eventually, the distribution of the particle between the two chambers will be independent of its initial location.}
    \label{fig:partitionBilliard}
\end{figure}
One simple example of a system with multiple sectors is a billiard table with two chambers (see figure \ref{fig:partitionBilliard}). We can start the particle in some simple wavepacket $\psi_i$ located in chamber $i$ (in this case $i$ is either 1 or 2, but the logic applies equally well to systems with many sectors). We will choose a wavepacket to have energy approximately $E$. 
After time $T$, we check what sector the ball is in. The probability of sector $j$ is $p_j(T|\psi_i)$. Importantly, the fact that the particle reaches equilibrium so fast within a chamber means that $p_j(T|\psi_i)$ depends only on which chamber $i$ you start with and the energy, not on any of the details of the initial state $\psi_i$. We can thus define $p_{i\to j}(T;E)$, the probability that a particle will be in sector $j$ at time $T$ given that it started in sector $i$ at time $0$.

In terms of the $p_{i \to j}$s, the TRP is
\begin{equation}
    \trp(T;E)=\sum_i p_{i \to i}(T;E)
    \label{eq:TRPDef}
\end{equation}
At $T=0$ $p_{i \to i}$ is 1 regardless of $i$, and the TRP just counts the number of sectors. At long times, once the system is in equilibrium, $p_{i\to i}$ is just the equilibrium probability for $i$. This implies that the TRP will be exactly one. If there is some other conserved quantity in the system, such as charge, momentum, or spin, the TRP counts the number of charge sectors for this quantity. Systems with a conserved charge will have an enhanced ramp at arbitrarily long times, a consequence of the completely lack of level repulsion between energy eigenstates in different charge sectors.

As the TRP interpolates from its large intial value to its final value, the system equilibrates. (Under certain circumstances the TRP can actually fall below one. This is symptomatic of some sort of oscillatory nature to the system, and is explored in chapter \ref{chapter:soundPole}).

It is important to clarify that the TRP can depend on the temperature or energy window of interest. For instance in the billiard example above, the velocity of the ball scales as $\sqrt(E)$, meaning that the time-scale to penetrate from one chamber to the other goes approximately as $E^{-1/2}$. This means that at a fixed time, $\trp(T;E)$ will decrease with decreasing $E$. Despite this, when convenient (such as in the subsection below), we will omit the dependence of the TRP on $E$. 

\subsection{Properties of the TRP}
In this subsection, we clarify a number of properties of the TRP that are only left implicit in the literature. The TRP is closely related to the dynamical zeta function in mathematics \cite{artinMazur,Ruelle2002DynamicalZF}. We will prove two properties of the TRP: it is multiplicative for independent systems, and it is independent of how the sectors are defined as long as they are `fine enough'.

First, the proof of the multiplicative property. Let's decompose our systems into subsystems $A$ and $B$, so each sector $i$ is defined by a pair $i_{A},i_{B}$. An example of this might be two distinguishable non-interacting balls moving around the stadium in figure \ref{fig:partitionBilliard}. The TRP can be written 
\begin{equation}
    \trp(T)=\sum_i p_{i \to i}(T)=\sum_{i_A, i_B} p_{i_A \to i_A}(T)p_{i_B \to i_B}(T)=
    \trp_A(T)*\trp_B(T)
\end{equation}
Thus, the TRP behaves in some sense like a partition function, albeit one that measures dynamics instead of thermodynamics. Like the usual partition function, we expect to see that for a system of size $N$, the TRP will be exponentially large in $N$.

\begin{figure}
    \centering
    \includegraphics[scale=0.3]{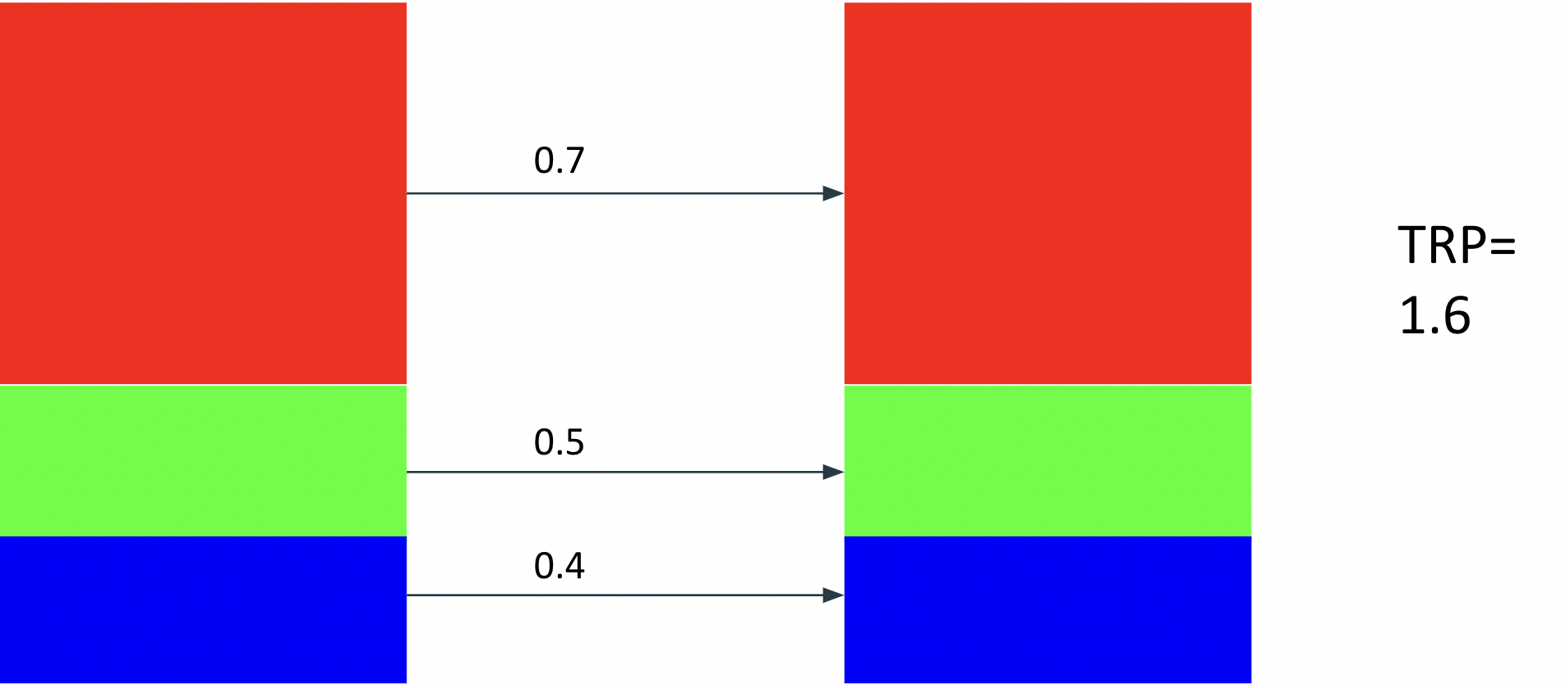}\\
    \vspace{1cm}
    \includegraphics[scale=0.3]{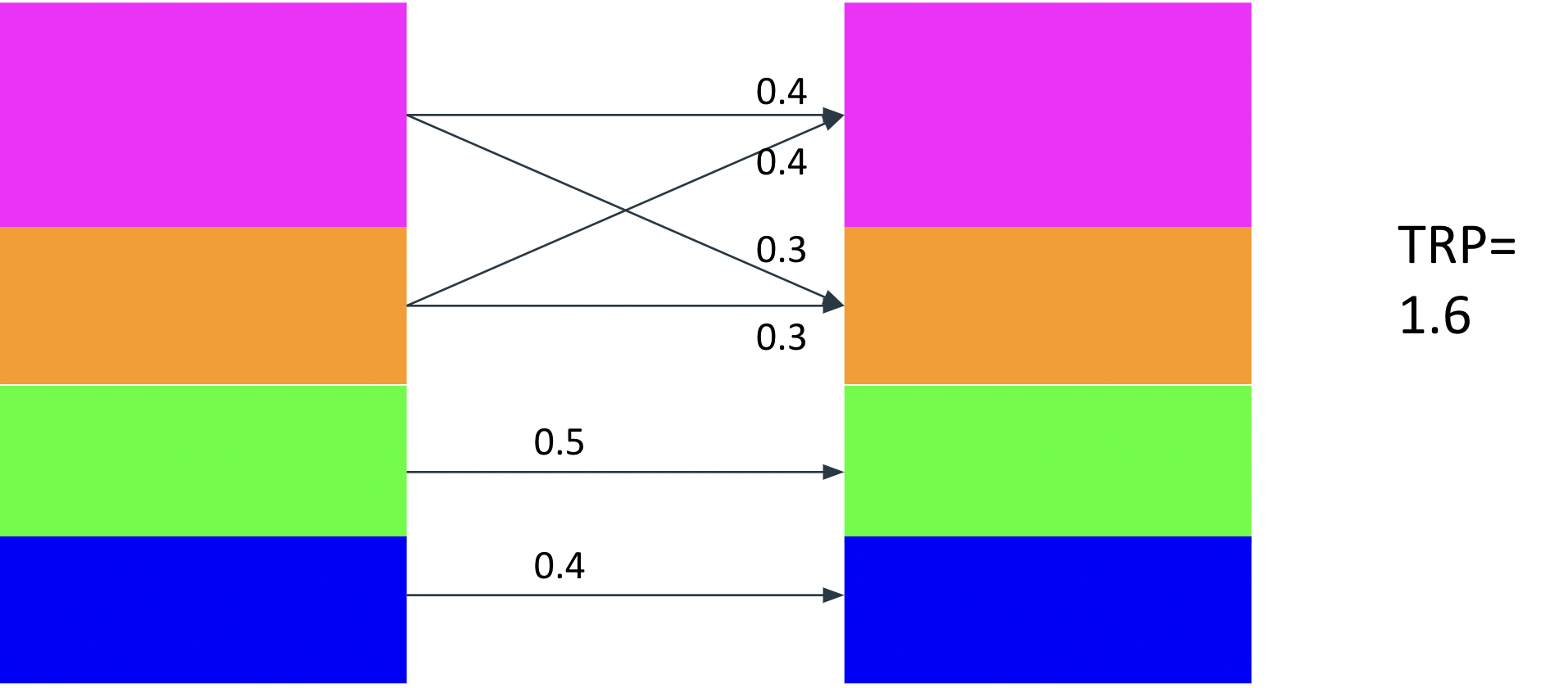}
    \caption{The same evolution described two different ways. In the top description there are three sectors, in the bottom description the red sector is partitioned into pink and orange. In both cases the TRP is 1.6.}
    \label{fig:trpResiliant}
\end{figure}

Our second theorem is that the TRP does not depend on how exactly we define the sectors, as long as the sectors are fine enough, or the time is long enough. More precisely, our condition is that there is no simple operator at time $T$ that provides evidence for where in a sector we started in at time 0. In the example of figure \ref{fig:trpResiliant}, there may be operators at time $T$ that can distinguish whether we started in the red or green sector, but nothing can distinguish between the orange and pink subsectors of the red sector.

To get a feel for this, we first show that given this condition is satisfied, further breaking a sector in two doesn't change the TRP. We will follow the example in figure \ref{fig:trpResiliant}.

We start with several sectors (in this case three) and their return probabilities (in this case 0.7, 0.5, and 0.4). We imagine breaking a sector (in this case red) into two subsectors (in this case pink and orange). We have assumed that no operator at time $T$ can distinguish between these two sectors, otherwise our initial sector definitions were too granular. This includes the operator that indicates what sector we are in at time $T$. Thus 
\begin{equation}p_{\textrm{pink}\to i}(T)=p_{\textrm{orange}\to i}(T)=p_{\textrm{red}\to i}(T)\end{equation}
for all sectors or subsectors $i$. This implies 
\begin{equation}
\begin{split}
   p_{\textrm{red}\to\textrm{red}}(T)=p_{\textrm{pink}\to\textrm{red}}(T)=p_{\textrm{pink}\to\textrm{pink}}(T)+p_{\textrm{pink}\to\textrm{orange}}(T)=\\
   p_{\textrm{pink}\to\textrm{pink}}(T)+p_{\textrm{orange}\to\textrm{orange}}(T)
\end{split}
\end{equation}
Thus the TRP remains the same under the partitioning of a single (fine enough) sector. We can use this to show that all fine enough partitions of phase space into $P_1$ and $P_2$ must yield the same TRP, by partitioning both of them into a set of sectors finer than both.
\subsection{The TRP In Physical Systems}
One of the first lines of research into the spectral properties of thermal systems was studying the single-level energy statistics of an electron in a disordered grain \cite{Altland_2000,altshuler1986repulsion,PhysRevB.60.3944}. In this class of papers, an electron in $d$-dimensional space is subject to some sort of disordered Hamiltonian
\begin{equation}
    \hat H=\frac{\hat p^2}{2m}+\hat V
\end{equation}
Where $\hat V$ is a weak randomly chosen potential. The electron will exhibit diffusive motion with some diffusion constant $D$. If we divide space into cubes of length $\ell$, the probability that the electron ends up in the cube it started in is $\left(\frac{\ell}{2\sqrt{\pi D T}}\right)^d$. If the system has volume $V$ the total number of sectors is $\frac{V}{\ell^d}$ meaning the TRP is exactly 
\begin{equation}
    \trp(T)=\frac{V}{(2\sqrt{\pi D T})^d}
\end{equation}
This goes to one after a time of approximately $V^{2/d}/D$, the timescale it takes for a particle to diffuse across the system. This timescale, the time necessary to completely forget the initial sector, is often called the \textbf{Thouless time}.

\section{Fluctuating Hydrodynamics}
In this thesis, special attention will be given to two types of slowly thermalizing systems: hydrodynamic systems and glassy systems. \textbf{Hydrodynamics} is the study of systems with local conservation laws. Its earliest roots go back to the study of fluids, which have five locally conserved currents (energy, three directions of momentum, and particle number). But more modern theories can take into account any number of conservation laws, including non-Abelian symmetries \cite{Glorioso_2021,Fern_ndez_Melgarejo_2017}, dipole conservation laws \cite{Gromov_2020,Glorioso_2022}, higher form symmetries \cite{Grozdanov:2016tdf}, and even integrable systems \cite{Essler_2023,Doyon_2020}.

Hydrodynamic theories are effective field theories (EFTs) written using conservation laws to guide the IR physics. The most important dynamical variables are typically densities of conserved quantities (number density, energy density, momentum density), or else closely related quantities (temperature, local velocity). Hydrodynamics has a very deep relationship with thermodynamics. In a word, it is thermodynamics with locality. In conventional thermodynamics, we study the maximal entropy state of a system given a global constraint on the energy or charge. Hydrodynamics takes this to the next level: on reasonable timescales charge isn't just conserved globally, it is also conserved locally. The total energy of the universe is conserved exactly, the total energy in a room changes on the order of hours, the total energy of a molecule is randomized trillions of times a second.

In other words, local conservation laws create slow modes, and locality slows down thermalization. Each smoothed-out density configuration $\rho(x)$ is its own sector. This enormous number of sectors- exponential in system size- allows for an enormous TRP and an enormous SFF. If we can calculate the total return probability of a hydrodynamic system, we know its spectral form factor. In order to accomplish this, we need a precise technology for making claims about unlikely events in hydrodynamic systems.

One particularly useful formulation of this problem is the Closed Time Path (CTP) formalism. A good introduction  can be found in see~\cite{glorioso2018lectures}. For more details see ~\cite{crossley2017effective,Glorioso_2017}. Other formulations of fluctuating hydrodynamics are explored in ~\cite{Grozdanov_2015,Kovtun_2012,Dubovsky_2012,Endlich_2013}. 

The CTP formalism begins on the \textbf{Schwinger-Keldysh contour}. The central element of the Schwinger-Keldysh method is the introduction of a time contour that loops back on itself, often visualized as preparing an equilibrium state at an intial time ($-\infty$ for our purposes), evolving to a final time ($+\infty$), and then looping back.

We begin with a system with one conserved current operator $j_\mu$ satisfying $\partial^\mu j_\mu=$ as an operator equation. In this case the CTP formalism calculates the following partition function on a Schwinger-Keldysh contour: 
\begin{equation}
    Z[A^\mu_1(t,x),A^\mu_2(t,x)]=\tr \left( e^{-\beta H} \mathcal{P} e^{i\int dt d^d x A^\mu_1j_{1\mu}} \mathcal{P} e^{-i\int dt d^d x A^\mu_2 j_{2\mu}}\right),
\end{equation}
where $\mathcal P$ represents path ordering on the Schwinger-Keldysh contour.

For $A_1=A_2$, this is exactly the thermal partition function at inverse temperature $\beta$. Differentiating $Z$ with respect to the $A$s generates insertions of the conserved current density $j_\mu$. Differentiating with respect to $A_1$ inserts on the forward leg, differentiating with respect to $A_2$ puts them on the later backwards leg. $Z[A^\mu_1(t,x),A^\mu_2(t,x)]$ and its derivatives generate all possible contour-ordered correlation functions of current density operators. The many constraints imposed on these correlation functions by unitarity and conservation laws become constraints on $Z$.

We enforce the conservation law $\partial^\mu j_{\mu}=0$ on both legs by expressing $Z$ as 
\begin{equation}
    \begin{split}
         Z[A^\mu_1,A^\mu_2]=\int \mathcal D \phi_1\mathcal D \phi_2 \exp\left(i\int dt dx W[B_{1\mu},B_{2\mu}]\right),\\
         B_{i\mu}(t,x)=\partial_\mu \phi_i(t,x)+A_{i\mu}(t,x).
    \end{split}
    \label{eq:AbelianB}
\end{equation}
We have introduced new fields $\phi$ on each leg to represent the slow fluctuating modes of the system. Insertions of the currents are still obtained by differentiating $Z$ with respect to the background gauge fields $A_{i\mu}$. From this one can derive that $\partial_\mu \frac{\delta Z}{\delta A_{i\mu}} = 0$.

There are additional constraints on the functional $W$. The key assumption of hydrodynamics is that the action $W$ is local. If the functional is non-local, that means that more fields need to be introduced. Moreover, when expressed in terms of
\begin{equation}
\begin{split}
    B_a=B_1-B_2,\\
    B_r=\frac{B_1+B_2}{2},
\end{split}
\end{equation}
there are additional conditions which follow from unitarity:
\begin{itemize}
    \item $W$ terms all have at least one power of $B_a$, that is $W=0$ when $B_a=0$. This ensures that $Z$ is the partition function when $A_1=A_2$.
    \item Terms odd (even) in $B_a$ make a real (imaginary) contribution to the action.
    \item A KMS constraint \cite{doi:10.1143/JPSJ.12.570,PhysRev.115.1342} deriving from the fact that $Z[A^\mu_1(t),A^\mu_2(t)]=Z[A^\mu_2(-t),A^\mu_1[i\beta-t]]^*$
\end{itemize}
As a case study, we will consider a system with only energy conservation. One simple Lagrangian consistent with these conditions and rotational invariance is
\begin{equation}
    L=iT^2cD \sum_j B_{aj}^2+cT(B_{a0}B_{r0}-D\sum_jB_{aj}B_{rj}).
\end{equation}
where $j$ ranges over spatial indices, $T=\beta^{-1}$ is temperature, $D$ is the diffusion constant and $c$ is the system's specific heat. Setting the external gauge fields $A$s to zero, the Lagrangian simplifies to
\begin{equation}
    L=\phi_a\left(\partial_r \epsilon-D\Delta \epsilon\right)+iT^2cD (\partial_j \phi_a)^2,
\label{eq:diffusiveLagrangian}
\end{equation}
with $\epsilon=cT\partial_t\phi_r$. We can show that $\epsilon$ is equal to the density of our conserved energy.

Examining equation \ref{eq:diffusiveLagrangian}, we see a part proportional to $\phi_a$ and a part proportional to $\phi_i^2.$ Focusing on just the first part, we see that $\phi_a$ is a Lagrange multiplier enforcing the diffusion equation $\partial_r \epsilon-D\Delta \epsilon.$ The quadratic part of equation \ref{eq:diffusiveLagrangian} introduces fluctuations into our dissipative system, turning a deterministic process into a probabilistic one. The KMS condition relates the strength of the fluctuations to the strengths of the dissipation, allowing us to recover the famous fluctuation-dissipation theorems.
\section{Thermalization in Glassy Systems}
Glasses provide a stark contrast to hydrodynamic systems. In a hydrodynamic systems, the sparseness of the interaction graphs means that conserved quantities like energy or charge take time to spread across the system, forcing us to develop a theory of local (in real space) equilibrium. By constrast, a \textbf{glass} is a system whose non-local interactions create impassible barriers in phase space, forcing us to develop a theory of local (in phase space) equilibrium.

\begin{figure}
    \centering
    \includegraphics[scale=0.5]{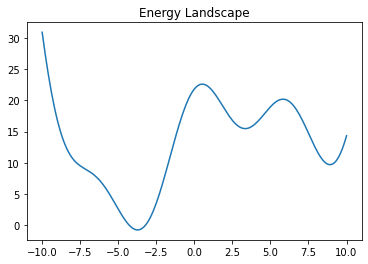}\\
    \includegraphics[scale=0.5]{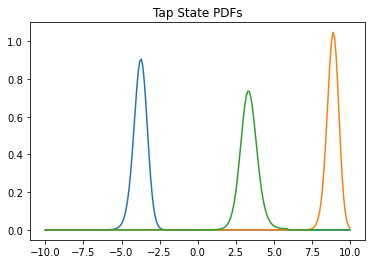}
    \caption{(Top) A complicated energy landscape with several local minima. (Bottom) Locally Gibbs-like states that approximate $p(x)~e^{-\beta V(x)}$ where $p$ is large.}
    \label{fig:1DGlass}
\end{figure}

By necessity, glassy systems have many degrees of freedom interacting in complicated ways, making it difficult to picture their phase spaces. In order to gain some intuition, we will look at the rugged energy landscape in figure \ref{fig:1DGlass}.
On the top, there is some complicated potential with many local minima, separated by walls of height much more than one. If we imagine putting a particle near one of these minima, and coupling it to a heat bath at temperature 1, thermal fluctuations would cause it to spread out. The entropy would increase, but the particle would take an exponentially long time to cross the potential energy barriers between the wells. There is an (almost) steady state which has density proportional to $e^{-\beta V(x)}$ in some region around a local minimum and is asymptotically zero away from that minimum. These locally-consistant long-lived states are called Thouless-Anderson-Palmer or \textbf{TAP states} \cite{Thouless1977Solution,CAG,Crisanti1995ThoulessAndersonPalmerAT}. Recalling the total return probability, we see that the TRP at intermediate times (long enough to thermalize within a TAP state, but too short to tunnel between TAP states) exactly counts the number of stable local minima. In chapter \ref{chapter:glass}, we calculate the spectral form factor of a simple glass model analytically, and show that the SFF's enhancement factor agrees with known calculations of the number of TAP states.
\section{Plan For The Rest Of This Thesis}
The body of this thesis consists of four chapters \ref{chapter:hydro,chapter:ssb,chapter:soundPole,chapter:glass}, adapted from refs \cite{winerprx,Winer_2022,Winer:2022gqz,Winer:2022ciz} by the author of this thesis and others. 

Chapter \ref{chapter:hydro} is a general overview of how slow thermalization affects the spectral form factor. It begins with systems with true conservation laws, and shows how Hamiltonians with conserved charge sectors have enhanced SFFs. It then discusses Hamiltonians of a nearly-bock form, deriving the fact that the ramp portion of the SFF is enhanced by the TRP, and showing the case with perfect conservation laws is a special case of this. Finally it makes the connection to hydrodynamics, arguing that hydrodynamic systems have a large number of approximate conservation laws (the density in each region of space changes only slowly) and deriving the exponentially large enhancement factor in a number of contexts, including diffusive hydrodynamics, subdiffusive hydro, and even an interacting theory. However, everything analyzed in this chapter is a Markovian process such as diffusion, we never explicitly discuss systems modeled by higher-order differential equations such as sound or Goldstone modes.

Chapter \ref{chapter:ssb} is a followup to chapter \ref{chapter:hydro}. In this work, we focus on systems with broken symmetries. We progress from $Z_n$ symmetries to finite non-Abelian groups to systems with broken $U(1)$ symmetries (superfluids) and finally to systems with spontaneously broken non-Abelian continuous symmetries. The results are surprising: while symmetries make the SFF larger, spontaneously broken symmetries make the SFF larger still. The essential reason for this is that the broken symmetries cause correlations in the level spacings of different charge sectors, enhancing fluctuations in the level density. Chapter \ref{chapter:ssb} focuses only on mean field systems, not on systems which have any spatial extent.

Chapter \ref{chapter:soundPole} is the next paper in this series. Chapter \ref{chapter:hydro} concluded by studying Markovian systems with oscillatory character, and chapter \ref{chapter:ssb} studied mean-field systems with Goldstone modes. Chapter \ref{chapter:soundPole} studies systems with sound propagating through space. Scientifically, this mostly takes the form of applying formulas in chapter \ref{chapter:hydro}, especially equation \ref{eq:diffShape}. However, we find this formula gives quantitatively new results in the context of oscillating hydrodynamic modes. In particular, the results depend on the spacings of the frequencies of the hydrodynamic modes. This puts us in the amusing situation of trying to calculate the spectral form factor of a complicated hydrodynamic system, and having our results depend on the spectral form factor of some much simpler system (the operator giving the evolution of the sound modes in the linearized effective theory).

Chapter \ref{chapter:glass} takes our story in a different direction, studying the spectral form factor of a specific system: the quantum $p$-spherical model, one of the simplest solvable models of a spin glass. We solve a path integral exactly in the limit of large system size $N$, and derive an expression for the connected SFF during the ramp phase. Our result is linear in $T$ just like the SFF for random matrix theory, but the coefficient in front is exponentially large in $N$. We show that this coefficient agrees with the known results counting the number of long-lived TAP states in the $p$-spherical model, and thus show that in this explicit model, the SFF is proportional to the total return probability as initially argued in chapter \ref{chapter:hydro}. These results open an entirely new method for counting TAP states in quantum spin glasses, a previously very difficult task.

\renewcommand{\thechapter}{2}
\newcommand{\sgn}{\textrm{sgn}}
\newcommand{\fstretch}{f_{\textrm{stretch}}}
\newcommand*\mean[1]{\overline{#1}}
\newcommand{\sff}{\textrm{SFF}}
\chapter{Hydrodynamic Theory of the Connected Spectral Form Factor}
\label{chapter:hydro}
\textbf{Authors:} \textit{Michael Winer, Brian Swingle}

\textbf{Abstract:} One manifestation of quantum chaos is a random-matrix-like fine-grained energy spectrum. Prior to the inverse level spacing time, random matrix theory predicts a `ramp' of increasing variance in the connected part of the spectral form factor. However, in realistic quantum chaotic systems, the finite time dynamics of the spectral form factor is much richer, with the pure random matrix ramp appearing only at sufficiently late time. In this article, we present a hydrodynamic theory of the connected spectral form factor prior to the inverse level spacing time. We start from a discussion of exact symmetries and spectral stretching and folding. We then derive a general formula for the spectral form factor of a system with almost-conserved sectors in terms of return probabilities and spectral form factors within each sector. Next we argue that the theory of fluctuating hydrodynamics can be adapted from the usual Schwinger-Keldysh contour to the periodic time setting needed for the spectral form factor, and we show explicitly that the general formula is recovered in the case of energy diffusion. We also initiate a study of interaction effects in this modified hydrodynamic framework and show how the Thouless time, defined as the time required for the spectral form factor to approach the pure random matrix result, is controlled by the slow hydrodynamics modes.

\section{Introduction}

There has been a surge of recent interest~\cite{haake2010quantum,PhysRevLett.52.1,mehta2004random} in the statistics of energy levels of chaotic quantum systems. Quantum chaos in this loose sense is typically invoked when quantizing a classically chaotic system and in the context of quantum systems that thermalize. It is widely believed that ensembles of such chaotic systems have the same spectral statistics as ensembles of random matrices, with examples from nuclear systems~\cite{doi:10.1063/1.1703775,wigner1959group} to condensed matter systems~\cite{bohigas1984chaotic,dubertrand2016spectral,PhysRevLett.121.264101} to holographic theories~\cite{Cotler2017,Stanford2018}. In fact, it is now common to take random matrix spectral statistics as one definition of quantum chaos.

Such a definition must be applied with care, however, since a particular chaotic quantum system will typically only have random matrix-like spectral features at sufficiently long times after features like spatial locality have been washed out. In this paper we present a hydrodynamic theory of the intermediate time spectral properties of such quantum chaotic systems. Symmetries and hydrodynamics are an inescapable part of the story because time-independent Hamiltonian systems always have at least time translation symmetry and energy conservation. Here we consider both exact and approximate symmetries, including the important case of slow modes arising, for example, from energy conservation. These results allow us to precisely characterize how the imprint of spatial locality on the energy spectrum gives way to pure random matrix statistics at long time. To setup a statement of our main results, we first review the basics of random matrix theory and the observables of interest.

A random matrix ensemble is characterized by two pieces of data. The first datum is the type of matrix (orthogonal, unitary, symplectic) and corresponding Dyson index $\boldsymbol\beta=1,2,4$. In physical terms, this relates to the number and nature of antiunitary symmetries. The second datum is a potential $V(E)$, where we choose matrix $H$ with probability $dP\propto \prod_{ij}dH_{ij} \exp \left( -\tr V(H)\right)$. These data give a joint probability for the eigenvalues $\{E_i\}$ equal to 
\begin{equation}dP=\frac 1 {\mathcal Z}\prod_{i<j} |E_i-E_j|^{\boldsymbol \beta} \prod_i e^{-V(E_i)},
\label{eq:RMTpdf}
\end{equation}
where $\mathcal{Z}$ is a normalization. 

This probability distribution can be conveniently interpreted in terms of a ``Coulomb gas'' of eigenvalues as follows. Eq.~\ref{eq:RMTpdf} has the form of a Boltzmann distribution at unit temperature for a gas of 1d particles at positions $E_i$ with logarithmic Coulomb interactions $U_{ij}=\bbeta \log |E_i-E_j|$ subject to an external potential $V$~\cite{mehta2004random}. In this way of thinking, the correlations of the density of particles/eigenvalues,
\begin{equation}
\rho(E)=\sum_{i}\delta(E-E_i),
\end{equation}
form a natural set of obserables. The most basic of these observables is the density of states, $\overline{\rho(E)}$, where the overline denotes the disorder average. For example, in a Gaussian random matrix ensemble in which the potential $V$ is quadratic, this average is well approximated by the famous Wigner semi-circle law. The simplest observable that probes spectral correlations is the 2-point function of the density, $\overline{\rho(E_1) \rho(E_2)}$. 

It is common~\cite{heusler2004universal}\cite{brezin1997spectral} to package this 2-point function into an object in the time domain called a spectral form factor, defined here to include a filter function $f$,
\begin{equation}
\sff(T,f)=\overline{|\tr f(H)e^{iHT}|^2}=\overline{\sum_{i,j}f(E_i)f(E_j)e^{i(E_i-E_j)T}}.
\label{eq:SFFdef}
\end{equation}
Very often we choose $f(H)=\exp(-\beta H)$, which we call the SFF at inverse temperature $\beta$. (In this paper bold $\bbeta$ is the Dyson index and $\beta$ is inverse temperature. $T$ is time and never temperature.) Another useful choice for $f$ will be a Gaussian function zeroed in on a part of the spectrum of interest. The SFF is then simply the squared magnitude of the $T$-component of the Fourier transform of $\rho_f(E) = f(E)\rho(E)$,
\begin{equation}
    \rho_f(T) = \int_{-\infty}^{\infty} dE \rho_f(E) e^{i E T}.
\end{equation}

\begin{figure}
\centering
\includegraphics[scale=0.6]{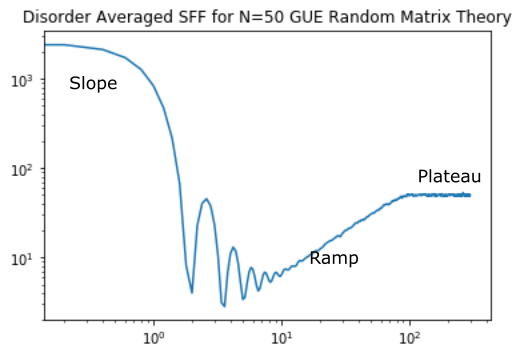}
\caption{The SFF of a simple random matrix system displaying slope, ramp, and plateau behaviors.}
\label{fig:BasicSFF}
\end{figure}

The SFFs of chaotic systems traditionally break into three regimes. First, a slope region, where Eq.~\ref{eq:SFFdef} is dominated by the disconnected part of the 2-point function of $\rho$. Once the system reaches the Thouless time $t_{\text{Th}}$ when all macroscopic degrees of freedom have relaxed, we reach a new stage.
This second state is the ramp, where the disorder-averaged SFF is linear in $T$. The ramp continues until times of order the level spacing (called the Heisenberg time), long enough that the off-diagonal terms in equation \ref{eq:SFFdef} average to zero. After this time, the disorder averaged SFF is flat, and we have a plateau.
An example log-log plot of a random matrix SFF is shown in Fig.~\ref{fig:BasicSFF}.

\begin{figure}
\centering
\includegraphics[scale=0.6]{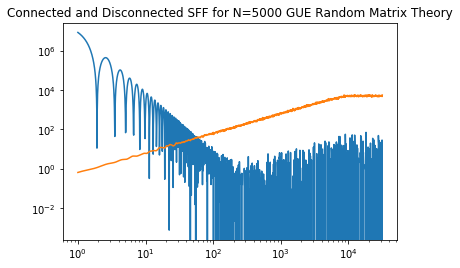}
\caption{The connected (orange) and disconnected (blue) parts of the Spectral Form Factor plotted on a log-log plot. The ramp come and plateau come entirely from the connected bit, and the slope entirely from the disconnected bit. This graph appears to show some small ramp-like behavior for the connected part of the SFF, but that's just because the sample variance (controlled by the connected SFF) is going up so the distribution of values gets wider.}
\label{fig:twoSFF}
\end{figure}

It is further useful to decompose the SFF into connected and disconnected pieces. In terms of the partition function,
\begin{equation}
    Z(T,f) = \sum_i f(E_i) e^{i E_i T},
\end{equation}
the SFF is 
\begin{equation}
    \sff(T,f) = \overline{Z(T,f) Z^*(T,f)} = \sff_{\text{conn}} + \sff_{\text{disc}},
\end{equation}
where 
\begin{equation}
    \sff_{\text{disc}} = |\overline{Z(T,f)}|^2
\end{equation}
and
\begin{equation}
    \sff_{\text{conn}} = \sff - \sff_{\text{disc}} = \overline{\left(Z(T,f)-\overline{Z(T,f)}\right)\left(Z(T,f)-\overline{Z(T,f)}\right)^*}.
\end{equation}
Fig.~\ref{fig:twoSFF} shows the very different behaviors of these two pieces of the SFF. The disconnected part is controlled just by the density of states, so we can more cleanly access the spectral correlations by focusing on the connected part.

One comment about notation is in order. The ramp typically refers to the linear in time part of the connected spectral form factor. In a many-body system of $N$ degrees of freedom with no symmetries or slow modes, the ramp is expected to onset after a short relatively short time of order $\log N$.\footnote{This is the time it takes for an exponentially decaying mode of the form $e^{-\lambda t}$ to reach a $1/N$ suppressed amplitude provided the rate $\lambda$ is not $N$-dependent.} The main topic of this article is the modification of the random matrix ramp due to slow modes and non-random matrix features of the system. One could conceivably speak about a `time-dependent ramp coefficient', but we prefer to consider the time period prior to the pure random matrix ramp as distinct regime. In this view, there are four time periods: (1) the very early regime, prior to a time of order $\log N$, when all the details matter, (2) the hydrodynamic regime, when the spectral form factor is determined by the symmetries and slow modes of the system, but is insensitive to other details, (3) the pure random matrix ramp regime, and (4) the plateau regime. Given this characterization, we define the Thouless time to be the time it takes for the SFF to come close the pure random matrix ramp. Later, we will derive an expression relating the connected SFF to return probabilities in equation \eqref{eq:bigAnswer}, giving precise meaning to the notion that RMT behavior takes over when the system has had time to fully explore Hilbert space~\cite{Schiulaz_2019}.

Given this background, we can now state our main results. We study the connected spectral form factor in the pure random matrix ramp regime and the hydrodynamic regime. First, in Section \ref{sec:RMT} we review the random matrix theory calculation of the ramp, focusing on its coefficient. We observe that the predicted coefficient agrees with analytical results in the SYK model and with numerical results in a variety of spin models. It is therefore natural to conjecture that both the linear-$T$ behavior and the precise coefficient are universal across chaotic systems. 
Second, in Section \ref{sec:Conserve} we show how symmetries and folding modify modify the coefficient of the ramp by breaking the Hamiltonian up into decoupled sectors. The random matrix theory prediction is again shown to agree with results in various models. 
Third, in Section \ref{sec:nearBlock} we discuss the case of approximate symmetries which correspond to slowly decaying modes. We show in such cases that the connected spectral form factor can be computed in terms of return probabilities for the slow modes.
Finally, in Section \ref{sec:Hydro} we argue that the theory of fluctuating hydrodynamics, conventionally formulated on the Schwinger-Keldysh contour, can be adapted to the periodic time contours defining the spectral form factor. Focusing on the case of energy diffusion, we show that this periodic CTP formalism recovers the ramp at late time and the return probability formula. At quadratic level, the formulas agree with previous results obtained in Floquet models; we also discuss novel effects arising from hydrodynamic interactions. 

To give some context for our work, we start by noting that there is a very large literature on quantum chaos extending back many decades. One key paper is \cite{original-thouless} which showed that the variance of the number of single particle energy levels in a band was random-matrix-like for energies smaller than the inverse Thouless time. This time originally arose in the context of mesoscopic transport as a measure of the sensitivity of the system to boundary conditions, but it has come to refer to the timescale beyond which quantum dynamics looks random matrix like. Other prior investigations of the Thouless time in a many-body setting include \cite{Schiulaz_2019,Gharibyan_2018,Friedman_2019,Altland_2018,Nosaka_2018}. It should be noted that the Thouless time can depend on the observable used to define it, for example, the spectral form factor versus some correlation function. There are also a growing number of exact diagonalization studies and analytic results on many-body spectral statistics and spectral form factors including~\cite{PhysRevLett.121.264101,Stanford2018,Garc_a_Garc_a_2017,altland2020late}. The theory of fluctuating hydrodynamics has been developed in a series of papers including~\cite{Dubovsky_2012,PhysRevD.91.105031,Haehl_2018,crossley2017effective,Jensen_2018}. One useful recent review on various aspects of quantum chaos is \cite{doi:10.1080/00018732.2016.1198134}.

\section{A Simple Ramp from Random Matrix Theory }
\label{sec:RMT}
As reviewed above, the spectral form factor $\sff(T,f)$ is the expectation over disorder of the square of the magnitude of $\rho_f(T)$. Like any expected value of a square, it has two parts: a square of an expected value and a variance. For $T$ small compared to the width of the distribution of $\rho_f$, the square of the expected value dominates and we have the slope portion of the SFF. When the variance part dominates, we have the ramp and plateau. We focus on the variance part by subtracting the disconnected part of the SFF. We thus consider
\begin{equation}
\sff(T,f)_{\text{conn}}=\int dE_1 dE_2 \,\text{cov} \left(\rho(E_1), \rho(E_2)\right) f(E_1)e^{iTE_1}f(E_2)e^{-iTE_2}
\label{eq:rampPlateau} 
\end{equation}

It is a classic result of the random matrix theory \cite{doi:10.1063/1.1703775} of GUE matrices that far from the edges of the spectrum, where the average level density is given by $\bar \rho$, the connected two point function of density is given by
\begin{equation}
    R(E_1,E_2)=\langle \rho(E_1)\rho(E_2)\rangle -\bar \rho^2=\left(\frac{\sin(\pi\bar \rho(E_2-E_1))}{\pi(E_2-E_1)}\right)^2-\bar \rho \delta(E_2-E_1)
\end{equation}
This quantity can be Fourier transformed to get the connected SFF contribution.
\begin{equation}
    \sff_{conn}(T,f)=\int dE_1dE_2 R(E_1,E_2)f(E_1)f(E_2)e^{iT(E_1-E_2)}
\end{equation}
Assuming both $\bar \rho$ and $f$ vary much more slowly than $e^{iTE}$, we have
\begin{equation}
\begin{split}
    \sff_{\text{conn}}(T,f)=\int dE f^2(E) F_{\text{GUE}}(T,\bar \rho),\\
    F_{\text{GUE}}(T,\bar \rho)=
    \begin{cases} 
      \frac{T}{2\pi} & 0\leq T\leq 2\pi \bar \rho, \\
      \bar \rho & 2\pi \bar\rho\leq T.
   \end{cases}
\end{split}
\end{equation}
There are analogous expressions for GOE and GSE matrices. All have the properties inherited from a coulomb gas that for $T\ll\bar \rho$, $F_{GXE}(T,\bar\rho)\approx \frac{T}{\pi \bbeta}$, and for $T\gg\bar \rho$, $F_{GXE}(T,\bar\rho)\approx \bar \rho$.
 Since $\rho$ is exponential in system size, the variance is proportional to $T$ for a very wide range of times. This is the famous ramp found in both random matrix theory and a plethora of chaotic systems. Setting $f=1$ and assuming $T\ll \rho$, the infinite temperature SFF ramp depends only on the spectral width~\cite{Stanford2018}, whether the random matrix ensemble is Gaussian or has a more exotic potential. With a general filter function, one obtains
\begin{equation}
\sff(T,f)_{\text{ramp}}= \int dE  f^2(E) \frac{T}{\pi \bbeta}.
\label{eq:bigFResult}
\end{equation}
An important commend about equation \eqref{eq:bigFResult} is that the slope doesn't depend on details of the Hamiltonian except the bounds of the spectrum. If we choose an $f$ which is extremely small or zero near the bounds of the spectrum (for more on such $f$s see section \ref{sec:filtering}), the prediction is that the disorder-averaged ramp for chaotic systems is actually invariant under any perturbation.

\begin{figure}
\centering
\includegraphics[scale=0.6]{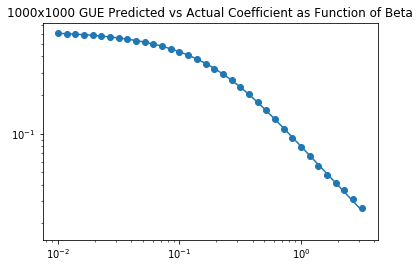}
\caption{This log-log plot shows the predicted results (line) and numerical results (dots) for a GUE matrix with ground state energy shifted to $0$ and maximum energy $4$. For $\beta \ll 1$, the $\beta$ dependence is slow, but as $\beta$ becomes larger the coefficient has an inverse relationship with $\beta$.}
\label{fig:GUEgraph}
\end{figure}

If we consider the SFF at inverse temperature $\beta$, the answer is given by the variance of $\rho_\beta(T)=\int dE e^{-\beta E}e^{-iTE}\delta \rho$. For $\beta \ll T \ll \rho$ this is just going to be the result obtained by plugging $f(E)=\exp(-\beta E)$ into equation \eqref{eq:bigFResult}.

\begin{equation}
\sff(T,\beta)_{\text{ramp}}=\int dE  \exp(-2\beta E) \frac{T}{\pi \bbeta}= 
\frac{T}{2 \pi \beta\bbeta}(e^{-2\beta E_{\min}}-e^{-2\beta E_{\max}})
\label{eq:bigTResult}
\end{equation}
Figure~\ref{fig:GUEgraph} compares the numerically extracted ramp coefficient to the formula derived above for a GUE ensemble with ground state energy shifted to zero. 

\subsection{Filtering and the Microcanonical SFF}
\label{sec:filtering}
\begin{figure}
\centering
\includegraphics[scale=0.6]{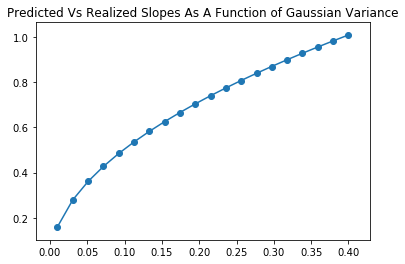}
\caption{Equation \eqref{eq:filterRamp} compared to numerics for a Gaussian filter on a GUE ensemble.}
\label{fig:filter}
\end{figure}

So far we have largely focused on the case $f(E)=\exp(-\beta E)$ or $f(E)=\exp(-\beta (E-E_{\min}))$. Another very useful choice is $f(E)=\exp(\frac{(E-\bar E)^2}{4 \sigma^2})$. This allows us to investigate the contribution to the SFF from energies within a small window of width $\sigma$ centered on $\bar E$. Using equation \eqref{eq:bigFResult} we have a coefficient of
\begin{equation}
\sff=\int dE \frac{T}{\pi \bbeta}\exp(\frac{(E-\bar E)^2}{2 \sigma^2})=\frac{\sqrt{2\pi}\sigma T}{\pi \bbeta}
\label{eq:filterRamp}
\end{equation} 
whenever $\bar E$ is well within $(E_{\min},E_{max})$. Figure \ref{fig:filter} shows the match between theory and numerics.

\begin{figure}
\centering
\includegraphics[scale=0.5]{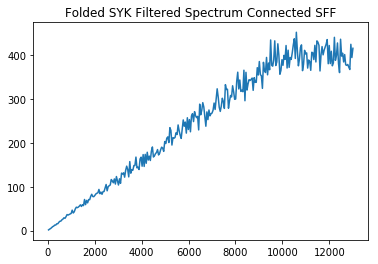}
\includegraphics[scale=0.5]{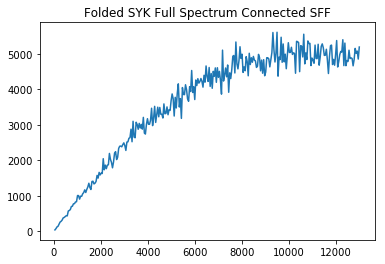}
\caption{A picture of the connected SFF of a stretched GUE matrix. The first picture is filtered to the energy range near $0$, the second is unfiltered. Notice the clearer transition near the Heisenberg time in the first picture.}
\label{fig:cleanHeisenberg}
\end{figure}

This microcanonical SFF has a number of uses. For instance, in systems with something other than uniformly chaotic behavior, it allows us to 'scan' the SFF for transitions to some other phase. Also, for systems with a wide range of Heisenberg times, it allows us to zoom in on a particular range of the spectrum and get a clearer picture of the ramp-plateau transition. To illustrate this, figure \ref{fig:cleanHeisenberg} shows the transition for a GUE ensemble stretched to $f(H)=H + H^3/10$, enough to display a wide range of Heisenberg times. For thermodynamic systems, the density of states, and thus the Heisenberg time, can vary by many orders of magnitude throughout the spectrum, and it can be even more important to filter.

\subsection{Ramp Coefficient For SYK}
\label{sec:SYK}
In the case of the SYK model~\cite{Maldacena_2016,Rosenhaus_2019,kobrin2020manybody,Stanford2018}, one can analytically obtain the same result as equation \eqref{eq:bigFResult}. As a reminder, the SYK is a disordered 0+1d system made of Majorana fermions with $q$-body interactions ($q$ is even). The SYK Hamiltonian is given by
\begin{equation}
    H[\psi] = i^{q/2}\sum_{1 \leq j_1 <...< j_q \leq N} J_{j_1j_2...j_q}\psi^{j_1}\psi^{j_2}...\psi^{j_q},
    \label{eq:SYKHamiltonian2}
\end{equation}
where $\psi^i, i=1,...,N$ represents the Majorana fermions and satisfy the anticommutation relation $\{\psi^i,\psi^j\}=\delta_{ij}$, and each $J_{j_1...j_q}$ is a Gaussian variable with mean zero and variance $\langle J_{j_1...j_q}^2 \rangle=\frac{J^2 (q-1)!}{N^{q-1}}$. 

It is often convenient to perform a series of exact manipulations on Hamiltonian \eqref{eq:SYKHamiltonian2} to get a mean-field Lagrangian description of the SYK model in terms of bilocal variables consisting of a Green's function $G$ and self-energy $\Sigma$. In particular, one can write an expression for the imaginary temperature partition function of the SYK Model as
\begin{equation}
Z(iT)=\int DG D\Sigma \exp N \Big[ \frac12 \Tr \log( \partial_t - i\Sigma)\\
    + \frac 12\int dt_1 dt_2 (i \Sigma(t_1,t_2) G(t_1,t_2) - \frac{J^2 }{q} G^q(t_1,t_2) ) \Big].
\end{equation}
The SFF can be thought of as a partition function of a doubled system living on two contours, with one contour running forward in time (corresponding to $e^{-i H T}$ in the SFF) and one contour running backward in time (corresponding to $e^{i H T}$ in the SFF). Generalizing the result for $Z(iT)$, one can write the SFF as
\begin{equation}
\begin{split}
   \sff(T,\beta=0) = \int DG D\Sigma\exp N \Big[ \frac12 \Tr \log( \partial_t - i \hat\Sigma)\\
    + \frac{1}2 \sum_{\alpha,\beta=1,2}\int dt_1 dt_2 (i \Sigma_{\alpha \beta} G_{\alpha\beta} - \frac{J^2 }{q}  (-1)^{\alpha+\beta} G_{\alpha\beta}^q ) \Big] , 
\label{eq:Action}
\end{split}
\end{equation}
where a hat above a variable signals a matrix representation, $(\hat \Sigma)_{\alpha\beta} \equiv \Sigma_{\alpha\beta}$. Because of the antiperiodic boundary conditions on the fermions, both $G_{\alpha \beta}$ and $\Sigma_{\alpha \beta}$ are antiperiodic under time shifts by $T$. Note also that the measures $DG$ and $D\Sigma$ each integrate over the space of two-index functions of two variables.

The authors of \cite{Stanford2018} study the SFF of the SYK model for all $q$ and show that the ramp comes from a family of semiclassical solutions.
At intermediate times, the path integral \eqref{eq:Action} is dominated by ``wormhole'' solutions derived from a thermofield double (TFD) solution. One takes $G$ and $\Sigma$ on the two contours just as they'd be in the bulk of a solution on a Schwinger-Keldysh contour for temperature $\beta_{aux}$. For any choice of $\beta_{\text{aux}}$, this is a saddle point of \ref{eq:Action}, up to exponentially small error. In particular, $\beta_{\text{aux}}$ can take any value, unrelated to the externally applied $\beta$. It is often convenient to replace $\beta_{\text{aux}}$ with the related parameter $E_{\text{aux}}$, where $E_{\text{aux}}$ is the energy of one copy of the SYK system at temperature $\beta_{\text{aux}}$. $E_{\text{aux}}$ ranges from $E_{\min}$ to $E_{\max}=-E_{\min}$

The other number parameterizing saddle points of equation \eqref{eq:Action} is $\Delta$, a relative time shift between the two contours. Because there are nonzero correlations between the two legs and both contours have time-translation symmetry, one can choose any point $t=\Delta$ on contour $2$ to line up with $t=0$ on contour 1. Because of the antiperiodic boundary conditions, the manifold of all possible $\Delta$s is a circle of circumference $2T$. The authors show the measure along this saddle manifold is $dE d\Delta/2\pi$. There is also a hidden symmetry, $Z_2$ for $q\equiv 2\textrm{ mod 4}$ and $Z_4$ for $q\equiv 0\textrm{ mod 4}$, that multiplies the number of saddle points by 
\begin{equation}
m_q=\begin{cases} 
      2 & q\equiv 2\textrm{ mod 4} \\
      4 & q\equiv 0\textrm{ mod 4}.
   \end{cases}
\end{equation}
Thus, \cite{Stanford2018} finds an infinite temperature ramp
\begin{equation}
\sff(T,\beta=0)=m_q\int_{0}^{2T} d\Delta \int_{E_{\min}}^{E_{\max}}dE_{\text{aux}} \frac{1}{2\pi}
\label{eq:SSSresult}
\end{equation}
The authors of \cite{Stanford2018} go through the $q\equiv 0\textrm{ mod 4}$ and $q\equiv 2\textrm{ mod 4}$ cases separately, and deal with varying $\bbeta$s and degeneracies for the corresponding matrix ensembles to show that \eqref{eq:SSSresult} gives the same answer as RMT for all $q$ and $N$.
  
\begin{figure}
\centering
\includegraphics[scale=0.6]{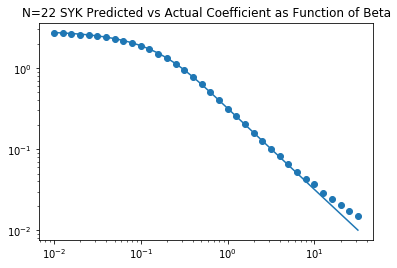}
\caption{A comparison of numerically extracted ramp coefficients versus the predictions of \eqref{eq:bigTResult} for the $N=22$ SYK model at various $\beta$s.}
\label{fig:SYKfigure}
\end{figure}

If we introduce an $f^2(E)$ insertion, then the path integral \eqref{eq:Action} is replaced with
\begin{equation}
\begin{split}
    \sff(T,f) = \int DG D\Sigma f(E_1)f(E_2)\exp N \Big[ \frac12 \Tr \log( \partial_t - i \hat\Sigma)\\
    + \frac{1}2 \sum_{\alpha,\beta=1,2}\int dt_1 dt_2 (i \Sigma_{\alpha \beta} G_{\alpha\beta} - \frac{J^2 }{q}  (-1)^{\alpha+\beta} G_{\alpha\beta}^q ) \Big] , 
\label{eq:Action2}
\end{split}
\end{equation}
where $E_1$ and $E_2$ are the energies of the field configuration on contours 1 and 2, respectively. The modified path integral is still dominated by the old saddle point manifold. However, only a few points on this manifold are still saddles, namely those where $f'(E_{\text{aux}})$ is zero. That being said, when we integrate we get the same $2 T m_q \int dE f^2(E)/(2 \pi)$. Figure \ref{fig:SYKfigure} shows the comparison between this analytical result and the ramp coefficient extracted from exact diagonalization for $q=4$, $N=22$.

\subsection{Other Models}
\label{sec:OtherModels}
It is straightforward to study the ramp coefficients of a wide variety of other models using exact diagonalization. We would like to put forward the explicit claim that in all systems with hydrodynamic behavior, at times large enough to equilibrate (which is the Thouless time) but less than the Heisenberg time, the SFF is a linear ramp with coefficient given by the pure RMT prediction \eqref{eq:bigFResult} or its generalizations to be discussed in the next section. The simplest justification for this is that the argument in \ref{app:SSS} ultimately requires nothing but a hydrodynamic description and that $T$ be large enough to forget everything not conserved. We also show in section \ref{sec:Hydro} that microcanonical ramp coefficients are invariant under small deformations in hydrodynamics once the Thouless time has been reached.

The first model we will consider will be a random all-to-all spin model analogous to the Sherrington-Kirkpatrick model~\cite{PhysRevLett.35.1792,Panchenko_2012}, which can be thought of as the SYK model with fermions replaced by spins. The Hamiltonian can be written as
\begin{equation}
H=\sum_{1\leq j_1\leq...\leq j_q<N}^{i_1,...,i_q=1,2,3} J_{j_1,...,j_q}^{i_1,...,i_q}\sigma_{i_1}^{j_1}...\sigma_{i_q}^{j_q}
\end{equation}
Often the choice of $q=2$ is made, as in our numerical analysis. The SK model is different from the SYK model in that it forms a spin glass near the top and bottom of the spectrum.
\begin{figure}
\centering
\includegraphics[scale=0.5]{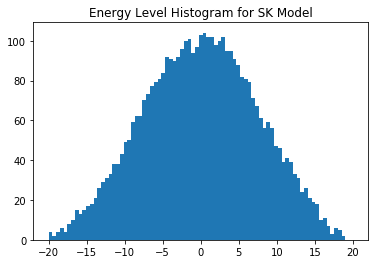}
\includegraphics[scale=0.5]{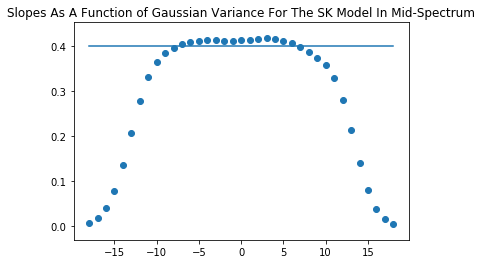}

\caption{Energy level density and a plot comparing theoretical versus realized slopes. The SK model seems to obey RMT near the center of the spectrum, but not further out. In particular, the 'ramp' developes a substantial y-intercept as its slope decreases, something pure RMT can't explain.}
\label{fig:skgraphs}
\end{figure}
This spin glass isn't fully understood, especially at comparatively small $N$. But it is clear from figure \ref{fig:skgraphs} that RMT breaks down when we aren't near the center of the spectrum.
\begin{figure}
\centering
\includegraphics[scale=0.5]{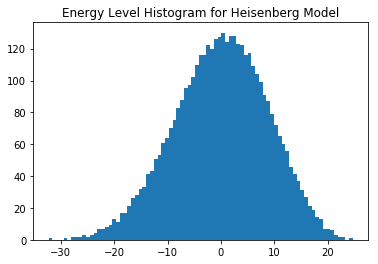}
\includegraphics[scale=0.5]{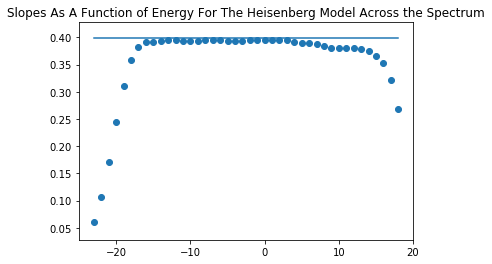}
\caption{Energy level density and a plot comparing theoretical versus realized slopes in the Heisenberg model. Near the center of the spectrum, the match with pure RMT is good, but there are visible deviations around $E=10$.}
\label{fig:heisgraphs}
\end{figure}

We can also consider a disordered Heisenberg model with disorder on both the bond strength and the field~\cite{Gubin_2012,phdthesis}. The Hamiltonian is
\begin{equation}
H=\sum_{j=1}^L \sum_{i=1,2,3} J_{ij}\sigma_j^i\sigma_{j+1}^i+F_{ij}\sigma_j^i
\end{equation}
Where each $J_{ij}$ is drawn from a $N(1,0.2)$ and $F_{ij}$ is drawn from $N(0,1)$. The parameters were chosen to break all symmetries except time translation, and to ensure that the system isn't integrable. For larger field strengths, or low temperatures, the Heisenberg model is in an MBL phase and no ramp is present~\cite{phdthesis}. We used a spin chain of length $L=12$ for our numerical analysis, which is shown in figure \ref{fig:heisgraphs}.

\section{Block Hamiltonians}
\label{sec:Conserve}

In this section, we discuss how the filtered SFF is modified when the Hamiltonian has a block structure such that it breaks up into disconnected pieces. Such a structure can arise, for example, due to symmetries or due to an imposed folding of the spectrum. We briefly consider both cases here.

\subsection{Random matrices with unitary symmetries}

One can impose additional conservation laws such as a $U(1)$ charge conservation on random matrix theory. In doing so, we break the Hamiltonian into different `sectors' labelled by their charge. There is eigenvalue repulsion within each sector, but no repulsion for eigenvalues in different sectors. This means that the eigenvalue densities in the different sectors are essentially uncorrelated. This, in turn, implies that the variances in eigenvalue densities, and thus the ramps, simply add together. If the total $U(1)$ charge is denoted $Q$, then we have
\begin{equation}
\sff(T)=\sum_Q \int dE \frac{ f^2(E,Q)}{\pi\bbeta}T.
\label{eq:sumSector}
\end{equation} 
We can also include conserved charge in the filter function, for example, taking $f=\exp(-\beta(E-\mu Q))$ for some chemical potential $\mu$.

\subsection{Charged SYK}

The above formula can be analytically obtained in the charged SYK model~\cite{Davison_2017,Gaikwad_2020,Sorokhaibam_2020}, with Hamiltonian given by
\begin{equation}
    H[\psi] = i^{q/2}\sum_{1 \leq j_1 <...< j_q/2 \leq N/2,1 \leq j_{q/2+1} <...< j_q \leq N/2} J_{j_1j_2...j_q}\psi^{j_1}\psi^{j_2}\psi^{j_{q/2}}\bar\psi^{j_{q/2+1}}...\bar\psi^{j_q},
    \label{eq:ChagedSYKHamiltonian}
\end{equation}
This Hamiltonian has a symmetry where the first $N/2$ fermions have charge $-1$ and the last $N/2$ fermions have charge +1. It displays very similar physics to the SYK model, including a holographic dual and maximal chaos.

\begin{figure}
\includegraphics[scale=0.5]{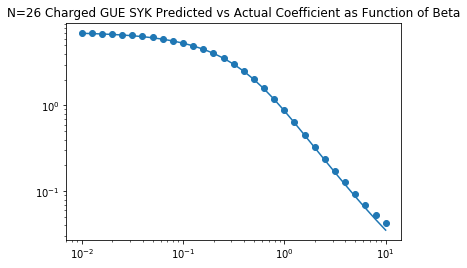} 
\includegraphics[scale=0.5]{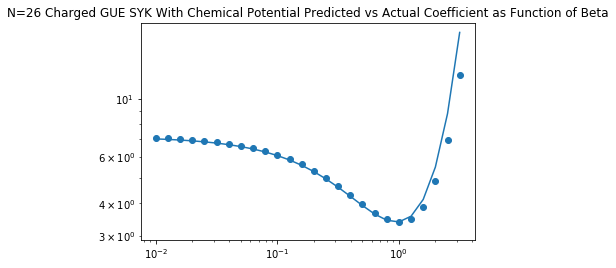}
\caption{Plots showing predicted versus realized slopes for $f=\exp(-\beta(E-\mu q))$ for $\mu=0$ and $\mu=0.1$.}
\label{fig:chargeSYK}
\end{figure}

As in the conventional SYK model, we can use logic basically identical to that of \cite{Stanford2018} to derive the RMT result with a semiclassical analysis. The main difference is that solutions are parameterized by a charge $Q_{\text{aux}}$ and a potential difference $\Delta_{\mu}$, in addition to $E_{\text{aux}}$ and a time difference $\Delta$. Instead of integrating over just $dE_{\text{aux}}d\Delta/2\pi$, the measure also has a factor of $dQ_{\text{aux}}d \Delta_{\mu}/2\pi$. If the quantum of charge is $q$, then the range of integration for $\Delta_\mu$ is $2\pi/q$ (assuming all charges are multiples of $q$, any gauge transformation with phase $2\pi/q$ is the identity). Integration over the saddle point manifold then gives the same result as equation \eqref{eq:sumSector} with $\bbeta=2$ (GUE). Figure \ref{fig:chargeSYK} shows two plots comparing the predicted and empirical values for filter functions $f=\exp(-\beta E)$ and $f=\exp(-\beta(E-\mu Q))$.

\subsection{Folded Spectra}
\label{sec:folded}

We will now examine the case of stretched and folded spectra. To start with, consider a random Hamiltonian $H'=\fstretch(H)$, where $H$ is a random matrix chosen from distribution \eqref{eq:RMTpdf} and $\fstretch$ is a smooth function with everywhere positive derivative. The quantum mechanics of such deformations have been considered recently in \cite{Gross_2020}. Another motivation to study folded spectra comes from the eigenstate thermalization hypothesis (ETH). ETH asserts that any local observable $ O$ can be written as a sum of a smooth function of energy $f_{O}(H)$ (related to the microcanonical expectation value) and a random-like erratic part $R$~\cite{PhysRevE.50.888,PhysRevA.43.2046}. Under this hypothesis, the SFF of a Hamiltonian perturbed by a local operator is then equivalent to the SFF of a stretched spectrum plus a random matrix, $H' = H + \epsilon O \sim H + \epsilon f_O(H) + R$.

Studies of the SFFs of folded systems are common~\cite{Gharibyan_2018,Gubin_2012,Vahedi_2016,phdthesis}. One reason is that a folding procedure (often called `unfolding') can be used to get semicircle statistics out of other level distributions in order to more easily compare numerical results with RMT. In this section we show analytically that non-singular folds indeed leave ramps invariant. For a comparison of folding versus filters as a way to look at parts of the spectrum see \cite{Nosaka_2018}.

Returning to $H' =\fstretch(H)$, there is generically no $V'$ such that $H'$ is distributed according to \eqref{eq:RMTpdf}. Rather, the pdf for $H'$ is given by
\begin{equation}dP=\frac 1 {\mathcal Z}\prod_{i<j} |\fstretch^{-1}(\lambda_i)-\fstretch^{-1}(\lambda_j)|^{\boldsymbol \beta} \prod_i e^{-V(\fstretch^{-1}(\lambda_i))+\log \fstretch^{-1'}(\lambda_i)}.
\label{eq:RMTpdfMod}
\end{equation}
Nonetheless, the spectral statistics of $H'$ are very similar to those given by \eqref{eq:RMTpdf}. 
This is because nearby eigenvalues still repel with repulsion term $(\fstretch^{-1}(E'_1)-\fstretch^{-1}(E'_2))^\bbeta$ which is roughly proportional to $(E'_1-E'_2)^\bbeta$. As such, the ramp still exists with coefficient given by \eqref{eq:bigFResult}. 

Another way to see this is to consider a Gaussian filter function. If the variance $\sigma$ in the filter function is small compared to the scale of variation in $f_{\text{stretch}}$\footnote{For example, if $f_{\text{stretch}}$ is a slowly varying function of the energy density.}, then for the small window around $\bar E$, the stretching simply rescales all the differences between eigenvalues by $f_{\text{stretch}}'(\bar E)$, which is a trivial change. The effect on SFFs with broader filter functions can be obtained by integrating over $\bar E$.

\begin{figure}
\begin{tabular}{cc}
\includegraphics[scale=0.5]{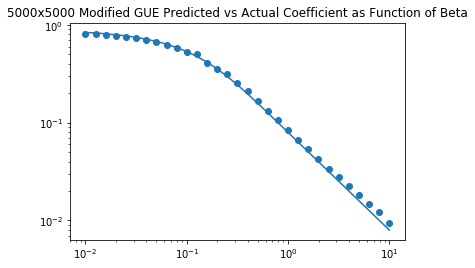}&\includegraphics[scale=0.5]{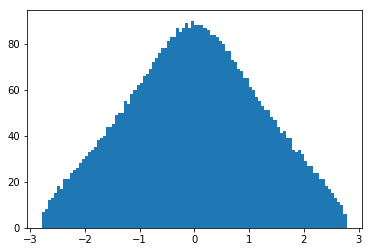}\\
\includegraphics[scale=0.5]{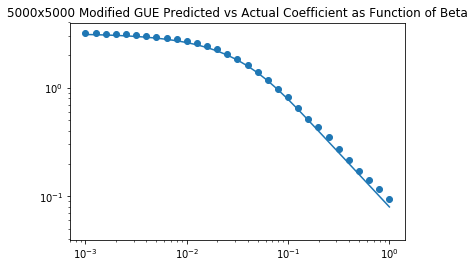}&\includegraphics[scale=0.5]{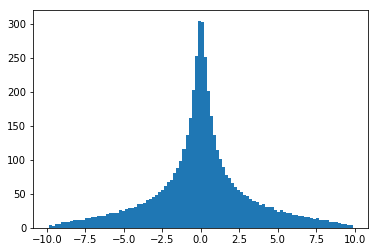}
\end{tabular}
\caption{Ramp coefficient plots and spectral densities for $\fstretch(E)=E+0.1E^3$ and $\fstretch(E)=E+E^3$.}
\label{fig:stretch}
\end{figure}

Figure \ref{fig:stretch} shows coefficient plots of  5000 by 5000 GUE matrices after transformations $\fstretch(E)=E+0.1E^3$ and $\fstretch(E)=E+E^3$, accompanied by histograms of their spectral density

The next natural is question to ask is what happens when we choose a function $\fstretch$ which doubles back on itself, for instance $\fstretch(E)=E-E^3$. In these cases we can have multiple `species' of eigenvalues near $E'$, corresponding to which branch of $\fstretch^{-1}$ the original $E$ lies on. There is almost no repulsion between different species of eigenvalue, so the ramp part of the SFF is given by
\begin{equation}
\int dE f^2(E) \frac T {\pi\bbeta} (\# \textrm{ of species at }E)
\label{eq:species}
\end{equation}

\begin{figure}
\begin{tabular}{cc}
\includegraphics[scale=0.5]{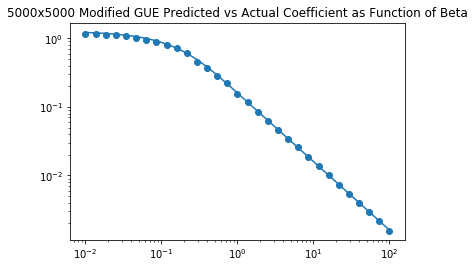}&\includegraphics[scale=0.5]{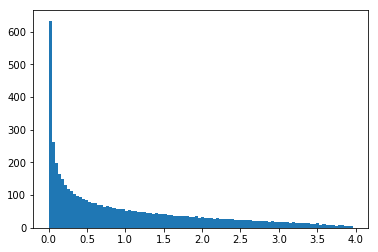}\\
\includegraphics[scale=0.5]{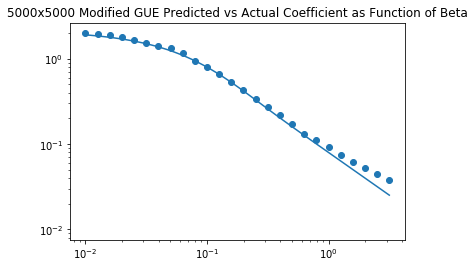}&\includegraphics[scale=0.5]{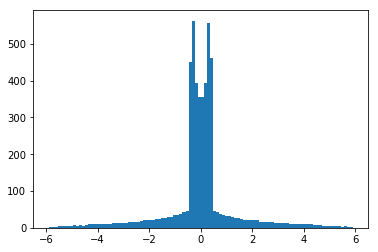}
\end{tabular}
\caption{Ramp coefficient plots and spectral densities for $\fstretch(E)=E^2$ and $\fstretch(E)=E-E^3$.}
\label{fig:stretchSpecies}
\end{figure}

Figure \ref{fig:stretchSpecies} shows the matching between \eqref{eq:species} and numerical experiment for $\fstretch(E)=E^2$ and $\fstretch(E)=E-E^3$.
One question that remains open is what the behavior is like near the turning points, characterized by $\frac{d}{dE}\fstretch(E)=0$, where the repulsion term becomes singular. Might there be a strong enough contribution to change the overall behavior? 

\section{Nearly Block Hamiltonians}
\label{sec:nearBlock}
Having developed the theory for SFFs with conserved quantities or decoupled sectors, it is time to turn our attention to SFFs for systems with one or more almost-conserved quantity. Suppose the Hamiltonian decomposes into two pieces, $H=H_0+V$, such that $H_0$ breaks into $\Omega_0$ decoupled blocks and $V$ causes transitions between the blocks. Suppose the $V$-induced transitions are slow, so that the $H_0$ blocks are random matrix like.

To compute $Z(iT) = \tr(e^{-i H T})$, we want to sum over all return amplitudes. Consider a basis for the Hilbert space labelled by the pair $(\alpha,i)$ where $\alpha$ denotes the block and $i$ indicates a basis vector within a block. Given an initial state $(\alpha,i)$, write its time development as
\begin{equation}
    |\psi_{(\alpha,i)}(T) \rangle = \sum_{\beta=1}^{\Omega_0} \sqrt{p_{\alpha\rightarrow \beta}(T)} |\phi_{\beta,(\alpha,i)}(T) \rangle,
\end{equation}
where $p_{\alpha\rightarrow \beta}(T)$ is the probability to transition to sector $\beta$ after starting in sector $\alpha$ (assumed to be independent of the within-sector label $i$) and $|\phi_{\beta,(\alpha,i)}(T)\rangle$ is the normalized state in sector $\beta$ originating from $\psi_{(\alpha,i)}$. The return amplitude is
\begin{equation}
    \langle \psi_{(\alpha,i)}(0) | \psi_{(\alpha,i)}(T)\rangle = \sqrt{p_{\alpha\rightarrow \alpha}(T)} \langle \psi_{(\alpha,i)}(0) | \phi_{\alpha,(\alpha,i)}(T)\rangle.
\end{equation}

The SFF is assembled by summing these amplitudes, taking the squared magnitude, and then averaging. Now, since the dynamics within each sector is random matrix like at the timescales of interest, the diagonal terms should reduce to the within-sector SFF and the off-diagonal terms should be small, 
\begin{equation}
 \sum_{i,j} \overline{\langle \psi_{(\alpha,i)}(0) | \phi_{\alpha,(\alpha,i)}(T)\rangle \langle \psi_{(\beta,j)}(0) | \phi_{\beta,(\beta,j)}(T)\rangle^*} = \delta_{\alpha,\beta} \sff_\alpha(T).
\end{equation}
Hence, the filtered SFF reduces to
\begin{equation}
    \sff(T,f) = \sum_{\alpha} f(E_\alpha)^2 p_{\alpha\rightarrow \alpha}(t) \sff_\alpha(t).
\end{equation}
When $\sff_\alpha$ is just a linear ramp with a known coefficient, the evaluation of the $\sff$ reduces to summing over the return probabilities. 

To understand the return probabilities in more detail and introduce a useful rate-matrix formalism, consider the instructive example of a particle stuck in one of $k$ potential wells, in a kinematic space complicated enough that the Hamiltonian within each well is well-approximated by a random matrix. The single almost-conserved quantity is an index ranging from one to $k$.

\begin{figure}
\centering
\includegraphics[scale=0.6]{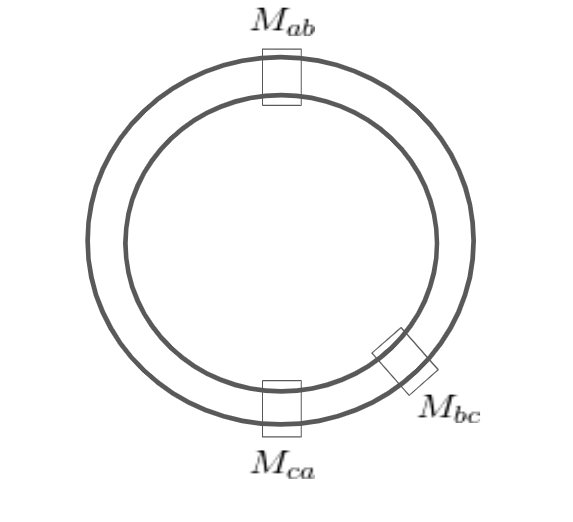}
\caption{The SFF is calculated on a doubled contour for the system. In this configuration, there are three instantons, one taking from well $a$ to well $b$, one shortly after going from $b$ to $c$, then eventually one taking the system from $c$ back to $a$. In between wells the system is well-described by the dynamics within a single sector. }
\label{fig:circleDiagram}
\end{figure}

We can solve this using doubled-system wormhole techniques like those in \cite{Stanford2018}, reviewed in appendix \ref{app:SSS}. Lets introduce some collective variables to denote a particle's state within a well, as well as the discrete variable $i$ denoting which well the particle is stuck in. The simplest solutions to the equations of motion in a doubled system are ones where $i$ is constant over the entire doubled contour.

There are also tunneling events which take the system from well to well, and we can put all their amplitudes into a transition rate matrix $M(E)$ ($M(E)$ also has elements on the diagonals to make sure probability is conserved). Because these tunneling events happen on a doubled system, their amplitudes have natural interpretations as probabilities for a single copy of the system. An illustration of one path which contributes to the path integral is given in figure \ref{fig:circleDiagram}. Note that $M$ is not a Hermitian matrix. It has all negative eigenvalues, except for one zero eigenvalue whose left eigenvector is $(1,1,1...)$ corresponding to conservation of probability). 

To get from the transition matrix to the SFF, the key point is that the same instanton gas that gives us the probability of transfer also shows up in a wormhole-like path integral calculation of the SFF. We start with out in thermofield double (TFD) for the various approximately disconnected sectors of the Hamiltonian. At each timestep from $t$ to $t+dt$, there is some amplitude (probability from the point of view of a single copy of the system) that the system will go from sector $i$ to sector $j$. This is just $M_{ij}dt$. Multiplying over all timesteps, and requiring that the doubled system start and end in the same sector gives 
\begin{equation}
\textrm{Factor from approximate symmetries}=\tr \prod_1^{T/dt} (I+Mdt) =\tr e^{MT}
\end{equation}
This means that the ramp is given by
\begin{equation}\sff=\int dE \frac{T}{\pi \bbeta} f^2(E) \tr \exp(M(E)T),
\label{eq:bigAnswer}
\end{equation} 
where the $T$ in front still comes from an overall displacement of one side relative to the other. The coefficient of the ramp in \eqref{eq:bigAnswer} starts out as $k$ for $k$ wells and goes down to $1$ at long time. It is also worth noting that if there are truly conserved quantities, formula \ref{eq:bigAnswer} will still give correct results. For instance if there are two blocks, each of which has two approximate subblocks of the same size, at long times the return probabilities will be $1/2$, and the factor will be $4 \times \frac 12 =2$, which is what we would get if we didn't know about the approximate subblock structure at all. 

It is also interesting to point out that \eqref{eq:bigAnswer} can be thought of as a more precise version of the claim that one gets the pure RMT result once enough time has passed for a state to explore all of Hilbert space \cite{Schiulaz_2019}.

\begin{figure}
\includegraphics[scale=0.6]{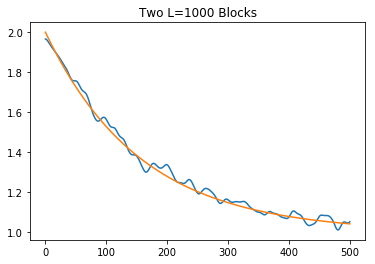}
\includegraphics[scale=0.6]{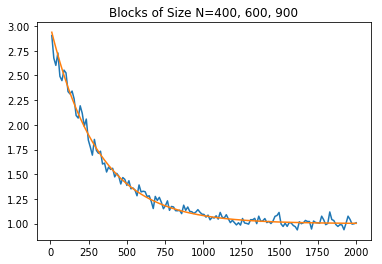}
\includegraphics[scale=0.6]{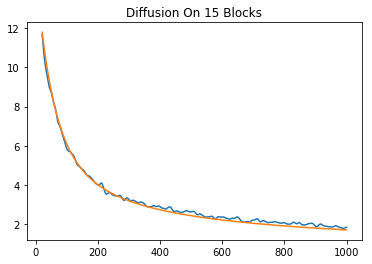}
\caption{Comparison of $\tr e^{MT}$ (orange) vs numerical realization of $\sff/\left(\int dE \frac{T}{\pi \bbeta} f^2(E)\right)$ where $f$ is chosen to be a tightly bunched Gaussian.}
\label{fig:threeGraphs}
\end{figure}

To illustrate the working of formula \eqref{eq:bigAnswer}, suppose we have a random Hermitian matrix of the following form: a $2N\times 2N$ complex symmetric matrix, decomposed into $N\times N$ blocks, with elements of size $J^2$ on the diagonal blocks and $k^2J^2$ on the off-diagonal blocks. We can use Fermi's Golden Rule to get transition rates: $\bra{i}H\ket{f}^2$ is just $k^2J^2$ and the density of states to transition to is $\rho(E)=\frac{2\sqrt{NJ^2-E^2}}{2\pi J^2}$. So the overall rate is $k^2{2\sqrt{NJ^2-E^2}}$. Figure \ref{fig:threeGraphs} shows three increasingly complicated scenarios. In the first one, there are two blocks connected with $k=0.04$. In the second, there are three blocks of different sizes. In the third, a chain of blocks where only neighboring blocks are connected. This is analogous to a particle slowly diffusing, where its position is approximately conserved. In each graph, we show the realized ratio of the connected SFF to the predicted single block SFF, and also $\tr e^{M(E)T}$.

\section{Hydrodynamics}
\label{sec:Hydro}

As the theory of a system's slow modes, hydrodynamics provides a natural framework in which to evaluate the return probabilities entering the general formula in equation \eqref{eq:bigAnswer}. For simplicity, we focus on the case of energy diffusion here. This theory is interesting not only as a simple test case, but because it is very generic: any spatially local Hamiltonian system which thermalizes and which does not have additional conserved quantities (the generic case, e.g. due to disorder breaking translation symmetry) is described by this theory at long time/distance. We also want to point out that the hydrodynamic formulation is attractive, since it connects observable manifestations of chaos, like thermalization, to a more fine grained characterization of quantum chaos in terms of energy level statistics.

At a given time $T$, such a hydrodynamic system has an extensive set of approximate conservation laws. For linear diffusion, the amplitude of a long-wavelength energy fluctuation with wavevector $k$ decays at rate $D k^2$, where $D$ is the energy diffusion constant. In this case, all modes with wavevector less than $k_T \sim (DT)^{-1/2}$ have not appreciably decayed. In spatial dimension $d$, the number of such modes is 
\begin{equation}
    N_T \sim \sum_k \theta(k_T - |k|) \sim V \int \frac{d^d k}{(2\pi)^d} = \frac{V S_d}{(2\pi)^d}\frac{k_T^d}{d},
\end{equation}
which is extensive in the system size $V$. Hence, in this case we can label the nearly decoupled sectors by the amplitudes of energy fluctuations with wavevector less than $k_T$. Hence, we are exactly in the situation considered in section \ref{sec:nearBlock}.

This is a problem in fluctuating and dissipative hydrodynamics~\cite{Kovtun_2012,Dubovsky_2012,Grozdanov_2015,Endlich_2013}. The particular toolset we use is a modification of the the closed time path (CTP) formalism~\cite{crossley2017effective,glorioso2018lectures}, which itself a special case of the Schwinger-Keldysh formalism\cite{Haehl_2017,Kamenev_2009,CHOU19851}. We include a lightning review of this formalism in appendix \ref{app:CTP}, and it is described in great detail in the references. In brief, this is a theory that formulates hydrodynamics in terms of an effective field theory of conserved quantities on a Schwinger-Keldysh contour. The CTP action can also often be written in terms of Langevin-like stochastic differential equations. The two contours correspond to forward, $e^{-i H T}$, and backward, $e^{i H T}$, time evolution, however the boundary conditions in our case are different owing to the separate traces in the definition of the SFF.

There are two ways of looking at the role of the CTP formalism in terms of SFFs. One is in terms of the return probability picture in equation \eqref{eq:bigAnswer}. Considering again the example of energy diffusion, we can say that
\begin{equation}
    \tr e^{MT} = \int \mathcal D \epsilon(x,t=0) \Pr(\epsilon(x,t=T)=\epsilon(x,t=0)),
\end{equation}
where $\epsilon(x,t)$ is the energy density at position $x$ and time $t$ and we exclude the spatial zero mode. This can be converted into a path integral over all periodic histories, 
\begin{equation}
    \tr e^{MT} \propto \int \mathcal D \epsilon(x,t)\mathcal D\phi_a(x,t) e^{iS[\epsilon,\phi_a]},
    \label{eq:returnCTP}
\end{equation}
where $\phi_a$ is the anti-symmetric counterpart of $\epsilon$ in the CTP formalism. Note that $\epsilon$ is an $r$-type variable, meaning symmetric between the two contours, while $\phi_a$ is an $a$-type variable, meaning antisymmetric between the two contours. Another way to look at equation \eqref{eq:returnCTP} is to view it a path integral which will have wormhole-like solutions as in \cite{Stanford2018} (see appendix \ref{app:SSS}). In particular, it is a path integral over two contours going in opposite directions and it focuses on a set of states that (locally) look like an equilibrium thermal state. We call this periodic time modification of the CTP formalism the doubled periodic time (DPT) formalism.

To be completely explicit, here are the assumptions underlying the following analysis of the DPT formalism. Consider a system with `bare' hydrodynamic action $S_{\text{hydro}} = \int d^d x dt L_{\text{hydro}}$ defined on the conventional CTP contour. By bare action we mean that we have integrated out all the fast modes, above some energy scale $\Lambda_{\text{fast}}$, but we have not integrated over any slow modes. Then we assume the following:
\begin{itemize}
    \item First, that the same bare hydro action on the CTP contour can be placed on the SFF contour by simply changing the boundary conditions in time, up to corrections of order $e^{-\Lambda_{\text{fast}}T}$. Physically, the expectation is that the fast modes cannot wrap efficiently around the thermal circle, and hence the action obtained from integrating them out is not sensitive to the boundary conditions. Note that this statement can only apply to the bare action: once we integrate out modes which can effectively wrap the time circle, then we can get new terms in the action. 
    \item Second, that the bare CTP action with SFF boundary conditions gives the dominant saddle point / phase for the connected SFF for a wide window of time. Specifically, it should be the dominant saddle after times of order $\Lambda_{\text{fast}}^{-1} \log(\text{system size})$ and before the inverse many-body level spacing time. Note that we are relying on the thermodynamic limit to evaluate the SFF by finding the dominant saddle point and computing fluctuations around it.
    \item Third, that there is some averaging over disorder which effectively connects the decoupled SFF contours and rationalizes the interactions between contours in the hydro action. Such averaging is required to make sense of the SFF as a smooth function of time, otherwise one would find an erratic time-dependence. While this disorder average is certainly required, it remains somewhat mysterious from the hydro point of view since the disorder doesn't explicitly appear in the hydro action. Note that the CTP contour already has connectivity between the contours due at least to the future boundary condition, so averaging is not required there if the observables of interest are self-averaging.
\end{itemize}
Applied to the case of energy diffusion, these assumptions yield the linear ramp at late time and recover the return probability formula. Moreover, we can treat interactions on top of the quadratic hydro theory giving linear diffusion. In essence, the technical point is that the CTP action with modified boundary conditions gives a candidate saddle point for the SFF path integral. If this saddle point dominates, then the ramp follows. In this sense, hydrodynamics implies quantum chaos in the spectral sense.

\subsection{Hydrodynamics, Wormholes, and the Thermofield Double}

Here we elaborate on the connection to the thermofield double and wormholes. As is pointed out in \cite{Stanford2018} and summarized in appendix \ref{app:SSS}, there are two significant saddle points of a path integral on the SFF contour. One is where the two circles host two decoupled saddle points. The other set derive from thermofield double solutions, which are correlated between the two contours and which exhibit a free relative time shift $\Delta$ and a free total energy. Though this phenomenon is general, in the case of holographic systems these TFD solutions also have an interpretation as wormholes. As such, they are literally the ``connected'' part of the SFF. In this context, hydrodynamics appears naturally because it can be viewed as the theory of expanding around a thermofield double solution. Moreover, it is necessary to use hydrodynamics to get a quantitative 1-loop or higher understanding of the size of these contributions.

To calculate the SFF, we need to do a saddle-point expansion for a thermofield double solution on a forward and backwards contour. In this subsection, we discuss the spatial zero modes of the hydrodynamic action. This can be viewed as a theory of zero dimensional systems (such as those with all-to-all interactions) or as the late time limit of a finite-dimensional system in finite volume. The path integral with a quadratic action, in terms of the relative time shift $\Delta$ and total energy $E_{\text{aux}}$, is
\begin{equation}
   \sff=\int \frac{\mathcal D E_{\text{aux}}\mathcal D \Delta}{2\pi}f^2(E_{\text{aux}})\exp(- i\int dt \Delta(t) \partial_t E_{aux}(t)).
\end{equation} 
The integrals over nonzero frequency modes yield delta functions which enforce energy conservation from moment to moment, while the integral over the zero frequency modes give the linear ramp:
\begin{equation}
    \frac{T}{2\pi} \int dE_{\text{aux}}f^2(E_{\text{aux}}).
\end{equation}

It is instructive to compare this answer with the traditional path integral on the CTP contour. In the CTP case, the zero frequency relative time shift is constrained to be zero due to the future boundary condition connecting the contours, but in the DPT case, this relative time shift is naturally unconstrained. Similarly, the total energy integral is weighted by a thermal factor (or the energy distribution of the initial state) in the CTP case, but it is unconstrained (apart from the imposed filter function) in the DPT case.

In the case of a time-reversal invariant Hamiltonian with GOE symmetries, an extra factor of two comes from the possibility of reversing time for one of the contours relative to another, so time $t$ on contour 1 maps to time $-t$ on contour 2. For physical Hamiltonians with GSE symmetries, these cannot be realized without the SFF picking up at least one factor of two in the numerator from degeneracies or blocks, and then we get the GUE answer.

Similar logic can be applied to higher order moments of $Z(T,f)$ with respect to the disorder average, with the assumption that the relevant saddle points are copies of the dominant DPT saddle point. There are actually two slightly different cases. In the most generic case, $Z$ is a complex number and the moments of interest are $Z^k Z^{*k}$. There are $k$ forward contours and $k$ backwards contours. Thus there are $k!$ ways to connect the forward and backwards contours into pairs. Once this is done, each one has a free $E$ and $\Delta$. Thus the $2k$-th moment of $Z$ (assuming there are no additional symmetries) is
\begin{equation}
   \overline{Z^k(T,f)Z^{*k}(T,f)}=k!\left(\frac{T}{\bbeta \pi}\int dE f^2(E)\right)^k.
\end{equation}
In another case, there is an operator $O$ which anticommutes with $H$, and the spectrum has $E\leftrightarrow -E$ symmetry. Provided $f$ is even, $Z$ is always real, and there is no difference between forwards and backwards contours of the SFF. So there are $(2k)!!$ pairings and the answer is
\begin{equation}
    \overline{Z^k(T,f)Z^{*k}(T,f)}=(2k)!!\left(\frac{T}{\bbeta \pi}\int dE f^2(E)\right)^k.
\end{equation}
These are exactly the moments one would get for complex or real Gaussian variables respectively, which is also what one would get from RMT. More surprisingly, we got this without specifying the type of disorder, which indicates that hydrodynamics knows about universal features of disordered systems provided the disorder is not so strong that it changes the structure of the hydro theory. 

\subsection{Diffusive Hydrodynamics}

We will now evaluate the quantity $\exp(MT)$ for the linear theory of energy diffusion. As we saw above, the ramp comes from the spatial zero modes, and the sum over return probabilies comes from the other spatial modes. Most of the calculation about to be shown is generic for any diffusing substance, but for concreteness we continue to use the language of energy diffusion. In the CTP framework, the theory of linear diffusion is given by a Lagrangian of the form
\begin{equation}
L=-\phi_a\left(\partial_t\epsilon-D\nabla^2\epsilon\right)+i\beta^{-2}\kappa(\nabla \phi_a)^2, 
\label{eq:CTPLag}
\end{equation}
where $D$ is the diffusion constant, $\kappa$ is the thermal conductivity, $\nabla^2$ is the Laplacian. One can also define the specific heat $c=\kappa/D$, in terms of which $\epsilon = c \beta^{-1} \partial_t \phi_r$. Note that since these physical properties typically vary with temperature/energy density, they should be regarded as functions of the zero mode $E_{\text{aux}}$ and we must integrate the final result over energy. For the analysis in this subsection, we consider the total energy of the system to be fixed and known.

Now, because the action is quadratic, the path integral breaks up into a product over different spatial wavevectors. Hence, the sum over return probabilities is
\begin{equation}
\tr e^{M T}=\prod_k \int d\epsilon_{k,\textrm{init}} p(\epsilon_{k,\textrm{final}}=\epsilon_{k,\textrm{init}})
\label{eq:return}
\end{equation}
Looking at a particular wavevector $k$, let the amplitude at time $t=0$ be $\epsilon_k$. At time $t=T$, the amplitude is given by some probability distribution with mean $e^{- \gamma_k T} \epsilon_k$, where $\gamma_k$ is the decay rate, and variance $\sigma^2(T)$. For the linear theory above, this distribution is a Gaussian,
\begin{equation}
    p(\epsilon_{k,\textrm{final}},T) = \frac{\exp\left( - \frac{(\epsilon_{k,\textrm{final}} - e^{-\gamma_k T} \epsilon_k)^2 }{2 \sigma^2(T)}\right)}{\sqrt{2\pi \sigma^2(T)}},
\end{equation}
although the precise shape turns out not to matter. The return probability integrated over the initial condition is
\begin{equation}
    \int d\epsilon_k p(\epsilon_{k,\textrm{final}}=\epsilon_k,T) = \frac{1}{1-e^{-\gamma_k T}},
\end{equation}
independent of the variance $\sigma^2(T)$.

For the DPT theory above with periodic boundary conditions, $\gamma_k = D k^2$, and the allowed values of $k$ are $k \in (2\pi/L)\mathbb{Z}^d$ where $L$ is the linear size, so that $V=L^d$. This implies that
\begin{equation}
    \tr e^{M T} = \prod_{ k \in (2\pi/L)\mathbb{Z}^d} \frac{1}{1 - e^{- Dk^2 T}}.
\end{equation}
For more general shapes, the decay rates are given by the  eigenvalues $\lambda$ of the Laplacian $\nabla^2$ (which are non-positive). Hence, the general formula is 
\begin{equation}
\tr e^{M T}=  \prod_{\lambda \in \text{spec}(\nabla^2)} \frac{1}{1-e^{D\lambda T}}
\label{eq:diffShape}
\end{equation}
Equation \eqref{eq:diffShape} can also be derived by directly computing the path integral taking into account the periodic boundary conditions in time, an exercise we do appendix \ref{app:direct}. Note that the zero mode, $k=0$, requires special attention. It corresponds to the exactly conserved quantity, and the divergence in $(1-e^{-\gamma T})^{-1}$ when $\gamma=0$ should be replaced with a sum over the allowed values of the conserved charge, as follows from the trace formula. This is just what we discussed in the previous subsection.

Considering times that are short enough that many modes have not decayed, so that we may ignore the discreteness of the spectrum of $\nabla^2$, the result for a box of volume $V$ is 
\begin{equation}
\log \tr e^{M T} =\frac{V}{(2\pi)^d}\int d^d k \sum_{j=1}^\infty \frac {\exp(-jDk^2T)}j=
\frac{V}{(2\pi)^d}\sum_j \frac 1j \left(\frac \pi {jDT}\right)^{d/2}=V\left(\frac{1}{4\pi DT}\right)^{d/2} \zeta(1+d/2).
\label{eq:diffusionSFF}
\end{equation}
When specialized to one dimension, this agrees exactly with the result in \cite{Friedman_2019} obtained for a particular Floquet model where the diffusing substance was a conserved $U(1)$ charge. At longer times, we see that the slowest modes control the approach to the linear pure random matrix ramp. In particular, when $T$ is large compared to the Thouless time $t_{\text{Th}} \sim V^{2/d}/D$, the trace is exponentially close to unity, $\tr e^{M T} \sim 1 + \mathcal{O}(e^{-T/t_{\text{Th}}})$.

\subsection{Subdiffusive Hydrodynamics}

As an aside, for some systems, such as fracton systems with multipole conservation~\cite{Gromov_2020,Nandkishore_2019} or systems near a localization transition, one can get subdiffusive dynamics of a conserved density. This can be taken into account by replacing $\nabla^2$ with $\nabla^{2n}$. In this case, the analogue of equation \eqref{eq:diffusionSFF} is
\begin{equation}
\begin{split}
\log \tr e^{M T} =\frac{V}{(2\pi)^d}\int d^d k \sum_{j=1}^\infty \frac {\exp(-jD_nk^{2n}T)}j\\
= \frac{V}{(2\pi)^d}\frac{S_d}{2n}\Gamma\left(\frac{d}{2n}\right) \sum_j \frac 1j \left(\frac 1 {jD_nT}\right)^{d/2n}\\
= V{(2\pi)^d}\frac{S_d}{2n}\Gamma\left(\frac{d}{2n}\right)\left(\frac{1}{D_nT}\right)^{d/2n} \zeta(1+d/2n),
\label{eq:subdiffusionSFF}
\end{split}
\end{equation}
which agrees with the result obtained in \cite{Moudgalya_2021} for such a system.

\subsection{Deformations}
\label{subsec:defs}

Here we discuss how deformations of the Hamiltonian manifest in the DPT formalism. This serves as a useful introduction to the new features arising due to the periodic temporal boundary conditions. Consider a Hamiltonian $H=H_0 + g  \delta H$. The derivative of $Z(iT) Z(-iT)$ with respect to $g$ is
\begin{equation}
    \delta |Z|^2 = \tr \left( i\delta H e^{iHT} \right)\tr e^{-iHT}-\tr e^{iHT}\tr \left( i\delta He^{-iHT}\right).
\end{equation}
Viewing the two contours as two copies of the system, this expression can be thought of as inserting $i \delta H \otimes I - I \otimes i \delta H$  into the DPT path integral. Because it is anti-symmetric between the two contours, it corresponds to an $a$ variable in DPT formalism, so the expression for $\delta |Z|^2$ is like the expectation value of an $a$ variable. In the standard CTP case, such an expectation value would be exactly zero. But in the DPT case, one can get a non-zero result. This had to be so, given our result above, since such a perturbation can certainly change the value of the diffusion constant, and thus the overall answer.

\begin{figure}
    \centering
    \includegraphics[scale=0.5]{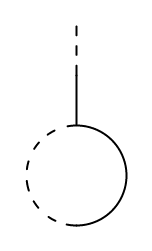}
    \caption{An $arr$-type vertex with an $a$-type variable (dashed line propagator) and an $r$-type variable (solid line propagator) contracted together. In the traditional CTP formulation this would be impossible, but with periodic time there is a contribution from one or more wrappings around the time circle. At long times, contributions from non-trivial wrappings are suppressed by factors of $e^{-T/t_{\text{Th}}}$.}
    \label{fig:tadpole}
\end{figure}

To show how this comes about in the formalism, consider an $arr$-type interaction. This vertex allows for diagrams such as in figure \ref{fig:tadpole}, which give a nonzero imaginary expectation value to $a$-type variables due to propagators that wrap around the $T$ circle. When $T$ is less than the Thouless time, such wrapping effects are not suppressed and the DPT formalism predicts that the SFF is sensitive to the deformation. However, at times long compared to the Thouless time, the effects of the periodic identification are exponentially small and the formalism predicts that $a$-type variables should have approximately zero average. This observation is how the hydrodynamic DPT formalism encodes the universality of the pure random matrix ramp at late time.

\subsection{Hydrodynamics With Periodic Time}

Before we discuss the effects of interactions, let's talk in more detail about how one would set up hydrodynamics on the SFF contours. It might not seem natural at first, but we are justified in using the same Lagrangian we would get in the conventional CTP formalism. 

At a structural level, one issue is that not all the rules of the usual Schwinger-Keldysh contour have to be obeyed. Because we are considering two contours, $\tr e^{-i H T} \tr e^{i H T}$, instead of a single folded contour, $\tr\left( e^{- i H T} e^{i H T}\right)$, automatic cancellations in CTP that arise from unitary are not guaranteed to occur. However, at sufficiently large $T$, the local physics of the two contours is identical, so integrating out fast modes should not generate novel terms not present in the usual CTP action. On the other hand, the slow fields are sensitive to the periodicity of time, and we will analyze these effects below. But the `bare' action obtained from integrating fast modes should be the same.

At a physical level, one issue is the arrow of time. The CTP formalism has an arrow of time from dissipation, instilled by the thermal loop in the past. The DPT contour is time symmetric, so how do dissipative terms arise? The reason this works mathematically is that we start with a sum over microstates in a possibly non-equilibrium macrostate in the doubled system, evolve it for time $T$, then take the amplitude with the phase. During this evolution, like (almost) all microstates in a non-equilibrium macrostate, the system feels an arrow of time. We can decide arbitrarily whether the initial state is at time $0$ or time $T$, and implant different arrows of time, but the end result will be the same.

There is also a nice interpretation in terms of the return probabilities in equation \eqref{eq:bigAnswer}. We are just integrating over all macrostates, and trying to calculate the probability that the system will return to the same macrostate. Again, the problem spontaneously introduces an arrow of time.

Turning now to the DPT formalism, recall that in the usual CTP case, $ra$ correlators are strictly causal. That means that $\langle \phi_a(0)\epsilon_r(t) \rangle$ is zero unless $t>0$. Doing perturbation theory on a time circle, this condition no longer makes sense since there will no longer be a coherent definition of future and past. Instead, we will wrap the propagator around the circle, so we have (where $a$ subscripts denote $\phi_a$ and $r$ subscripts denote the $r$-type variable $\epsilon$) 
\begin{equation}
\begin{split}
G^{\text{DPT}}_{ar}(t)=\sum_{n=-\infty}^{\infty} G^{\text{CTP}}_{ar}(t+nT),\\
G^{\text{DPT}}_{rr}(t)=\sum_{n=-\infty}^{\infty} G^{\text{CTP}}_{rr}(t+nT).
\end{split}
\label{eq:wrap}
\end{equation}

Considering again the example of energy diffusion with Lagrangian \eqref{eq:CTPLag}, the conventional CTP propagators are
\begin{equation}
\begin{split}
G^{\text{CTP}}_{ar}(t,k)=i\theta_+(t)e^{-Dk^2t},\\
G^{\text{CTP}}_{rr}(t,k)=\kappa\frac {Dk^2}{2}e^{-Dk^2|t|}.
\end{split}
\end{equation}
Here $\theta_+(t)$ denotes a step function with $\theta_+(0)=0$. Wrapping $G_{ar}$ around the circle and taking $t \in (0,T]$, we find
\begin{equation}
    G^{\text{DPT}}_{ar}(t) = i\frac{e^{- D k^2 t} }{1- e^{-D k^2 T}},
\end{equation}
while for $t=0$ we find
\begin{equation}
    G^{\text{DPT}}_{ar}(0) = i\frac{e^{- D k^2 T} }{1- e^{-D k^2 T}}.
\end{equation}
Note the $1/k^2$ IR divergance at low wavevector. A similar formula can be obtained for the wrapped $rr$ propagator, but the overall $k^2$ factor renders that object less IR divergent. 

With this setup, we are now ready to discuss interactions. In the next section, we highlight a few qualitative effects arising from time periodicity. Here we quickly recall the basic story without time periodicity. Taking again the example of linear diffusion above, interactions arise at the very least because the parameters of theory, like the diffusion constant, are themselves functions of the energy density. Taking the scaling from the quadratic theory~\eqref{eq:CTPLag}, one finds that interactions are irrelevant by power counting. For this reason, one argues that they can be treated in perturbation theory. The leading 1-loop diagrams can then be computed to exhibit a variety of corrections to parameters and other non-analytic features. 

\subsection{Interaction Effects in a Toy Model of Diffusive Hydro}

We now compute some novel effects of interactions that arise due to time periodicity in an interacting version of~\eqref{eq:CTPLag}. We focus on a simple 1-loop effect, treated self-consistently and resummed, but it is an open question whether this is sufficient to capture all the relevant physical effects.

\begin{figure}
\centering
\includegraphics[scale=1]{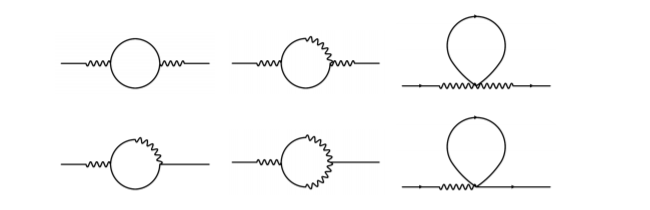}
\caption{The leading Feynman diagrams modifying the propagator. Wiggly lines are $\phi_a$s and straight lines are $\epsilon$s. Time wraps around horizontally, so both diagrams are zero-point functions and in each of them one propagator wraps around time $T$ some integer number of times. Figure taken from \cite{Chen_Lin_2019}.}
\label{fig:fdiagrams}
\end{figure}

Following the conventions in \cite{Chen_Lin_2019}, we consider the Lagrangian
\begin{equation}
L = i\beta^{-2}\kappa(\nabla \phi_a)^2-\phi_a\left(\partial_t\epsilon-D\Delta\epsilon\right)
+\frac{\lambda}{2}\Delta \phi_a\epsilon^2+\frac{\lambda'}{3}\Delta \phi_a\epsilon^3+ic\beta^{-2} (\grad\phi_a)^2(\tilde \lambda \epsilon+\tilde\lambda' \epsilon^2)
\end{equation}
as the simplest model of interacting hydrodynamics. In the case of conventional hydrodynamics, the dominant diagrams are the ones given by figures \ref{fig:fdiagrams}.
The UV cutoff is not strongly modified by the time periodicity, since high momentum legs are exponentially suppressed when they wrap around the circle. As discussed in \cite{Chen_Lin_2019}, such modes give a correction of the form
\begin{equation}
\begin{split}
\Sigma_{ar}(k)=-\frac{S_d}{d}\frac{cT^2}{2D\ell^d}\lambda_Dk^2 \\
\lambda_D=\lambda^2+\lambda\tilde \lambda+2\lambda'D,
\end{split}
\end{equation}
where $\ell$ is a cutoff scale.

\begin{figure}
\centering
\includegraphics[scale=0.5]{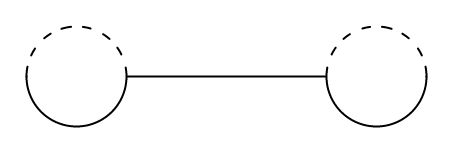}
\caption{Two $\lambda$ vertices. The propagator along the dumbell is infinite, and the diagram diverges unless $\lambda=0$ and thus we are at a local extremum of diffusivity.}
\label{fig:dumbbell}
\end{figure}

Our interest is in the new classes of diagrams allowed by periodic time. One interesting diagram is the dumbbell in figure \ref{fig:dumbbell}. This diagram diverges unless $\lambda=0$. This condition is equivalent to requiring we expand around an energy density which is an extremum of diffusivity. This makes perfect sense in light of the SFF formula in equation \eqref{eq:diffusionSFF}, which suggests that the dominant contribution should come from the minimum of diffusivity. If instead we add a filter function which localizes the total energy integral around some $\bar E$, then this effectively adds a mass to the zero mode and removes the divergence.

\begin{figure}
\centering
\includegraphics[scale=0.5]{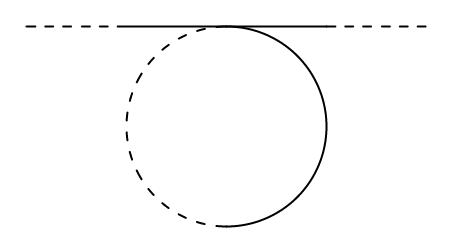}
\caption{A diagram contributing to the $rr$ self energy. In the case of normal hydrodynamics, there would be no $rr$ self energy.}
\label{fig:selfEn}
\end{figure}

Another important diagram is the one in figure \ref{fig:selfEn}.
This diagram would normally vanish in the CTP setup, but here it contributes a non-vanishing $rr$ self energy. We treat this term self-consistently by adding an undetermined self energy to the action and fixing it self-consistently. Since $\phi_a$ self-interactions still need to have a derivative in front of them by CTP rules, an constant $rr$ self energy $\Sigma$ is indeed the IR-biggest term we can add. Then the propagators are
\begin{equation}
\begin{split}
G_{rr}(\omega,k)=\frac{D\kappa k^2}{\Sigma D\kappa k^2+(D^2k^4+\omega^2)}\\
G_{ar}(\omega,k)=\frac{i\omega+Dk^2}{\Sigma D\kappa k^2+(D^2k^4+\omega^2)}\\
G_{aa}(\omega,k)=\frac{\Sigma}{\Sigma D\kappa k^2+(D^2k^4+\omega^2)}.
\end{split}
\end{equation}  
To solve for the self energy, we will need to sum $G_{ar}$ over all frequencies $\omega=2\pi n/T$. The sum is 
\begin{equation}
\sum_{\omega=2\pi n/T} G_{ar}(\omega,k)=\frac{Dk^2}{\sqrt{D^2k^4+\Sigma D\kappa k^2}}\frac{\exp(-\sqrt{D^2k^4+\Sigma D\kappa k^2}T)}{1-\exp(-\sqrt{D^2k^4+\Sigma D\kappa k^2}T)},
\end{equation}
which gives an expression for $\Sigma$:
\begin{equation}
\Sigma=\lambda' \sum_n\int \frac{d^d k}{(2\pi)^d} k^2 \frac{Dk^2}{\sqrt{D^2k^4+\Sigma D\kappa k^2}}\exp(-n\sqrt{D^2k^4+\Sigma D\kappa k^2}T)
\end{equation}
There isn't an IR divergence on the right hand side, so to leading order in large $T$, the $\Sigma$ dependence on the right hand side can be dropped. We are left with
\begin{equation}
\Sigma=\lambda' \sum_n\int \frac{d^d k}{(2\pi)^d} k^2 \exp(-nDk^2T)=\lambda'\frac{d}{2DT} \sqrt{\frac{1}{4\pi DT}}^d\zeta(1+d/2),
\end{equation}
where for the last equality we work in the time regime where the wavevector may be treated as continuous. This result also has an intuitive interpretation. At a minimum of diffusivity, the result \eqref{eq:diffusionSFF} gets a quadratic dependence on $\epsilon$ which is exactly the self-energy.

If we add $\Sigma$ to the action and take a determinant we get
\begin{equation}
\det \begin{pmatrix}
\Sigma & (i\omega+Dk^2)/2\\
(-i\omega+Dk^2)/2 & D\kappa k^2
\end{pmatrix}= D^2k^4+\Sigma D\kappa k^2+\omega^2.
\end{equation}
Using the results in appendix \ref{app:direct}, it follows that the coefficient of the ramp is modified to 
\begin{equation}
\log \text{coeff}(T)=\frac{V}{(2\pi)^d}\int d^dk \sum_{j=1}^\infty \frac {\exp(-j\sqrt{D^2k^4+\Sigma D\kappa k^2}T)}j
\end{equation}
For $d>2$, this is dominated by the $D^2$ bit, and we get the free answer for long times. For $d=2$, both contributions are important and there is no obvious way to simplify the integral. For $d=1$ with $V = L$, keeping just the $\Sigma$ term gives
\begin{equation}
\log \text{coeff}(T)=\frac{L}{2\pi}\frac{2}{T\sqrt{\Sigma D\kappa}}\frac{\pi^2}{6}.
\label{eq:d1result}
\end{equation}
Plugging in the expression for $\Sigma$, we see that the time dependence is now $T^{-1/4}$ instead of $T^{-1/2}$ in $d=1$. At later time, the self energy gets exponentially small past the Thouless time set by the lowest mode, so one still has an exponential late time approach to the pure random matrix ramp.

\section{Discussion}

In this paper we developed a theory of connected SFFs using tools from both RMT and hydrodynamics. This framework provides a number of key results, including formulas like equation \eqref{eq:bigAnswer} in cases with nearly conserved quantities, exemplified by equation \eqref{eq:returnCTP}. This allows us to rederive results previously only obtained for Floquet systems \cite{Friedman_2019,Moudgalya_2021} using general hydrodynamic principles. We are also able to give new formulas that include nonlinear effects as in equation \eqref{eq:d1result}. Such nonlinearities are typically associated with long-time tails in hydrodynamics \cite{PhysRevA.4.2055}\cite{Kovtun_2003} and here we see them manifest in the spectral form factor. Our results shed light on how spatial locality in Hamiltonians interacts with ergodicity. Finally, while we focused on the simplest case of energy diffusion for simplicity, analogous results can be obtained in a wide variety of hydrodynamic theories. Quite generally, the Thouless time can be read off from the decay rates of the slowest hydrodynamic modes. And in a system without slow modes, the Thouless time should scale like the logarithm of the system size, since any mode with a system-size-independent decay rate will have a system size suppressed amplitude after logarithmic time.

One emerging lesson highlighted by our work is that quantum chaos should be viewed as a robust phase of matter. In particular, the emergence of a pure random matrix ramp after the Thouless time is a feature that is stable to small perturbations. In fact, as we emphasized, the linear growth with time as well as the exact coefficient of the ramp are seemingly universal. What evidence is there for this? First, as discussed in section~\ref{subsec:defs}, when considering deformations $H=H_0+g \delta H$, the derivative of the SFF with respect to $g$ is an expectation of an $a$-type variable and such expectations are suppressed by factors of $e^{-T/t_{\text{Th}}}$. This means the SFF is unaffected up to exponentially small corrections. Second, as discussed in section~\ref{sec:folded}, the addition of an ETH-obeying perturbation to the Hamiltonian corresponds to a stretching of the spectrum plus the addition of a random matrix. Hence, if the system had a linear ramp without this perturbation, it will also have one with the perturbation. Third, the basic phenomenon of the linear ramp comes from a symmetry breaking effect arising from the spontaneous breakdown of the relative time translation between the two SFF contours. Because this relative time translation symmetry cannot be explicitly broken by any time-independent Hamiltonian perturbation (i.e. without completely changing the problem), the corresponding symmetry broken phase should be both distinct from the unbroken phase and stable. 

When we glimpse different manifestations of quantum chaos like hydrodynamics and ETH connecting to the emergence of RMT, it suggests to us that a larger synthesis may be possible. Certainly there are many connections between chaos, random matrix statistics, and eigenstate thermalization, e.g. as reviewed in \cite{doi:10.1080/00018732.2016.1198134}, as well as connections to notions of complexity, e.g.~\cite{Roberts_2017}. However, work remains to understand how all the different timescales obtained from various manifestations of chaos fit together, e.g.~\cite{dymarsky2018bound}. We hope to elaborate on these points in future work.

There are several issues that are still not fully understood, leaving room for further work. One is whether hydrodynamic methods can derive plateau behaviors in SFFs. Such a path integral derivation would need be be very unusual to reproduce the fact that plateau behavior is non-perturbative in the Heisenberg time, $T_{\textrm{Heisenberg}} \sim e^S$. Perhaps inspiration could be taken from other path-integral derivations of plateaus such as \cite{saad2019late,altland2020late,M_ller_2005}. Another question is the role of disorder. It seems that hydrodynamic SFFs naturally spit out values consistent with disorder averaging, despite there being no explicit disorder-averaging in the definition of the CTP formulation. Certainly for non-periodic times, the CTP formulation does not require disorder averaging to get correct real-time dynamics \cite{Blake_2018,glorioso2018lectures}. One possible resolution of this puzzle is that we need some disorder in order to make sense of the CTP action on the SFF contours, but this disorder can be small when the system size is large so that no intensive quantities, like transport parameters, are modified.

Finally, it is important to fully understand the possible effects of interactions in our modified CTP formalism. In the conventional CTP context, power counting indicates that interactions are irrelevant in the renormalization group sense. Interactions do generate novel effects not present in the Gaussian fixed point theory, but the claim is that these effects can be accurately captured in perturbation theory. We showed that there are new effects arising from time periodicity, but perhaps these effects can still be accurately captured by resumming few loop contributions. It would be interesting to formulate a generalized renormalization group analysis to better understand the situation.

\section{Acknowledgements}

We thank Subhayan Sahu and Christopher White for helpful discussions throughout this process. This work is supported in part by the Simons Foundation via the It From Qubit Collaboration (B. S.) and by the Air Force Office of Scientific Research under award number FA9550-17-1-0180 (M.W.). M.W. is also supported by the Joint Quantum Institute.

\renewcommand{\thechapter}{3}

\chapter{Spontaneous Symmetry Breaking, Spectral Statistics, and the Ramp}
\label{chapter:ssb}
\textbf{Authors:} \textit{Michael Winer, Brian Swingle}

\textbf{Abstract:} Ensembles of quantum chaotic systems are expected to exhibit energy eigenvalues with random-matrix-like level repulsion between pairs of energies separated by less than the inverse Thouless time. Recent research has shown that exact and approximate global symmetries of a system have clear signatures in these spectral statistics, enhancing the spectral form factor or correspondingly weakening level repulsion. This paper extends those results to the case of spontaneous symmetry breaking, and shows that, surprisingly, spontaneously breaking a symmetry further enhances the spectral form factor. For both RMT-inspired toy models and models where the symmetry breaking has a description in terms of fluctuating hydrodynamics, we obtain formulas for this enhancement for arbitrary symmetry breaking patterns, including broken Abelian symmetries $Z_n$ and $U(1)$, and partially or fully broken non-Abelian symmetries.
\newpage

\section{Introduction}

This paper is concerned with the statistical properties of energy levels of chaotic quantum systems exhibiting spontaneous symmetry breaking (SSB). The phenomenon of SSB can occur whenever a system possesses a symmetry and a suitable thermodynamic limit. This limit can be achieved either with a system extended in space or, in the cases examined in this paper, a zero-dimensional system with a large number of degrees of freedom. SSB is said to occur at a given energy density (energy per degree of freedom) if, after first taking the thermodynamic limit, a vanishingly small symmetry breaking perturbation in the Hamiltonian leads to a non-symmetric equilibrium state (reviews include \cite{Beekman_2019,Hidaka_2020}). Because the system's symmetry constrains the structure of Hamiltonian, SSB also manifests as a reorganization of the energy levels as a function of energy density. The purpose of this paper is to understand how this reorganization manifests in the correlations between energy levels and how the symmetry unbroken case is recovered in a finite size system.

More precisely, the theory developed here computes the so-called spectral form factor (SFF), the simplest version of which is the Fourier transform of the 2-point correlation between pairs of energy levels. Because it is a Fourier transform, the SFF naturally lives in the time domain. For a quantum chaotic system, the expectation is that, after a time known as the Thouless time, the SFF will approach a random matrix form determined by the symmetry of the Hamiltonian. This paper develops a theory of the SFF in zero-dimensional system exhibiting SSB. In particular, the theory explains how the SFF deviates from the appropriate symmetric random matrix form at early time and how random matrix behavior is recovered at late time.  

Although the SFF is understood in a wide variety of regimes \cite{saad2019semiclassical,PhysRevResearch.3.023118,PhysRevResearch.3.023176}, including systems with symmetries \cite{Friedman_2019}, there is yet to be a systematic study of the SFF in systems with spontaneously broken systems. This paper fills this gap in the zero-dimensional case, with the case of spatially extended systems left to future work. There have also been a few other studies of the interplay of quantum chaos, eigenstate thermalization, and spontaneous symmetry breaking including~\cite{Zhao_2014,Fratus_2015,fratus2017eigenstate}; see~\cite{D_Alessio_2016} for a review of notions of quantum chaos.

We analyze the spectral correlations in these systems as a function of energy density. At high energy density, one typically finds symmetric random-matrix-like energy levels, a hallmark of quantum chaos ~\cite{bohigas1984chaotic,berry1977level,doi:10.1063/1.1703775,haake2010quantum}. SSB then sometimes occurs as the energy density is lowered, in which case it corresponds to a breaking of ergodicity in the thermodynamic limit. Our theory quantitatively explains how ergodicity and symmetry are restored at finite system size from the point of the view of the energy spectrum. 

Because symmetry restoration is a long-time process, the theory must deal with special slow dynamical processes associated with the order parameter of the broken symmetry for which a hydrodynamic-like effective theory is the right description~\cite{crossley2017effective,Glorioso_2017,Grozdanov_2015,Kovtun_2012,Dubovsky_2012,Endlich_2013} (see~\cite{kamenev_2011} for an accessible introduction to the Schwinger-Keldysh technique underlying these effective theories). In previous work~\cite{winerprx}, we showed how the field theory formulation of fluctuating hydrodynamics could be adapted to predict a random-matrix-like spectral form factor at late times and to compute finite time corrections to random matrix theory (RMT). The present paper can be viewed as an extension of the earlier theory to systems with the additional physics of spontaneous symmetry breaking~\cite{Lallouet_2003,Hurtado_2011}.

\subsection{Setup and observables}

We now describe the setup and observables in more detail. A quantum system has symmetry group $G$ if (1) $G$ acts on the Hilbert space of the system by some faithful (but typically reducible) representation $\mathcal{U}$ and (2) the Hamiltonian of the system commutes with every representative, $\mathcal{U}(g) H = H \mathcal{U}(g)$. We focus on systems where the representation of $G$ acting on the full Hilbert space is unitary and linear. For disordered systems corresponding to an ensemble of Hamiltonians, we require that every element of the ensemble commute with the same $\mathcal U$s. Because it commutes with the Hamiltonian, such a symmetry constrains the energy spectrum by forbidding certain matrix elements in the Hamiltonian. More precisely, the Hamiltonian breaks into decoupled blocks labelled by the irreducible representations (irreps) of $G$.

In the simplest quantum chaotic case, each irrep block will consist of a number of copies (equal to the dimension of the irrep) of a system-specific matrix with random-matrix-like level correlations~\cite{bohigas1984chaotic,berry1977level,doi:10.1063/1.1703775,haake2010quantum}. Furthermore, the matrices for different irreps will be independent of each other. We quantify the correlations in these spectra using a filtered form of the spectral form factor (SFF)~\cite{brezin1997spectral} which zooms in on the particular energy density where SSB occurs. The spectral form factor with filter function $f$ is
\begin{equation}
    \SFF(T,f) = \overline{ | \text{Tr}[U(T) f(H)]|^2},
\end{equation}
where $U(T) = e^{-i H T}$ is the time evolution operator and overline denotes a disorder average over an ensemble of Hamiltonians where each representative is symmetric. We will say more about this averaging shortly; it is necessary in order to render the SFF a smoothly varying function of time $T$.
\begin{figure}
\centering
\includegraphics[scale=0.5]{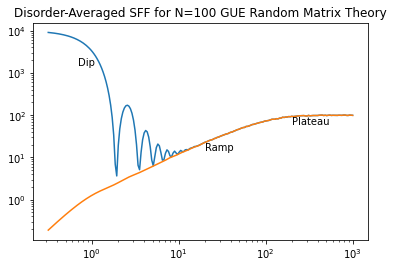}
\caption{SFF (blue) of a random matrix exhibiting the dip-ramp-plateau structure. The connected SFF only exhibits the ramp and plateau.}
\end{figure}

It is also useful to define the connected and disconnected contributions to the SFF. The disconnected part, 
\begin{equation}\SFF_{\text{dis}}(T,f)=|\overline{\text{Tr}[U(T) f(H)]}|^2,
\end{equation}
is distinguished by the full SFF because the squaring happens after the averaging. The connected part is the difference
\begin{equation}\SFF_{\text{con}}(T,f)=\SFF(T,f)-\SFF_{\text{dis}}(T,f)
\end{equation}
It can be shown that both the connected and disconnected parts are positive definite. In the standard dip-ramp-plateau picture, the dip comes from the disconnected component, and the ramp and plateau come from the autocorrelation of $\tr U(T) f(H)$ found in the connected component of the SFF.

For a quantum chaotic system with no symmetry, the expectation is that, after a Thouless time the SFF will agree with the random matrix result \cite{bohigas1984chaotic,berry1977level,doi:10.1063/1.1703775,haake2010quantum}. Up to the Heisenberg time, which is proportional to the level density, the RMT result is~\cite{mehta2004random} the linear ramp,
\begin{equation}
    \SFF(T,f) = \int dE f^2(E) \frac{T}{\pi \bbeta},
\end{equation}
where $\bbeta =1,2,4$ is the Dyson index from random matrix theory determined by the presence or absence of antiunitary symmetries. If we choose $f$ to be a Gaussian filter of the form $f = \exp\left( -\frac{(E-E_0)^2}{4\sigma^2}\right)$, then the result is
\begin{equation}
    \SFF(T,f) = \frac{\sqrt{2\pi} \sigma T}{\pi \bbeta}
\end{equation}
provided $E_0$ sits within the spectrum of $H$. 

If instead we have a system with $G$ symmetry which is unbroken at energy $E_0$, then the Gaussian filtered SFF will be
\begin{equation}
    \SFF(T,f) = \sum_{R} |R|^2 \frac{\sqrt{2\pi} \sigma T}{\pi \bbeta},
    \label{eq:nonabelianBase}
\end{equation}
where the sum is over irreps $R$ appearing in the spectrum and $|R|$ is the dimension of $R$. As discussed above, each irrep block is composed of $|R|$ identical copies of an independent random matrix (one for each state in $R$), so the $|R|^2$ factor arises because all the subblocks are perfectly correlated. See appendix \ref{app:nonAb} for more details.

The new feature associated with SSB at energy $E_0$ is that the different irrep blocks will no longer be effectively independent. At large but finite system size, the energy eigenstates will be `cat states' which transform in different representations of the symmetry group but have nearly-identical energies. This means that the spectra of different blocks will be strongly correlated, and this additional correlation will cause the SFF to take a larger value at early to intermediate times. Then at fixed system size (and for typical forms of SSB), the system will crossover to the unbroken behavior at very long time provided the order parameter fluctuates rapidly compared to the Heisenberg time of each block.

The remainder of the paper is organized as follows. In section \ref{sec:Zn} we consider the case of spontaneously broken $Z_n$ symmetry in various toy models, obtaining analytic and numerical results in excellent agreement. Next, in section \ref{sec:discreteG}, we do the same for more general finite groups $G$. Next, we consider a hydrodynamic calculation of the symmetry-breaking SFF in sections \ref{sec:AbelianHydro} and \ref{sec:NonAbelianHydro}, focusing first on Abelian Lie groups and then on general Lie Groups.

\section{$Z_n$ SSB With Zero Spatial Dimensions}
\label{sec:Zn}

We will start with the simplest possible case of discrete SSB: the case of a $Z_n$ symmetry and a charge-$1$ order parameter. Since any discrete Abelian group is a product of $Z_n$ factors (with possibly different $n$s), this case captures most of the interesting physics of discrete Abelian SSB. We also restrict attention to the case of zero spatial dimensions, so the reader should have in mind a cluster or similar sort of system where a large number of degrees of freedom can interact without geometric restrictions.

The charge-$1$ order parameter is described by a basis $\ket \phi \in \Phi$ with $\phi$ an integer from $0$ to $n-1$. Although in this chapter the symmetry and the corresponding order parameter are discrete, we will occasionally call $\phi$ a Goldstone mode when analogy with the continuous case would be helpful.

In addition to $\phi$, we take the other degrees of freedom to be described by a state $\psi$ in an $N$-dimensional Hilbert space $\Psi$, for some large $L$. In a physical system, $L$ would be exponential in the system size. A state in the total Hilbert space will be an $nL$-dimensional superposition of states of the form $\ket \phi \otimes \ket \psi$. In this section, we will require that $\ket \psi$ transform trivially under $G$. We briefly consider more general behavior in appendix \ref{app:IntCharge}.

The Hamiltonian is built from a collection of operators $H_k$ acting on $\Psi$ that are associated with (possibly trivial) transitions of the order parameter from sector $\phi$ to sector $\phi+k$. Using the shift operator $M_k$ defined as $M_k \ket{\phi} = \ket{\phi + k}$, we write this decomposition as
\begin{equation}
\begin{split}
    H=\sum_k M_k\otimes H_k\\
    (M_k)_{ij}=\delta_{i,j+k},
\end{split}
\end{equation}
with the arguments of the delta function in $M_k$ all taken mod $n$. Hermiticity of $H$ requires that $H_k^\dagger = H_{n-k}$ since $M_k^\dagger = M_{n-k}$. For instance, the $n=4$ Hamiltonian written out in block matrix form is
\begin{equation}
\begin{split}
    H=\begin{pmatrix}
    H_0&H_3&H_2&H_1\\
    H_1&H_0&H_3&H_2\\
    H_2&H_1&H_0&H_3\\
    H_3&H_2&H_1&H_0\\
    \end{pmatrix}\\
    \text{with } H_0=H_0^\dagger, H_1=H_3^\dagger, H_2=H_2^\dagger.\\
\end{split}
\end{equation}

At this point, we must ask what sorts of matrices $H_0$ and $H_k$ make good models of the sorts of systems we see in real life. As a simple model, consider the case where each $H_k$ is chosen independently consistent with the constraints imposed by Hermiticity. This should be a reasonable description of the spectral properties of generic chaotic systems after all other modes have decayed. Each block has matrix elements with variance $J_k^2/L$, and the physics of SSB is modeled by the condition $J_0\gg J_{k\neq 0}$. This leads to a diffusive motion in the order parameter space.  

The $H_k$ could also have more structure. For example, one could have $H_k=g_k(H_0)$ for some simple slowly-varying matrix function $g$. A salient case is where $g_k$ is roughly constant over the energy range of $H_0$, and is $0$ unless $k=\pm 1$.  In this world, the Hamiltonian can be roughly written as $I \otimes H_0 + K \otimes g(H_0)$, where $K_{ij}=\delta_{i,j+1}+\delta_{i,j-1}$. We can simultaneously diagonalize $H_0$ and $K$ to diagonalize this matrix. Given an eigenstate $\ket \psi$ of $H_0$ with $H_0\ket \psi=E_\psi \ket \psi$ and an eigenstate $\ket q$ of $K$ with $K\ket q=\lambda_q \ket q$ the energy of $\ket q \otimes\ket \psi$ is $E_\psi+g(E_\psi)\lambda_q$. Especially when $n$ is large, it makes sense to talk about a dispersion relation depending on $q$. We can think of the state as a particle with internal degrees of freedom propagating ballistically in order parameter space. In a real life system, the most realistic choice of $g$ would simply be an identity matrix, corresponding to a temperature/energy-independent kinetic term for the Goldstone mode. More general $g$s allow more complicated dependence. And most realistic systems add an ordered matrix such as the identity in this paragraph to a disordered random collection as in the previous paragraph. For instance, \textcolor{red}{Better example?} in the continuous case one can consider a bound state of interacting atoms in a physical Mexican hat potential. The system has Goldstone mode $\theta$, the angle to the center of mass of the atoms. The transition matrix is $\frac{1}{2I} \partial_\theta^2$, where $I$ is the moment of inertia of the system. By the eigenstate thermalization hypothesis, this moment of inertia has some rough slow dependence of $E$, but also some RMT contributions.

\subsection{The SFF With a Purely Random Kinetic Term}

Consider first the case of purely random $H_k$. To calculate the SFF, it is convenient to work with two copies of the system with total Hamiltonian. If the Hamiltonian of a single copy of the system is $H_{sys}$,
\begin{equation}
H_{\text{tot}}=H_{\text{sys}}\otimes I-I\otimes H^*_{\text{sys}}.
\end{equation}
The SFF of a single copy of the system is then
\begin{equation}
    \text{SFF}(T,1)=\overline{\tr \exp(-iH_{\text{tot}}T)}.
\end{equation}
Note that the complex conjugation in the definition of $H_{\text{tot}}$ is added for convenience.

\begin{figure}
    \centering
    \includegraphics[scale=0.07]{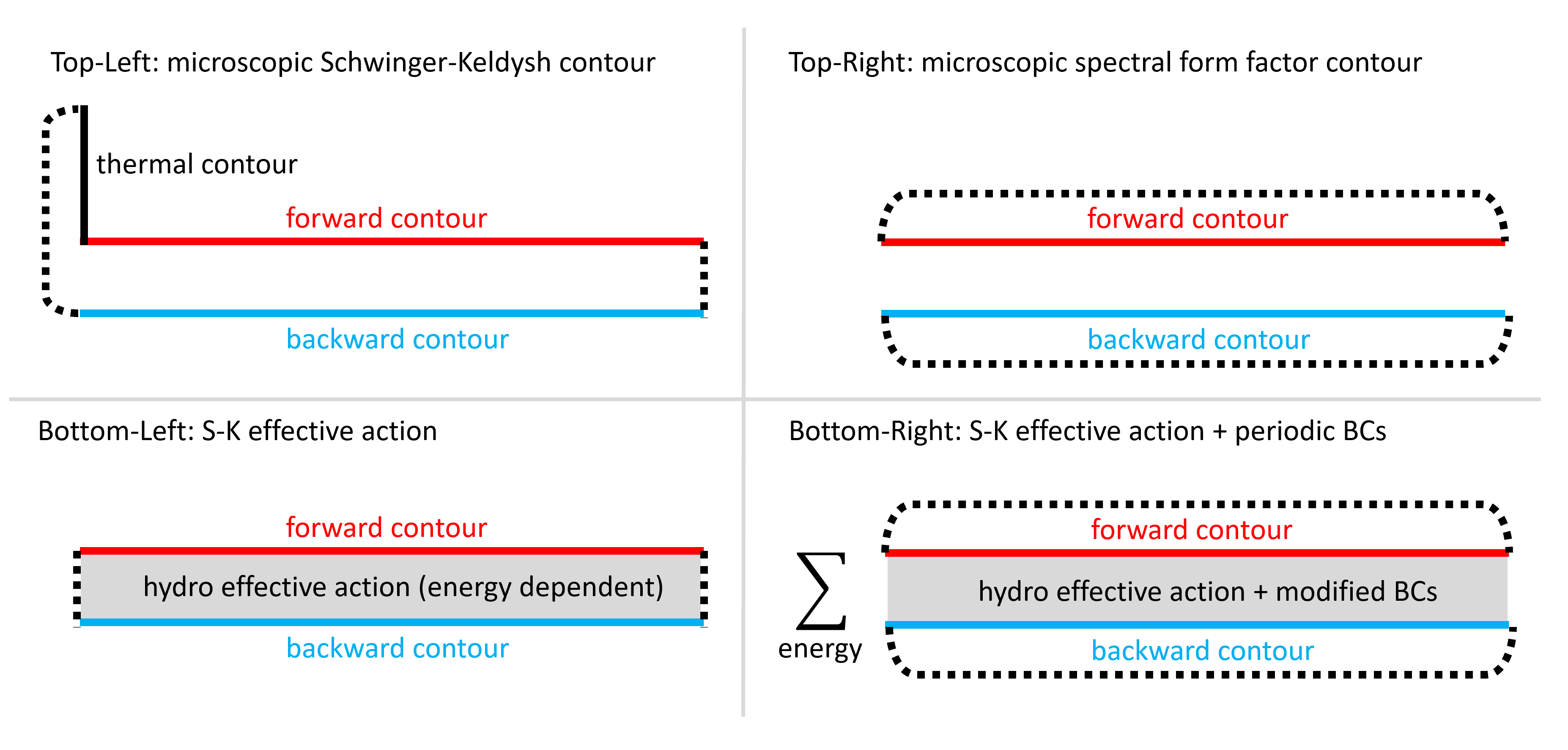}
    \caption{A pictorial representation of how one goes from an exact expression for the SFF (top right) to an effective action on periodic boundary conditions (bottom right). The process causes an effective coupling between same-time variables on opposite contours, analogous to how performing this procedure on on a Schwinger-Keldysh contour also introduces coupling between the two legs. In fact, the two procedures will produce the same action, up to exponentially small corrections.}
    \label{fig:masterpiece}
\end{figure}
\begin{figure}
\centering
\begin{tabular}{cc}
\includegraphics[scale=0.5]{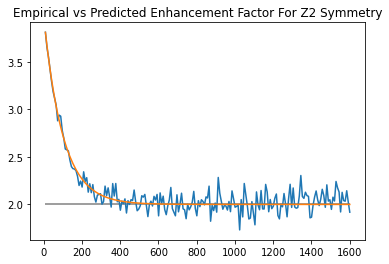}\\
\includegraphics[scale=0.5]{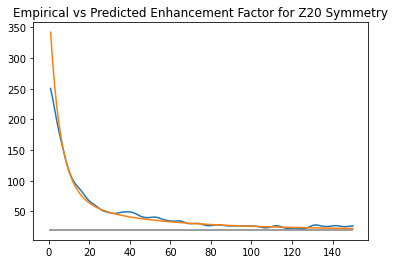}\\
\end{tabular}
\caption{The predicted (orange) vs displayed (blue) enhancements for two $Z_n$-symmetric Hamiltonians. The pure RMT prediction for a matrix with that symmetry group is in gray.}
\label{fig:ZnGraph}
\end{figure}

As a reminder, if all the $H_{k \neq 0}$ are set to zero, then the order parameter is frozen and the SFF is controlled by the diagonal $H_0$ blocks. $H_0$ is a random matrix, so its SFF is given by the random matrix result, but since each block is identical and there are $n$ blocks, the total SFF of the system is $n^2$ times the random matrix result. Nonzero $H_{k \neq 0}$ blocks cause the order parameter to fluctuate in time. We want to calculate the modifications to the SFF by summing over quantum trajectories of the order parameter. This calculation is viewed as a kind of path integral calculation on the doubled system with Hamiltonian $H_{\text{tot}}$. The averaging operation discussed in the introduction corresponds here to averaging over the elements of the $H_k$s.

Let's take a moment to consider what sort of formula we should expect for this sum. As with other ramp-related quantities, it is helpful to juxtapose it with Schwinger-Keldysh/CTP contour $e^{-\beta H}e^{iTH}e^{-iTH}$. We can generally factor the Hilbert space into slow degrees of freedom $\Phi$ and fast degrees $\Psi$. In this paper $\Phi$ is the order parameter of a symmetry and the various $\Phi$ sectors will be related by that symmetry. But $\Phi$ could just as well include conserved quantities, gauge bosons, or fermion degrees of freedom with a chiral symmetry.

At any point along the contour we have some density matrix $\rho_{\Phi\Psi}\propto e^{-\beta H}$ indicating the density for both order parameter $\phi$ and the microscopic degrees of freedom $\psi$ (equivalently, this can be viewed as a state of the two-replica system). If we trace out the $\Psi$ modes, $\rho_{\Phi}$ evolves by multiplication by a time-dependent superoperator, $e^{-\text{Trans}(E) T}$, which is generated from a sort of transfer matrix $\text{Trans}(E): \Phi^2\to \Phi^2$ that acts on two replicas of the order parameter space. More precisely,
\begin{equation}
    \rho_{\Phi}(T)=\int dE e^{-\text{Trans}(E)T} P_E\rho_{\Phi}(0)
\end{equation}
where $P_E$ projects down to an energy window around $E$.

If it weren't for the symmetry relating the different $\phi$s, the superoperator $\text{Trans}(E)$ would approximately annihilate doubled-system states $\ket{\phi_1}\ket{\phi_1'}$ with $\phi_1\neq \phi_1'$. One intuitive reason for this is that with no similarities between the different $\phi$ sectors we would be free to add a phase to one and not the others. The only terms that would be invariant under this are $\phi_1\phi_2\to\phi_1\phi_2$ or $\phi_1\phi_1\to\phi_2\phi_2$. Averaging over disorder, the first of these has to have zero amplitude in $\exp(-\textrm{Trans}(E)T)$ because we can add some small amount of energy to one sector but not another and rotate the term by a relative phase. As such, without symmetry between sectors we could essentially replace
\begin{equation}
    \text{Trans}(E)_{\phi_1 \phi'_1 \phi_2 \phi'_2}\rightarrow \delta_{\phi_1 \phi'_1} \delta_{\phi_2 \phi'_2} \text{Trans}(E)_{\phi_1 \phi_2}.
    \label{eq:WrongAssumption}
\end{equation} 
This is the case considered in \cite{winerprx}. But here the symmetry allows for constructive interference, and we need to trace over $\text{Trans}(E)$ in its full glory. Equation \ref{eq:WrongAssumption} is no longer valid.
\begin{figure}
    \centering
    \includegraphics[scale=0.5]{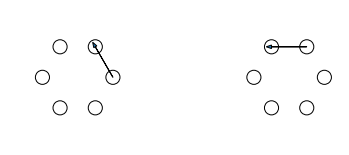}\\
    \includegraphics[scale=0.5]{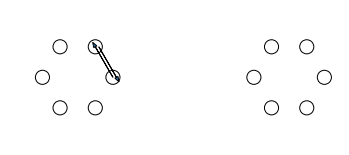}
    \caption{The two types of processes. In the top figure, the first and second copies of the system both have identical jumps. In the bottom figure, the first system jumps and jumps back.}
    \label{fig:processes}
\end{figure}
If the matrix elements of $H_0$ and $H_{k \neq 0}$ are independent random numbers, then there are only two perturbative processes that contribute to the path integral and thus to $\text{Trans}(E)$, shown in figure \ref{fig:processes}. The first is when both systems go from (possibly distinct) $\psi_i$s to (possibly distinct) $\psi_f$s using the same $H_k$. The second is when one of the two systems goes from $\psi_i$ to $\psi_f$ to $\psi_i$, using $H_k$ and $H_k^\dagger$. The both-sides-jump perturbation is parameterized by the two times $t_1,t_2$ of the two jumps, and by the energies $E_{\text{final}}$ after the jump. The amplitude for any one of these processes for any given $\psi_f$ depends on the square of the magnitude of the matrix element. The double integral over time gives one ``center of mass'' time and an energy delta function, 
\begin{equation}
   \int dt_1 dt_2 |\bra{\psi_f}H_k\ket{\psi_i}|^2e^{i(E_{\text{final}}-E_{\text{init}})(t_2-t_1)} \propto T \delta(E_{\text{final}}-E_{\text{init}}).
\end{equation}

Disorder-averaging over the matrix elements and converting the sum over final states into an integral, the total amplitude for this process to happen once after time $T$ is a quantity depending on the root-mean-square (rms) matrix elements of $H_k$, which can be simplified using Fermi's golden rule.
\begin{equation}
\begin{split}
\sum_{E_{\text{final}}}\int dt_1dt_2 |H_k\textrm{ rms matrix element}|^2e^{i(E_{\text{final}}-E_{\text{init}})(t_2-t_1)}\\
=r_k(E_{\text{init}}) T
\end{split}
\end{equation}
where the rate $r_k$ is
\begin{equation}
    r_k(E) = 2\pi |H_k\textrm{ rms matrix element}|^2 \rho(E)
\end{equation}
with $\rho(E)$ the density of states; this is just the Fermi's golden rule. This is thus an amplitude $r_k(E)$ per unit time for a process that multiplies the states of both replicas by $M_k$.

For the $\psi_1\to\psi_2\to\psi_1$ processes, we have the condition $t_2>t_1$, which produces a factor of $\frac 12$ relative to when the two jumps happen on separate contours. However there is a factor of 2 because the two jumps can happen in either replica of the system. The net result of these two jumps is that both subsystems are in the same sector, so the total amplitude is $r_k(E)$ for a process that multiplies the states of both replicas by $I$.

Summing all processes we get
\begin{equation}
\begin{split}
\text{Trans}(E)=  \sum_{k\neq 0} r_k(E)\left(  I \otimes I - M_k\otimes M_k \right) =\\
\frac{1}{2} \sum_{k\neq 0} r_k(E)\left( 2 I \otimes I - M_k\otimes M_k - M_{-k} \otimes M_{-k} \right)
\end{split}
\label{eq:transferEq}
\end{equation}
The second equality follows from $r_{k} = r_{n-k}$ and $M_{-k} = M_{n-k} = M_k^\dagger$. Note that this is a tensor product of different spaces, two factors of $\Phi$, rather than a factor of $\Phi$ and a factor of $\Psi$. This means that the transfer matrix has four $\Phi$ indices, i.e. it is a superoperator. The enhancement factor to the SFF is just $\int dE f^2(E) \tr e^{-\text{Trans}(E) T}$. We can see this prediction borne out in Figure~\ref{fig:ZnGraph}.

Note that the late time enhancement of the SFF follows from the number of zero modes of $\text{Trans}(E)$. The general spectrum of the transfer matrix is obtained from states of the form
\begin{equation}
    \sum_{\phi \phi'} e^{i q \phi + i q' \phi'} \ket \phi \ket{\phi'}, 
\end{equation}
where $q,q'$ are $\frac{2\pi}{n}$ times an integer $\in \{0, \cdots, n-1\}$. The $q,q'$ state has eigenvalue
\begin{equation}
    \sum_{k\neq 0} r_k (1 - \cos k (q+q') ).
\end{equation}
The set of zero modes is given by $q=-q'$, hence, there are $n$ zero modes of $\text{Trans}(E)$. This implies that the late time SFF enhancement is $n$, as expected in a situation where the symmetry has been restored.

\subsection{The SFF With A Mixed Kinetic Term}

\begin{figure}
\centering
\includegraphics[scale=0.5]{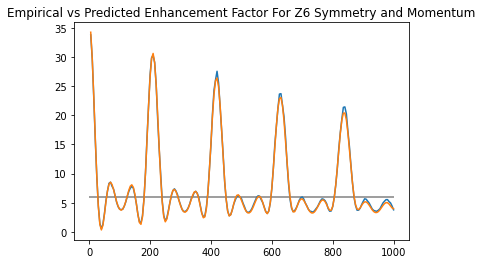}
\caption{Each $H_0$ is a 500 by 500 matrix, with consecutive blocks connected by $c_1R+c_2I$,  where $R$ is a random complex matrix,  $c_1$=0.02, and $c_2$=0.03. The numerical results are in blue, the analytic predictions are red. The pure RMT prediction for a matrix with that symmetry group is in gray.}
\label{fig:ZnMomentum}
\end{figure}

The effect of adding in a slowly varying kinetic term like $H_k = g_k(H_0)$ is to add in additional processes which can contribute to transfer matrix. More precisely, let's break our Hamiltonian into
\begin{equation}
H_{\text{sys}}=\sum_k M_k \otimes H_k  + \sum_k M_k \otimes g_k(H_0),
\end{equation}
with the $H_k$s fully independent of $H_0$ and the $g_k$s reasonably slowly varying analytic functions of small enough value to be treated by perturbation theory. 
\begin{figure}
    \centering
    \includegraphics[scale=0.5]{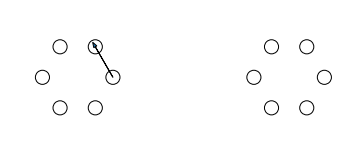}
    \caption{An additional figure, where one system jumps from one sector to another, with a coherent amplitude $g_k(E)$}
    \label{fig:processesLast}
\end{figure}
We will get an additional contribution to the transfer matrix of
\begin{equation}
\text{Trans}(E)_{\text{kinetic}} = - i \sum_k g_k(E) \left( M_k\otimes I-I \otimes M_k \right),
\label{eq:transfer2}
\end{equation}
which is just the amplitude of a single jump caused by the perturbation $f_j$. Here again the tensor product refers to two factors of $\Phi$. We see this prediction confirmed in Figure~\ref{fig:ZnMomentum}.

\section{General Discrete SSB}
\label{sec:discreteG}

In this section we will generalize from the simple Abelian case from last section and consider an arbitrary (finite) non-Abelian symmetry group $G$. Apart from its intrinsic interest, one motivation for considering $Z_n$ SSB is that we can approach $U(1)$ SSB by taking the limit $n\rightarrow \infty$. Unfortunately, there is no way to make arbitrarily good approximations of continuous non-Abelian groups with finite groups. Nonetheless, such discrete non-Abelian groups do appear in nature and can be spontaneously broken. A standard example is a valence bond solid, which preserves the $SU(2)$ spin symmetry but breaks the crystal point group symmetry.

We will start with some discrete symmetry group $G$, and some order parameter $\phi\in \Phi$. It will be helpful to think of $\phi$ not necessarily as a number, but as a thing which transforms under $G$. For instance in a ferromagnet, $\phi$ is a unit vector on the unit sphere $S^2$ indicating the direction of polarization. It transforms under the rotation group $G=SO(3)$. In effect, each element $g\in G$ induces a function $f_g(S^2)\to S^2$, where $f_{g_1}\circ f_{g_2}=f_{g_2g_1}$. Mathematicians call this a group action~\cite{artin}. We say that $\Phi$ is the orbit of $\phi$ under $G$. There is some (possibly trivial) subgroup $G'$ which maps $\phi$ to itself. This is called the stabilizer of $\phi$, and we have $|G'||\Phi|=|G|$. 

In addition to the order parameter space $\Phi$,  we will also have an 'internal' state $\ket \psi \in \Psi$, which transforms trivially under $G$. Once again we will set $\Phi$ to be $N$-dimensional. There is again some Hamiltonian random $H_0$ (variance $J_0^2/N$) within each of those $\phi$ subspaces, and a number of different matrices $\{H_i\}$ connecting different sectors. Let's again choose each element independently, with variance $J_i^2/N$, and where $J_{i}\ll J_0$. Because of the symmetry, each of the $H_i$s will show up in more than one place. 

Consider the example of the group $D_4$, which can be realized as the symmetry group of a square. This is a discrete group with $8$ elements generated by $90^{\circ}$ counterclockwise rotations ($a$) and reflections about one diagonal ($b$). These obey the relations $a^4=b^2 =e $ ($e$ is the identity) and $ab = b a^{-1}$. The center is $\{e,a^2\}$. The group has four one-dimensional irreps and one two-dimensional irrep. 

The order parameter corresponds to a choice of one corner of the square, so it takes $4$ values. Suppose that the only allowed jump of the order parameter corresponds to the rotation $a$ and the inverse rotation $a^{-1} = a^3$. Let the corresponding operators acting on $\Psi$ be $H_a$ and $H_{a^3}$. Invariance under the reflection $b$ requires $H_a = H_{a^3}$. The block matrix form of the system Hamiltonian is thus
\begin{equation}
\begin{split}
H=\begin{pmatrix}
H_0&H_a&0&H_a\\
H_a&H_0&H_a&0\\
0&H_a&H_0&H_a\\
H_a&0&H_a&H_0
\end{pmatrix}
\end{split}
\end{equation}

More generally, the system Hamiltonian is given by
\begin{equation}
H_{\text{sys}}=I \otimes H_0+\sum_i M_i\otimes H_i+M_i^\dagger\otimes H_i^\dagger
\label{eq:HRandom}
\end{equation}
where the $M_i$s are matrices which are mostly $0$s with a few $1$s connecting pairs of $\phi$s related by the same group action.  More precisely, the $M_i$s are indexed by $\Phi^2/G$. That is, two ordered pairs of $\phi$s share the same $M_i$ if their pairs are connected by an element of $G$.

In the non-Abelian case, we only consider the situation where the $H_i$ matrices are completely random and uncorrelated with $H_0$ and each other. In this case, the walk over the order parameter space is a random walk. This is a good model for, say, a particle stuck in one of $k$ identical wells (with enough internal structure to be chaotic), which jumps between wells using diffusive Poisson-frequency instantons. If the wells are placed on the vertices of a $k$-gon, the relevant group would be dihedral group $D_k$. If they were the vertices of a $k-1$-simplex, the relevant group would be the symmetric group $S_k$. This paradigm can be mathematically modeled.

\subsection{The SFF With a Purely Random Kinetic Term}

\begin{figure}
\centering
\begin{tabular}{cc}
\includegraphics[scale=0.5]{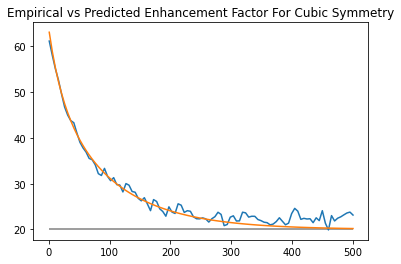}\\
\includegraphics[scale=0.5]{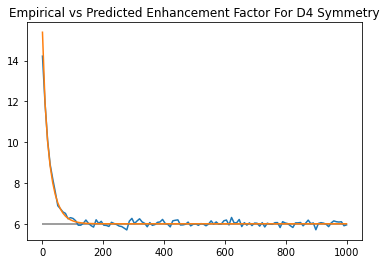}\\
\includegraphics[scale=0.5]{Chapters/chapter3/plots/Z2.png}\\
\includegraphics[scale=0.5]{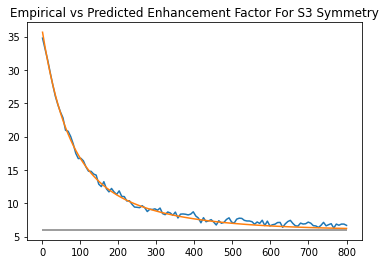}
\end{tabular}
\caption{The predicted (orange) vs displayed (blue) enhancements for a variety of group actions: Cubic Symmetry acting on the vertices of a cube, $D_4$ acting on the corners of a square,  $Z_2$ acting on two points, $S_3$ acting by multiplication on elements of $S_3$.}
\label{fig:nonAbelian}
\end{figure}
As in the Abelian case, to calculate the SFF, we want to have two copies of the system with total Hamiltonian
\begin{equation}
H_{\text{tot}}=H_{\text{sys}}\otimes I-I\otimes H^*_{\text{sys}}
\end{equation}
The SFF of a single copy of the system is still
\begin{equation}
    \text{SFF}(T,1)=\overline{\tr \exp(-iH_{\text{tot}}T)}.
\end{equation}
The analysis proceeds as in the Abelian case, in particular, we still have a path integral picture and treat the order parameter jumps in perturbation theory.


For each jump type $M_i$, there is a corresponding rate $r_i$. The transfer matrix is
\begin{equation}
\begin{split}
\text{Trans}(E)=\\
\frac 12 \sum r_i(E)\Big( \{ M_i, M_i^T \} \otimes I + I \otimes \{ M_i, M_i^T\}\\
-2M_i\otimes M_i-2M_i^T\otimes M_i^T\Big)\\
= \frac 12 \sum r_i(E) \Big((M_i\otimes I-I\otimes M_i^T)(M_i^T\otimes I-I\otimes M_i)\\
+(M_i^T\otimes I-I\otimes M_i)(M_i\otimes I-I\otimes M_i^T)\Big),
\end{split}
\label{eq:transferEqNonAbelian}
\end{equation}
where we used the fact that the $M_i$ have real matrix elements. The enhancement factor to the SFF is $\tr e^{-\text{Trans}(E) T}$. Figure \ref{fig:nonAbelian} shows the predicted and realized enhancements for various choices of $G$ and $\Phi$. In appendix \ref{app:longTime} we show that at long times this gives the RMT result consistent with symmetry $G$.

\section{Abelian SSB Hydro and the SFF}
\label{sec:AbelianHydro}

We now turn to the description of SFFs in systems with spontaneously broken continuous symmetries. We begin with the Abelian case, focusing again on the simplest case of $U(1)$ symmetry and a charge $1$ order parameter. Here the full power of the corresponding Schwinger-Keldysh hydro effective theory is revealed, so we first review that theory and then describe its modification to treat the SFF as in~\cite{winerprx}.

\subsection{Quick Review of Hydrodynamics}
\label{subsec:CTP}

At the broadest level, hydrodynamics is the program of creating effective field theories (EFTs) for systems based on the principle that macro-physics should be driven primarily by conservation laws. We will utilize the technology of the CTP formalism. For an accessible introduction, see~\cite{glorioso2018lectures}, and for more details see ~\cite{crossley2017effective,Glorioso_2017}. Additional information about fluctuating hydrodyamics can be found in~\cite{Grozdanov_2015,Kovtun_2012,Dubovsky_2012,Endlich_2013}. We first discussed the application of hydro EFTs for non-SSB spectral statistics in~\cite{winerprx}.

The CTP formalism starts with the following partition function of a Schwinger-Keldysh contour: 
\begin{equation}
    Z[A^\mu_1(t,x),A^\mu_2(t,x)]=\tr \left( e^{-\beta H} \mathcal{P} e^{i\int dt d^d x A^\mu_1j_{1\mu}} \mathcal{P} e^{-i\int dt d^d x A^\mu_2 j_{2\mu}}\right),
\end{equation}
where $\mathcal P$ is a path ordering on the Schwinger-Keldish contour.

For $A_1=A_2=0$, this is just the thermal partition function. Differentiating with respect to the $A$s generates insertions of the conserved current density $j_\mu$ along either leg of the Schwinger-Keldysh contour. Thus $Z$ is a generator of all possible contour ordered correlation functions of current operators.

One way to enforce the conservation law $\partial^\mu j_{i\mu}=0$ is to require 
\begin{equation}
    \begin{split}
         Z[A^\mu_1,A^\mu_2]=\int \mathcal D \phi_1\mathcal D \phi_2 \exp\left(i\int dt dx W[B_{1\mu},B_{2\mu}]\right),\\
         B_{i\mu}(t,x)=\partial_\mu \phi_i(t,x)+A_{i\mu}(t,x).
    \end{split}
    \label{eq:AbelianB3}
\end{equation}
Here the fields $\phi_i$ have been ``integrated in'' and represent the slow fluctuating modes of the system. Insertions of the currents are obtained by differentiating $Z$ with respect to the background gauge fields $A_{i\mu}$. A single such functional derivative gives a single insertion of the current, and so one presentation of current conservation is the identity $\partial_\mu \frac{\delta Z}{\delta A_{i\mu}} = 0$.

This is derived as follows. Since $Z$ only depends on $A_{i\mu}$ via the combined field $B_{i\mu}$, the functional derivative of $Z$ reduces to a functional derivative of the action,
\begin{equation}
    \frac{\delta Z}{\delta A_{i\mu}} = \int \mathcal D \phi_1\mathcal D \phi_2 \frac{\delta }{\delta (\partial_\mu \phi_i)}\exp\left(i\int dt dx W\right).
\end{equation}
Acting with $\partial_\mu$ and suppressing integration variables, we get
\begin{equation}
    \int \partial_\mu \frac{i \delta W}{\delta (\partial_\mu \phi_i)} e^{i \int W}.
\end{equation}
Because $W$ does not depend explicitly on $\phi_i$ (only on its derivatives), this is the function integral of a functional total derivative and hence vanishes.

The functional $W$ is not arbitrary. The key assumption of hydrodynamics is that the action $W$ is local. Moreover, when expressed in terms of
\begin{equation}
\begin{split}
    B_a=B_1-B_2,\\
    B_r=\frac{B_1+B_2}{2},
\end{split}
\end{equation}
there are several constraints which follow from unitarity:
\begin{itemize}
    \item $W$ terms all have at least one power of $B_a$, that is $W=0$ when $B_a=0$,
    \item Terms odd (even) in $B_a$ make a real (imaginary) contribution to the action,
    \item A KMS constraint imposing fluctuation-dissipation relations.
\end{itemize}
When calculating SFFs, one typically sets the external sources $A$ to zero, so the action can be written purely in terms of the derivatives of the $\phi$s.

The $\phi$s have a physical interpretation as phases transforming under the $U(1)$ symmetry. Performing a symmetry operation corresponds to adding a constant to $\phi$. If the symmetry is compact, this requires that adding $2\pi R$ to $\phi$ is actually a gauge transformation that doesn't change the state at all.
\subsection{The Hydrodynamic SFF}
As discussed in~\cite{winerprx} and illustrated in figure \ref{fig:masterpiece}, in order to calculate a spectral form factor, one performs the hydrodynamic integral with periodic boundary conditions in time. For instance, in a system with only energy conservation, the simplest hydro Lagrangian is
\begin{equation}
    L= C \partial_t \phi_a\partial_t \phi_r.
\end{equation}
One can show by taking derivatives with respect to the $A$s that the energy is $E=\int d^d x C \partial_t \phi_r$. So the SFF becomes
\begin{equation}
    \text{SFF}(T,f)=\int \frac {\mathcal D E \mathcal D \phi_a}{2\pi} f^2(E)\exp(-i\int dt \phi_a \partial_t E)
\end{equation}
This is a purely Gaussian integral. One can show that (subsection \ref{subsec:AbelianLagrangian}) that integrating out the nonzero frequencies with proper regularization leaves us with a prefactor in front of the measure of $(2\pi)^{-1}$. This leaves us with just an integral over the zero modes,
\begin{equation}
    \text{SFF}(T,f)=\int \frac {d E d \phi_a}{2\pi} f^2(E)  =\frac T{2\pi}\int dE f^2(E),
\end{equation}
the well-known result for GUE systems (for GOE there is an extra factor of two because one can reverse time on one contour relative to the other).

\subsection{Hydro and the Symmetry-Broken Spectral Form Factor}

Before we get into details, let's ask the most basic question: why should we expect spontaneous symmetry breaking to have any effect whatsoever on the spectral hydrodynamic path integral? For systems with spatial extent, SSB allows novel terms in the hydro action\cite{Glorioso_2017}. But it is impossible to write such terms in 0+1d. However, we know from previous sections that in the case of discrete symmetry, SSB has a clear signature in the connected SFF at short times.

The answer is the in the details, and in particular, the boundary conditions. In conventional fluctuating hydrodynamics, when we break our field into $\phi_a$ and $\phi_r$, it is standard practice to consider an overall constant addition to $\phi_r$ as a gauge symmetry. This is because phases aren't observable, only differences in phases like $\phi_a$ or $\phi_r(t_1)-\phi_r(t_2)$ are. But once a symmetry is spontaneously broken, phases do become observable. Water is invariant under a total rotation, but for ice you get a distinct quantum state.

What does this mean for boundary conditions? When calculating the SFF using conventional fluctuating hydrodynamics, it is sufficient to ensure that $\phi_a$ and $\partial_t \phi_r$ are periodic in time. But for SSB SFF, $\phi_r$ itself needs to be periodic. We can think of restoring the symmetry as gauging away $\phi_r$.

Of course, it is itself a phase defined on a circular manifold, and the path integral can wrap around that manifold any integer number of times. This introduces a summation into our calculation. We will handle the details below.

\subsection{The SSB Hydro Action}

We can write the Lagrangian as  
\begin{equation}
L=M\partial_t \phi_a (\partial_t \phi_r+b\partial_t^2\phi_r)+iM\frac{b}{2\beta} (\partial_t \phi_a)^2
\end{equation}
plus terms with more derivatives and/or fields. We show in appendix \ref{app:HydroDerive} that this Lagrangian can be derived as a continuous limit of the discrete $Z_n$ model in the previous section. The nature of spontaneous symmetry breaking means that in order to study it in 0+1d, we need some sort of large-$N$ limit. Thus, in our case, we are justified in dropping nonlinearities and interactions.

\subsection{Hamiltonian Approach}
\label{subsec:hamApproach}

We have formulated the SFF as a path integral involving a hydro-like effective action. In this subsection, we evaluate this path integral by converting to a corresponding Hamiltonian description. In the next two subsections, we give a direct Lagrangian calculation. Again, the contribution of charge to the SFF is 
\begin{equation}
\begin{split}
Z=\int \mathcal D\phi_1\mathcal D \phi_2 \exp(i\int dt L)f(q_1)f(q_2),\\
L=M \partial_t \phi_a (\partial_t \phi_r+b\partial_t^2\phi_r)+iM\frac{b}{2\beta} (\partial_t \phi_a)^2,
\end{split}
\label{eq:hydroPathIntegral}
\end{equation}
where $f$ is some function (perhaps a Gaussian) to regularize over the infinite sum over charges/momenta.

We make use of the fact that a path integral in one dimension is equivalent to a quantum mechanics problem. The Hilbert space is just a suitable space of functions on the configuration space of the path integral, in this case a basis is given by $U(1)^2$.

Viewed as an effective field theory, this path integral arrives from integrating out a number of microscopic modes. But like any Lagrangian path integral it is also equivalent to a Schrodinger-like evolution on the square of the Goldstone manifold (the imaginary $\phi_a^2$ part of the action corresponds to a non-unitarity for the fictitious Schrodinger dynamics).

To go from a Lagrangian quadratic in velocities to a Hamiltonian quadratic in momenta essentially involves inverting the Lagrangian. It is convenient to convert back to $\phi_1$ and $\phi_2$. Dropping the dissipative term in the equations of motion, the Lagrangian is
\begin{equation}
    L = \frac{M}{2} (\partial_t \phi_1)^2 - \frac{M}{2} (\partial_t \phi_2)^2 + i \frac{M b }{2\beta} (\partial_t \phi_1 - \partial_t \phi_2)^2.
\end{equation}
The corresponding canonical momenta are
\begin{equation}
    \pi_1 = \frac{\partial L}{\partial (\partial_t \phi_1)} = M \partial_t \phi_1 + i \frac{Mb}{\beta} (\partial_t \phi_1 - \partial_t \phi_2)
\end{equation}
and
\begin{equation}
     \pi_2 = \frac{\partial L}{\partial (\partial_t \phi_1)} = - M \partial_t \phi_2 - i \frac{Mb}{\beta} (\partial_t \phi_1 - \partial_t \phi_2).
\end{equation}
The Hamiltonian for this fictitious particle on the doubled system is then
\begin{equation}
    H_{eff} = \pi_i \partial_t \phi_i - L = \frac{\pi_i^2}{2M} - \frac{\pi_2^2}{2M} - i \frac{b}{2\beta M} (\pi_1 + \pi_2)^2.
\end{equation}
$H_{eff}$ is not the Hamiltonian of any underlying system, rather it is a mathematical trick that comes from evaluating the Lagrangian of an effective field theory.

Since each $\phi_i$ is periodic with period $2\pi R$, the eigenvalues of $\pi_i$ are quantized in terms of integer charges $q_i$ to be $\frac{q_i}{R}$. Hence, the eigenstates of the Hamiltonian are
\begin{equation}
\begin{split}
H_{eff}\psi_{q_1,q_2}=E_{q_1,q_2}\psi_{q_1,q_2},\\
E_{q_1,q_2}=(q_1^2-q_2^2)\frac{1}{2MR^2}-i\frac{b}{2\beta M R^2}(q_1+q_2)^2,\\
\psi_{q_1,q_2}(\phi_1,\phi_2)=e^{iq_1\phi_1/R}e^{iq_2\phi_2/R}.
\end{split}
\label{eq:Eexpression}
\end{equation}
The non-decaying solutions correspond to $q_1=-q_2, E_{q_1,q_2}=0$.  At long times, these are the only remaining contributions, and the long-time SFF is just equal to the number of charge sectors allowed in the sum. At general times, the overall sum is
\begin{equation}
    Z=\tr \left[f(q_1)f(q_2)e^{-iH_{eff}T}\right]=\sum_{q_1,q_2}f(q_1)f(q_2)\exp(-iE_{q_1,q_2}T).
    \label{eq:abCoeff}
\end{equation}

It is worth taking a moment to point out that in the large-$N$ limit necessary for 0d spontaneous symmetry breaking, the fictitious $E$s in expressions \eqref{eq:Eexpression} and \eqref{eq:abCoeff} are $1/N$ quantities, meaning the slowest decay time is actually extensive in the system size.

\subsection{Lagrangian Approach}
\label{subsec:AbelianLagrangian}

In this section, we will start be developing technology for arbitrary Gaussian SFFs. We have our path integral of $\exp\left(i \int \phi_a D\phi_r+\phi_a D_{aa}\phi_a \right)$, where $D=a \prod_{k=1}^{K} (\partial_t-\lambda_k)$, with $\textrm{Re}(\lambda_k)<0$ for all $k$, is some differential operator. The form of the action means that the determinant depends only on $D$, and we will set $D_{aa}$ to zero throughout this section.  

We proceed via spacetime discretization, and say that $\mathcal D\phi_a\mathcal D \phi_r=\prod_{j=0}^{T/\Delta t-1} \frac c{2\pi} d\phi_{aj}d\phi_{rj}$, the factor $c$ in the measure is to be determined. We compute the path integral by going to the Fourier basis
\begin{equation}
    \tilde{\phi}_i(\omega) = \sum_{j} \frac{e^{i \omega t_j}}{\sqrt{T/\Delta t}} \phi_i(t_j),
\end{equation}
where $t_j = j \Delta t$ and $\omega = 2\pi n/T$ for $n=0,\cdots,T/\Delta t -1$. The action becomes
\begin{equation}
\begin{split}
    \sum_j \Delta t \phi_a(t_j) [D \phi_r](t_j) = \\
    \sum_\omega \Delta t \tilde{\phi}_a(-\omega) a \prod_{k=1}^K \left( \frac{e^{- i \omega \Delta t} -1}{\Delta t} - \lambda_k\right) \tilde{\phi}_r(\omega).
\end{split}
\end{equation}
With only the $ra$ cross terms, the $\tilde{\phi}(-\omega)$ integral just gives $2\pi$ times a delta function of $\Delta t  D \tilde{\phi}_r(\omega)$. 

Hence, the path integral evaluates to 
\begin{equation}
   Z= \prod_{\omega} \frac{c}{a \Delta t} \prod_{k=1}^{K} \left(\frac{e^{-i \omega \Delta t}-1}{\Delta t}-\lambda_k \right)^{-1}.
\end{equation}
What constant $c$ in the measure keeps the integral from blowing up? This is pretty clearly a UV question. We want some $c$ such that
\begin{equation}
\begin{split}
Z =\prod_{n=0}^{T/\Delta t-1}  \frac{c}{a \Delta t} \prod_{k=1}^{K} \left(\frac{e^{-2\pi i n \Delta t/T}-1}{\Delta t}-\lambda_k \right)^{-1} =\\ \prod_{n=0}^{T/\Delta t-1}  \frac{c}{a \Delta t^{1-K}} \prod_{k=1}^{K} \left(e^{-2\pi i n \Delta t/T} - 1 -\lambda_k \Delta t \right)^{-1}
\end{split}
\end{equation}
doesn't blow as $\Delta t \rightarrow 0$. If we switch the order of the two products we get 
\begin{equation}
Z=\left( \frac{c}{a \Delta t^{1-K}}\right)^{T/\Delta t} \prod_{k=1}^{K} \left(1-e^{\lambda_kT}\right)^{-1}
\end{equation}
For this not to blow up, we need to regularize with $c=a \Delta t^{1-K}$.

Now that we have a general technology for regularizing CTP integrals, let's apply it to the specific case of the hydro Lagrangian $L = M\partial_t \phi_a (\partial_t \phi_r+b\partial_t^2\phi_r)+iM\frac{b}{2\beta} (\partial_t \phi_a)^2$. The corresponding differential operator is
\begin{equation}
D= - M b \partial_t^2 (\partial_t + b^{-1}).    
\end{equation}
In this case, $a= - Mb$ (although only the magnitude of $a$ is relevant), $\lambda_1 = 0$, $\lambda_2 = 0$, and $\lambda_3 = -b^{-1}$. However, there are two things keeping the hydro path integral from being a vanilla Gaussian integral: the zero-frequency modes where one integrates over the hydro  manifold, and the topological aspect of trajectories that wind around the manifold.  For now, we will focus on the saddle point with no winding.

In the non-winding sector, the non-zero-frequency modes are treated as Gaussian variables, while the integral zero frequency modes are integrated over exactly. The zero modes $\tilde{\phi}_i(0)$ range over a period of length $2\pi R\sqrt{T/\Delta t}$, and there are two of them. The path integral in the order parameter sector is thus
\begin{equation}
\begin{split}
Z_{\text{no-winding}}=\\
\frac {c}{2\pi}(2\pi R)^2 \frac{T}{\Delta t} \prod_{n=1}^{T/\Delta t-1} \frac{c}{a \Delta t^{1-K}} \prod_k \left(e^{-2\pi i n\Delta t/T}-1-\lambda_k \Delta t \right)^{-1}.
\end{split}
\end{equation}
The $\lambda_1=\lambda_2 =0$ terms in the product over $k$ give $\left( \frac{\Delta t}{T} \right)^2$, and the $\lambda_3=-1/b$ term gives $\frac{\Delta t}{b(1-e^{-T/b})}$. These pieces combine to give
\begin{equation}
    Z_{\text{no-winding}}=\frac{c}{2\pi} \frac{\Delta t}{T} (2\pi R)^2 \frac {\Delta t}{b(1-e^{-T/b})}.
\end{equation}
After inserting in our expression for $c=Mb\Delta t^{-2}$ we have 
\begin{equation}
Z_{\text{no-winding}}=\frac {1}{2\pi}\frac{M}{T}(2\pi R)^2 \frac{1}{(1-e^{-T/b})}.
\label{eq:coeffProd}
\end{equation}
If we wanted, we could add in higher derivatives to our original action. This would just result in more decaying terms corrections like the $1-e^{-T/b}$ in the denominator.

But there is a complication to equation~\eqref{eq:coeffProd}! The periodicity of $\phi$ actually means that there is an infinitude of saddle-points. In particular, the saddle-points of the action are parameterized by $n_1,n_2$,  winding numbers of $\phi_1,\phi_2$ around the circle. As a function of $n_r=\frac {n_1+n_2}2,n_a=n_1-n_2$, the winding contribution to the action is
\begin{equation}
\Delta S_{\text{winding}} (n_r,n_a)=M \frac{(2\pi R)^2}{T}\left( n_an_r+\frac{ib}{2\beta}n_a^2\right)
\end{equation}
The full path integral is obtained by summing over these modes, and since each winding sector has the same Gaussian part, we get
\begin{equation}
\begin{split}
Z=\\
\frac {1}{2\pi}\frac{M(2\pi R)^2}{T(1-e^{-T/b})}\sum_{n_r,n_a}\exp\left(i M\frac{(2\pi R)^2}{T}n_an_r-M\frac{(2\pi R)^2}{T}\frac{b}{2\beta}n_a^2\right).
\end{split}
\label{eq:WeirdSum}
\end{equation}

\subsection{Dealing With the Discrete Sum}

Equation~\eqref{eq:WeirdSum} is a divergent sum that needs to be regulated. In particular, there are an infinite number of $n_a=0$ solutions that add up. To regularize them, we need to remember that $n_r$ is proportional to the momentum/charge of the system, and so to get a finite enhancement we should only be summing over a finite set of charges, rather than all charges from $-\infty$ to $\infty$ (or the allowed set of microscopic charges). Then we can evaluate the path integral with an $f(Q)$ insertion, where $f$ is a regulating function. (For now we will assume it can be naively inserted and removed at will, and will analyze it more carefully in the next subsection.)
We get
\begin{equation}
\begin{split}
Z=\frac {1}{2\pi}\frac{M(2\pi R)^2}{T(1-e^{-T/b})}\sum_{n_r,n_a}f(Q_1)f(Q_2)\times\\
\exp\left(iM\frac{(2\pi R)^2}{T}n_an_r-M\frac{(2\pi R)^2}{T}\frac{b}{2\beta}n_a^2\right).
\end{split}
\end{equation}
Here charge $Q_i$ is related to $n_i$ by
\begin{equation}
Q_i=2\pi M  R^2 n_i/T.
\end{equation}
Moreover, in writing Eq.~\eqref{eq:WeirdSum} we implicitly assumed that $f$ is slowly varying so that the Gaussian part of the path integral is not appreciably modified.

Note that this expression for $Q$ in terms of winding numbers is such that, if we use it inside the path integral with an $f(Q)$ insertion, then we recover the correct sum over discrete charges of $f(Q)$ in the final answer. However, the quantization of $Q$ in the path integral arising from the winding sectors is different than the microscopic quantization. Either way of doing the sum, direct microscopic calculation or path integral calculation, will give the same answer, and they are related, roughly speaking, by a Poisson resummation. The situation is similar to the comparison between the Hamiltonian and Lagrangian descriptions of a particle on a ring.

For long times, we can rewrite $n_a$ in terms of $Q_a$ to see that the $\propto - Q_a^2 T$ term in the exponent suppresses $Q_a \neq 0$ contributions. The path integral is thus
\begin{equation}
\frac {1}{2\pi}\frac{M(2\pi R)^2}{T(1-e^{-T/b})}\sum_{n_r}f^2(2\pi M  R^2 n_r/T)=\int dQ f^2(Q),
\end{equation}
exactly what we want. Again, this integral over charge $Q$ approximates the discrete sum on the left hand side and is not a sum over the microscopic charges, but in the limit of slowly varying $f$ where Eq.~\eqref{eq:WeirdSum} is valid, we obtain the same answer.

Note that the functional substitution $Q=M\partial_t \phi_r$ would give us the exact same action as in the non-SSB case. The reason we get a more complicated short-time behavior is that we are evaluating the path integral in a different way. One way to look at the change is as a change in boundary conditions. Before breaking the symmetry, we treated an overall addition to $\phi_r$ as a gauge symmetry, and allowed arbitrary additions of it over the period. Whereas now we require that $\phi_r$ change by a quantized amount equal to $2\pi R n_r$.

\subsection{A More Careful Accounting of $f$}

We can express the filter function as 
\begin{equation}
 f(q) = \int_{0}^{2\pi R} dx \tilde f(x)e^{iqx}.   
\end{equation}
Inserting any operator $f(Q)$ into the path integral is thus equivalent to a double integral over basic insertions of the form $e^{i Q_1 x_1} \otimes e^{ i Q_2 x_2}$. Since $Q_i$ is conjugate to $\phi_i$, this insertion when acting on a state of definite $\phi_i$ shifts the value by $ - x_i$. When placed after the final resolution of the identity in the path integral, one gets a delta function $\delta(\phi_i(0) - \phi_i(T)+x_i)$ that effectively shifts the boundary condition to $\phi_i(T)=\phi_i(0)+x_i+2\pi R n_i $. The path integral \eqref{eq:hydroPathIntegral} can thus be written as
\begin{equation}
\begin{split}
Z=\int dx_1dx_2 \int \mathcal D_{x_1}\phi_1\mathcal D_{x_2} \phi_2 \exp(i\int dt L)\tilde f(x_1)\tilde f(x_2),\\
L=M \partial_t \phi_a (\partial_t \phi_r+b\partial_t^2\phi_r)+iM\frac{b}{2\beta} (\partial_t \phi_a)^2,
\end{split}
\label{eq:hydroPathIntegralBC}
\end{equation}
where $\mathcal D_{x}$ is the path integral measure with twisted periodic boundary conditions $\phi_i(T)=\phi_i(0)+x_i+2\pi R n_i$. 

If we repeat the analysis in the last two sections, we find that these twisted boundary conditions don't change the Gaussian part of the path integral at all. Defining $x_r=\frac{x_1+x_2}{2}$, $x_a=x_1-x_2$, equation \eqref{eq:WeirdSum} becomes 

\begin{equation}
\begin{split}
Z=\frac{1}{2\pi} \frac{M(2\pi R)^2}{T(1-e^{-T/b})}\int_{0}^{2\pi R} dx_1\int_{0}^{2\pi R}dx_2\sum_{n_r,n_a}\tilde f(x_1)\tilde f(x_2)  \exp( \Delta S)
\end{split}
\end{equation}
with
\begin{equation}
\begin{split}
\Delta S = iM\frac{\left((2\pi R)n_r+x_r\right)\left((2\pi R)n_a+x_a\right)}{T}-\\
M\frac{\left((2\pi R)n_a+x_a\right)^2}{T}\frac{b}{2\beta},\\
Z =\frac {1}{2\pi}\frac{M(2\pi R)^2}{T(1-e^{-T/b})}\int_{-\infty}^{\infty} dx_1\int_{-\infty}^{\infty} dx_2\tilde f(x_1)\tilde f(x_2)\\
\times \exp\left(iM\frac{x_rx_a}{T}-M\frac{x_a^2}{T}\frac{b}{2\beta}\right),
\end{split}
\label{eq:WeirderSum}
\end{equation}
where in the last equation we extend $\tilde f$ to be a period function with period $2\pi R$. 
This formula is fully general for any regulating function $f$ no matter how rapidly it varies. 

Let's investigate its behavior at short times. At short times we can consider the $\tilde f$s roughly constant compared to the Gaussian integral. Performing this integral and factoring in the determinant $2\pi \frac{T}{M}$ from the delta function, we have the short-time formula
\begin{equation}
    Z_{\textrm{short time}}=\frac{1}{(1-e^{-T/b})}(2\pi R)^2 \tilde f(0)^2=\left(\sum_{-\infty}^\infty f(q)\right)^2.
\end{equation}

More generally, we can replace $\tilde f(x_1)=\frac 1{2\pi R}\sum_{q'_1}e^{-iq'_1x_1}$. If we do the same for $x_2$, and define $q'_r=\frac{q'_1+q'_2}{2},$ $q'_a=q'_1-q'_2$, we have a Gaussian integral that works out to
\begin{equation}
    Z=\frac{1}{(1-e^{-T/b})}\sum_{q'_1,q'_2}f(q'_1)f(q'_2) \exp(i\frac{T}{M}q'_aq'_r+\frac{2b}{\beta}\frac{T{q'}_a^2}{M}).
\end{equation}
If we make the $T\gg b$ assumption of subsection \ref{subsec:hamApproach}, this is the exact same formula as \eqref{eq:abCoeff}. For long $T$ the $q'_a\neq 0$ terms become zero and we have 
\begin{equation}
    Z_{\textrm{long time}}=\sum_{q'}f^2(q').
\end{equation}

\section{Non-Abelian Hydro and the SFF}
\label{sec:NonAbelianHydro}

In this section, we treat the case of spontaneous symmetry breaking of continuous non-Abelian symmetries. As in the continuous Abelian case, a modified version of the hydro theory of non-Abelian SSB provides a useful way to formulate the SFF. 

\subsection{Overview of Non-Abelian Hydro}

In Abelian $U(1)$ hydrodynamics, the phase field $\phi$ can be regarded as an element of the group $U(1)$ by exponentiating it. In non-Abelian hydrodynamics~\cite{nonAbelian}, we replace the exponentiated phase field with a field $g$ that takes values in the (non-Abelian) global symmetry group $G$. The hydro generating function is expressed as 
\begin{equation}
\begin{split}
    Z_{\text{hydro}}[A^\mu_1(t,x),A^\mu_2(t,x)]=\\
    \int \mathcal D g_1\mathcal D g_2 \exp\left(i\int dt d^d x W[B_{1\mu},B_{2\mu}]\right),\\
    B_\mu=A_\mu + i g^{-1}\partial_\mu g,
\end{split}
\end{equation}
a non-Abelian generalization of Eq.~\eqref{eq:AbelianB3}. This action and generating function will always have a $G$ premultiplication symmetry. Note that in the case of $G=U(1)$, we recover the original Abelian hydro formalism.
For the purposes of this paper we will focus on the unsourced $A=0$ case, and also restrict to zero spatial dimensions. 

Note that any action constructed out of $B$s will have a $G$ premultiplication symmetry for both the right and left replica, for a total symmetry group of $G_L\times G_R$. In conventional hydro, this symmetry is broken to the diagonal by the future and past boundary conditions, but these boundary conditions are not present in the SFF case. A representative Lagrangian for $0$d non-Abelian hydro would be
\begin{equation}
L=B_{ta}(M+Mb\partial_t)B_{tr}+iM\frac{b}{2\beta}B_{ta}^2
\label{eq:nonAbelianAc}
\end{equation} 
where the $B$s are in the adjoint representation of the Lie algebra of $G$, and there is an implied summation over the representation indices.

Action (\ref{eq:nonAbelianAc}) has two time derivatives. Since we are working in 0+1d QFT, this means there is a quantum mechanics interpretation, just as in the Abelian case there was an interpretation in terms of non-unitary evolution on $U(1)^2$.

\subsection{Hamiltonian Approach for Full SSB}

We can use a Legendre transform to go from the Lagrangian \eqref{eq:nonAbelianAc} to a Hamiltonian,
\begin{equation}
H_{eff}=\frac{1}{2M}\sum L_{i1}^2-\frac{1}{2M}\sum L_{i2}^2-i\frac{b}{2\beta M}\left(\sum L_{i1}+L_{i2}\right)^2,
\end{equation}
where we sum over group generators $L_i$, which are canonically conjugate to the velocity components of $i g^{-1} \partial_t g$. The sum $\sum_i L_i^2$ is called the Casimir operator $L^2$, and has a number of important properties. Intuitively, it plays the same role as a Laplacian, but on a group manifold. Just as the Laplacian operator commutes with any momentum operator, $L^2$ commutes with all elements of the group $G$. As such it can be shown to be constant within any irreducible representation of $G$. For the Abelian case $G=U(1)$, where irreducible representations $R$ are parameterized by integer charges $q$, $L^2(R)=q^2$.

Let's evaluate
\begin{equation}
    Z=\tr \left[ e^{-iH_{eff}T}f(L^2_1)f(L^2_2) \right],
\end{equation}
where we promoted the filter function $f(Q)$ from the Abelian case to $f(L^2)$ in the non-Abelian case. Wavefunctions transform in the square of the regular representation of $R_{reg}$ of $G$. The regular representation has the contains $|R_i|$ copies of representation $R_i$. $R_{reg}\times R_{reg}$ can be decomposed as
\begin{equation}
\begin{split}
    R_{reg}\times R_{reg}=\bigoplus_{\bar R} \bar R^{\oplus n(\bar R)_{R_{reg}R_{reg}}}\\
    n(\bar R)_{R_{reg}R_{reg}}=\sum_{R_1 R_2}|R_1||R_2| n(\bar R)_{R_1R_2},
\end{split}
\end{equation}
where $n(\bar R)_{R_1R_2}$ is the number of times $\bar R$ appears in $R_1\times R_2$. For instance if $R_1$ and $R_2$ are the spin 1/2 and spin 1 representations of $SU(2)$ and $\bar R$ is the spin 1/2 representation, then $n(\bar R)_{R_1 R_2}=1$.

The Hamiltonian depends on the Casimir of each $R_i$ and on the composite system Casimir $(L_1+L_2)^2$.  The enhancement factor is thus
\begin{equation}
\begin{split}
Z=\sum_{R_1,R_2,\bar{R}} f(L^2(R_1))f(L^2(R_2))|R_1||R_2||\bar R|\times\\
\exp\left(- i \frac{[L^2(R_1)-L^2(R_1)]T}{2M} -\frac{b L^2(\bar R)T}{2\beta M}\right)n(\bar R)_{R_1R_2},
\label{eq:tripleSum}
\end{split}
\end{equation}
Here $R_1$ and $R_2$ are representations of $G$ on the first and second replica of the system, whereas $\bar R$ is a representation that lives in the Hilbert space of $H_{eff}$ on the full doubled system.

The longtime behavior is given by the $\bar R$ trivial case where $L^2(\bar R)=0$. This requires $\bar R$ to be the trivial 1D representation. If $\bar R$ is trivial,  $n(\bar R)_{R_1R_2}$ is zero unless $R_1$ is the complex conjugate of $R_2$,  in which case it is one. So the long time value is 
\begin{equation}
Z(T\rightarrow \infty)=\sum_{R_1} f^2(L^2(R_1))|R_1|^2 .
\end{equation}
This is exactly what one would predict from random matrix theory. We can also make use of $|R_1||R_2|=\sum_{\bar R}n(\bar R)_{R_1R_2} |\bar{R}|$ to show that in the short time limit this is
\begin{equation}
   Z(T \rightarrow 0) =\sum_{R_1}\sum_{R_2} f(L^2(R_1))f(L^2(R_2))|R_1|^2|R_2|^2.
\end{equation}
Since the regular representation has $|R_i|$ copies of $|R_i|$, this is essentially saying all states of all charges constructively interfere.

\subsection{Partial SBB}

If a subgroup $G'$ of $G$ is unbroken, this means overall $G'$ transformations that affect both replicas are unphysical. Thus we should gauge them out. In general, one gauges out a group by inserting a projector
\begin{equation}
P=\frac 1 {\textrm{Vol }G'}\int_{G'} dg' g'
\label{eq:projection}
\end{equation}
into the path integral. If we break representations $\bar R$ of $G$ into representations of $\bar R=\bigoplus \bar R'$ $G'$, this projects out all nontrivial representations of $G'$. For instance if $G'=G$, this will remove all the nontrivial $\bar R'$ leaving us with the result for RMT with symmetry group $G$. If $G'$ is trivial, nothing is projected out and equation \eqref{eq:tripleSum} still holds. In general, different symmetry-breaking patterns can be thought of as different degrees of freedom being observable, which means different projection operators $P$ are needed as boundary conditions.

For example, let's imagine a system with $SO(3)$ symmetry, which is broken down to $SO(2)$ by an order parameter. We are interested in some term in equation \eqref{eq:tripleSum}, for example $R_1=\textbf 5, R_2=\textbf 3, \bar R=\textbf 3$. We can verify that $n(\bar R)_{R_1R_2}=1$, meaning that when we multiply our representations we get one copy of $\bar R=\textbf 3$. What happens to this representation under the projection in equation \eqref{eq:projection}? To answer this, let's break of the representation of $G$ into representations of $G'$. The vector representation of $SO(3)$ breaks into a scalar and vector representation of $SO(2)$. Integrating $g'$ over the vector representation (as over any nontrivial representation of any group) we get zero. Integrating over the trivial representation we of course get one. So the projection operator projects down from three dimensions to one, and the $|\bar R|$ in equation \eqref{eq:tripleSum} becomes 1 instead of 3.

\section{Discussion}
In this paper we extended the understanding of quantum chaotic level repulsion to include systems with spontaneous symmetry breaking. We started with toy models with discrete symmetries, solved them, and confirmed our solutions with exact diagonalization. Next we used hydrodynamics, extending known techniques for unbroken symmetry to the case of spontaneous breaking. The technique is powerful enough to prove the correct longtime behavior, and flexible enough to handle any possible symmetry breaking pattern. Interestingly, we found that SSB typically enhances the SFF beyond that of a system with unbroken symmetry. In terms of the spectral form factor we have schematically
\begin{equation}
    \textrm{No Symmetry}<\textrm{Symmetry}<\textrm{Spontaneously Broken Symmetry}.
\end{equation}
One is left wondering how gauge symmetry might fit into that hierarchy.

In terms of SSB, the next step would be to handle higher dimensional systems. At least two interesting phenomena would reveal themselves in this case. First, the presence of sound poles associated with Goldstone modes. In higher dimension spontaneous symmetry breaking allows new terms in the hydrodynamic Lagrangian consistent with unitarity, such as $\phi_a\partial_x^2\phi_r$. This would allow the hydrodynamic variables to have sound poles, leading to a potentially rich new phenomenology in the SFF.

Higher dimensions also allows the possibility of topological effects. For instance, in a periodic system the Goldstone mode could wrap around the manifold several times. This new topological charge would lead to an expansion in the number of sectors and an additional enhancement to the SFF that could last for exponentially long times until the system tunnels into a topologically uncharged state. More exotic Goldstone manifolds and spatial manifolds would result in even more exciting topological concerns.

Finally, there is the issue of the plateau structure, entirely ignored in this paper. The lack of the hydrodynamic description of plateau behavior is made all the more striking by the fact that certain systems with some sort of resonant behavior (the peaks in figure \ref{fig:ZnMomentum}) can have `ramp' values of the SFF exceeding the final plateau value.

\section{Acknowledgements}

This work is supported in part by the Simons Foundation via the It From Qubit Collaboration (B. S.) and by the Air Force Office of Scientific Research under award number FA9550-17-1-0180 (M.W.). M.W. is also supported by the Joint Quantum Institute.

\renewcommand{\thechapter}{4}

\chapter{Emergent Spectral Form Factors in Sonic System}
\label{chapter:soundPole}
\textbf{Authors:} \textit{Michael Winer, Brian Swingle}

\textbf{Abstract:} We study the spectral form factor (SFF) for hydrodynamic systems with a sound pole, a large class including any fluid with momentum conservation and energy conservation, or any extended system with spontaneously broken continuous symmetry. We study such systems in a finite volume cavity and find that the logarithm of the hydrodynamic enhancement to the SFF is closely related to the spectral form factor of a quantum particle moving in the selfsame cavity. Depending upon the dimensionality and nature of the effective single-particle physics, these systems exhibit a range of behaviors including an intricate resonance phenomenon, emergent integrability in the SFF, and anomalously large fluctuations of the SFF.
\newpage
\section{Introduction}
The statistics of the energy spectrum is one of the most important diagnostics of quantum chaos~\cite{haake2010quantum,PhysRevLett.52.1,mehta2004random}. There is strong evidence that ensembles of chaotic systems have the same Hamiltonian spectral statistics as ensembles of Gaussian random matrices, with examples including nuclear systems~\cite{doi:10.1063/1.1703775,wigner1959group}, mechanical systems \cite{berry1977level,McDonald,BERRY1981163}, condensed matter systems~\cite{bohigas1984chaotic,altshuler1986repulsion,andreev1995spectral,dubertrand2016spectral,PhysRevLett.121.264101,Wittmann_W__2022,speck,bunin2023fisher,kamenev_1999,PhysRevResearch.3.L012019}, holographic models~\cite{Cotler2017,saad2019semiclassical,PhysRevD.98.086026,saad2019late,saad2021wormholes,saad2022convergent} and (generalizing to time-dependent Hamiltonians) circuit models \cite{shivam_2023,Chan_2018,Chan_2018Min,Chan_2019,Liao_2022}. A central quantity in spectral statistics in the Spectral Form Factor (SFF). Among other interesting properties \cite{PhysRevLett.127.230602,CCRM}, this quantity diagnoses whether energy levels repel as they do in random matrices~\cite{bohigas1984chaotic,Kos_2018,Flack_2020}, have independent Poissonian statistics~\cite{berry1977level}, or have some more exotic behavior~\cite{2020Prosen,PhysRevLett.125.250601,winerprx,Li_2021}. The SFF has spawned relatives including the partial SFF \cite{Joshi_2022}, the entanglement SFF \cite{ma2022quantum,2021,shaffer_2014}, and the Loschmidt SFF \cite{WinerLoschmidt,Weidenmuller_2005,guhr1998random}.

The SFF can be written as
\begin{equation}
    \SFF(T,f) =\overline{ | \text{Tr}[U(T) f(H)]|^2},
\end{equation}
where $U(T) = e^{-i H T}$ is the time evolution operator, $f$ is a filter function (perhaps a Gaussian centered on a specific energy band), and overline denotes a disorder average over an ensemble of Hamiltonians. These Hamiltonians can differ either through microscopic disorder or, as we will see, through the overall shape of the region containing the system. For a discussion of how different the Hamiltonians must be to constitute ``sufficient coverage'', see~\cite{WinerLoschmidt,Weidenmuller_2005,guhr1998random,https://doi.org/10.48550/arxiv.2205.12968,PhysRevLett.70.4063}. This disorder average is necessary because while the disorder-averaged SFF has smooth behavior as seen below, the SFF of a single Hamiltonian is a highly erratic function of time. A central problem in the field of quantum chaos is thus to calculate the SFF for a variety of physical systems in order to understand the general conditions under which random matrix behavior emerges.

In ~\cite{winerprx, Winer_2022}, the authors set forth a hydrodynamic theory of the spectral form factor. This effective theory predicts pure random matrix behavior at late time and computes corrections at earlier times due to slow modes. Intuitively, the presence of slow modes is related to atypically small matrix elements in the Hamiltonian, and atypically small matrix elements suppress the level repulsion which is characteristic of a random matrix. These corrections in the SFF persist until the system's ``Thouless time'', after which the random matrix behavior is recovered. In particular, approximate symmetries enhance the ramp by an amount exponential in system size, consistent with \cite{Chan_2018,Friedman_2019,Moudgalya_2021,kos_2021} . This enhancement factor was calculated for simple models of a system with conserved modes and with spontaneous symmetry breaking. These calculations included tree-level effects and diagrammatic corrections not seen in traditional hydrodynamics. \cite{winerprx} dealt exclusively with theories first order in time, and \cite{Winer_2022} dealt with systems with no spatial extent. However, the problem of hydrodynamic systems with sound poles was left unstudied. In this work, we fill in this gap, pointing out that sound pole effects can lead to new phenomena such as exponentially increasing SFFs and spectral form factors with sensitive qualitative dependence on the precise shape of the system.

\subsection{Summary of Results and Sketch of Paper}
\label{subsec:SumSketch}
The theory in \cite{winerprx,Winer_2022} expressed the enhancement of the SFF in linear hydrodynamics as a product over all modes. Here hydrodynamics (hydro) refers to a effective theory of a system's long-wavelength and long-time dynamics which incorporates the effects of conservation laws and other general constraints while coarse-graining over physics at more microscopic scales. \cite{winerprx} considered only cases where the hydro equations were first-order in time while \cite{Winer_2022} only studied systems with no spatial extent. In contrast, this paper uses the methods of hydrodynamics to study SFFs in systems with sound poles. This framework is meant to apply to systems for which the underlying microscopic theory is quantum chaotic and the long-wavelength physics is described by oscillatory sound modes. We treat the sound modes at the quadratic level within hydrodynamics (except for an appendix discussing some effects of hydrodynamic interactions), but this quadratic approximation still allows for the the modes to decay and is consistent with an underlying microscopic chaotic dynamics.

For the SFF enhancement, our hydrodynamic theory predicts that an oscillatory mode with angular frequency $\omega$ decaying at rate $\lambda$ contributes to the enhancement like $\left|\frac{1}{1-e^{i\omega T-\lambda T}}\right|^2$. This expression, which is obtained in Section 2, rapidly oscillates with angular frequency $\omega$, and the product of many such terms can generate much richer time-dependence than is possible with purely decaying modes (such as diffusive modes). The work of Sections 3, 4, and 5 is dedicated entirely to the study of these products. For this purpose, the spectrum of allowed $\omega$s is very important, and we find a variety of different behaviors depending on the statistical properties of the $\omega$s. This spectrum is determined by Laplacian of the region, or ``cavity'', in which the system resides. In particular, it is crucial to understand whether the spectrum of sound modes admits constructive or destructive interference in the product over $\omega$s.

In the case of 1d linear hydrodynamics, we find that patterns of interference impart an intricate fractal structure on the graph of the spectral form factor. There is an exponentially tall peak at every rational time in units of the system length. Unfortunately, this result is delicate; at the very least, hydrodynamic interactions destroy the sound pole in 1d~\cite{spohn_2020}, with the dynamics flowing to to the KPZ universality class \cite{KPZ,Krug1997OriginsOS,krajnik_2020}. However, we still include this discussion as a particularly simple example of the dramatic effects of oscillations and because the sound pole can survive in some special situations, such as when the local Hilbert space dimension is large.

More generally, for cavities in higher dimensions, one sees an enhancement factor that is qualitatively exponential in the single-particle SFF for a billiard of the same shape as the cavity. This surprising statement means that even though the fundamental system is made out of many microscopic degrees of freedom undergoing aperiodic motion, the full system SFF depends very strongly on the motion of wavepackets of sound. We distinguish between ``integrable'' cavities, in which the spectrum of the Laplacian is Poissonian, and ``chaotic'' cavities in which the spectrum of the Laplacian is random matrix like (although we stress that this integrable or chaotic adjective refers to the structure of the cavity modes, not to the many-body levels, which are always ultimately random-matrix-like in our models).

For example, local disordered systems in integrable cavities such as tori or ellipsoids see an exponential (in volume) enhancement to the many-body SFF corresponding to the Poissonian statistics of sound wavepackets, whereas disordered systems in a chaotic cavity, such a two-dimensional region in the shape a Bunimovich stadium billiard, see an exponentially growing enhancement to their SFF corresponding to the ramp in the SFF of a sonic wavepacket.

The rest of the paper is organized as follows. In the remainder of the introduction, we briefly review the SFF. In Section 2, we review the hydrodynamic theory of \cite{winerprx} and discuss its application to systems with a sound pole. In Section 3, we treat a one-dimensional system where the SFF exhibits an intricate fractal structure. In Section 4, we move to higher dimensions and discuss the case of a sound pole when the system is confined to an integrable cavity. In Section 5, we discuss the analogous problem in a chaotic cavity. In Section 6, we discuss interaction effects, with the preceeding discussion all at the quadratic level. This is followed by a brief discussion of the results; appendix A also contains a comparison to our prior work on SYK2.

\subsection{Review of the form factor}
\label{subsec:SFFReview}
Before proceeding to the derivations of these results, we first review the SFF in more detail. The starting point for any understanding of the spectral form factor is level repulsion. If we take the well-known joint probability density function for the $N$ eigenvalues of an $N$ by $N$ random Hermitian (GUE) matrix, we have
\begin{equation}
\begin{split}
    P(\lambda_1,\lambda_2,\lambda_3...\lambda_N)=\frac{1}{Z_N}\exp(\frac{N}{2}\sum_i \lambda_i^2)\prod_{ij}(\lambda_i-\lambda_j)^2=\\
    \frac{1}{Z_N}\exp\left(\int_{\lambda_1>\lambda_2} 2\log (\lambda_1-\lambda_2)\rho(\lambda_1)\rho(\lambda_2)d\lambda_1d\lambda_2-\int \frac{N}{2}\lambda^2\rho(\lambda) d\lambda\right).
    \label{eq:JPDF}
\end{split}
\end{equation}
It is the last term in the product on the first line that is responsible for level repulsion. Intuitively, if we zoom in on a near-degeneracy involving two levels, then the $2$ by $2$ submatrix involving those two levels only has an exact degeneracy if the coefficients of each basis matrix (Pauli $\sigma_x$, $\sigma_y$, and $\sigma_z$) are all zero, so near-degeneracy requires a special degree of fine-tuning.

The SFF is closely related to the correlation function of the density of states.
Formally, the (filtered) density of states is given by
\begin{equation} \label{eq:general_density_of_states_definition}
\rho(E, f) \equiv \sum_n f(E_n) \delta(E - E_n) = \textrm{Tr} f(H) \delta(E - H),
\end{equation}
where $n$ labels the eigenstate of $H$ with eigenvalue $E_n$, and its correlation function is
\begin{equation} \label{eq:density_of_states_correlation_definition}
C(E, \omega, f) \equiv \mathbb{E} \left[ \rho \left( E + \frac{\omega}{2}, f \right) \rho \left( E - \frac{\omega}{2}, f \right) \right].
\end{equation}
We have that
\begin{equation} \label{eq:SFF_density_of_states_relationship}
\begin{aligned}
\textrm{SFF}(T, f) &= \mathbb{E} \Big[ \textrm{Tr} f(H) e^{-iHT} \textrm{Tr} f(H) e^{iHT} \Big] \\
&= \int dE d\omega \, e^{-i \omega T} \mathbb{E} \left[ \textrm{Tr} f(H) \delta \left( E + \frac{\omega}{2} - H \right) \textrm{Tr} f(H) \delta \left( E - \frac{\omega}{2} - H \right) \right] \\
&= \int d\omega \, e^{-i \omega T} \int dE \, C(E, \omega, f).
\end{aligned}
\end{equation}
The SFF is simply the Fourier transform of the correlation function with respect to $\omega$. 

The SFF can be split into two contributions:
\begin{equation} \label{eq:general_SFF_decomposed}
\textrm{SFF}(T, f) = \big| \mathbb{E} \textrm{Tr} f(H) e^{-iHT} \big|^2 + \bigg( \mathbb{E} \left[ \big| \textrm{Tr} f(H) e^{-iHT} \big|^2 \right] - \big| \mathbb{E} \textrm{Tr} f(H) e^{-iHT} \big|^2 \bigg).
\end{equation}
The first term, the disconnected part of the SFF, comes solely from the average density of states. It is the absolute value square of the density's Fourier Transform.

It is the second term, the connected part of the SFF, that contains interesting information on the correlation between energy densities.
``Random matrix universality''~\cite{PhysRevLett.52.1,Altland1997} is the principle that an ensemble of quantum chaotic Hamiltonians will generically have the same \textit{connected} SFF as the canonical Gaussian ensembles of random matrix theory~\cite{mehta2004random,tao2012topics}.
This behavior is illustrated in Fig.~\ref{fig:SFFgraph}, which plots the disorder-averaged SFF of the Gaussian unitary ensemble (one of the aforementioned canonical ensembles).
The graph shows the three regimes of the random matrix theory SFF:
\begin{figure}
\centering
\includegraphics[width=0.9\textwidth]{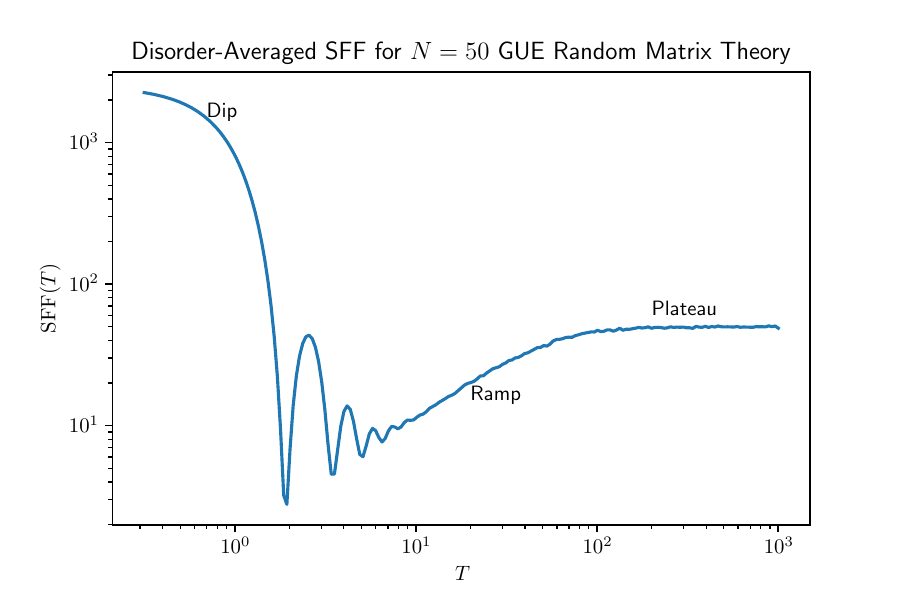}
\caption{The disorder-averaged SFF for the Gaussian unitary ensemble (GUE). The matrices in this ensemble have dimension $N = 50$. The SFF was computed numerically by averaging over ten thousand realizations. The three regimes --- dip, ramp, plateau --- are labeled.}
\label{fig:SFFgraph}
\end{figure}
\begin{itemize}
\item The ``dip'', occurring at early times, comes from the disconnected piece of the SFF (and thus its precise shape is non-universal and depends on the details of $f$ and the thermodynamics of the system).
Its downward nature reflects a loss of constructive interference --- the different terms of $\textrm{Tr} e^{-iHT}$ acquire different phase factors as $T$ increases.
\item The ``ramp'', occurring at intermediate times, is arguably the most interesting regime.
In canonical random matrix ensembles, the ramp follows from the result~\cite{mehta2004random}
\begin{equation} 
\mathbb{E} \left[ \rho \left( E + \frac{\omega}{2} \right) \rho \left( E - \frac{\omega}{2} \right) \right] - \mathbb{E} \left[ \rho \left( E + \frac{\omega}{2} \right) \right] \mathbb{E} \left[ \rho \left( E - \frac{\omega}{2} \right) \right] \sim -\frac{1}{\bbeta \pi^2 \omega^2},
\label{eq:repulsion}
\end{equation}
where $\bbeta = 1$, $2$, $4$ for the orthogonal, unitary, and symplectic ensembles respectively \cite{mehta2004random}.
The fact that the right hand side is negative is known as level repulsion in quantum chaotic systems \cite{wigner1959group}. Since the energy levels of a system repel each other, fluctuations in the density of states are suppressed. Long-wavelength (short time) fluctuations are suppressed the most, short-wavelength (long time) fluctuations are suppressed the least. 
Taking the Fourier transform of equation \ref{eq:repulsion} respect to $\omega$ gives a term proportional to $T$ for the connected SFF.
Such a linear-in-$T$ ramp is often taken as a defining signature of quantum chaos.
\item The ``plateau'', occurring at late times, is fundamentally a consequence of the discreteness of the spectrum.
At times much larger than the inverse level spacing or ``Heisenberg time'', all off-diagonal terms in the double-trace of the SFF average to zero, meaning that
\begin{equation} \label{eq:SFF_plateau_derivation}
\textrm{SFF}(T, f) \approx \sum_{mn} e^{-i(E_m - E_n)T} f(E_m) f(E_n) \sim \sum_n f(E_n)^2.
\end{equation}
For integrable systems, the plateau is reached very quickly with little to no ramp regime \cite{berry1977level,PhysRevLett.125.250601,winerprl}. For chaotic systems it isn't reached until a time exponential in system size. Nonetheless, it is the long-term fate of any system without a degenerate spectrum.
\end{itemize}
This paper will focus on the connected SFF at time scales towards the beginning of the ramp, scaling polynomially in system size instead of exponentially. At this time scale, physical systems do not behave exactly like random matrices, and exciting new phenomena can be observed. The timescale separating these non-universal phenomena from the linear ramp of RMT is known as the Thouless time, see e.g. \cite{Gharibyan_2018,Chan_2018} for recent discussions in the many-body context.

Before we specialize to the case of hydrodynamics, a word on filter functions $f$. One common choice of $f$ is $f(E)=e^{-\beta E}$. It is a matter of Fourier analysis to show that this would give a ramp looking like $\int_{E_{\min}}^{E_{\max}} dE e^{-2\beta E}\frac{T}{2\pi}$. This integral would be dominated by the lowest-energy region of the spectrum. Importantly, the region of the spectrum which in thermodynamics corresponds to inverse temperature $\beta$ is often very heavily suppressed. The physical reason why the coefficient of the ramp doesn't depend on the density of states (and hence why the densest region of the spectrum cannot dominate) can be seen from the equation \eqref{eq:JPDF}. The distribution of the eigenvalues can be interpreted as a Boltzmann distribution where the eigenvalues are particles in some confining potential $NV(x)$ ($V = \frac{x^2}{2}$ for a Gaussian random matrix) that repel each other with potential $2\log|\lambda_i-\lambda_j|$. If we approximate $\rho$ as a continuous density instead of a train of $\delta$ functions, then there is some saddle point density which can be solved from an integro-differential equation involving $V$. The connected SFF, however can be seen as the fluctations in $\rho$. At short times and large energies, where this approximation is valid, we see that the energy function we are trying to minimize is purely quadratic in the $\rho$s, so fluctuations shouldn't depend on the background $\rho$. A longer discussion on this point can be found in \cite{haake2010quantum}.

Because canonical SFFs select only dynamics near the ground state, we need a different approach. If we are interested in spectral statistics far from the ground state, we must instead choose an $f$ like $f(E)=e^{-(E-E_0)^2/4\sigma^2}$, which samples around the energy window of interest. This well-known but unusual property puts spectral statistics in contrast with thermodynamics, where canonical and microcanonical ensembles are equivalent.

In this paper, we discuss the connected SFF. We restrict ourselves to timescales long enough for hydrodynamic effects to kick in. We pay attention only to times much less than the many-body Heisenberg time, after which generic systems have no interesting spectral properties. Our results are non-trivial only at times less than the decay-time for the slowest sound mode, which is quadratic in system length. For a more careful discussion of different timescales in this work, see the end of section \ref{sec:freeStadium}.

\section{Overview of the Hydrodynamic Spectral Form Factor}
\label{section:DPT}

Our approach to the problem is to formulate an effective theory of the SFF contour. We do this by comparing two different but related sets of contours and arguing for a relationship between their effective descriptions. We will suppress the time $T$ and filter function $f$ in many of the expressions below.

\begin{figure}
    \centering
    \includegraphics[scale=0.12]{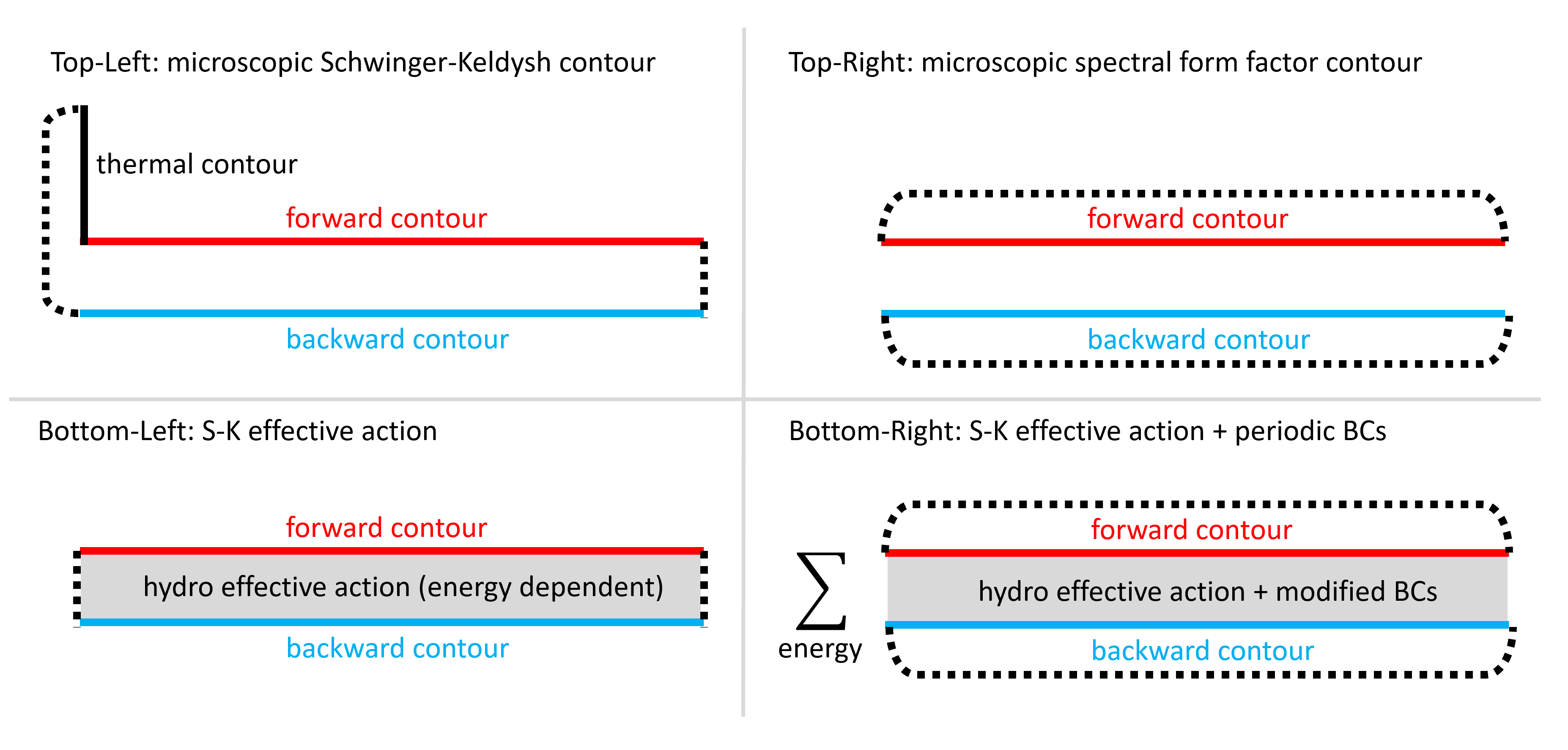}
    \caption{The Schwinger-Keldysh (S-K) contour (left) has a thermal circle of inverse temperature $\beta$ and forward and backwards time evolution legs. By integrating out local microscopic degrees of freedom (bottom), one arrives at an effective theory at the same temperature. The integration couples the left and right legs of the contour. The SFF contour (right) consists of two totally disconnected legs each with periodic boundary condition. By integrating out local microscopic degrees of freedom (bottom), one arrives at the same effective theory as in the Schwinger-Keldysh case (at least up to loop corrections). Remarkably, this integration results in a coupling between the two otherwise decoupled legs. This figure first appeared in \cite{winerprx}}
    \label{fig:masterpiece4}
\end{figure}

Case 1: The Schwinger-Keldysh (S-K or Kel) contour (Figure~\ref{fig:masterpiece4}, left), which computes
\begin{equation}
    Z_{\text{Kel}} = \text{tr}( U[A_1] \rho U[A_2]^\dagger),
\end{equation}
where $\rho$ is the initial state and the $A_i$ are backgrounds fields along the forward ($A_1$) and backward ($A_2$) legs of the contour. In this case, we have a single trace and an initial state which sets the background value of the energy and other conserved charges. 

Case 2: The form factor contour (Figure~\ref{fig:masterpiece4}, right), which computes 
\begin{equation}
    \text{SFF} = \tr(f(H) U[A_1]) \tr(f(H) U[A_2]^\dagger),
\end{equation}
where again we allow for different background fields on the forward and backward contours. In this case, we have two traces and no initial state, although we can consider the addition of a filter function $f$ which can be used to select part of the energy spectrum of interest. 

It is important to note that the SFF will be an erratic function of time; we must include some implicit averaging, for example, over time or over an ensemble of Hamiltonians, to give a smooth function of time. We can also compute the same ensemble average of the Schwinger-Keldysh contour, but we will here assume that the SK contour computes self-averaging observables that are only weakly affected by the ensemble average. Henceforth, we compare ensemble averaged versions of Case 1 and Case 2.

The Schwinger-Keldysh contour (Case 1) computes a generating function that gives access, via derivatives with respect to the background fields, to correlators and response functions. It is typically used as a tool to compute dynamical properties out of equilibrium. For our purposes here, we have in mind the Schwinger-Keldysh contour as a tool to compute the long wavelength hydrodynamic response of the system. As we review below, one can formulate an effective theory using the degrees of freedom on the SK contour which can be used to describe the hydrodynamic response of the system.

The SFF contour (Case 2) computes a generalization of the spectral form factor, with the usual form factor being recovered when the background fields are set to zero on each contour. At late times, our expectation is that the SFF will approach that of an appropriate random matrix, but at early times, especially in a system with slow hydrodynamic modes, one can expect systematic deviations from the random matrix result. We are interested in formulating an effective theory for the SFF contour, analogous to the hydrodynamic effective theory of the SK contour.

In a prior work, we gave a proposal for an effective theory of the SFF contour. We now review that proposal and the arguments in favor of it. The basic observation is that the Schwinger-Keldysh and SFF contours can both be viewed as different ways summing matrix elements of the same unitary operator, $U[A_1] \otimes U[A_2]^*$, defined on two copies of the system. For the Schwinger-Keldysh contour (Case 1), we connect the two copies to each other via a link at the initial time which encodes the initial state and a link at the final time which encodes the single trace. For the SFF contour (Case 2), we connect each copy to itself via a link from the initial time to the final time which encodes the two traces.

Based on this clue, we proposed that an effective theory of the Schwinger-Keldysh contour should also give an effective theory of the SFF contour provided the boundary conditions are modified appropriately. As a reminder, this proposal should be viewed as a statement about the ensemble averaged Schwinger-Keldysh contour and the ensemble averaged SFF contour. For the Schwinger-Keldysh contour, the boundary conditions set the value of the total energy (initial state) and introduce a fine-grained correlation between the final states of the two legs (final trace). For the SFF contour, there is no built in correlation between the fine-grained states of the two legs, but the ensemble average produces such a correlation. The SFF contour also does not have a set value of the total energy (or other conserved quantities), so we must integrate over them.

The final proposal is then as follows. Suppose, for a given total energy $E_0$, we have some effective fields $\phi_{1,2}$ which compute $Z_{\text{Kel}}$ as
\begin{equation}
    Z_{\text{Kel}}[A] = \int \mathcal{D}\phi e^{i \int dt d^d x L_{\text{Kel}}[\phi,A;E]},
\end{equation}
where the effective action $W_{SK}$ depends on the background energy $E_0$. Then, we can write a similar effective theory to compute the $\text{SFF}$,
\begin{equation}
    \text{SFF} = \int dE_0 \int  \mathcal{D}' \phi e^{i \int dt d^d x L_{\text{SFF}}[\phi,A;E_0]},
\end{equation}
where 
\begin{itemize}
    \item we now integrate over the total energy $E_0$ (and other conserved charges),
    \item we modify the boundary conditions of $\phi$ to be periodic in time on each contour, 
    \item and we have $L_{\text{SFF}} \approx L_{\text{Kel}}$.
\end{itemize}
The argument for this final point is that the fast decaying degrees of freedom which have been integrated out to yield $L_{\text{Kel}}$ are not significantly affected by the change in boundary conditions provided the time duration $T$ is much larger than the lifetime $\tau_{\text{fast}}$ of any integrated out mode. Performing the same integrating out procedure on the SFF contour thus gives 
\begin{equation}
    L_{\text{SFF}} = L_{\text{Kel}} + \mathcal{O}(e^{-T/\tau_{\text{fast}}}).
\end{equation}

In subsection \ref{subsec:CTP4} we quickly review the closed time path (CTP) formalism which deals with the Schwinger-Keldysh contour. In subsection \ref{subsec:DPT} we discuss the doubled periodic time (DPT) effective theory, a cousin of CTP which deals with the SFF contour. Finally in subsection \ref{subsec:soundPole} we discuss a special modification of the CTP and DPT for systems with oscillatory modes.

\subsection{Review of Closed Time Path Formalism}
\label{subsec:CTP4}

Hydrodynamics is the program of creating effective field theories (EFTs) for systems based on the principle that long-time and long-range physics is driven primarily by conservation laws and other protected slow modes. One particularly useful formulation is the CTP formalism explained concisely in ~\cite{glorioso2018lectures} and in more detail in ~\cite{crossley2017effective,Glorioso_2017,gao2018ghostbusters}. Other approaches to fluctuating hydrodyamics can be found in~\cite{Grozdanov_2015,Kovtun_2012,Dubovsky_2012,Endlich_2013}.

The CTP formalism lives on the Schwinger-Keldysh contour, pictured in figure \ref{fig:masterpiece4}. Its central object of study is the partition function
\begin{equation}
    Z_{\text{Kel}}[A^\mu_1(t,x),A^\mu_2(t,x)]=\tr \left( e^{-\beta H} \mathcal{P} e^{i\int dt d^d x A^\mu_1j_{1\mu}} \mathcal{P} e^{-i\int dt d^d x A^\mu_2 j_{2\mu}}\right),
\end{equation}
where $\mathcal P$ is a path ordering on the Schwinger-Keldish contour. The $j$ operators are local conserved currents. 

For $A_1=A_2=0$, $Z_{\text{Kel}}$ reduces to thermal partition function at inverse temperature $\beta$. Differentiating $Z_{\text{Kel}}$ with respect to the $A$s generates insertions of the conserved current density $j_\mu$ along either leg of the Schwinger-Keldysh contour. Thus $Z_{\text{Kel}}$ is the generating function of all possible contour-ordered correlation functions of current operators. In particular, for systems with a conserved energy, the energy density operator can always be extracted from the hydrodynamic action.

One can write $Z_{\text{Kel}}$ as 
\begin{equation}
      Z_{\text{Kel}}[A^\mu_1,A^\mu_2]=\int \mathcal D \phi^i_1\mathcal D \phi^i_2 \exp\left(i\int dt d^d x L_{\text{Kel}}[A_{1\mu},A_{2\mu},\phi^i_1,\phi^i_2]\right),
\end{equation}
for some collection of local fields $\phi$s. The fundamental insight of hydrodynamics is that at long times and distances, any massive $\phi$s can be integrated out. All that's left over is one $\phi$ per contour to enforce the conservation law $\partial^\mu j_{i\mu}=0$. Our partition function can be written 
\begin{equation}
    \begin{split}
         Z_{\text{Kel}}[A^\mu_1,A^\mu_2]=\int \mathcal D \phi_1\mathcal D \phi_2 \exp\left(i\int dt d^dx L_{\text{Kel}}[B_{1\mu},B_{2\mu}]\right),\\
         B_{i\mu}(t,x)=\partial_\mu \phi_i(t,x)+A_{i\mu}(t,x).
    \end{split}
    \label{eq:abelianB4}
\end{equation}
Insertions of the currents are obtained by differentiating $Z_{\text{Kel}}$ with respect to the background gauge fields $A_{i\mu}$. A single such functional derivative gives a single insertion of the current, and so one presentation of current conservation is the identity $\partial_\mu \frac{\delta Z_{\text{Kel}}}{\delta A_{i\mu}} = 0$.

The effective Lagrangian $L_{\text{Kel}}$ satisfies a number of constraints. The most important is locality. There are no slow modes besides $\phi$, and at long enough distance and time scales integrating out fast modes should yield a local Lagrangian depending on $B_{1,2}$ and their derivatives. There are several additional constraints following from unitarity. They are best expressed in terms of
\begin{equation}
\begin{split}
    B_a=B_1-B_2,\\
    B_r=\frac{B_1+B_2}{2}.
\end{split}
\end{equation}
The key constraints, which will not be proven here but are proven in, say, \cite{crossley2017effective}, are:
\begin{itemize}
    \item All terms in $L_{\text{Kel}}$ have at least one factor of $B_a$, that is $L_{\text{Kel}}=0$ when $B_a=0$.
    \item Terms odd (even) in $B_a$ make a real (imaginary) contribution to the action.
    \item All imaginary contributions to the action are positive imaginary (or zero).
    \item Any correlator in which the chronologically last variable has $a$-type will evaluate to 0 (known as the last time theorem or LTT).
    \item A KMS constraint imposing fluctuation-dissipation relations.
    \item Unless the symmetry is spontaneously broken, all factors of $B_r$ have at least one time derivative. This condition will be lifted in this paper, as we will be considering systems with a spontaneously broken symmetry.
\end{itemize}
For many applications, including calculating SFFs, one typically sets the external sources $A$ to zero, so the action can be written purely in terms of the derivatives of the $\phi$s.

The $\phi$s often have a physical interpretation depending on the precise symmetry in question. In the case of time translation, the $\phi$s are the physical time corresponding to a given fluid time (and are often denoted $\sigma$). In the case of a U(1) internal symmetry, they are local fluid phases. One simple quadratic action which is consistent with the above rules and which describes an energy-conserving system exhibiting diffusive energy transport is (with $\phi_{a,r} \rightarrow \sigma_{a,r}$)
\begin{equation}
    L_{\text{Kel}}^{\text{(diffusion)}}=\sigma_a\left(\kappa \beta^{-1}\partial_t^2\sigma_r-D\kappa \beta^{-1}\nabla^2\partial_t \sigma_r\right)+i\beta^{-2}\kappa(\nabla \sigma_a)^2. 
    \label{eq:lhydro1}
\end{equation}
Here $\kappa$ and $D$ can all be viewed as functions of the background energy $E_0$ which is set by the temperature $\beta$. Later, it will be convenient to also view the $\beta$ appearing in \eqref{eq:lhydro1} as a function of the background energy $E_0$.

\subsection{Review of the Doubled Periodic Time Formalism}
\label{subsec:DPT}
We now explain in more detail how our DPT theory~\cite{winerprx}, which is built from the CTP formalism, can be used to compute SFFs. We focus for concreteness on the diffusive action \eqref{eq:lhydro1} as an example. The hydro approach to the SFF predicts that the SFF can be obtained by evaluating the following path integral (still with $\phi_{a,r} \rightarrow \sigma_{a,r}$),
\begin{equation}
    \textrm{SFF}(T,f) = \int_{\mathcal C} \mathcal{D} \sigma_1 \mathcal{D} \sigma_2 f(E_1)f(E_2)e^{i \int dt d^d x L_{\text{Kel}}(\sigma_1,\sigma_2)}.
    \label{eq:SFFhydro}
\end{equation}
Here $\mathcal C$ represents the SFF contour seen on the right of figure \ref{fig:masterpiece4}. This contour has two disconnected legs each with real-time periodicity $T$. $\sigma_{1,2}$ represent time reparameterization modes on the two legs of the contour.

The contour in equation \eqref{eq:SFFhydro} should be contrasted with the Schwinger-Keldysh contour. These contours both have two long legs, but very different boundary conditions. The fact that an action initially written to calculate two-point functions in fluctuating hydrodynamics on the SK contour can---when evaluated with different boundary conditions---calculate the spectral form factor is the surprising result of \cite{winerprx}.

The key change is in the boundary conditions of the fields. To see this, we first define
\begin{equation}
\begin{split}
    \rho=\frac {\partial L_{\text{Kel}}}{\partial(\partial_t \sigma_a)} = - \kappa \beta^{-1} \partial_t \sigma_r ,
\end{split}
\end{equation}
which can be thought of as the average energy density on the two contours. Consider the spatial- and time-zero-modes of $\rho$ and $\sigma_a$, $\rho_0$ and $\sigma_{a,0}$. $\rho_{0}$ is nothing but the total energy $E_0$ while $\sigma_{a,0}$ is the total relative time shift between the two contours. In the SK contour, $\rho_0$ is set by the initial state, which gives a strongly peaked probability distribution for $\rho_0$. Similarly, in the SK contour, $\sigma_{a,0}$ is fixed to be zero since the times on the two contours are fixed to agree in the far future. By contrast, on the SFF contour the overall energy is not constrained by an initial state. The filter functions can select an energy, and they contribute to the path integral as $f(\rho_0)^2$. (Contributions in which the energies on the two contours are substantially different are suppressed.) Similarly, on the SFF contour the overall time shift is not fixed to be zero, so $\sigma_{a,0}$ should now be integrated over. The domain of integration is periodic since time is identified. In this way, the zero modes produce the expected random matrix ramp,
\begin{equation}
    \text{SFF} \sim \int d\sigma_{a,0} \int dE_0 = T \int d E_0.
\end{equation}

Now we review the calculation in more detail, including both zero- and non-zero-modes. Using the definition of $\rho$, we can rewrite equation \eqref{eq:lhydro1}
\begin{equation}
   L_{\text{Kel}}^{\text{(diffusion)}}=-\sigma_a\left(\partial_t\rho-D\nabla^2\rho\right)+i\beta^{-2}\kappa(\nabla \sigma_a)^2. 
    \label{eq:lhydro2}
\end{equation}
Since the action is entirely Gaussian, we can evaluate the path integral exactly. We first break into Fourier modes in the spatial directions. The integral becomes
\begin{equation}
    \prod_{k}\int\mathcal D\epsilon_k\mathcal D\sigma_{ak}f(E_1)f(E_2) \exp\left(-i\int dt \sigma_{ak}
    \partial_t\rho_k+Dk^2\sigma_{ak}\rho_k-\beta^{-2} \kappa k^2 \sigma_{ak}^2\right) 
\end{equation}
For $k\neq 0$, breaking the path integral into time modes gives an infinite product which evalutes~\cite{winerprx} to $\frac{1}{1-e^{-Dk^2T}}$ (and there is indeed no explicit $\kappa$ dependence). For $k=0$, we just integrate over the full manifold of possible $\sigma_a$s and $\rho$s to get $\frac T{2\pi} \int f^2(E_0) dE_0$, up to an overall factor that depends on the measure. So the full connected SFF for systems with a single diffusive mode is
\begin{equation}
   \textrm{SFF}= \frac T{2\pi} \int f^2(E_0) dE_0 \prod_k \frac{1}{1-e^{-D(E) k^2T}} .
   \label{eq:SFFProduct}
\end{equation}
We emphasize again that $D(E_0)$ depends on the background energy $E_0$ and we are integrating over $E_0$. 

Let us now suppose we work in the thermodynamic limit and use a Gaussian filter $f$ to select a particular background energy $E_0$ (up to subextensive fluctuations). In this large-volume limit, the product over non-zero modes can be evaluated by taking the log and then Taylor expanding $\log(1-e^{-Dk^2T})$. The result is
\begin{equation}
    \log \frac{\textrm{SFF}}{\text{SFF}_{\text{zero-modes}}} \approx V\left(\frac{1}{4\pi DT}\right)^{d/2} \zeta(1+d/2),
    \label{eq:diffusiveSFF}
\end{equation}
where $\zeta(s)=\sum_{n=1}^\infty \frac{1}{n^s}$ is the famous Riemann zeta function. This approximation breaks down at times near the Thouless time of the system $T=1/(DL^2)$, where $L$ is the characteristic length. At this point the product over modes in equation \eqref{eq:SFFProduct} can be better approximated as just 1, and the ramp takes on the value one would expect from conventional random matrix theory. The form of \eqref{eq:diffusiveSFF} agrees with the expression derived in~\cite{Friedman_2019} in the limit of large local Hilbert space dimension; the hydro theory predicts \eqref{eq:diffusiveSFF} just given the diffusive dynamics and it can be used to show that that the leading perturbative correction due to hydrodynamic interactions (non-quadratic terms in $L_{\text{Kel}}$) is small when the volume is large.

The for a huge array of systems (including many with sound modes), the product $\prod_k \frac{1}{1-e^{-D(E) k^2T}}$ actually has a physical interpretation as the Total Return Probability (TRP)~\cite{winerprx}. If one partitions the configuration space into sectors labeled by $i,j$, then one can define $p_{i\to j}(T)$ as the probability that a system starting in sector $i$ is in sector $j$ at time $T$. The total return probability is
\begin{equation}
    \trp(T)=\sum_{i}p_{i\to i}(T).
\end{equation}
Remarkably, if the sectors are small enough that one can't tell where in a sector the system started after time $T$, then this quantity is very resilient to how exactly sectors are chosen. For instance cutting a sector $i$ into $i', i''$ will replace $p_{i\to i}$ with $p_{i'\to i'}+p_{i'\to i'}=p_{i\to i'}+p_{i\to i''}=p_{i \to i}$. If one chooses to have each sector be a single configuration $\psi$ then the TRP can be written 

\begin{equation}
    \trp=\sum_{\psi} \left|\expval{e^{-iHT}}{\psi}\right|^2.
\end{equation}
The TRP has an interpretation as measuring how much a system still remembers after time $T$. If a system has not spread through configuration space, the TRP will be large, whereas if the system has forgotten its initial configuration the TRP will be one. In purely dissipative systems, the TRP will always be greater than one, and the connected SFF will always be larger than the RMT result. But for systems with some oscillatory character then TRP's behavior can be much more complicated.

That completes our review of the DPT formalism in the context of diffusive dynamics. For a general quadratic hydro theory, the frequencies $\omega=iDk^2$ will be replaced by a more general set of modes $\omega_j(k)$. The parameters specifying this dispersion may also depend on the background energy and the background values of other conserved quantities. The general SFF predicted by the DPT formalism is then
\begin{equation}
    \textrm{SFF}= \frac T{2\pi} \int f^2(E_0) dE_0 \prod_{\textrm{Slow Modes }\phi_j}\prod_{k}\frac{1}{1-e^{i\omega_j(k)T}}.
\end{equation}
We again assume for simplicity that $f$ is chosen to select a particular background energy for which the parameters of $\omega_j(k)$ take some particular value. Then in terms of the random matrix form factor,
\begin{equation}
     \textrm{SFF}_{\textrm{GUE}} =  \frac T{2\pi} \int f^2(E) dE,
\end{equation}
the ramp is enhanced by a factor of 
\begin{equation}
    \Zs = \frac{ \textrm{SFF}}{ \textrm{SFF}_{\textrm{GUE}}} = \prod_{\textrm{Slow Modes }\phi_j}\prod_{k}\frac{1}{1-e^{i\omega_j(k)T}}.
    \label{eq:enhancementDiff}
\end{equation}
The rest of this paper will focus on the computation the product in \eqref{eq:enhancementDiff} for various systems with sound poles. We find our results depend in detail on the geometry of the system, in stark contrast with the diffusive result in equation \ref{eq:diffusiveSFF}, which depends on on volume.

\subsection{Doubled Periodic Time Formalism with a Sound Pole}
\label{subsec:soundPole}
We are finally ready to tackle the problem of sound poles SFF hydrodynamics. To do so, we simple need to specify the frequencies $\omega_j$ entering into \eqref{eq:enhancementDiff}. We use a simple model of sound poles described by the hydro Lagrangian
\begin{equation}
    L_{\text{Kel}}^{\text{(sonic)}} =\frac 12 \phi_a\left(\partial_t^2+\frac{2\Gamma}{c^2} \partial_t^3-c^2\partial_\mu^2\right)\phi_r+\frac{2i\Gamma}{\beta c^2}\phi_a\partial_t^2\phi_a +\textrm{higher derivative terms}.
\end{equation}
This sort of Lagrangian might arise in a superfluid, where $\phi$ plays the role of an order parameter. Alternatively this can be taken as a minimal schematic for a system with a more conventional sound pole, such as familiar fluid systems with energy and momentum conservation.

This system has a characteristic length scale $\ell=\Gamma/c$. Physically, this is the scale below which hydrodynamics breaks down, and corresponds roughly to the mean free path of the constituent molecules.

In general this Lagrangian is cubic in frequency, and the equations of motion gives us three solutions for $\omega$. For small $k$ such that $k\ell\ll 1$, these solutions are  $\omega\approx \pm ck+i\Gamma k^2$, and $\omega \approx i\frac{2\Gamma}{c^2}$. This last mode is fast, and can be ignored in the infrared. Note that the solutions for $i\omega$ are either real or come in complex-conjugate pairs, so the enhancement in \eqref{eq:enhancementDiff} is real as it must be.

Ignoring the fast decaying mode, we finally have our formula for the SFF enhancement in the presence of a sound pole,
\begin{equation}
\log \Zs=-\sum_{k, \mathfrak{s}=\pm}\log(1-\exp\{(i\mathfrak{s}ck-\Gamma k^2)T\}).
\label{eq:enhancement}
\end{equation}
where sums over $k$ are always over the numbers $k>0$ such that $k^2$ is an eigenvalue of the Laplacian in the system holding our fluid.

By Taylor expanding the log on the right, we also have the alternate formula
\begin{equation}
\log \Zs=\sum_{j=1}^\infty \sum_{k^2,\mathfrak{s}}\frac{\exp(j(i\mathfrak{s}ck-\Gamma k^2)T)}j.
\label{eq:enhancementExpanded}
\end{equation}
The presence of the complex exponentials in equation \eqref{eq:enhancementExpanded} leads to qualitatively new behavior not seen in purely diffusive hydro. While systems without sound poles see $\Zs$ decay monotonicly as $T$ increases, in sonic systems we see an intricate interplay between the positive and negative terms in \eqref{eq:enhancementExpanded}.

We will now compute the enhancement in a variety of sonic scenarios. Corrections to the formula \eqref{eq:enhancement} arise from higher-derivative terms and non-Gaussian terms not included in $L_{\text{Kel}}^{\text{(sonic)}}$. Corrections can also arise from the finite width of the filter function.

\section{Babylon: The 1D Sound Pole}

\begin{figure}
\centering
    \includegraphics[scale=0.5]{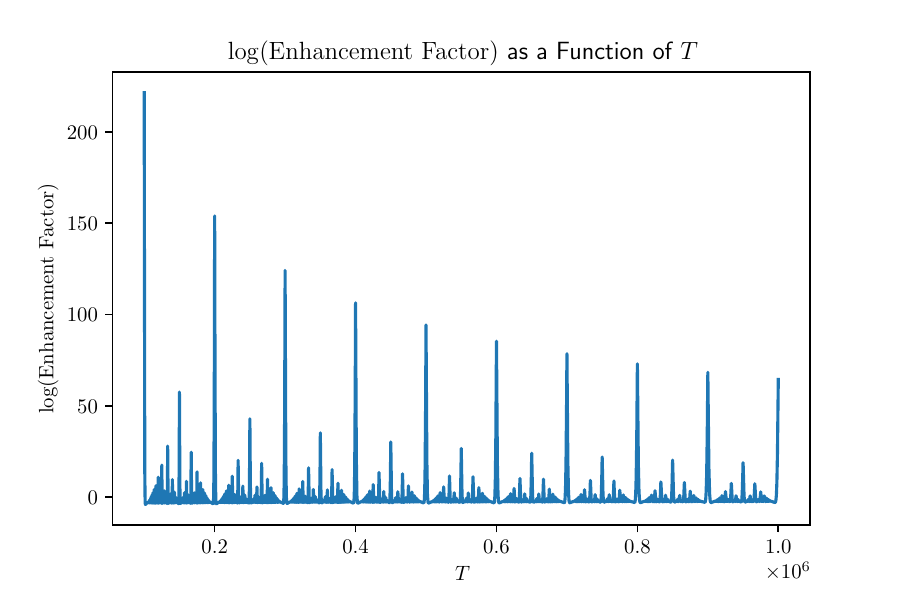}\includegraphics[scale=0.5]{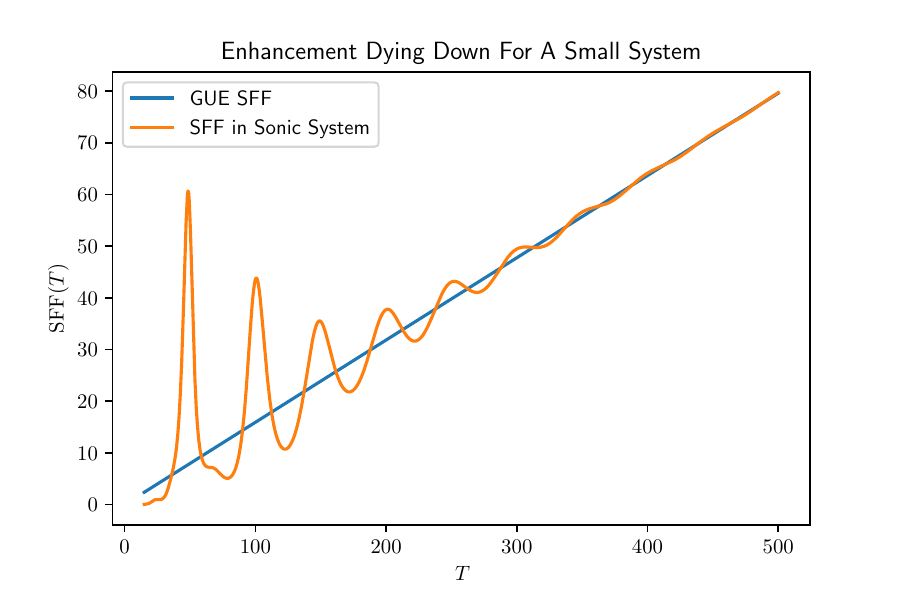}
    \caption{The first image shows the log of the SFF enhancement factor for a system of length $L=10^5$ with $c=1$, $\Gamma=1$, as calculated in equation \ref{eq:SFF1Dproduct}. The graph cuts off at a lower bound of $T=9\times 10^4$, otherwise the $T=0$ divergence would overwhelm the rest of the image. The second graph shows a more sedate choice of $L=50$ $c=1$, $\Gamma=1$. At this smaller scale, only fifty times the mean free path, fewer terms in product \ref{eq:enhancementDiff} and the enhancement is small enough that the sonic SFF and the (connected) GUE SFF can be plotted on the same axes. Note the oscillatory behavior, and the fact that at certain times the sonic SFF is actually smaller than the pure random matrix system.}
    \label{fig:blowup}
\end{figure}

In this section we will concern ourselves with a very specific problem: the SFF enhancement for a 1D system with periodic boundary conditions. In this case, the $k$s are just $2\pi n/L$. In order to get our bearings, let's first evaluate equation \eqref{eq:enhancement} numerically. For a particular parameter choice, the results are shown in figure \ref{fig:blowup}.

\begin{figure}
    \centering
    \includegraphics{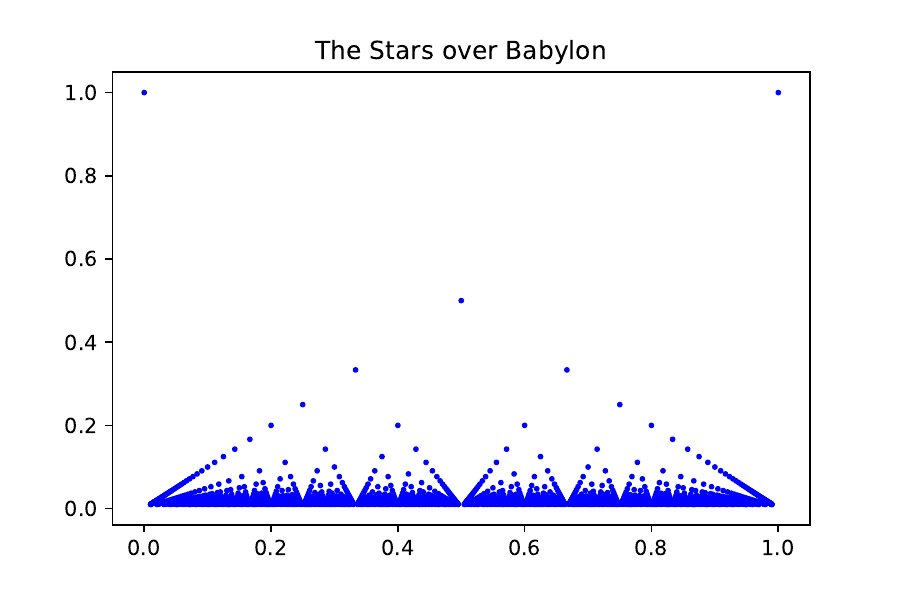}
    \caption{The Stars over Babylon. The Stars over Babylon function in this range corresponds to one period in figure \ref{fig:blowup}.}
    \label{fig:babylon}
\end{figure}

Before we can understand this fascinating picture, we need to discuss the function
\begin{equation}
    f(x) = 
\begin{cases}
  \frac{1}{n} &\text{if }x = \tfrac{m}{n}\quad (x \text{ is rational), with } m \in \mathbb Z \text{ and } n \in \mathbb N \text{ coprime}\\
  0           &\text{if }x \text{ is irrational.}
\end{cases}
\label{eq:babylon}
\end{equation}
This function is known by many fanciful names including the popcorn function, the raindrop function, the countable cloud function, and, best of all, the Stars over Babylon. The function is graphed in figure \ref{fig:babylon}.

What does this fairytale function have to do the sums in equations \eqref{eq:enhancement}? The answer is that there is a huge enhancement whenever the system is in resonance, that is whenever $cT/L$ is rational. Setting $k= 2\pi \q/L$ and $cT/L=m/n$ the sum in equation \eqref{eq:enhancement} becomes
\begin{equation}
\begin{split}
    \log \Zs=-2 \sum_{\q>0, \mathfrak{s}=\pm}\log(1-\exp\left\{\left(2\pi i\mathfrak{s} \frac{c\q}{L} -\Gamma \frac{4\pi^2\q^2}{L^2}\right)T\right\})\\
    =-2 \sum_{\q>0, \mathfrak{s}=\pm}\log(1-\exp\left\{\left(2\pi i\mathfrak{s} \frac{m \q}{n} -\Gamma \frac{4\pi^2q^2}{L^2}\right)T\right\}).
    \label{eq:SFF1Dproduct}
\end{split}
\end{equation}
so every $n$th term in the sum over $q$ contributes a term $-\log(1-\exp(-\Gamma k^2 T))$, which is very close to $-\log 0$ when $k$ is small. Since only a $1/n$ fraction of the modes contribute, and contributions like this swamp out any others, we get a structure like equation \eqref{eq:babylon}. A more careful accounting in the next subsection will reveal that the Babylon formula is actually modified to $n^{-3/2}$ from $1/n$. Here $n$, as the denominator in a Babylon-like function, plays the role of a sort of order. Smaller $n$ spikes are more resilient to dissipation while larger $n$ spikes get wiped out more quickly.

\subsection{Details on the 1D Spectral Form Factor}

\begin{figure}
\begin{center}
    \includegraphics[scale=0.8]{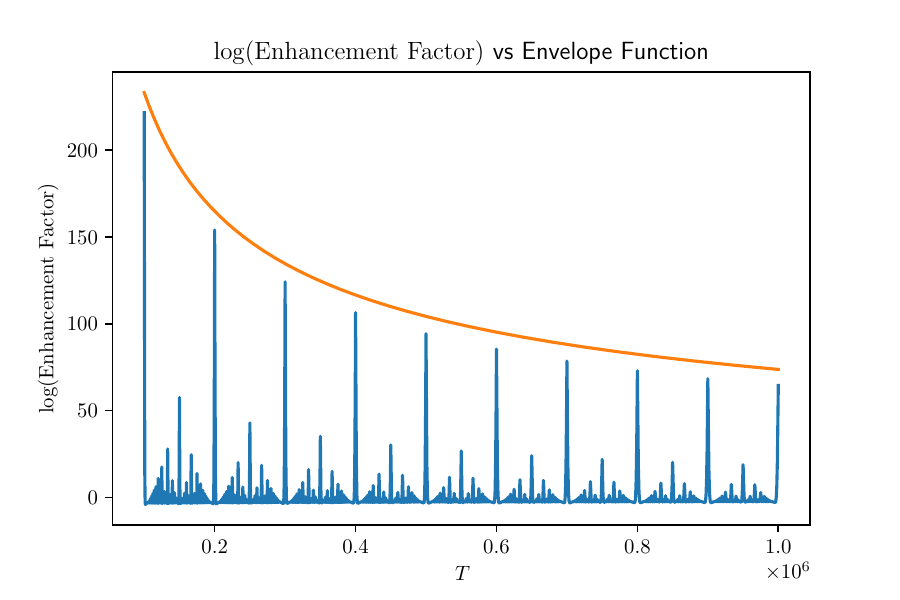}
    \end{center}
    \caption{Equation \eqref{eq:evelope1} does a good job estimating the envelope for $L=1,000,000$, $c=1$, $\Gamma=1$. The source of the discrepancy comes from approximating the sum in equation \eqref{eq:envelopeSum} as an integral. This approximation is valid in the limit $cL\gg \Gamma$ where many terms contribute to the sum.}
    \label{fig:envelope1}
\end{figure}

Not content to observe the Stars over Babylon pattern, let's work out some quantitative details. For instance, how tall should one expect the peaks at full resonance to be? These are the times $T$ satisfying $cT/L=m$ for some integer $m$. So we have
\begin{equation}
    \log \Zs(T=mL/c)=-2\sum_{\q\in \mathbb N/\{0\}}\log(1-\exp(-\Gamma \left(\frac{2\pi \q}{L}\right)^2T))
    \label{eq:envelopeSum}
\end{equation}
If we replace $\Gamma$ with $D$, we see that at these special resonant times, this is also the enhancement factor for normal diffusion. So for times $T\ll L^2/\Gamma$ we can thus use the same methods leading to \eqref{eq:diffusiveSFF} (namely Taylor expanding the log and approximating the sum over $q$s with an integral) to get
\begin{equation}
    \log \Zs \left(T=\frac{mL}c\right)=2 L\left(\frac{1}{4\pi \Gamma T}\right)^{1/2} \zeta(3/2).
    \label{eq:evelope1}
\end{equation}
For a picture of this envelope versus the actual function, see figure \ref{fig:envelope1}.
We can also evaluate an envelope for the shorter peaks for $n>1$. In these cases it is best to break the sum into sets of $n$ consecutive terms. We can write this exactly as 
\begin{equation}
    \log \Zs\left(T=\frac{mL}{nc}\right)=-\sum_{\q\in \mathbb N/\{0\},\mathfrak{s}=\pm}\sum_{\q_2=0}^{n-1}\log(1-\exp(-\Gamma \left(\frac{2\pi (n \q-\q_2)}{L}\right)^2T+\mathfrak s 2\pi i \q_2/n)).
\end{equation}
In the thermodynamic limit $c L \gg \Gamma$ we can treat the Gaussian as constant within each inner sum. We can then use the identity
\begin{equation}
    \sum_{\q=0}^{n-1} \log(1-a\exp(2\pi i\q/n))=\log( 1-a^n)
\end{equation}
to perform the inner sum. We have
\begin{equation}
    \log \Zs\left(T=\frac{mL}{nc}\right)=-\sum_{\q\in \mathbb N/\{0\}}\sum_{\q_2=0}^{n-1}\log(1-\exp(-n\Gamma \left(\frac{2\pi n\q}{L}\right)^2T)).
\end{equation}
This is the same sum again but with $\Gamma$ replaced with $n^3$ $\Gamma$. This means that we have
\begin{equation}
    \log \Zs\left(T=\frac{mL}{nc}\right)=L\frac 1 {n^{3/2}} \left(\frac{1}{4\pi \Gamma T}\right)^{1/2} \zeta(3/2).
    \label{eq:envelopen}
\end{equation}
This $n^{-3/2}$ is interesting and surprising. Again, in the original Babylon function, there is just an $1/n$. We justified this by saying that $1/n$ of the terms contribute exploding positive contributions to the sum. However, the other $n-1$ terms can be shown to on average contribute slightly negative suppressing terms. This provides the intuition for the $n^{-3/2}$ in equation \eqref{eq:envelopen}. 

A final word about when these approximations become valid. Based on numerically evaluating equation \eqref{eq:enhancement}, it seems that the spikes at integers don't become clearly visible until this system size is a thousand relaxation lengths, and the full fractal structure isn't visible until ten thousand. Figures \eqref{fig:blowup} and \eqref{fig:envelope1} use a system size of a million. Needless to say, this is quite beyond the realm of any foreseeable exact diagonalization calculation or experimental technology.

\subsection{A Fourier Perspective and the Width of the Peaks}
\label{subsec:Fourier}
The analysis in the above subsection tells us the height of the peaks, but doesn't tell us anything about their breadth or shape. We can extract this information by taking the Taylor expansion of $\log (1-x)=\sum_{j=1}^\infty -\frac 1j x^j$
\begin{equation}
    \log \Zs=2 \sum_{\q>0, \mathfrak{s}=\pm}\sum_{j=1}^\infty \frac 1j\exp\left(2\pi i\q \mathfrak{s}\frac{cT}{L}j \right) \exp\left(-j\frac{4\pi^2 \Gamma \q^2}{L^2}T\right),
\end{equation}
For each $j$ this is the Fourier series representation of a chain of tight peaks (missing a $\q=0$ term). The peaks have Gaussian shape, and recur after time $\frac L{cj}$. Thus we see that the $j$th term in this sum corresponds only to the peaks at $T=\frac{mL}{jc}$, or in other words we can identify $j$ with $n$ in the previous section.

The total area under the Gaussian from the $j$th sum is given by the amplitude of the missing $\q=0$ term times the period. This is just an amplitude  of $\frac 1j$ times a period of $\frac L {cj}$ for a total area of $\frac{L}{cj^2}$.

Combined with the results of the previous subsection, we know that the width of the Gaussian is something like $\# \frac 1c \sqrt{\frac{\Gamma T}{n}}$. This implies that at sufficiently large $n$, the width of the peaks does become greater than the area between the peaks, which means that the lovely Stars over Babylon structure does not have infinite resolution.

\subsection{Instability to Interactions}
\label{subsec:stardestroyer}
It is also worth investigating how higher derivative terms and interactions would affect the qualitative result of the Babylon-shaped enhancement factor. In the thermodynamic limit, the higher derivative terms don't substantially affect the pattern. The higher derivative terms only affect larger values of $k$, which are already suppressed since they have factors of $e^{-\Gamma k^2 T}$.

Interactions, however, can have a more noticeable impact. A more detailed discussion will need to wait until appendix \ref{app:Interaction}; we will quote the results of that appendix here. An important point is that while in the CTP formalism the velocity would merely be renormalized by a smooth amount, in DPT the  velocity can be renormalized by an amount depending sensitively on frequency. We hypothesize that this irregularity would substantially derange the intricate pattern, likely in favor of a more erratic pattern of spikes.

If there were some large number of local degrees of freedom, in the tradition of \cite{Chan_2018,Friedman_2019,Moudgalya_2021,Chan_2022}, then these interactions would be heavily suppressed. But a large local number of degrees of freedom combined with a large system size means a truly large Hilbert space dimensions, making the empirical observation of these patterns a remote possibility for the foreseeable future.

Moreover, in the Schwinger-Keldysh case, 1d sonic hydrodynamics is known to be unstable \cite{spohn_2020}, and flows to the KPZ universality class \cite{KPZ,Krug1997OriginsOS,krajnik_2020}. For SFF hydrodynamics, these concerns are modified. The periodic time changes the significance of diagrammatic corrections in complicated ways \cite{winerprx}, and it is unclear in what dimensions weakly coupled hydrodynamics is stable in the IR limit. This will be discussed more in the outlook.


That said, it is far from obvious that the KPZ scaling completely destroys the Stars. This is because the ballistic propagation of sound, which is responsible for the basic resonance structure, is still present in the KPZ case. Said differently, one can view the flow to the KPZ universality class as modifying the dispersion from $\omega = c k - i \Gamma k^2 + \cdots$ to $\omega = c k - i \tilde{\Gamma} k^{3/2} + \cdots$. This suggests that the KPZ physics most strongly modifies the envelope function and may have a weaker effect on the resonances. We emphasize again that this requires further study and is a non-trivial problem in the DPT formalism, one that combines relevant interactions with periodic time.

However, we can sketch out one starting point for the analysis. One way to interpret $Z_{\text{enh}}$ for the unstable sonic fixed point is via a pair of biased diffusion equations, one for the left movers and one for the right movers. These equations are
\begin{equation}
    \partial_t \rho_\pm = \pm c \partial_x \rho_\pm + \Gamma \partial_x^2 \rho_\pm  + \xi_\pm,
\end{equation}
where $\pm$ refers to the left and right movers and we included a stochastic force $\xi_\pm$ to describe hydrodynamic fluctuations. The $Z_{\text{enh}}$ is obtained as the probability that a given initial condition $\rho_\pm(x,0)$ is recovered at later time at later time $T$. Focusing on the $+$ component, we have
\begin{equation}
    Z_{\text{enh}}(T) = \int D \rho_+(0) D \xi_+ P(\xi_+) \delta[\rho_+^{\text{diff}}(T;\xi_+) - \rho_+(0)],
\end{equation}
where $D\rho_+(0) D\xi$ denotes a functional integral over the noise and the initial condition and $\rho_+^{\text{diff}}(T;\xi)$ solves the biased diffusion equation with noise $\xi$.

The resonance condition, $m L = n cT$, arises as follows. If we ignore the $\Gamma$ term and the stochastic term, then a given profile $\rho_+(x,0)$ is merely translated by the dynamics to $\rho_+(x,T)=\rho_+(x+cT,0)$. Hence, for a localized wavepacket the probability to return is zero unless
\begin{equation}
    m L = c T
\end{equation}
for some integer $m$. The full set of resonances arises by considering perturbations of specific wavelenths, e.g. the $n=2$ case corresponds to perturbations of wavelength $\lambda = L/2, L/4, L/6, \cdots$ that need only be translated by half the length of the system to return to themselves. The effect of the noise and $\Gamma$ term is to broaden these sharp features as discussed above.

Now consider the KPZ case. Because of the peculiarities of 1+1d kinematics, the decomposition into left and right movers still provides an approximate starting point for the analysis. Focusing again on $\rho_+$, the simplest equation which captures the relevant effects is 
\begin{equation}
    \partial_t \rho_+ = c \partial_x \rho_+ + \frac{c'}{2} \partial_x \rho_+^2 + D \partial_x^2 \rho_+ + \xi_+. \label{eq:kpz}
\end{equation}
The new term is the $c'$ term, which turns out to be relevant in the scaling sense. Before proceeding, we emphasize again that it is not clear how the time periodicity modifies the standard analysis and whether the displayed terms are sufficient to capture all the physics of interest in our case. 

With that caveat, the enhancement takes the same form,
\begin{equation}
    \Zs(T) = \int D \rho_+(0) D \xi_+ P(\xi_+) \delta[\rho_+^{\text{KPZ}}(T;\xi_+) - \rho_+(0)],
\end{equation}
where now $\rho_+^{\text{KPZ}}(T;\xi_+)$ solves \eqref{eq:kpz} instead of the biased diffusion equation. We see immediately that the translating effect of the $c \partial_x \rho_+$ term is still present, so a localized wavepacket will still have vanishing return probability unless $m L = c T$. Of course, the other terms are crucial, but as in the diffusive case, their effect is plausibly to broaden these primary resonances rather than to destroy them. On the other hand, the subleading peaks, which arose from profiles with special wavelengths that enjoyed an enhanced translation symmetry, e.g., by $L/2$, are harder to analyze in the non-linear theory given by \eqref{eq:kpz}. Indeed, we cannot analyze the physics wavelength-by-wavelength since the non-linear term couples different wavelengths. We think it is plausible that there could still be enhancements in $\Zs$ corresponding to the subleading resonances, but we cannot say for sure without a more complete analysis of the KPZ return probability.


\section{Sound In An Integrable Cavity}
\label{sec:intStadium}
\begin{figure}
    \centering
    \includegraphics{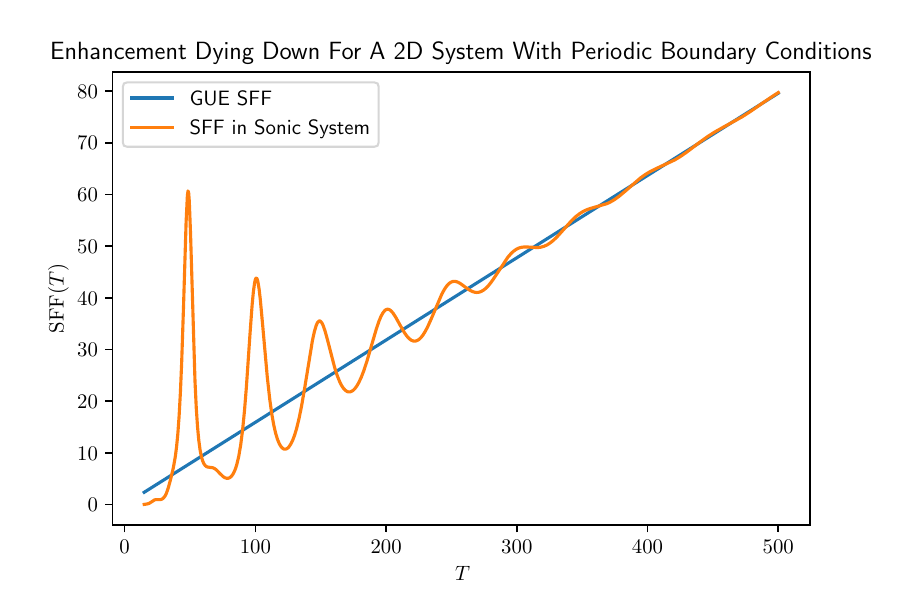}
    \caption{(Blue) Unenhanced ramp for GUE-like SFF. (Orange) the connected SFF for a $c=\Gamma=1$ 2d hydrodynamic system of size $5 \times 5$ with periodic boundary conditions. If the size were significantly larger, the enhancement would be prohibitively large and the two SFFs couldn't share the same plot.}
    \label{fig:SFFs2D}
\end{figure}

We will now make the jump from quadratic hydrodynamics in 1D to higher dimensions. In this case we are now faced with the choice of what shape our system should take: spherical, toroidal, or something more exotic. The particular case of a square system with periodic boundary conditions is plotted in figure \ref{fig:SFFs2D}.

While in most cases---including the diffusive hydro SFF---the answers to hydrodynamic questions do not depend on this sort of choice, the sonic hydrodynamic SFF will care about the precise shape we choose.
We see that equation \eqref{eq:enhancement} depends sensitively on the detailed eigenvalues of the Laplacian in our system. In this section we will examine the case where the Laplacian operator in our cavity has Poissonian spectral statistics. This might occur, for instance, if we are doing hydrodynamics in a torus, a rectangular prism, or an ellipsoid. More generally, Poissonian spectral statistics for the Laplacian is thought to describe the case where the cavity, when viewed as a stadium/billiard table, gives rise to integrable dynamics for a particle moving inside the stadium~\cite{berry1977level}. For this reason, we will call this case an integrable cavity. But it is important to remember that while the emergent dynamics of an individual sound mode may be integrable, our hydrodynamic assumption requires that the full many-body dynamics be chaotic in terms of microscopic degrees of freedom.

For a general hydrodynamic system, equation \eqref{eq:enhancement} can be rewritten as 
\begin{equation}
    \Zs=\prod_{k}\frac 1{(1-\exp\{(ick-\Gamma k^2)T\})(1-\exp\{(-ick-\Gamma k^2)T\})}.
    \label{eq:intCoeffFormula}
\end{equation} 
Using the Poissonian statistics, we can calculate the expected value of this enhancement factor. In any Poissonian system, the density of Laplacian ``energy'' eigenvalues in the interval $[E_1,E_1+d E_1]$ is independent of that in region $[E_2,E_2+dE_2]$ for $E_1 \neq E_2$. This independence implies a similar independence for $k\sim\sqrt E$. So we can evaluate the expected value of equation \eqref{eq:intCoeffFormula} by partitioning the spectrum of $k$ into non-overlapping regions $[k_i,k_i+\delta k]$, finding the expected value of the product over eigenvalues in that region, and then multiplying the results together. 

For concreteness, we will treat the values of $k$ as a Poisson process with intensity $\bar \rho(k)$. This means that in any given region $[k,k+dk]$ the product is $[(1-\exp\{(ick-\Gamma k^2)T\})(1-\exp\{(-ick-\Gamma k^2)T\})]^{-1}$ with probability $\bar \rho dk$ (eigenvalue present) and $1$ with probability $1-\bar \rho dk$ (eigenvalue absent). The expected value is thus \begin{equation}
    1+\left(\frac 1{(1-\exp\{(ick-\Gamma k^2)T\})(1-\exp\{(-ick-\Gamma k^2)T\})}-1\right)\bar \rho dk.
\end{equation}
This means that in expectation we can write
\begin{equation}
    \langle{\Zs}\rangle=\exp \left(\int_0^\infty dk \left(\frac 1{(1-\exp\{(ick-\Gamma k^2)T\})(1-\exp\{(-ick-\Gamma k^2)T\})}-1\right)\bar \rho(k)  \right),
    \label{eq:TruePoisson}
\end{equation}
where the angle brackets represent an average over different cavity configurations.

\begin{figure}
    \centering
    \includegraphics{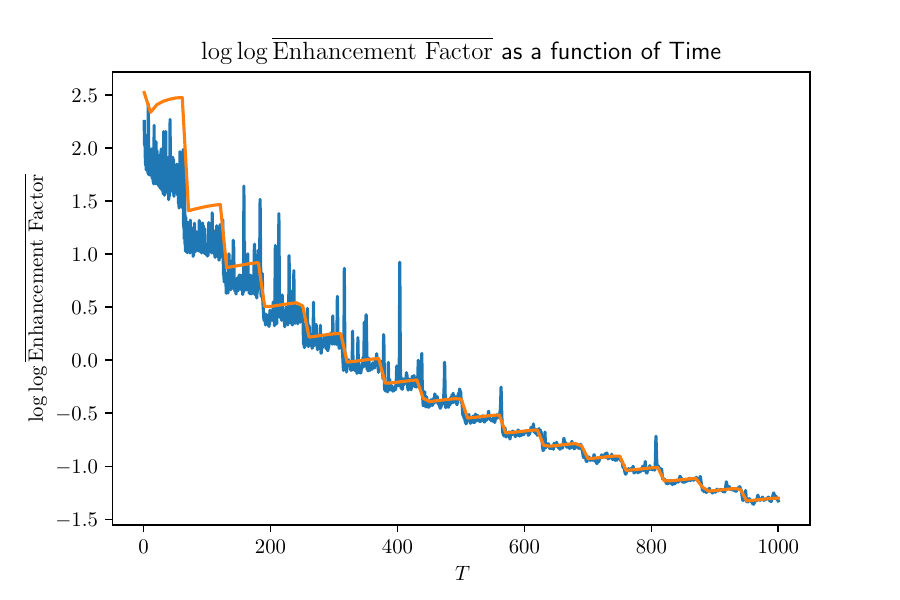}
    \caption{A graph illustrating the agreement of numerical (blue) vs theoretical (orange) predictions for equation \eqref{eq:TruePoisson}. We chose $c=1,\Gamma=0.1$, $\bar \rho(k)=1$. Because equation \eqref{eq:TruePoisson} has a divergence at low $k$ we had a cutoff of $k=0.1$. This cutoff is responsible for the interesting steplike behavior seen in both functions. The blue line is an average of 1971459 samples.}
    \label{fig:TruePoisson}
\end{figure}

This formula isn't the result of an expansion, it assumes only a purely Poissonian density for $k$. Figure \ref{fig:TruePoisson} shows numerics backing up this prediction. To create figure \ref{fig:TruePoisson}, we used the most Poissonian process possible: independent random numbers. It does not correspond to a Laplacian on any particular cavity shape.

It is worth noting that equation \eqref{eq:intCoeffFormula} is a product over many terms. As such, depending on context, the mean value might not be a good representation of typical values, for the same reason that log-normal distributions are not well-clustered around their mean. The expected value of the log of the coefficient can be calculated straightforwardly by taking the log of equation \eqref{eq:intCoeffFormula}.

Another caveat is that real billiard systems, even integrable ones, do not have exactly Poissonian spectral statistics. The clearest exhibition of this is in the case of periodic orbits. To explain how this affects the result, consider a torus of dimensions $2\pi\times 4\pi$. The eigenfunctions of the Laplacian are parameterized by a wavevector of the form $(n_x,n_y)$ where $n_x$ is integer valued and $n_y$ can also take half-integer values. For example, one possible Laplacian eigenstate has wavevector $(1,\frac 12)$, which gives $k^2=\frac 54, k=\frac{\sqrt 5}{2}$. There is another eigenstate with wavevector $(2,1)$, which corresponds to $k=\sqrt 5$. In general, there will be states with $k=n\frac{\sqrt 5}{2}$ for all positive integers $n$. When we multiply the enhancement factors for all of these states, we get the same intricate pattern seen in figure \eqref{fig:blowup}. 

We think that this same effect can exist more generally in any cavity in which a classical particle can take a closed periodic path. To show this fact about the eigenvalues of the Laplacian, we imagine that we start a quantum wavepacket moving under a fictitious Hamiltonian $H_{\text{fict}}=-\frac{1}{2m_{\text{fict}}}\grad^2$ at velocity $v$ around a periodic path of length $L$. The wavepacket does not have a definite energy, but instead a spread of energies well-centered on $E_{\text{fict}}=\frac{k^2}{2m_{\text{fict}}}=\frac{m_{\text{fict}}v^2}{2}$. If we choose a $k$ very large compared to the inverse system size (well into the semiclassical limit) then all of the energy eigenvalues contributing to this wavepacket are close to $E_{\text{fict}}$. 

Because the classical motion is periodic, the wavefunction is going to be approximately periodic in time with period $L/v$.  This means that the wavepacket has overlap with $H_{\text{fict}}$ eigenstates with energies that differ by integer multiples of $\frac{2\pi v}{L}$. So there will be values of $k=\sqrt{2m_{\text{fict}}E_{\text{fict}}}$ spaced out, separated by integer multiples of $\frac{2\pi}{L}$. These evenly spaced out modes violate our assumption of Poissonness, and will lead to contributions like in figure \eqref{fig:blowup}. These effects rely only on the existence of periodic orbits in the cavity, which are are present in both integrable or chaotic cavities.

It is likely, however that these effects will drown out after a comparatively short time once the wavefunction has time to spread out. For chaotic systems this time is known as the Ehrenfest time, and is on the order of $\frac{\log \frac{\textrm{Phase Space Volume}}{h^d}}{\lambda}$, where $\lambda$ is the Lyapunov exponent. For integrable systems the time is given by the same star destroying effects as in subsection \eqref{subsec:stardestroyer}. For a generic integrable system, $\omega$ will depend on $d$ commuting quantum numbers. The dependence will be well approximated as linear, but any higher-derivative terms or interactions in the hydro theory will break that perfect interference.

\section{Sound in a Chaotic Cavity}
\label{sec:freeStadium}
\begin{figure}
    \centering
    \includegraphics[scale=0.8]{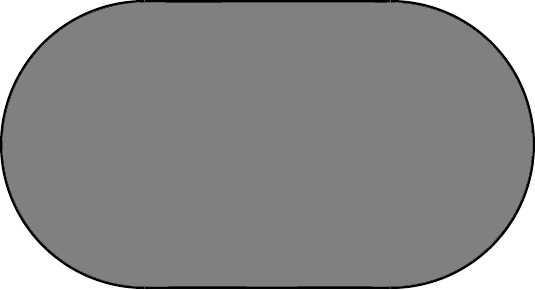}
    \includegraphics[scale=0.8]{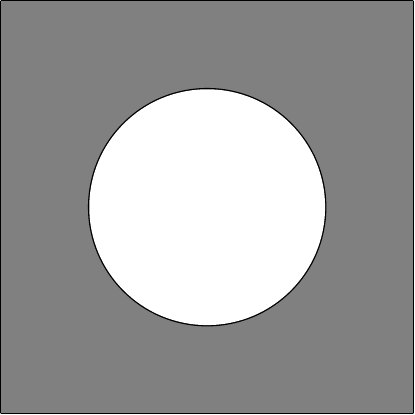}
    \caption{The eigenvalues of the Laplacian are known to have Wigner-Dyson/RMT-like statistics in both Bunimovich Stadium (left) and the Sinai Billiard (right). The fluids fill up the gray regions in these two shapes, meaning that sound modes in the fluid have RMT-like spectra.}
    \label{fig:bunSinai}
\end{figure}

In this section we will turn our attention to the case of quadratic hydrodynamic enhancements in a chaotic cavity. As a reminder, the expression for the enhancement from sound poles in a generic cavity is
\begin{equation}
\begin{split}
\log \Zs=-\sum_{k^2, \mathfrak{s}=\pm}\log(1-\exp\{(i\mathfrak{s}ck-\Gamma k^2)T\})\\
=\sum_{j=1}^\infty\sum_{k, \mathfrak{s}=\pm} \frac{1}{j}\exp((i\mathfrak{s}ck-\Gamma k^2)jT)
\end{split}
\label{eq:stadium}
\end{equation}
Let's imagine the shape of the cavity is a Bunimovich stadium or a Sinai billiard (figure \ref{fig:bunSinai}). Then the $k^2$s become the eigenvalues of a level-repelling chaotic Hamiltonian of the Gaussian Orthogonal Ensemble (GOE) universality class. The $k$s are stretched eigenvalues, but still exhibit GOE-type level repulsion.
\begin{figure}
    \centering
    \includegraphics[scale=0.8]{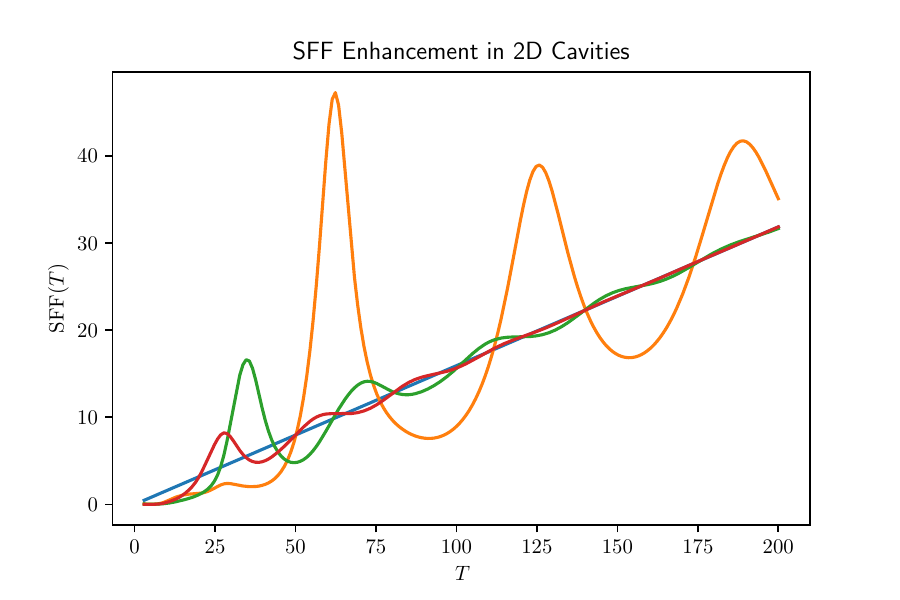}
    \caption{RMT Result (Blue) versus sound-enhanced results in 3 different chaotic cavities with area 40 (Orange, Green, and Red). Notice how the system with one very slow mode (Orange) seems to have fairly different behavior than the faster decaying Green and Reds. As system sizes become larger compared to the mean free path, and more modes can oscillate without dying, interference between various oscillations matters more than the magnitude of any one slow mode.}
    \label{fig:chaosSound}
\end{figure}
We can view the right hand side of equation \eqref{eq:stadium} as a sum over partition functions with imaginary temperature $x$ associated with the $k$ spectrum, 
\begin{equation}
    Z(x,f)=\int dk \rho(k) e^{i k x} f(k)
\end{equation}
where $\rho$ is the exact density of states and $f$ is a filter function. In particular, we have
\begin{equation}
    \log \Zs=\sum_{j=1,\, \mathfrak{s}=\pm}^\infty \frac 1j Z\left[j\mathfrak{s}cT,f_j=\exp(-j\Gamma Tk^2)\right].
    \label{eq:wdef}
\end{equation}
If we fix particular stadium, then we expect that the enhancement factor will be an erratic function of time. Some examples of these functions are plotted in figure \ref{fig:chaosSound}.

To get a smooth result which is calculable, we average over a number of configurations for our stadium (being careful to remain in the GOE universality class) instead of calculating the logarithm of $\Zs$ for a given realization of the stadium. Denoting by $W$ the quantity in equation \eqref{eq:wdef}, we have $\log \langle \Zs \rangle=\log \langle e^W \rangle$, which is by definition the sum of the cumulants of $W$.

Doing a cumulant expansion, we get
\begin{equation}
    \log \langle \Zs \rangle =\sum_\ell \frac 1{\ell!}c_\ell = \mathbb{E}\left[\sum_{j,\mathfrak{s}} \frac 1j Z(j\mathfrak{s}cT,f_j)\right]+\frac 12 \textrm{var}\left[ \sum_{j,\mathfrak{s}} \frac 1j Z(j\mathfrak{s}cT,f_j)\right]+\dots
    \label{eq:basicCumulantExpansion}
\end{equation}
For times smaller than the Heisenberg time of the single-particle system, we can safely approximate the cumulant expansion with just the first two terms, knowing all subsequent terms will suppressed by factors of the cavity volume. We will consider these two terms in turn.

The first cumulant is obtained from the average density of states. Our system has some density of states $\rho(k)$ which fluctuates depending on the precise shape of the cavity. Averaging over cavity shapes we get that $\rho(k)$ fluctuates about some $\bar \rho(k)$. $\bar \rho$ can be calculated in a semiclassical approximation. If the cavity has $d$-dimensional volume $V$, we can expect a density of states
\begin{equation}
    \bar \rho(k)=\frac {V}{(2\pi)^d} S_{d-1}k^{d-1},
\end{equation}
where $S_{d-1}=\frac{2\pi^{\frac{d-1}2}}{\Gamma(\frac {d-1}2)}$ is the surface area of a $d-1$ sphere.

So the expected value of the imaginary-time partition function is
\begin{equation}
    \mathbb{E} \left[\frac 1j Z(j\mathfrak{s}cT,f_j=\exp(-j\Gamma Tk^2))\right]=\frac {V S_{d-1}} {(2\pi)^dj}\int _{0}^\infty dk k^{d-1}\exp(ij\mathfrak{s}cTk)\exp(-j\Gamma Tk^2).
\end{equation}
After including the sum over $\mathfrak{s}$ and relabeling $k \rightarrow -k$ in the $\mathfrak{s}=-$ term, we have
\begin{equation}
     \sum_{\mathfrak{s}} \mathbb{E} \left[\frac 1j Z(j\mathfrak{s}cT,f_j)\right]=\frac {V S_{d-1}} {(2\pi)^dj}\int _{-\infty}^\infty dk |k|^{d-1}\exp(ijcTk)\exp(-j\Gamma Tk^2)
\end{equation}
Right away, we notice that the behavior is qualitatively different for even versus odd $d$. For odd $d$, we have the Fourier transform of an analytic function, while for even $d$, there is a non-analyticity at $k=0$. 

If we evaluate the expression, we get
\begin{equation}
   \sum_{\mathfrak{s}} \mathbb{E} \left[\frac 1j Z(j\mathfrak{s}cT,f_j)\right] =
\begin{cases}
  \frac {V S_{d-1}} {(2\pi)^dj} \sqrt{\frac{2\pi }{j\Gamma T}}(jT)^{1-d}\partial_c^{d-1}\exp(-j\frac{c^2}{4\Gamma}T)&\text{if }d \textrm{ odd},\\
  \frac {2V S_{d-1}} {(2\pi)^dj} \frac{(d-1)!}{(ijcT)^d}+O(T^{-d-1})          &\text{if }d \textrm{ even}.
\end{cases}
\label{eq:dip}
\end{equation}
As we see, in odd dimensions this is an exponential decay, while in even dimensions it is a power-law decay. This is related to the fact that sound waves have sharp edges in odd dimensions and soft edges in even dimensions. The terms in \eqref{eq:dip} can plugged into \eqref{eq:basicCumulantExpansion} and the sum over $j$ computed to get
\begin{equation}
    \log \langle{\Zs}\rangle\supset \begin{cases}
  \frac {V S_{d-1}} {(2\pi)^d} \sqrt{\frac{2\pi }{\Gamma T}}(T)^{1-d}\partial_c^{d-1}\exp(-\frac{c^2}{4\Gamma}T)&\text{if }d \textrm{ odd},\\
  \zeta(d+1)\frac {2V S_{d-1}} {(2\pi)^d} \frac{(d-1)!}{(icT)^d}+O(T^{-d-1})          &\text{if }d \textrm{ even}.
\end{cases}
\label{eq:1ParticleDip}
\end{equation}

The odd dimension result is zero at times greater much than $\frac{\Gamma}{c^2}$. Unfortunately, before this time hydrodynamics is dominated by higher derivative corrections, so in odd dimensions equation \ref{eq:1ParticleDip} must be approached with caution. This fast decay of the exponential-in-volume enhancement factor can be contrasted with the diffusive result quoted in equation \ref{eq:diffusiveSFF}, which decays slowly in all dimensions.
In contrast the rapid decay in odd dimensions, the even-dimensional part of equation \ref{eq:1ParticleDip} is large until times extensive in the system length, and could, in principle, be observed. But it still decays far faster than equation \ref{eq:diffusiveSFF}.

If we stopped here, at the first term in the cumulant expansion, we would be making the approximation $\log \langle{\Zs}\rangle=\langle{\log \Zs}\rangle$. By analogy with the glass literature, we refer to $\log \langle{\Zs}\rangle$ as the annealed average and $\langle{\log \Zs}\rangle$ as the quenched average. The above approximation is thus analogous to the approximation that quenched equals annealed. Typically, the quenched value of a partition function is the value for a typical realization of disorder, whereas the unquenched or annealed value is dominated by more extreme terms. If one were to take a small number of cavity shapes and evaluate the SFF enhancement coefficients, most of the coefficients would look like equation \eqref{eq:1ParticleDip}. This is because we have $\log \Zs \sim \rho$ sample by sample, so the quenched average contains only a linear-in-$\rho$ term whereas the annealed average has contributions from higher cumulants. Equation \eqref{eq:1ParticleDip} also gives the quenched answer for an integrable system.

Now onto the second cumulant, which is
\begin{equation}
    \textrm{var}\left[\sum_{j,\mathfrak{s}} \frac 1j Z(j\mathfrak{s}cT,f_j)\right]=\sum_{j,j',\mathfrak{s},\mathfrak{s}'} \frac{1}{j j'}\int dk dk' \mathbb{E}(\rho(k)\rho(k'))_{\text{conn}} e^{i j k \mathfrak{s} c T + i j' k' \mathfrak{s}' c T} f_j(k) f_{j'}(k'),
\end{equation}
where $\mathbb{E}(\rho(k)\rho(k'))_{\text{conn}}$ is the connected pair correlation function. Further progress can be made under the assumption that the connected correlator has only weak dependence on $k+k'$. In this case, only terms with $j=j'$ and $\mathfrak{s}=-\mathfrak{s}'$ contribute, giving
\begin{equation}
    \textrm{var}\left[\sum_{j,\mathfrak{s}} \frac 1j Z(j\mathfrak{s}cT,f_j)\right] \approx \frac{1}{j^2} \sum_{j,\mathfrak{s}} \int dk dk' \mathbb{E}(\rho(k)\rho(k'))_{\text{conn}} e^{i j \mathfrak{s} (k-k')  c T} f_j(k) f_j(k').
\end{equation}

Now, this expression is essentially a sum over single-particle spectral form factors evaluated at the times $j c T$ and with filter function $f_j$. For each of these we may use the standard GOE result (assuming the single particle SFF is always in the ramp phase) to obtain
\begin{equation}
    \textrm{var}\left[\sum_{j,\mathfrak{s}} \frac 1j Z(j\mathfrak{s}cT,f_j)\right] \approx \sum_j \frac{2}{j^2} \int_0^\infty d\bar{k} \left( \frac{j c T}{\pi} \right) \exp( - 2 j \Gamma T \bar{k}^2).
\end{equation}
The integral is straightforward, and the total variance is thus
\begin{equation}
    \textrm{var}\left[\sum_{j,\mathfrak{s}} \frac 1j Z(j\mathfrak{s}cT,f_j)\right] \approx \sum_j  \frac{c T}{j \pi}\sqrt{\frac{\pi}{2 j \Gamma T}} = \zeta(3/2) \sqrt{ \frac{c^2 T}{2 \pi \Gamma}}. \label{eq:varTerm}
\end{equation}
This answer is eerily reminiscent of equation \eqref{eq:evelope1}. But most interestingly, it is a highly universal contribution to the ramp of a hydrodynamic system with a sound pole. It doesn't depend on any details of the shape of the system, or even its overall size. Just that the sound waves propagating around experience chaotic dynamics.

Equation \eqref{eq:varTerm} increases indefinitely, and it is worth knowing when exactly it starts to fail. The answer is that at times on the order of the single-particle Heisenberg time (inverse level spacing of the single-particle system) the second cumulant reaches a plateau, and cumulants after the second stop being negligible. After a few Heisenberg times, there is no longer any residue of the single-particle level repulsion and the enhancement for chaotic billiards should resemble that for integrable billiards. This, in turn, falls to 1 as the various sound modes die down, long before the full many-body Heisenberg time.

In summary, here is a list of important time scales for the connected sound pole SFF in a chaotic cavity:
\begin{itemize}
    \item Hydrodynamic scattering time of order $\Gamma/c^2$. This is the time-scale after which hydrodynamics becomes a valid approximation.
    \item Single particle Thouless time, on the order of $L/c$. This is the time scale at which equation \eqref{eq:varTerm} becomes a good approximation.
    \item The single particle Heisenberg scale on the order of $\left(\frac 1c V \Gamma^{-\frac{d-1}{2}}\right)^{\frac{2}{d+1}}$. Equation \eqref{eq:varTerm} is dominated by modes with $k$ on the order of $k\sim (\Gamma T)^{-1/2}$. The single particle Heisenberg scale is the time at which $T$ exceeds the density of states in this region. Above this time scale a chaotic billiard should behave like an integrable billiard.
    \item The lifetime of the slowest sound modes $L^2/\Gamma$. This can also be thought of as a many-body Thouless time. After this time the ramp looks like the pure random matrix theory result. 
    \item The many-body Heisenberg time at order $e^{S}$. This is the time scale at which the SFF of the full many-body system plateaus.
\end{itemize}

\section{Discussion and Outlook}

In this paper, we expanded the theory of the hydrodynamic spectral form factor to the more realistic case of hydrodynamics with a second-time derivative. This more sophisticated theory and its accompanying sound pole structure allowed a new phenomenon: interference in the emergent hydrodynamic SFF formalism. We discovered a variety of different interference patterns, all connecting intimately with the SFF of a single particle problem in various cavities. We started with the simplest case, the (necessarily integrable) dynamics of a sound mode in a single dimension, and got the remarkable function in figure \ref{fig:blowup} in the absence of dispersion. In the case of higher dimensional cavities, our results mostly concerned disorder-averages over cavities. Using Wigner-Dyson versus Poissonian statistics for the eigenvalues of the Laplacian of the cavity, we derived the results of sections \ref{sec:intStadium} and \ref{sec:freeStadium}. We derive equations \eqref{eq:TruePoisson} and \eqref{eq:basicCumulantExpansion} respectively, and graph the predicted enhancements for toy examples of the laplacian spectral statistics.


Throughout this work, we find that the expected spectral form factor depends sensitively on both the dimensionality of the system as well as whether the cavity supports chaotic or integrable billiard dynamics. For example, equation \eqref{eq:1ParticleDip} take on entirely different forms in odd versus even dimensions, for the same reason that sound has a non-analytic shockwave in odd but not even dimensions. We also note that, if momentum is exactly conserved, then there will be distinct blocks in the many-body Hamiltonian labelled by the many-body momentum (similar to the discussion of conserved quantities in \cite{Winer_2022}).

This work opens the gates for calculations in a wide array of settings, including CFTs (which necessarily have both momentum and energy conservation) and systems with spontaneously broken symmetry. In particular, CFTs on spheres have been shown~\cite{Freivogel_2012} to have eternal non-decaying hydrodynamic modes. This follows from the conformal symmetry, where certain ladder operators $L^+$ satisfy the commutation relation $[H,L^+]=\frac 1R L^+$. Just like in a simple harmonic oscillator, this leads to two-point functions which oscillate exactly, forever, even at times much longer than the Heisenberg time of the system.
Each primary operator with energy $\frac \Delta R$ sits at the bottom of a tower of states which contribute $\frac{e^{i\Delta T/R}}{(1-e^{iT/R})^D}$ to the SFF. These $\frac 1{(1-e^{iT/R})^D}$ factors would lead to strong enhancements in the SFF out to arbitrary times, even deep into the plateau region. It would be interesting to look for a hydrodynamic explanation for these effects, and relate them to the above formalism.

As our ability to numerically calculate the SFFs of quantum fields theories improves \cite{luca}, one might look for at least the leading peaks in a numerical spectral form factor. One highly optimistic possibility is that signatures of these results could be found in the original home of Wigner-Dyson statistics: atomic nuclei. These nuclei are hydrodynamic systems with vibrational modes, these modes might have a signature in the spectral form factor.

\begin{figure}
    \centering
    \includegraphics[scale=0.5]{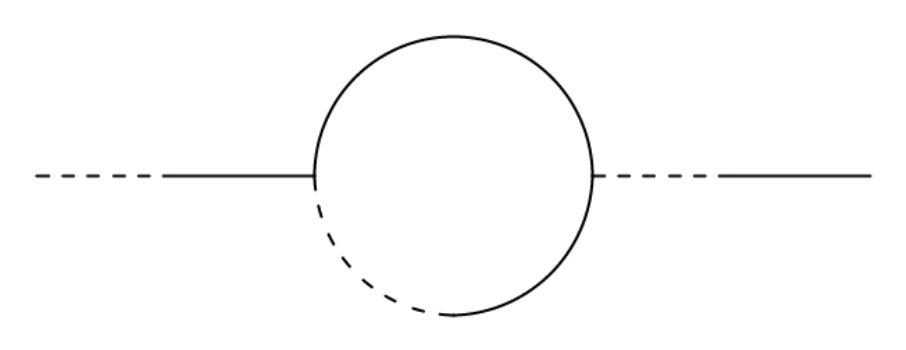}
    \caption{The loop integral in this diagram depends on the second moment of the spectral density.}
    \label{fig:twoPointDiagram}
\end{figure}

An important direction for future work is seeing whether this emergent spectral effect can give rise to enhanced 2-point functions in the CTP hydrodynamic theory. Figure \eqref{fig:twoPointDiagram} shows a diagram which depends on an internal integral over the frequencies of the two legs. If the two-point function of these frequencies is not analytic (as is the case for both integrable and Wigner-Dyson spectral statistics), that leads to a long-time tail (albeit suppressed by a factor of volume).
Finally, the fate of DPT hydro in the long-range limit is still an open question. It is known that in conventional hydrodynamics, the 1D theory is always strongly coupled. In the DPT theory, each leg in each Feynman diagram is modified due to each mode wrapping around the periodic time (see equation \ref{eq:GDPT}). One could imagine these modes destructively interfering at times $T$ incommensurate with the period of the sound modes, thus rescuing the diffusive theory. On the other hand, it is known \cite{winerprx} that for purely diffusive dynamics the additional wrapping makes periodic-time hydro strongly coupled even up to two dimensions. We leave this question for future work.

This work was supported by the Joint Quantum Institute (M.W.) and by the Air Force Office of Scientific Research under FA9550-19-1-0360 (B.G.S.).

\renewcommand{\thechapter}{5}

\chapter{Spectral Form Factor of a Quantum Spin Glass}
\label{chapter:glass}
\textbf{Authors:} \textit{Michael Winer, Richard Barney, Christopher L. Baldwin, Victor Galitski, Brian Swingle}

\textbf{Abstract:} It is widely expected that systems which fully thermalize are chaotic in the sense of exhibiting random-matrix statistics of their energy level spacings, whereas integrable systems exhibit Poissonian statistics. In this paper, we investigate a third class: spin glasses. These systems are partially chaotic but do not achieve full thermalization due to large free energy barriers. We examine the level spacing statistics of a canonical infinite-range quantum spin glass, the quantum $p$-spherical model, using an analytic path integral approach. We find statistics consistent with a direct sum of independent random matrices, and show that the number of such matrices is equal to the number of distinct metastable configurations---the exponential of the spin glass ``complexity'' as obtained from the quantum Thouless-Anderson-Palmer equations. We also consider the statistical properties of the complexity itself and identify a set of contributions to the path integral which suggest a Poissonian distribution for the number of metastable configurations. Our results show that level spacing statistics can probe the ergodicity-breaking in quantum spin glasses and provide a way to generalize the notion of spin glass complexity beyond models with a semi-classical limit.

\section{Introduction} \label{sec:introduction}



An isolated quantum many-body system which reaches an effective thermal equilibrium state starting from an out-of-equilibrium initial state is often called ``quantum chaotic." As commonly used, quantum chaos is a loose term referring to a family of phenomena that typically co-occur, including the ability of the system to serve as its own heat bath ~\cite{Deutsch1991,Srednicki1994,Rigol2008Thermalization}, hydrodynamic behavior of conserved quantities ~\cite{glorioso2018lectures,crossley2017effective,PhysRevD.91.105031,Haehl_2018,Jensen_2018}, and random-matrix-like energy eigenvalues~\cite{PhysRevLett.52.1,doi:10.1063/1.1703775,mehta2004random,guhr1998random}. Given this variety, it is crucial to understand the relationships between different manifestations of quantum chaos~\cite{DAlessio2016From,Santos2010}.

These relationships are complicated and interesting in large part because the systems in question have structure, such as locality and symmetry. For example, if the Hamiltonian has spatial locality, energy conservation implies the existence of slow hydrodynamic modes and an associated long time scale, the Thouless time, such that random-matrix behavior is only present for energy levels closer than the inverse Thouless time~\cite{Chan_2018,Moudgalya_2021}. Similarly, if the Hamiltonian possesses a symmetry, then it can be organized into blocks labelled by irreducible representations of the symmetry. One finds random-matrix statistics within each individual block, but full ergodicity is broken because matrix elements between different blocks are forbidden~\cite{2019Santos,PhysRevE.102.060202,winerprx,Winer_2022,Roy2022zig}.

It is natural to ask whether there are other ways in which ergodicity can be lost, and if so, what the resulting spectral statistics of the Hamiltonians are. In particular, we will better understand the relations between different measures of quantum chaos by understanding how they are lost and what replaces them. 

Quantum spin glasses provide one well-established context to explore these questions, since they exhibit a rich phenomenology associated with the inability to fully thermalize~\cite{Binder1986Spin,Mezard1987,Fischer1991,Nishimori2001,Castellani2005Spin,Mezard2009,Stein2013}
In this paper, we determine the spectral statistics of an analytically tractable spin glass model, the quantum $p$-spherical model. We find that up to times polynomial in the system size, the Hamiltonian can effectively be described as approximately block-diagonal. Each block behaves as a random matrix independent of the others, and the number of blocks depends on the energy per particle. At high energies, there is only one block and the system is ergodic. Below a critical energy density, the Hamiltonian breaks into exponentially many blocks --- the average number of blocks jumps discontinuously from the high energy regime and then decreases as the energy density decreases further. We establish these results via a path integral computation of the spectral form factor (SFF), which measures correlations between pairs of energy levels~\cite{saad2019semiclassical,saad2019late,winerprl,PhysRevE.72.046207,yiming}.

In the remainder of the introduction, we give some physical context by reviewing the spectral form factor and mean-field spin glasses, and then summarize our results.
In Sec.~\ref{sec:overview_PSM}, we review the $p$-spherical model in detail.
In Sec.~\ref{sec:ergodic_ramp}, we calculate the SFF of this model in the high-temperature ergodic regime, and in Sec.~\ref{sec:nonergodic_ramp}, we do so in the non-ergodic regime. Finally, in Sec.~\ref{sec:HigherMoments}, we investigate higher-moment analogues of the SFF.
We then discuss implications of these results and directions for future work in Sec.~\ref{sec:conclusion}.

\begin{figure}[t]
    \centering
    \includegraphics[width=.9\textwidth]{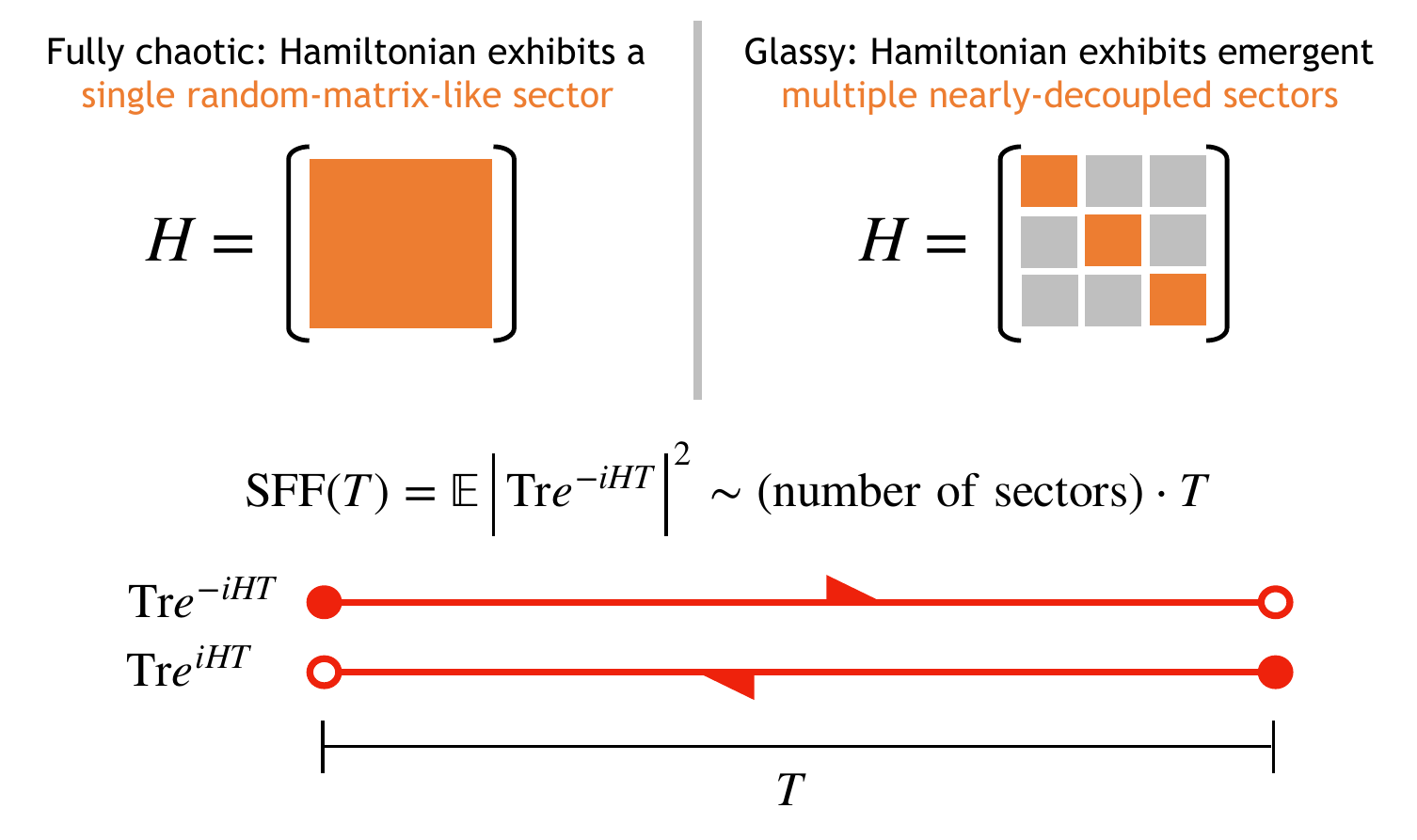}
    \caption{(Top left) Fully chaotic systems have energy levels that are statistically similar to a Gaussian random matrix, indicated by the orange block. (Top right) By contrast, quantum spin glasses in the non-ergodic phase have spectral statistics that resemble a collection of many nearly-decoupled random matrices (Bottom) Spectral statistics can be diagnosed via the spectral form factor, denoted $\textrm{SFF}(T)$, which consists of a path integral over a pair of real-time contours as indicated by the red lines. The universal part of $\textrm{SFF}(T)$, which is proportional to $T$, is enhanced by the number of effectively uncoupled sectors (other non-universal contributions are not indicated here).}
    \label{fig:my_label}
\end{figure}

\subsection{Review of the spectral form factor} \label{subsec:review_spectral_form_factor}

To study the spectral correlations of a Hamiltonian $H$, a standard tool is the spectral form factor (SFF) \cite{Cotler2017,brezin1997spectral}, defined as
\begin{equation} \label{eq:general_SFF_definition}
\textrm{SFF}(T) \equiv \big| \textrm{Tr} e^{-iHT} \big|^2.
\end{equation}
In situations where the spectrum is unbounded, or when one wishes to concentrate on a portion of the spectrum, the trace in Eq.~\eqref{eq:general_SFF_definition} is regulated by a filter function $f(H)$:
\begin{equation} \label{eq:filtered_SFF_definition}
\textrm{SFF}(T, f) \equiv \big| \textrm{Tr} f(H) e^{-iHT} \big|^2.
\end{equation}
One common choice is $f(H) = e^{-\beta H}$\cite{saad2019semiclassical,Papadodimas_2015}, and another is $f(H) = e^{-c(H - E_0)^2}$.
The latter allows one to study level statistics near a specified energy $E_0$.

For a single Hamiltonian, the SFF is an erratic function of time \cite{brezin1997spectral}.
Thus one usually considers an ensemble of Hamiltonians and defines the SFF as the average of Eq.~\eqref{eq:filtered_SFF_definition} over the ensemble.
Throughout this paper, we use the notation $\mathbb{E}[ \, \cdot \, ]$ to denote the ensemble average.

The SFF is closely related to the correlation function of the density of states.
Formally, the (filtered) density of states is given by
\begin{equation} \label{eq:general_density_of_states_definition5}
\rho(E, f) \equiv \sum_n f(E_n) \delta(E - E_n) = \textrm{Tr} f(H) \delta(E - H),
\end{equation}
where $n$ labels the eigenstate of $H$ with eigenvalue $E_n$, and its correlation function is
\begin{equation} \label{eq:density_of_states_correlation_definition5}
C(E, \omega, f) \equiv \mathbb{E} \left[ \rho \left( E + \frac{\omega}{2}, f \right) \rho \left( E - \frac{\omega}{2}, f \right) \right].
\end{equation}
We have that
\begin{equation} \label{eq:SFF_density_of_states_relationship5}
\begin{aligned}
\textrm{SFF}(T, f) &= \mathbb{E} \Big[ \textrm{Tr} f(H) e^{-iHT} \textrm{Tr} f(H) e^{iHT} \Big] \\
&= \int dE d\omega \, e^{-i \omega T} \mathbb{E} \left[ \textrm{Tr} f(H) \delta \left( E + \frac{\omega}{2} - H \right) \textrm{Tr} f(H) \delta \left( E - \frac{\omega}{2} - H \right) \right] \\
&= \int d\omega \, e^{-i \omega T} \int dE \, C(E, \omega, f).
\end{aligned}
\end{equation}
The SFF is simply the Fourier transform of the correlation function with respect to $\omega$, integrated over $E$ (although the filter function allows one to concentrate on an arbitrary subset of the spectrum).

\begin{figure}
\centering
\includegraphics[width=0.9\textwidth]{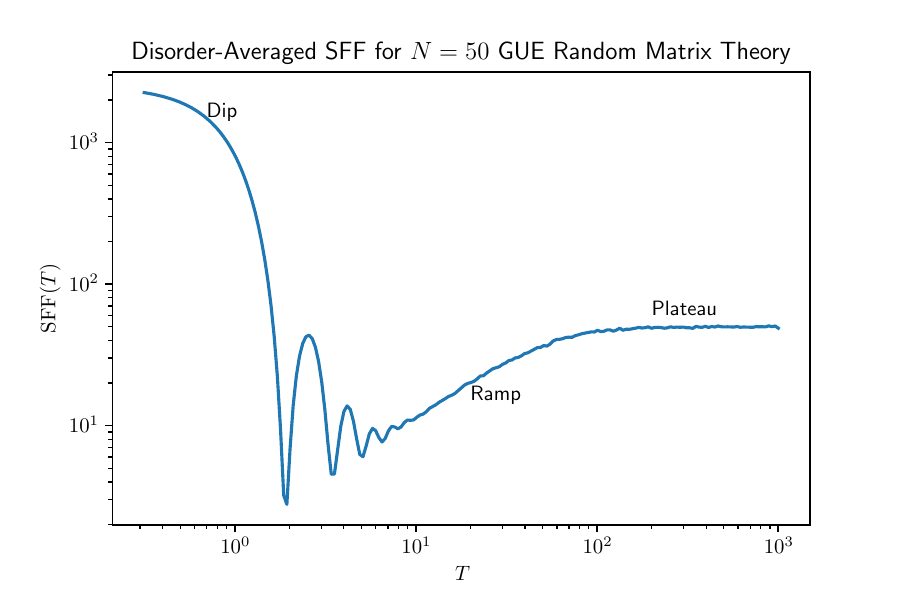}
\caption{The disorder-averaged SFF for the Gaussian unitary ensemble (GUE) of matrix dimension $N = 50$, computed numerically by averaging over ten thousand realizations. The three distinct regimes --- dip, ramp, plateau --- are indicated.}
\label{fig:5SFFgraph}
\end{figure}

It is conceptually useful to split the SFF into two contributions:
\begin{equation} \label{eq:general_SFF_decomposed5}
\textrm{SFF}(T, f) = \big| \mathbb{E} \textrm{Tr} f(H) e^{-iHT} \big|^2 + \bigg( \mathbb{E} \left[ \big| \textrm{Tr} f(H) e^{-iHT} \big|^2 \right] - \big| \mathbb{E} \textrm{Tr} f(H) e^{-iHT} \big|^2 \bigg).
\end{equation}
The first term, the disconnected piece of the SFF, comes solely from the average density of states.
It is the second term, the connected piece, that contains information on the correlation between energy levels.
The assertion of ``random matrix universality''~\cite{PhysRevLett.52.1,Altland1997} can be phrased as the statement that an ensemble of quantum chaotic Hamiltonians will generically have the same \textit{connected} SFF as the canonical Gaussian ensembles of random matrix theory~\cite{mehta2004random,tao2012topics}.
This conjectured universal behavior is illustrated in Fig.~\ref{fig:5SFFgraph}, which plots the disorder-averaged SFF of the Gaussian unitary ensemble (one of the aforementioned canonical ensembles).
Note the three distinct regimes:
\begin{itemize}
\item The ``dip'', occurring at short times, comes from the disconnected piece of the SFF (and thus its precise shape is non-universal).
It reflects a loss of constructive interference --- the different terms of $\textrm{Tr} e^{-iHT}$ acquire different phase factors as $T$ increases.
\item The ``ramp'', occurring at intermediate times, is arguably the most interesting regime.
In the canonical matrix ensembles, it is a consequence of the result\cite{mehta2004random}
\begin{equation} \label{eq:matrix_ensembles_ramp_result}
\mathbb{E} \left[ \rho \left( E + \frac{\omega}{2} \right) \rho \left( E - \frac{\omega}{2} \right) \right] - \mathbb{E} \left[ \rho \left( E + \frac{\omega}{2} \right) \right] \mathbb{E} \left[ \rho \left( E - \frac{\omega}{2} \right) \right] \sim -\frac{1}{\bbeta \pi^2 \omega^2},
\end{equation}
where $\bbeta = 1$, $2$, $4$ in the orthogonal, unitary, and sympletic ensembles respectively \cite{mehta2004random}.
The right-hand side being negative is a reflection of the well-known level repulsion in quantum chaotic systems \cite{wigner1959group}.
Taking the Fourier transform with respect to $\omega$ gives a term proportional to $T$ for the connected SFF.
Such a linear-in-$T$ ramp is often taken as a defining signature of quantum chaos.
\item The ``plateau'', occurring at late times, results from the discreteness of the spectrum.
At times much larger than the inverse level spacing, one expects that all off-diagonal terms in the double-trace of the SFF sum to effectively zero, meaning that
\begin{equation} \label{eq:5SFF_plateau_derivation}
\textrm{SFF}(T, f) = \sum_{mn} e^{-i(E_m - E_n)T} f(E_m) f(E_n) \sim \sum_n f(E_n)^2.
\end{equation}
As the plateau regime is both challenging to access analytically and not particularly informative, we shall not consider it further in this work.
\end{itemize}

The bulk of our analysis in this paper is devoted to calculation of the ramp in a well-known quantum spin glass model, the $p$-spherical model (discussed below).
The results can be understood via the elementary observation that when a Hamiltonian is block diagonal,
\begin{equation} \label{eq:block_diagonal_matrix}
H = \begin{pmatrix} H_1 & 0 & 0 \\ 0 & H_2 & 0 & \hdots \\ 0 & 0 & H_3 \\ & \vdots & & \ddots \end{pmatrix},
\end{equation}
then $\textrm{Tr} e^{-iHT} = \sum_k \textrm{Tr} e^{-iH_kT}$.
If the different blocks are independent, then the variance of $\textrm{Tr} e^{-iHT}$ is the sum of the variance of each $\textrm{Tr} e^{-iH_kT}$, i.e., \textit{the SFF is the sum of the SFF for each block.}
In particular, the coefficient of the universal linear-in-$T$ ramp is multiplied by the number of independent blocks.
Systems with only approximately block-diagonal Hamiltonians, for which there are small matrix elements between blocks, have this enhancement of the ramp up to the transition timescale between blocks.
For a more detailed analysis, see Ref.~\cite{winerprx}.

\subsection{Review of mean-field spin glasses} \label{subsec:review_mean_field_spin_glasses}

\begin{figure}[t]
\centering
\includegraphics[width=0.9\textwidth]{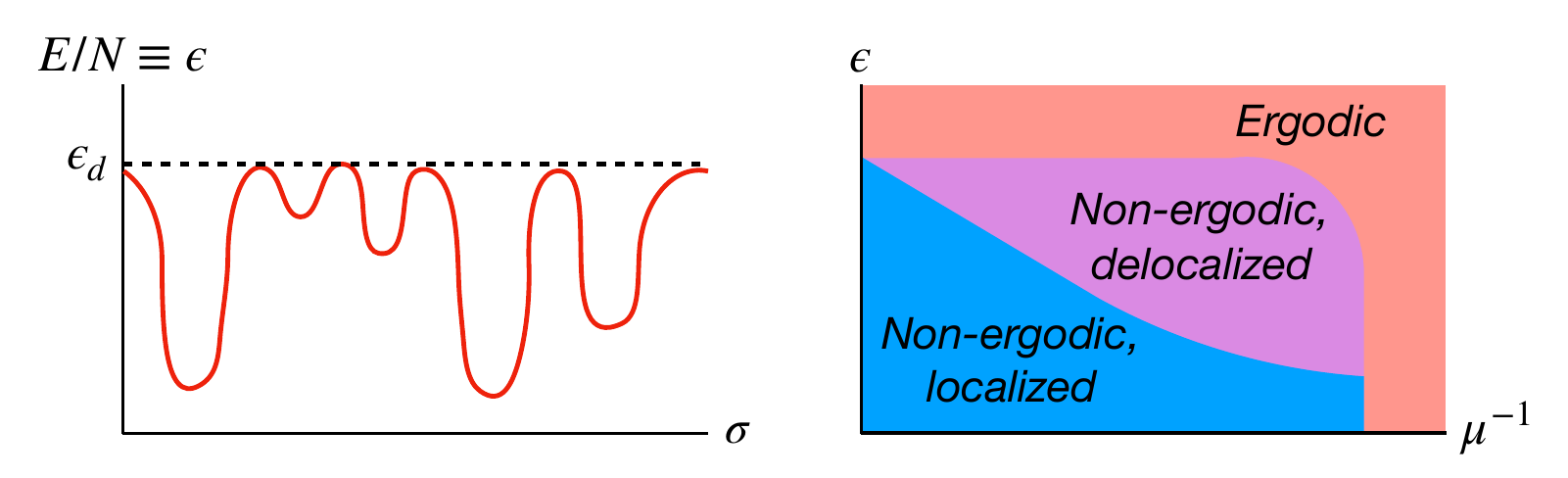}
\caption{(Left) Cartoon of the energy landscape in a 1RSB spin glass. The y-axis is energy per spin, $E/N$, where $E$ is energy and $N$ is the number of spins. Different points on the x-axis represent (very roughly, since the actual configuration space is $N$-dimensional) different spin configurations $\sigma$. The dashed line indicates the energy density $\epsilon_d$ below which the system is non-ergodic. (Right) Sketch of the dynamical phase diagram for a quantum 1RSB spin glass. The x-axis represents parameters controlling the strength of quantum fluctuations, and the y-axis is energy density. Note that many other types of phase transitions are also present, in particular equilibrium transitions, but are not indicated here. See, e.g., Refs.~\cite{Castellani2005Spin,Mezard2009,Anous:2021eqj} for more information.}
\label{fig:spin_glass_cartoons}
\end{figure}

Broadly speaking, spin glasses are systems in which the magnetic moments $\sigma_i$ are frozen but disordered at low temperatures.
However, this definition (much like that of ``quantum chaos'') encompasses a wide variety of phenomena which are in many ways quite distinct, as is made clear by the literature on the subject~\cite{Binder1986Spin,Mezard1987,Fischer1991,Nishimori2001,Castellani2005Spin,Mezard2009,Stein2013}.
In the present paper, we focus on what are known as ``one-step replica symmetry breaking'' (1RSB) spin glass phases~\cite{Mezard2009}.
We are specifically interested in quantum spin glasses, but we first review the corresponding classical case, for which configurations are labelled by a list $\sigma \equiv \{ \sigma_1, \cdots, \sigma_N \}$ and the Hamiltonian is simply a function of $\sigma$.

While the technical definition of 1RSB is somewhat involved, the qualitative physics is straightforward to understand and captured by the sketch in Fig.~\ref{fig:spin_glass_cartoons}.
The energy landscape, i.e., energy as a function of spin configuration, has many deep wells and steep barriers.
In particular, the number of wells is $e^{O(N)}$ and the heights of the energy barriers separating wells are $O(N)$, where $N$ is the number of spins. As a result, below a certain energy density $\epsilon_d$, the system is extremely non-ergodic: it remains trapped within an exponentially small fraction of the thermodynamically relevant configuration space until exponentially long timescales.
While the 1RSB phenomenon was originally studied in the context of stochastic classical dynamics~\cite{Kirkpatrick1987Dynamics,Crisanti1993Spherical,Cugliandolo1993Analytical,Barrat1996Dynamics}, it has recently been shown to imply exponentially long \textit{tunneling} timescales for isolated quantum dynamics as well~\cite{Altshuler2010Anderson,Bapst2013Quantum,Zhao2014Three,Baldwin2018Quantum,Smelyanskiy2020Nonergodic}.

TAP states (named after Thouless, Anderson, and Palmer~\cite{Thouless1977Solution}) provide a more quantitative description of such ``deep wells''.
Arguably the most general definition (see Ref.~\cite{Nishimori2001} for others) is in terms of the Legendre transform of the free energy with respect to local fields:
\begin{equation} \label{eq:TAP_Legendre_transform}
F \big( \{ m_i \} \big) = -\frac{1}{\beta} \log{\textrm{Tr} e^{-\beta H + \beta \sum_i h_i \sigma_i}} + \sum_i h_i m_i,
\end{equation}
where $H$ is the Hamiltonian of interest and the fields $\{ h_i \}$ are chosen so that $\langle \sigma_i \rangle = m_i$ (where $\langle \, \cdot \, \rangle$ indicates a thermal average).
TAP states are simply the local minima of $F(\{m_i\})$.
Physically, each corresponds to a different ``well'' of the energy landscape, including thermal fluctuations around the lowest point (thus TAP states do generically depend on temperature).
The partition function can be decomposed as a sum over TAP states:
\begin{equation} \label{eq:TAP_partition_decomposition}
Z \equiv \sum_{\sigma} e^{-\beta H(\sigma)} = \sum_{\alpha} \left[ \sum_{\sigma} \delta_{\sigma \in \alpha} e^{-\beta H(\sigma)} \right] \equiv \sum_{\alpha} Z_{\alpha},
\end{equation}
where $\alpha$ denotes a TAP state and $\delta_{\sigma \in \alpha}$ restricts the trace to only those states belonging to TAP state $\alpha$.
Note that in this discussion, $\sigma$ can refer to any set of degrees of freedom: Ising spins, vector spins, continuous coordinates, etc.
In all cases, Eqs.~\eqref{eq:TAP_Legendre_transform} and~\eqref{eq:TAP_partition_decomposition} can be interpreted accordingly.

Quantum generalizations of spin glasses are usually obtained by adding non-commuting terms to the Hamiltonian.
For example, with an Ising Hamiltonian, one often interprets $\sigma_i$ as the Pauli spin-$z$ operator $\sigma_i^z$ and includes an additional transverse field $\Gamma \sum_i \sigma_i^x$~\cite{Ishii1985Effect,Thirumalai1989Infinite,Goldschmidt1990,Buttner1990Replica}.
On the other hand, with systems having continuous degrees of freedom (including the one which we study in this paper), one can interpret $\sigma_i$ as a position coordinate and include the ``kinetic energy'' $\sum_i \pi_i^2 / 2\mu$, where $\pi_i$ is the momentum operator conjugate to $\sigma_i$~\cite{Cugliandolo1999RealTime,Cugliandolo2001}.
Generically, the resulting system has a frozen spin glass phase at low energy and small quantum fluctuations (the latter being controlled by $\Gamma$ and $\mu^{-1}$ respectively in the examples above), and has a paramagnetic phase at either high energy or large quantum fluctuations.
A sketch of the typical phase diagram is shown in Fig.~\ref{fig:spin_glass_cartoons}, with these two phases indicated by ``non-ergodic'' and ``ergodic''.

It has recently been noted that quantum 1RSB spin glasses can exhibit \textit{eigenstate} phase transitions which are distinct from the above~\cite{Laumann2014Many,Baldwin2017Clustering,Biroli2021Out}.
Qualitatively speaking, on the low energy/fluctuation side of the eigenstate phase boundary, each eigenstate of the Hamiltonian is localized on a single TAP state.
This implies that under the system's internal dynamics alone (i.e., as given by the Schrodinger equation), the system cannot tunnel between TAP states on \textit{any} timescale, even times exponential in the number of spins.
On the other side of the phase boundary, each eigenstate is delocalized over many TAP states in accordance with random matrix behavior.
As discussed in Ref.~\cite{Baldwin2018Quantum}, while this implies that the system does tunnel between TAP states, the timescale for tunneling is necessarily exponential in system size, analogous to the activation times under open-system dynamics.
Only when there exists a single TAP state can one identify the phase as genuinely thermalizing.
As a result, one finds phase diagrams like that sketched in Fig.~\ref{fig:spin_glass_cartoons}, with ``non-ergodic''/``ergodic'' indicating whether multiple TAP states exist and ``localized''/``delocalized'' referring to the eigenstate properties.


\subsection{Summary of results} \label{subsec:implications}

In this paper, we calculate the SFF for a particular ensemble of quantum spin glasses, the quantum $p$-spherical model (PSM)~\cite{Cugliandolo1998Quantum,Cugliandolo1999RealTime,Cugliandolo2001}. 
We find that in the ergodic phase, the connected part of the SFF agrees with the expectation from random matrix theory (Eq.~\eqref{eq:SFF_connected_contribution_ergodic_phase} below), while in the non-ergodic phase, it is enhanced by a factor which is precisely the number of TAP states (Eq.~\eqref{eq:SFF_final_result}).
Given the discussion in Secs.~\ref{subsec:review_spectral_form_factor} and~\ref{subsec:review_mean_field_spin_glasses}, this makes precise and validates the idea that each metastable state (i.e., TAP state) corresponds to a block of the Hamiltonian that is quantum chaotic on its own but is nearly decoupled from all others, thus making the system as a whole non-ergodic~\cite{Baldwin2017Clustering}.
This is the main result of the present work.

We also consider higher moments of the evolution operator and identify a set of saddle points (Eq.~\eqref{eq:higher_moment_special_final_expression}) which, in addition to confirming the picture that different TAP states have independent level statistics, suggest that at least at low complexity, the number of TAP states at a given energy is Poisson-distributed and independent of other energies.
Yet as we shall discuss, since our analysis does not consider the perturbative corrections around each saddle point, this does not constitute a complete calculation and serves more as motivation for future investigation.

\section{Real-time dynamics of the quantum $p$-spherical model} \label{sec:overview_PSM}

\subsection{The model} \label{subsec:model}

The classical $p$-spherical model (PSM)~\cite{Crisanti1992} is a disordered spin model with all-to-all $p$-body interactions.
It is defined by the classical Hamiltonian
\begin{equation} \label{eq:classical_Hamiltonian}
H_{\textrm{cl}} \equiv \sum_{(i_1 \cdots i_p)} J_{i_1 \cdots i_p} \sigma_{i_1} \cdots \sigma_{i_p},
\end{equation}
where the couplings $J_{i_1\dots i_p}$ are independent Gaussian random variables with mean zero and variance
\begin{equation} \label{eq:p_spin_coupling_variance}
\mathbb{E} {J_{i_1 \cdots i_p}}^2 = \frac{J^2(p-1)!}{C_{i_1 \cdots i_p} N^{p-1}}.
\end{equation}
Here and throughout, $\mathbb{E}$ indicates an average over couplings.
The notation $(i_1 \cdots i_p)$ denotes sets of $p$ indices such that $1 \leq i_1 \leq \cdots \leq i_p \leq N$. The sum in Eq.~\eqref{eq:classical_Hamiltonian} is over all such sets.
Our treatment differs from the standard convention by including a parameter $J$ for the overall strength of the disorder.
To recover the standard expressions, simply set $J^2=p/2$.
We also include the combinatorial factor $C_{i_1 \cdots i_p} = \prod_{1 \leq i \leq N} n_i!$, where $n_i$ is the number of indices set equal to $i$.
This term is almost always one, but its inclusion avoids $1/N$ corrections in the action. 

The $\sigma_i$ are real, continuous spin variables subject to the spherical constraint
\begin{equation} \label{eq:spherical_constraint}
\sum_{i=1}^N \sigma_i^2 = N,
\end{equation}
which ensures that the system has an extensive free energy.
It is apparent that this is a mean-field model without any spatial structure.
This allows for infinite free energy barriers around metastable states in the thermodynamic limit, making the model ideal for examining the impact of metastability on the spectral statistics of spin glasses.

In this work, we follow Refs.~\cite{Cugliandolo1998Quantum,Cugliandolo1999RealTime,Cugliandolo2001} in generalizing Eq.~\eqref{eq:classical_Hamiltonian} to a quantum Hamiltonian $H$.
We treat the $\sigma_i$ as commuting position operators, and define conjugate momentum operators $\pi_i$ which satisfy the commutation relations
\begin{equation} \label{eq:canonical_commutation_relations}
[\sigma_i, \pi_j] = i \delta_{ij}.
\end{equation}
The \textit{quantum} PSM simply includes a kinetic energy term in the Hamiltonian:
\begin{equation} \label{eq:quantum_Hamiltonian}
H = \sum_{i=1}^N \frac{\pi_i^2}{2\mu} + \sum_{(i_1 \cdots i_p)} J_{i_1 \cdots i_p} \sigma_{i_1} \cdots \sigma_{i_p}.
\end{equation}
The mass $\mu$ is an additional parameter controlling the strength of quantum fluctuations.
To incorporate the spherical constraint, we take the Hilbert space to be the subspace in which $\sum_i \sigma_i^2$ has eigenvalue $N$.

The quantum PSM may be interpreted as a soft-spin version of the Ising $p$-spin model in an external transverse field --- itself the subject of much study~\cite{Gardner1985,Goldschmidt1990,Nieuwenhuizen1998,Dobrosavljevic1990,DeCesare1996} --- where $\mu^{-1}$ is analogous to the transverse field.
Alternatively, if we think of $\sigma \equiv \{ \sigma_1, \cdots, \sigma_N \}$ as a position vector in $N$-dimensional space, the quantum PSM has a natural interpretation as a particle of mass $\mu$ moving on a hypersphere of radius $\sqrt{N}$.
This particle experiences the Gaussian random potential
\begin{equation} \label{eq:hypersphere_random_potential}
V(\sigma) = \sum_{(i_1 \cdots i_p)} J_{i_1 \cdots i_p} \sigma_{i_1} \cdots \sigma_{i_p},
\end{equation}
whose correlation function is
\begin{equation} \label{eq:random_potential_correlation}
\mathbb{E} V(\sigma) V(\sigma') = \frac{J^2 (p-1)!}{C_{i_1 \cdots i_p} N^{p-1}} \sum_{(i_1 \cdots i_p)} \sigma_{i_1} \sigma'_{i_1} \cdots \sigma_{i_p} \sigma'_{i_p} = \frac{J^2}{pN^{p-1}} \big( \sigma \cdot \sigma' \big)^p.
\end{equation}

Note that there is a very important difference between $p=2$ and $p>2$: the former is a Gaussian model, essentially (but for the spherical constraint) a system of linearly coupled harmonic oscillators.
It therefore has qualitatively different behavior than the $p>2$ models, which are genuinely interacting and serve as reasonable toy models for rugged energy landscapes.
In this work, we exclusively consider $p>2$.

\subsection{Schwinger-Keldysh path integral} \label{subsec:Schwinger_Keldysh_path_integral}

\begin{figure}[t]
\centering
\includegraphics[width=1.0\textwidth]{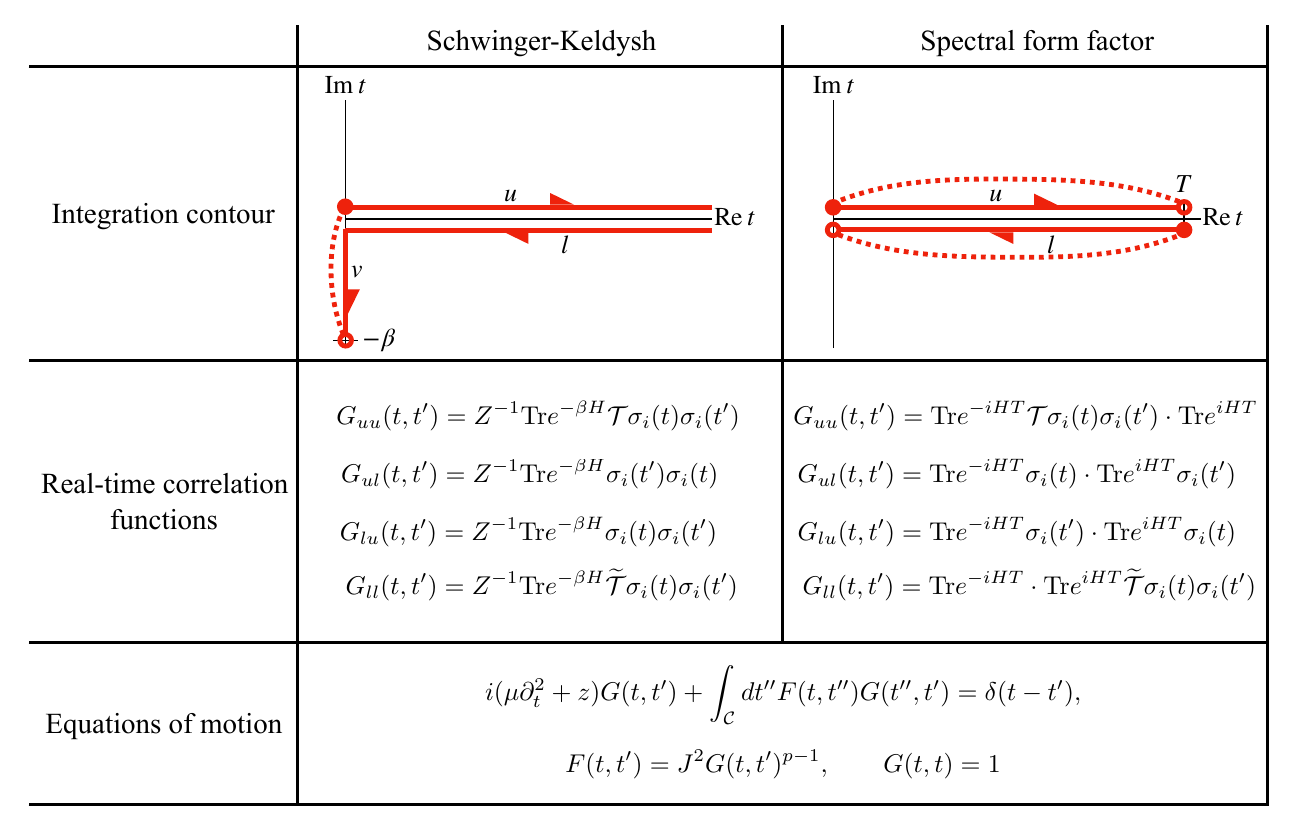}
\caption{Summary of the contours, order parameters, and (at least at high temperature) equations of motion considered in this work. The left column gives the quantities appropriate to the Schwinger-Keldysh path integral, and the right column to the spectral form factor (SFF) path integral. \\ (Top row) Contours for the respective path integrals. Each of the different branches is labelled, and directions are indicated by arrowheads. Points connected by dashed lines are identified, making the contours periodic. \\
(Middle row) Relationship between order parameters of the theory and observable quantities. $H$ and $Z$ are the $p$-spin Hamiltonian and partition function respectively. $\mathcal{T}$ and $\widetilde{\mathcal{T}}$ denote time ordering and anti-ordering. \\
(Bottom row) Equations of motion. These take the same form for both path integrals, differing only in the contour $\mathcal{C}$ being used.}
\label{fig:contour_summary}
\end{figure}

Just as other all-to-all models have a saddle-point/mean-field description at large $N$, so too does the PSM.
We start with the disorder-averaged (i.e., ``annealed'') path integral on the Schwinger-Keldysh contour at inverse temperature $\beta$, illustrated in the left column of Fig.~\ref{fig:contour_summary}.
While it is in general incorrect (often grossly) to disorder-average the path integral itself, it is known that the annealed approximation is accurate in the PSM as long as $\beta$ is less than a critical value $\beta_s$~\cite{Castellani2005Spin,Thomson_2020}.
We shall assume that this is true throughout.
The annealed path integral is
\begin{equation} \label{eq:Keldysh_path_integral_start_1}
\begin{aligned}
\mathbb{E} Z_{\textrm{SK}} &= \int \mathcal{D} \sigma^N \exp \left[ \int_{\mathcal{C}} dt \sum_i \left( \frac{i \mu}{2} \big( \partial_t \sigma_i(t) \big)^2 - \frac{i z(t)}{2} \big( \sigma_i(t)^2 - 1 \big) \right) \right] \\
&\qquad \qquad \quad \cdot \int dP(J) \exp \left[ -i \int_{\mathcal{C}} dt \sum_{(i_1 \cdots i_p)} J_{i_1 \cdots i_p} \sigma_{i_1}(t) \cdots \sigma_{i_p}(t) \right],
\end{aligned}
\end{equation}
where
\begin{equation} \label{eq:PSM_coupling_distribution}
dP(J) \propto \prod_{(i_1 \cdots i_p)} dJ_{i_1 \cdots i_p} \exp \left[ -\frac{N^{p-1} C_{i_1 \cdots i_p} J_{i_1 \cdots i_p}^2}{2(p-1)!J^2} \right].
\end{equation}
For brevity, we use $\mathcal{C}$ to denote the entire contour.
Thus $\int_{\mathcal{C}} dt$ indicates a contour integral within the complex-$t$ plane.
The Lagrange multiplier $z(t)$ is included to enforce the spherical constraint.
It can be interpreted as a time-dependent harmonic potential whose value is chosen such that $\sum_i \sigma_i(t)^2 = N$ at all times.
Thus the measure $\mathcal{D}\sigma^N$ is simply the product measure over each $\sigma_i$ independently.
From here, the same manipulations used to get Schwinger-Dyson equations for the SYK model will give us equations of motion for the PSM.

One can immediately perform the Gaussian integrals over the couplings to obtain
\begin{equation} \label{eq:disorder_averaged_partition_function}
\mathbb{E} Z_{\textrm{SK}} = \int \mathcal{D} \sigma^N e^{-NS'},
\end{equation}
where
\begin{equation} \label{eq:annealed_effective_action}
\begin{aligned}
NS' &\equiv \int_{\mathcal{C}} dt \sum_i \left( -\frac{i\mu}{2} \big( \partial_t \sigma_i(t) \big)^2 + \frac{iz(t)}{2} \big( \sigma_i(t)^2 - 1 \big) \right) \\
& \qquad \qquad + \frac{J^2 (p-1)!}{2 C_{i_1 \cdots i_p} N^{p-1}} \sum_{(i_1 \cdots i_p)} \int_{\mathcal{C}} dt dt' \sigma_{i_1}(t) \sigma_{i_1}(t') \cdots \sigma_{i_p}(t) \sigma_{i_p}(t') \\
&= \int_{\mathcal{C}} dt \sum_i \left( -\frac{i\mu}{2} \big( \partial_t \sigma_i(t) \big)^2 + \frac{iz(t)}{2} \big( \sigma_i(t)^2 - 1 \big) \right) + \frac{NJ^2}{2p} \int_{\mathcal{C}} dt dt' \left( \frac{1}{N} \sum_i \sigma_i(t) \sigma_i(t') \right)^p.
\end{aligned}
\end{equation}
Next introduce a ``fat unity'',
\begin{equation} \label{eq:partition_function_fat_unity}
\begin{aligned}
1 =& \int \mathcal{D}\mathcal{G} \prod_{tt'} \delta \Big( N\mathcal{G}(t, t') - \sum_i \sigma_i(t) \sigma_i(t') \Big) \\
=& \int \mathcal{D}\mathcal{G} \mathcal{D}\mathcal{F} \exp\left[ \frac{N}{2} \int_{\mathcal{C}} dt dt' \mathcal{F}(t, t') \left( \mathcal{G}(t, t') - \frac{1}{N} \sum_i \sigma_i(t) \sigma_i(t') \right) \right].
\end{aligned}
\end{equation}
The integral over the self-energy $\mathcal{F}(t, t')$ runs along the imaginary axis, making the second line simply the identity $\int dp e^{ipx} = 2\pi \delta(x)$ (we absorb factors of $2\pi$ into the measure $\mathcal{D}\mathcal{F}$).
However, when we ultimately evaluate the path integral by saddle point, we shall find that the saddle point value of $\mathcal{F}(t, t')$ is real.
Inserting Eq.~\eqref{eq:partition_function_fat_unity} into the path integral gives
\begin{equation} \label{eq:disorder_averaged_expanded_partition_function}
\mathbb{E} Z_{\textrm{SK}} = \int \mathcal{D}\mathcal{G} \mathcal{D}\mathcal{F} \int \mathcal{D} \sigma^N e^{-NS''},
\end{equation}
where
\begin{equation} \label{eq:annealed_expanded_effective_action}
\begin{aligned}
NS'' &\equiv -\frac{iN}{2} \int_{\mathcal{C}} dt z(t) + \frac{N}{2} \int_{\mathcal{C}} dt dt' \left( \frac{J^2}{p} \mathcal{G}(t, t')^p - \mathcal{F}(t, t') \mathcal{G}(t, t') \right) \\
&\qquad \qquad + \frac{1}{2} \sum_i \left[ \int_{\mathcal{C}} dt \left( -i \mu \big( \partial_t \sigma_i(t) \big)^2 + iz(t) \sigma_i(t)^2 \right) + \int_{\mathcal{C}} dt dt' \sigma_i(t) \mathcal{F}(t, t') \sigma_i(t') \right].
\end{aligned}
\end{equation}
We can now perform the integral over $\sigma_i$, resulting in
\begin{equation} \label{eq:disorder_averaged_final_partition_function}
\mathbb{E} Z_{\textrm{SK}} = \int \mathcal{D}\mathcal{G} \mathcal{D}\mathcal{F} e^{-NS_{\textrm{eff}}},
\end{equation}
where
\begin{equation} \label{eq:annealed_final_action}
S_{\textrm{eff}} \equiv -\frac{i}{2} \int_{\mathcal{C}} dt z(t) + \frac{1}{2} \int_{\mathcal{C}} dt dt' \left( \frac{J^2}{p} \mathcal{G}(t, t')^p - \mathcal{F}(t, t') \mathcal{G}(t, t') \right) + \frac{1}{2} \log{\textrm{Det}} \Big[ i(\mu \partial_t^2 + z) + \mathcal{F} \Big].
\end{equation}

At large $N$, the remaining path integral can be evaluated within the saddle point approximation.
The locations of the saddle points are determined by setting to zero the functional derivatives of Eq.~\eqref{eq:annealed_final_action}:
\begin{equation} \label{eq:Keldysh_EOM}
\begin{gathered}
i \big( \mu \partial_t^2 + z(t) \big) \mathcal{G}(t, t') + \int_{\mathcal{C}} dt'' \mathcal{F}(t, t'') \mathcal{G}(t'', t') = \delta(t - t'), \\
\mathcal{F}(t, t') = J^2 \mathcal{G}(t, t')^{p-1}, \qquad \mathcal{G}(t, t) = 1.
\end{gathered}
\end{equation}
Keep in mind that the time arguments in Eq.~\eqref{eq:Keldysh_EOM} are complex and range over the entire Schwinger-Keldysh contour.
In particular, although it is hidden in this compact notation, the infinitesimals $dt$ acquire different phases depending on the branch of the contour: $dt$ is a positive real infinitesimal on the upper (``forward'') real-time branch, a negative real infinitesimal on the lower (``backward'') real-time branch, and a negative imaginary infinitesimal on the thermal branch.

$\mathcal{G}(t, t')$ is the order parameter of this theory.
As is clear from the manner by which it was introduced (top line of Eq.~\eqref{eq:partition_function_fat_unity}), expectation values of $\mathcal{G}(t, t')$ within the path integral are equivalent to expectation values of $N^{-1} \sum_i \sigma_i(t) \sigma_i(t')$.
The latter are simply time-ordered correlation functions.
We shall focus on the real-time correlation functions, for which it is more transparent to explicitly indicate the branches by $\alpha \in \{u, l\}$ and have $t$ be simply a real variable.
Formally, we have that
\begin{equation} \label{eq:Keldysh_formal_expectation_values}
\begin{aligned}
\big< \mathcal{G}_{uu}(t, t') \big> &= \mathbb{E} \Big[ Z_{\textrm{SK}}^{-1} \textrm{Tr} e^{-\beta H} \mathcal{T} \sigma_i(t) \sigma_i(t') \Big], & \qquad \big< \mathcal{G}_{ul}(t, t') \big> &= \mathbb{E} \Big[ Z_{\textrm{SK}}^{-1} \textrm{Tr} e^{-\beta H} \sigma_i(t') \sigma_i(t) \Big], \\
\big< \mathcal{G}_{lu}(t, t') \big> &= \mathbb{E} \Big[ Z_{\textrm{SK}}^{-1} \textrm{Tr} e^{-\beta H} \sigma_i(t) \sigma_i(t') \Big], & \qquad \big< \mathcal{G}_{ll}(t, t') \big> &= \mathbb{E} \Big[ Z_{\textrm{SK}}^{-1} \textrm{Tr} e^{-\beta H} \widetilde{\mathcal{T}} \sigma_i(t) \sigma_i(t') \Big],
\end{aligned}
\end{equation}
where $\mathcal{T}$ denotes time ordering and $\widetilde{\mathcal{T}}$ denotes time anti-ordering.
Note that we can omit the sum over $i$ because the different spins (upon disorder-averaging) have equivalent behavior.

A number of formal properties of $\mathcal{G}_{\alpha \alpha'}(t, t')$ are evident from Eq.~\eqref{eq:Keldysh_formal_expectation_values}.
For one thing, $\mathcal{G}_{\alpha \alpha'}(t, t')$ clearly depends only on the time difference $t - t'$, and we shall often write $\mathcal{G}_{\alpha \alpha'}(t)$ with $t'=0$.
Since the four components differ only in time ordering, we see that for \textit{any} function $f(x)$,
\begin{equation} \label{eq:Keldysh_component_symmetry_identity}
f \Big( \mathcal{G}_{uu}(t) \Big) + f \Big( \mathcal{G}_{ll}(t) \Big) = f \Big( \mathcal{G}_{ul}(t) \Big) + f \Big( \mathcal{G}_{lu}(t) \Big).
\end{equation}
We can further express all four components in terms of a single complex-valued function (equivalently two real-valued functions).
For example, write $\mathcal{G}_{lu}(t)$ in terms of its real and imaginary parts as $\mathcal{G}^R(t) + i \mathcal{G}^I(t)$.
Since $\mathcal{G}_{lu}(t)^* = \mathcal{G}_{lu}(-t)$, $\mathcal{G}^R(t)$ is even and $\mathcal{G}^I(t)$ is odd.
One can easily confirm that
\begin{equation} \label{eq:Keldysh_correlation_relationships}
\begin{aligned}
\mathcal{G}_{uu}(t) &= \mathcal{G}^R(t) + i \textrm{sgn}[t] \mathcal{G}^I(t), & \qquad \mathcal{G}_{ul}(t) &= \mathcal{G}^R(t) - i \mathcal{G}^I(t), \\
\mathcal{G}_{lu}(t) &= \mathcal{G}^R(t) + i \mathcal{G}^I(t), & \qquad \mathcal{G}_{ll}(t) &= \mathcal{G}^R(t) - i \textrm{sgn}[t] \mathcal{G}^I(t).
\end{aligned}
\end{equation}

One of the most important features of $\mathcal{G}_{\alpha \alpha'}(t, t')$ is the limiting behavior at large $|t - t'|$, as a function of the inverse temperature $\beta$.
Numerical solution of Eq.~\eqref{eq:Keldysh_EOM} demonstrates that there is a critical value $\beta_d$ (which is less than $\beta_s$):
\begin{itemize}
\item For $\beta < \beta_d$, $\lim_{|t-t'| \rightarrow \infty} \mathcal{G}_{\alpha \alpha'}(t, t') = 0$.
We call this the ``ergodic'' phase ($\mathbb{E} \langle \sigma_i(t) \rangle = 0$ by symmetry regardless of temperature, and so in this phase $\mathbb{E} \langle \sigma_i(t) \sigma_i(t') \rangle \rightarrow \mathbb{E} \langle \sigma_i(t) \rangle \mathbb{E} \langle \sigma_i(t') \rangle$).
\item For $\beta_d < \beta < \beta_s$, $\lim_{|t-t'| \rightarrow \infty} \mathcal{G}_{\alpha \alpha'}(t, t') = q_{\textrm{EA}} > 0$.
We call this the ``non-ergodic'' phase.
The quantity $q_{\textrm{EA}}$ is referred to as the ``Edwards-Anderson'' order parameter.
\item For $\beta_s < \beta$, our initial annealed approximation is no longer valid.
The replica trick is required to obtain accurate results~\cite{Mezard1987,Fischer1991}, but (at least for finite-time dynamical properties) the behavior is qualitatively similar to that of the non-ergodic phase.
\end{itemize}

\subsection{TAP equations on the Schwinger-Keldysh contour}
\label{subsec:QTAP}

The dynamical calculation described above only hints at the complexity of the non-ergodic phase.
A more complete picture emerges from a generalization in the spirit of the TAP equations.
Our treatment follows that of Ref.~\cite{Biroli_2001}, which derived TAP equations on the thermal circle for the quantum PSM.
While the extension to real-time dynamics is straightforward, we are not aware of any explicit calculation in the literature.
Thus we present a detailed derivation of the following equations in App.~\ref{sec:TAP_derivation}.

As discussed in Sec.~\ref{subsec:review_mean_field_spin_glasses}, the TAP free energy (or Gibbs potential) is the Legendre transform of the free energy with respect to local fields.
It is therefore a function of the magnetization $m_i$ of each spin.
For the free energy of quantum systems, the magnetization should also have an imaginary time index $m_i(\tau)$.
The imaginary-time correlation function $\mathcal{G}(\tau, \tau')$ becomes an additional order parameter.

We define the TAP \textit{action} on the Schwinger-Keldysh contour analogously.
It is a function of the magnetizations $m_i(t)$ and the correlation function $\mathcal{G}(t, t')$, with $t$ again being complex-valued and ranging over the entire contour.
Specifically,
\begin{equation} \label{eq:Keldysh_TAP_action}
\begin{aligned}
iNS_{\textrm{TAP}}[m, \mathcal{G}] &\equiv \log{\int \mathcal{D}\sigma^N \exp \left[ i \sum_i S_i^0 - i \int_{\mathcal{C}} dt \sum_{(i_1 \cdots i_p)} J_{i_1 \cdots i_p} \sigma_{i_1}(t) \cdots \sigma_{i_p}(t) \right]} \\
&\qquad \qquad + \frac{iN}{2} \int_{\mathcal{C}} dt z(t) - i \int_{\mathcal{C}} dt \sum_i h_i(t) m_i(t) + \frac{iN}{2} \int_{\mathcal{C}} dt dt' \Lambda(t, t') \mathcal{G}(t, t'),
\end{aligned}
\end{equation}
where $\mathcal{C}$ denotes the Schwinger-Keldysh contour and
\begin{equation} \label{eq:Keldysh_TAP_noninteracting_action}
S_i^0 \equiv \int_{\mathcal{C}} dt \left( \frac{\mu}{2} \big( \partial_t \sigma_i(t) \big)^2 - \frac{z(t)}{2} \sigma_i(t)^2 + h_i(t) \sigma_i(t) \right) - \frac{1}{2} \int_{\mathcal{C}} dt dt' \Lambda(t, t') \sigma_i(t) \sigma_i(t').
\end{equation}
The fields $h_i(t)$ and $\Lambda(t, t')$ are \textit{not} independent parameters.
They are instead chosen so that $\langle \sigma_i(t) \rangle = m_i(t)$ and $N^{-1} \sum_i \langle \sigma_i(t) \sigma_i(t') \rangle = \mathcal{G}(t, t')$, just as $z(t)$ is again chosen to enforce $N^{-1} \sum_i \langle \sigma_i(t)^2 \rangle = 1$, where the expectation value is with respect to the action in Eq.~\eqref{eq:Keldysh_TAP_action}.

Due to the Legendre-transform structure of $S_{\textrm{TAP}}$, we have that
\begin{equation} \label{eq:quantum_TAP_Legendre_relations}
N \frac{\partial S_{\textrm{TAP}}}{\partial m_i(t)} = -h_i(t), \qquad \frac{\partial S_{\textrm{TAP}}}{\partial \mathcal{G}(t, t')} = \frac{1}{2} \Lambda(t, t').
\end{equation}
The TAP equations are those for $m_i(t)$ and $\mathcal{G}(t, t')$ which one gets by setting the right-hand sides of Eq.~\eqref{eq:quantum_TAP_Legendre_relations} to zero.
The solutions are therefore the values of magnetization and correlation function which the system can consistently possess ``on its own,'' without any external fields.
In this sense, each solution corresponds to a distinct metastable state.
There is no reason why there cannot be many self-consistent solutions, and indeed, spin glass models such as the PSM do have many at sufficiently low temperature.

We calculate the TAP equations in App.~\ref{sec:TAP_derivation}.
They are simplified by the fact that we can take $m_i(t) = m$ and $z(t) = z$.
We also define $q_{\textrm{EA}} \equiv N^{-1} \sum_i m_i^2$.
The equations come out to be (together with $\mathcal{G}(t, t) = 1$)
\begin{equation} \label{eq:Keldysh_TAP_EOM}
i \big( \mu \partial_t^2 + z \big) \Big( \mathcal{G}(t, t') - q_{\textrm{EA}} \Big) + J^2 \int_{\mathcal{C}} dt'' \Big( \mathcal{G}(t, t'')^{p-1} - q_{\textrm{EA}}^{p-1} \Big) \Big( \mathcal{G}(t'', t') - q_{\textrm{EA}} \Big) = \delta(t - t'),
\end{equation}
\begin{equation} \label{eq:Keldysh_TAP_magnetization_equation}
J^2 \int_{\mathcal{C}} dt' \Big( \mathcal{G}(t, t')^{p-1} - (p-1) q_{\textrm{EA}}^{p-2} \mathcal{G}(t, t') + (p-2) q_{\textrm{EA}}^{p-1} \Big) m_i = -izm_i - i \sum_{(i_1 \cdots i_p)} J_{i_1 \cdots i_p} \frac{\partial (m_{i_1} \cdots m_{i_p})}{\partial m_i}.
\end{equation}
Note that Eq.~\eqref{eq:Keldysh_TAP_magnetization_equation} is $N$ equations, one for each spin $i$, and that it holds equally for any value of $t$ due to time translation invariance.
Defining $\mathcal{F}(t, t') \equiv J^2 \mathcal{G}(t, t')^{p-1}$, Eq.~\eqref{eq:Keldysh_TAP_EOM} is quite similar to Eq.~\eqref{eq:Keldysh_EOM}.
The only difference is that Eq.~\eqref{eq:Keldysh_TAP_EOM} uses $\Delta \mathcal{G}(t, t') \equiv \mathcal{G}(t, t') - q_{\textrm{EA}}$ and $\Delta \mathcal{F}(t, t') \equiv \mathcal{F}(t, t') - J^2 q_{\textrm{EA}}^{p-1}$, which decay to zero at large $|t - t'|$, rather than $\mathcal{G}(t, t')$ and $\mathcal{F}(t, t')$ themselves.

Despite the more involved derivation, $\mathcal{G}(t, t')$ remains a contour-ordered expectation value.
Thus, returning to the notation in which $\alpha \in \{u, l\}$ labels branches and $t$ is real, $\mathcal{G}_{\alpha \alpha'}(t - t')$ possesses the same formal properties as discussed in the previous subsection (Eqs.~\eqref{eq:Keldysh_component_symmetry_identity} and~\eqref{eq:Keldysh_correlation_relationships}).
Of particular importance will be the Fourier transform of $\Delta \mathcal{G}_{\alpha \alpha'}(t)$ at zero frequency, denoted $\Delta \widetilde{\mathcal{G}}_{\alpha \alpha'}(0)$, as well as its (matrix) inverse, $\Delta \widetilde{\mathcal{G}}_{\alpha \alpha'}^{-1}(0)$.
Also define $L \equiv \int_{-\infty}^{\infty} dt \Delta \mathcal{G}^R(t)$ and $\Lambda \equiv \int_0^{\infty} dt \Delta \mathcal{G}^I(t)$.
Then from Eq.~\eqref{eq:Keldysh_correlation_relationships}, we see that
\begin{equation} \label{eq:Keldysh_zero_frequency_matrix}
\begin{pmatrix} \Delta \widetilde{\mathcal{G}}_{uu}(0) & \Delta \widetilde{\mathcal{G}}_{ul}(0) \\ \Delta \widetilde{\mathcal{G}}_{lu}(0) & \Delta \widetilde{\mathcal{G}}_{ll}(0) \end{pmatrix} = \begin{pmatrix} L + 2i \Lambda & L \\ L & L - 2i \Lambda \end{pmatrix},
\end{equation}
\begin{equation} \label{eq:Keldysh_zero_frequency_matrix_inverse}
\begin{pmatrix} \Delta \widetilde{\mathcal{G}}_{uu}(0) & \Delta \widetilde{\mathcal{G}}_{ul}(0) \\ \Delta \widetilde{\mathcal{G}}_{lu}(0) & \Delta \widetilde{\mathcal{G}}_{ll}(0) \end{pmatrix}^{-1} = \frac{1}{4 \Lambda^2} \begin{pmatrix} L - 2i \Lambda & -L \\ -L & L + 2i \Lambda \end{pmatrix}.
\end{equation}

The multiplicity of solutions to the TAP equations comes from Eq.~\eqref{eq:Keldysh_TAP_magnetization_equation}.
By use of Eqs.~\eqref{eq:Keldysh_TAP_EOM},~\eqref{eq:Keldysh_zero_frequency_matrix}, and~\eqref{eq:Keldysh_zero_frequency_matrix_inverse}, it can be written (associating $u$ with 0 and $l$ with 1)
\begin{equation} \label{eq:Keldysh_TAP_magnetization_alternate}
\begin{aligned}
\left[ (-1)^{\alpha} \sum_{\alpha'} \Delta \widetilde{\mathcal{G}}_{\alpha \alpha'}^{-1}(0) - (p-1)J^2 q_{\textrm{EA}}^{p-2} \sum_{\alpha'} (-1)^{\alpha'} \Delta \widetilde{\mathcal{G}}_{\alpha \alpha'}(0) \right] &m_i \\
= \left[ \frac{1}{2i \Lambda} - (p-1)J^2 q_{\textrm{EA}}^{p-2} 2i \Lambda \right] &m_i = -i \sum_{(i_1 \cdots i_p)} J_{i_1 \cdots i_p} \frac{\partial (m_{i_1} \cdots m_{i_p})}{\partial m_i}.
\end{aligned}
\end{equation}
Eq.~\eqref{eq:Keldysh_TAP_magnetization_alternate} is identical to that which appears and has been well-studied for the \textit{classical} PSM \cite{Crisanti1992,Crisanti1993Spherical,Crisanti1995ThoulessAndersonPalmerAT,Castellani2005Spin}.
Thus we simply quote the following results.
In addition to the inverse temperature $\beta$, solutions to Eq.~\eqref{eq:Keldysh_TAP_magnetization_alternate} are parametrized by the quantity
\begin{equation} \label{eq:normalized_energy_definition}
\mathcal{E} \equiv \frac{1}{NJq_{\textrm{EA}}^{p/2}} \sum_{(i_1 \cdots i_p)} J_{i_1 \cdots i_p} m_{i_1} \cdots m_{i_p},
\end{equation}
which can be interpreted as a ``normalized'' potential energy density: each magnetization has a value which is (very roughly) comparable to $q_{\textrm{EA}}^{1/2}$, and thus the natural scale for the interaction energy is $Jq_{\textrm{EA}}^{p/2}$.
The value of $q_{\textrm{EA}}$ for a given $\mathcal{E} < 0$ is given by the largest solution to
\begin{equation} \label{eq:Keldysh_TAP_overlap_closed_equation}
-2Jq_{\textrm{EA}}^{p/2 - 1} \Lambda = \frac{p}{2(p-1)} \left( -\mathcal{E} - \sqrt{\mathcal{E}^2 - \mathcal{E}_{\textrm{th}}^2} \right), \qquad \mathcal{E}_{\textrm{th}} \equiv -\frac{2\sqrt{p-1}}{p},
\end{equation}
where $\Lambda$ depends on $q_{\textrm{EA}}$ through Eq.~\eqref{eq:Keldysh_TAP_EOM}.
One can show that solutions to Eq.~\eqref{eq:Keldysh_TAP_overlap_closed_equation} exist only for $\beta > \beta_d$, with $\beta_d$ the same as defined in Sec.~\ref{subsec:Schwinger_Keldysh_path_integral}.
Furthermore, Eq.~\eqref{eq:Keldysh_TAP_overlap_closed_equation} only makes sense if $\mathcal{E} \leq \mathcal{E}_{\textrm{th}}$.
In that case, the number of solutions $\mathcal{N}(\beta, \mathcal{E})$ to Eq.~\eqref{eq:Keldysh_TAP_magnetization_alternate} --- in addition to the trivial solution $m_i = 0$ --- is exponential in system size: $N^{-1} \log{\mathcal{N}(\beta, \mathcal{E})} \sim \Sigma(\mathcal{E})$, with\footnote{
As written, Eq.~\eqref{eq:TAP_complexity} is a bit sloppy.
$\mathcal{N}(\mathcal{E})$ is given by Eq.~\eqref{eq:TAP_complexity} when the latter is non-negative and $\beta$ is such that solutions to Eq.~\eqref{eq:Keldysh_TAP_overlap_closed_equation} exist.
In all other cases, $\mathcal{N}(\mathcal{E}) = 0$.
}
\begin{equation} \label{eq:TAP_complexity}
\Sigma(\mathcal{E}) = \frac{1}{2} \left( 1 + 2\log{\frac{p}{2}} \right) - \frac{p \mathcal{E}^2}{2} + \frac{p^2}{8(p-1)} \Big( \mathcal{E} + \sqrt{\mathcal{E}^2 - \mathcal{E}_{\textrm{th}}^2} \Big)^2 + \log{\Big( -\mathcal{E} + \sqrt{\mathcal{E}^2 - \mathcal{E}_{\textrm{th}}^2} \Big)}.
\end{equation}
The exponent $\Sigma(\mathcal{E})$ is referred to as the ``complexity'' in the spin glass literature.

The connection between this TAP approach and the conventional Schwinger-Keldysh path integral lies in the fact that: i) the inverse temperature $\beta_d$ at which TAP states with non-zero magnetization appear is identical to that at which the autocorrelation function acquires a non-zero late-time limit; ii) the overlap determined by Eq.~\eqref{eq:Keldysh_TAP_overlap_closed_equation} is identical to the late-time value of the autocorrelation function.
This strongly suggests the following picture:
\begin{itemize}
\item For $\beta < \beta_d$ (the ``ergodic'' phase), there exists a single equilibrium state with zero magnetization, and the correlation function decays to zero on a finite timescale.
\item For $\beta_d < \beta$ (the ``non-ergodic'' phase), there exist exponentially many metastable states having non-zero magnetization.
The number of states is given by the exponential of the complexity $\Sigma(\mathcal{E})$.
Dynamically, in the $N \rightarrow \infty$ limit, a system prepared in one metastable state will remain in that state for all time.
At finite $N$, it is only on a timescale exponential in $N$ that the system can transition between states.
\end{itemize}
Much more can be said about these phases (in particular how the replica-symmetry-breaking transition at $\beta_s$ appears within the TAP approach), and a large body of literature is devoted to this topic.
We refer in particular to Ref.~\cite{Castellani2005Spin} as an excellent starting point.

In the present work, we determine the spectral statistics of the PSM in both the ergodic and non-ergodic phase.
Those of the former can be computed very much along the lines of Ref.~\cite{saad2019semiclassical}, which we do in Sec.~\ref{sec:ergodic_ramp}.
Those of the latter, however, require novel calculations which we present in Sec.~\ref{sec:nonergodic_ramp}.
Unsurprisingly, the properties of TAP states shall play an essential role.

\section{The semiclassical ramp in the ergodic phase}
\label{sec:ergodic_ramp}

To reiterate, we are evaluating
\begin{equation} \label{eq:generic_SFF_definition}
\textrm{SFF}(T, f) \equiv \mathbb{E} \big| \textrm{Tr} f(H) e^{-iHT} \big|^2 = \mathbb{E} \Big[ \textrm{Tr} f(H) e^{-iHT} \textrm{Tr} f(H) e^{iHT} \Big],
\end{equation}
where $H$ is the PSM Hamiltonian (Eq.~\eqref{eq:quantum_Hamiltonian}) and $f$ is a filter function as discussed in Sec.~\ref{subsec:review_spectral_form_factor}.
Here we consider the ergodic phase, for which the results are analogous to those of SYK~\cite{saad2019semiclassical}. We then consider the non-ergodic phase in Sec.~\ref{sec:nonergodic_ramp}.

\subsection{Effective action} \label{subsec:ergodic_effective_action}

The calculation begins by retracing the steps described in Sec.~\ref{subsec:Schwinger_Keldysh_path_integral}, only on a modified contour.
We still have upper and lower branches indicated by $\alpha \in \{u, l\}$ (with $u = 0$ and $l = 1$), but now each is separately periodic.
Furthermore, we no longer have a thermal branch.
See the right column of Fig.~\ref{fig:contour_summary}, as compared to the left column.
While some care is required to account for the filter functions (as discussed in Appendix~\ref{sec:filter_functions}), we ultimately arrive at an expression analogous to Eq.~\eqref{eq:annealed_final_action}:
\begin{equation} \label{eq:SFF_formal_path_integral}
\textrm{SFF}(T, f) = \int \mathcal{D}G \mathcal{D}F \, f \big( \epsilon_u[G] \big) f \big( \epsilon_l[G] \big) e^{-NS_{\textrm{eff}}[G, F]},
\end{equation}
\begin{equation} \label{eq:SFF_formal_action}
\begin{aligned}
S_{\textrm{eff}}[G, F] &= -\frac{i}{2} \int_0^T dt \sum_{\alpha} (-1)^{\alpha} z_{\alpha}(t) + \frac{1}{2} \int_0^T dt dt' \sum_{\alpha \alpha'} (-1)^{\alpha + \alpha'} \left( \frac{J^2}{p} G_{\alpha \alpha'}(t, t')^p - F_{\alpha \alpha'}(t, t') G_{\alpha \alpha'}(t, t') \right) \\
&\qquad \qquad + \frac{1}{2} \log{\textrm{Det}} \Big[ i (-1)^{\alpha} \delta_{\alpha \alpha'} \big( \mu \partial_t^2 + z_{\alpha} \big) + (-1)^{\alpha + \alpha'} F_{\alpha \alpha'} \Big],
\end{aligned}
\end{equation}
where the ``energy density'' $\epsilon_{\alpha}[G]$ is defined as ($0^+$ denotes a positive infinitesimal)
\begin{equation} \label{eq:energy_density_definition}
\epsilon_{\alpha}[G] \equiv -\frac{\mu}{2} \partial_t^2 G_{\alpha \alpha}(0^+, 0) - \frac{iJ^2}{p} \int_0^T dt \sum_{\alpha'} (-1)^{\alpha'} G_{\alpha \alpha'}(t, 0)^p.
\end{equation}
See App.~\ref{sec:filter_functions} for details.
The saddle point of $S_{\textrm{eff}}$ is given by the equations (compare to Eq.~\eqref{eq:Keldysh_EOM})
\begin{equation} \label{eq:SFF_general_saddle_point_equations}
\begin{gathered}
i \big( \mu \partial_t^2 + z_{\alpha}(t) \big) G_{\alpha \alpha'}(t, t') + \int_0^T dt'' \sum_{\alpha''} (-1)^{\alpha''} F_{\alpha \alpha''}(t, t'') G_{\alpha'' \alpha'}(t'', t') = (-1)^{\alpha} \delta_{\alpha \alpha'} \delta(t - t'), \\
F_{\alpha \alpha'}(t, t') = J^2 G_{\alpha \alpha'}(t, t')^{p-1}, \qquad G_{\alpha \alpha}(t, t) = 1.
\end{gathered}
\end{equation}

Denoting averages with respect to the path integral of Eq.~\eqref{eq:SFF_formal_path_integral} by $\langle \, \cdot \, \rangle$, the expectation value of $G$ is related to the original degrees of freedom as follows (we omit the filter functions here for brevity):
\begin{equation} \label{eq:SFF_order_parameter_spin_relationship}
\begin{aligned}
\big< G_{uu}(t, t') \big> &= \mathbb{E} \Big[ \textrm{Tr} e^{-iHT} \mathcal{T} \sigma_i(t) \sigma_i(t') \textrm{Tr} e^{iHT} \Big], &\qquad \big< G_{ul}(t, t') \big> &= \mathbb{E} \Big[ \textrm{Tr} e^{-iHT} \sigma_i(t) \textrm{Tr} e^{iHT} \sigma_i(t') \Big], \\
\big< G_{lu}(t, t') \big> &= \mathbb{E} \Big[ \textrm{Tr} e^{-iHT} \sigma_i(t') \textrm{Tr} e^{iHT} \sigma_i(t) \Big], &\qquad \big< G_{ll}(t, t') \big> &= \mathbb{E} \Big[ \textrm{Tr} e^{-iHT} \textrm{Tr} e^{iHT} \widetilde{\mathcal{T}} \sigma_i(t) \sigma_i(t') \Big],
\end{aligned}
\end{equation}
where $\mathcal{T}$ denotes time ordering and $\widetilde{\mathcal{T}}$ denotes time anti-ordering.
One immediately sees from Eq.~\eqref{eq:SFF_order_parameter_spin_relationship} that:
\begin{enumerate}[label=\roman*)]
\item all components of $\langle G_{\alpha \alpha'}(t, t') \rangle$ are time-translation invariant and have period $T$;
\item $\langle G_{uu}(t, t') \rangle$ and $\langle G_{ll}(t, t') \rangle$ are even functions of $t - t'$;
\item $\langle G_{ul}(t, t') \rangle$ and $\langle G_{lu}(t, t') \rangle$ are in fact independent of both time arguments;
\item $\langle G_{uu}(t, t') \rangle^* = \langle G_{ll}(t, t') \rangle$;
\item $\langle G_{ul}(t, t') \rangle^* = \langle G_{lu}(t, t') \rangle$.
\end{enumerate}
Solutions to Eq.~\eqref{eq:SFF_general_saddle_point_equations} do \textit{not} necessarily share all these properties, since some of the symmetries may be spontaneously broken.

However, one simple solution that obeys all of the above is to take $G_{ul}(t, t') = G_{lu}(t, t') = 0$.
The resulting action is precisely what one would get from averaging each factor of $\textrm{Tr} e^{-iHT}$ separately, i.e., this solution gives the disconnected contribution to the SFF:
\begin{equation} \label{eq:SFF_disconnected_contribution}
\mathbb{E} \Big[ \textrm{Tr} f(H) e^{-iHT} \textrm{Tr} f(H) e^{iHT} \Big] = \mathbb{E} \Big[ \textrm{Tr} f(H) e^{-iHT} \Big] \mathbb{E} \Big[ \textrm{Tr} f(H) e^{iHT} \Big] + \cdots,
\end{equation}
where $\cdots$ denotes the contribution to the path integral from non-zero $G_{ul}$ and/or $G_{lu}$.
Eq.~\eqref{eq:SFF_disconnected_contribution} holds equally well in the non-ergodic phase, and thus the remainder of this paper will be concerned with determining those additional contributions.

\subsection{Connected solutions} \label{subsec:ergodic_connected_solutions}

Following Ref.~\cite{saad2019semiclassical}, we construct approximate solutions to Eq.~\eqref{eq:SFF_general_saddle_point_equations} which become accurate at large $T$.
We first present the solutions and justify them afterwards.
Take $\mathcal{G}_{\alpha \alpha'}(t, t')$ to be the Schwinger-Keldysh correlation function at inverse temperature $\beta_{\textrm{aux}}$, exactly as given in Sec.~\ref{subsec:Schwinger_Keldysh_path_integral} (Eq.~\eqref{eq:Keldysh_formal_expectation_values} in particular).
Again define $\mathcal{F}_{\alpha \alpha'}(t, t') \equiv J^2 \mathcal{G}_{\alpha \alpha'}(t, t')^{p-1}$.
A solution to the SFF saddle point equations (up to terms which vanish at large $T$) is
\begin{equation} \label{eq:SFF_connected_solution_G}
G_{\alpha \alpha'}(t, t') = \sum_{n = -\infty}^{\infty} \mathcal{G}_{\alpha \alpha'}(t - t' + \delta_{\alpha \neq \alpha'} \Delta + nT),
\end{equation}
\begin{equation} \label{eq:SFF_connected_solution_Sigma}
F_{\alpha \alpha'}(t, t') = \sum_{n = -\infty}^{\infty} \mathcal{F}_{\alpha \alpha'}(t - t' + \delta_{\alpha \neq \alpha'} \Delta + nT).
\end{equation}
Here $\Delta$ can be any real number between 0 and $T$.
Thus Eqs.~\eqref{eq:SFF_connected_solution_G} and~\eqref{eq:SFF_connected_solution_Sigma} constitute a two-parameter family of solutions, the parameters being $\beta_{\textrm{aux}}$ and $\Delta$.
Every such solution contributes to the SFF.

As for the Lagrange multipliers $z_{\alpha}(t)$, they are independent of $t$ due to time translation invariance.
We further have that $z_u = z_l \equiv z$: both equal the value of the chemical potential needed to satisfy the \textit{equilibrium} spherical constraint, i.e., $N^{-1} \sum_i \textrm{Tr} Z_{\textrm{SK}}^{-1} e^{-\beta_{\textrm{aux}} H} \sigma_i^2 = 1$ (time translation invariance then implies that $\mathcal{G}_{\alpha \alpha}(t, t) = 1$ for all times and both branches).

To justify Eqs.~\eqref{eq:SFF_connected_solution_G} and~\eqref{eq:SFF_connected_solution_Sigma}, it is essential that $\mathcal{G}_{\alpha \alpha'}(t - t')$ decay exponentially to zero as $|t - t'| \rightarrow \infty$.
Thus these solutions only apply in the ergodic phase. 
With this in mind, the following comments together establish their validity:
\begin{itemize}
\item The sum over $n$ ensures that $G_{\alpha \alpha'}(t - t')$ has period $T$, even though $\mathcal{G}_{\alpha \alpha'}(t - t')$ does not.
\item Since $\mathcal{G}_{\alpha \alpha}(t - t')$ decays exponentially, $G_{\alpha \alpha}(0) \sim 1$ up to terms which are exponentially small in $T$.
\item The equation $F_{\alpha \alpha'}(t, t') = J^2 G_{\alpha \alpha'}(t, t')^{p-1}$ is satisfied up to exponentially small terms because, when raising Eq.~\eqref{eq:SFF_connected_solution_G} to the $p-1$'th power, all cross terms are exponentially small (as is the sum over them).
In other words,
\begin{equation} \label{eq:cross_terms_neglecting}
\left( \sum_n \mathcal{G}_{\alpha \alpha'}(t - t' + nT + \delta_{\alpha \neq \alpha'} \Delta) \right)^{p-1} \sim \sum_n \mathcal{G}_{\alpha \alpha'}(t - t' + nT + \delta_{\alpha \neq \alpha'} \Delta)^{p-1}.
\end{equation}
\item $\mathcal{G}_{\alpha \alpha'}(t, t')$ obeys Eq.~\eqref{eq:Keldysh_EOM}, written explicitly in terms of components as
\begin{equation} \label{eq:Keldysh_EOM_explicit}
\begin{aligned}
&i \big( \mu \partial_t^2 + z \big) \mathcal{G}_{\alpha \alpha'}(t - t') + \int_0^{\infty} dt'' \sum_{\alpha''} (-1)^{\alpha''} \mathcal{F}_{\alpha \alpha''}(t - t'') \mathcal{G}_{\alpha'' \alpha'}(t'' - t') \\
&\qquad \qquad \qquad \qquad \; \; \; - i \int_0^{\beta_{\textrm{aux}}} d\tau'' \mathcal{F}_{\alpha v}(t + i\tau'') \mathcal{G}_{v \alpha'}(-i\tau'' - t') = (-1)^{\alpha} \delta_{\alpha \alpha'} \delta(t - t'),
\end{aligned}
\end{equation}
where $v$ denotes the thermal branch of the contour.
For $t, t' \gg 1$ (which still allows $t - t'$ to take any value), $\mathcal{G}_{\alpha v}(t + i\tau)$ is exponentially small for all $\tau$ and the last term on the left-hand side can be neglected.
We can also take the lower limit of the $t''$ integral to $-\infty$.
Thus when checking whether Eq.~\eqref{eq:SFF_connected_solution_G} satisfies Eq.~\eqref{eq:SFF_general_saddle_point_equations}, we have that
\begin{equation} \label{eq:SFF_connected_solution_demonstration}
\begin{aligned}
&i \big( \mu \partial_t^2 + z \big) G_{\alpha \alpha'}(t - t') + \int_0^T dt'' \sum_{\alpha''} (-1)^{\alpha''} F_{\alpha \alpha''}(t - t'') G_{\alpha'' \alpha'}(t'' - t') \\
&\; \; \sim i \big( \mu \partial_t^2 + z \big) \mathcal{G}_{\alpha \alpha'}(t - t') + \int_{-\infty}^{\infty} dt'' \sum_{\alpha''} (-1)^{\alpha''} \mathcal{F}_{\alpha \alpha''}(t - t'') \mathcal{G}_{\alpha'' \alpha'}(t'' - t') \\
&\; \; \sim (-1)^{\alpha} \delta_{\alpha \alpha'} \delta(t - t'),
\end{aligned}
\end{equation}
again making use of the fact that $\mathcal{G}_{\alpha \alpha'}(t - t')$ is exponentially small when $|t - t'|$ is large.
The equation is indeed satisfied.
\item Finally, the off-diagonal components $G_{ul}(t, t')$ and $G_{lu}(t, t')$ contain the parameter $\Delta$ because they break the \textit{separate} time translation symmetries in $t$ and $t'$ (see property iii above).
Thus if any choice of $\Delta$ solves Eq.~\eqref{eq:SFF_general_saddle_point_equations}, so do all choices of $\Delta \in [0, T)$.
\end{itemize}

As noted above, we have thus identified a two-parameter family of solutions to the SFF saddle point equations.
In what follows it will be more convenient to parametrize the solutions by the equilibrium energy density $\epsilon(\beta_{\textrm{aux}})$ corresponding to inverse temperature $\beta_{\textrm{aux}}$.
We can express $\epsilon(\beta)$ in terms of $\mathcal{G}$ (and thus $G$) by inserting a factor of $H$ into the Schwinger-Keldysh contour.
Since $H$ clearly commutes with the evolution operator $e^{-\beta H} e^{iHt} e^{-iHt}$, it can be inserted at any point, in particular at a late time for which (again because $\mathcal{G}_{\alpha \alpha'}(t - t')$ decays exponentially) the thermal branch can be neglected.
By following the same steps as in Appendix~\ref{sec:filter_functions}, we find that $\epsilon(\beta)$ is given precisely by Eq.~\eqref{eq:energy_density_definition}, evaluated on either branch:
\begin{equation} \label{eq:thermal_energy_density_expression}
\begin{aligned}
\epsilon =& -\frac{\mu}{2} \partial_t^2 \mathcal{G}_{uu}(0^+) - \frac{iJ^2}{p} \int_{-\infty}^{\infty} dt \Big( \mathcal{G}_{uu}(t)^p - \mathcal{G}_{ul}(t)^p \Big) \\
=& -\frac{\mu}{2} \partial_t^2 \mathcal{G}_{ll}(0^+) + \frac{iJ^2}{p} \int_{-\infty}^{\infty} dt \Big( \mathcal{G}_{ll}(t)^p - \mathcal{G}_{lu}(t)^p \Big).
\end{aligned}
\end{equation}

\subsection{Contribution of connected solutions} \label{subsec:ergodic_connected_action}

Having demonstrated that Eqs.~\eqref{eq:SFF_connected_solution_G} and~\eqref{eq:SFF_connected_solution_Sigma} solve the SFF saddle point equations, it remains to calculate the action (Eq.~\eqref{eq:SFF_formal_action}) evaluated at the solutions.
First note that, since each solution obeys Eq.~\eqref{eq:SFF_general_saddle_point_equations}, we can rewrite the action as
\begin{equation} \label{eq:SFF_on_shell_action}
S_{\textrm{eff}} = -\frac{J^2 (p-1)T}{2p} \int_0^T dt \sum_{\alpha \alpha'} (-1)^{\alpha + \alpha'} G_{\alpha \alpha'}(t)^p - \frac{1}{2} \sum_{\omega} \log{\textrm{Det}} \widetilde{G}_{\alpha \alpha'}(\omega),
\end{equation}
where $\omega \in 2\pi \mathbb{Z}/T$ and $\widetilde{G}_{\alpha \alpha'}(\omega) \equiv \int_0^T dt e^{i \omega t} G_{\alpha \alpha'}(t)$.
Note that the Lagrange multiplier terms have dropped out since $z_u = z_l$.
Furthermore, since $\int dt G_{\alpha \alpha'}(t)^p \sim \int dt \mathcal{G}_{\alpha \alpha'}(t)^p$, the general relation in Eq.~\eqref{eq:Keldysh_component_symmetry_identity} implies that the first term of Eq.~\eqref{eq:SFF_on_shell_action} in fact vanishes.

For the second term, note that by Eq.~\eqref{eq:SFF_connected_solution_G},
\begin{equation} \label{eq:SFF_connected_solution_Fourier_transform}
\widetilde{G}_{\alpha \alpha'}(\omega) = e^{-i \delta_{\alpha \neq \alpha'} \omega \Delta} \widetilde{\mathcal{G}}_{\alpha \alpha'}(\omega), \qquad \widetilde{\mathcal{G}}_{\alpha \alpha'}(\omega) \equiv \int_{-\infty}^{\infty} dt e^{i \omega t} \mathcal{G}_{\alpha \alpha'}(t).
\end{equation}
The exponential decay of $\mathcal{G}_{\alpha \alpha'}(t)$ implies that $\widetilde{\mathcal{G}}_{\alpha \alpha'}(\omega)$ (and thus $\widetilde{G}_{\alpha \alpha'}(\omega)$) is an infinitely differentiable function of $\omega$.
Strictly speaking, since the path integral is regularized by a timestep $\Delta t \rightarrow 0$, $\widetilde{\mathcal{G}}_{\alpha \alpha'}(\omega)$ is furthermore periodic with period $2\pi/\Delta t$.
The same is true of $\log{\textrm{Det}} \widetilde{G}_{\alpha \alpha'}(\omega)$.
Thus the Euler-Maclaurin formula~\cite{Knuth1994} gives
\begin{equation} \label{eq:SFF_on_shell_action_second_term}
\sum_{n=-\pi/\Delta t}^{\pi/\Delta t} \log{\textrm{Det}} \widetilde{G}_{\alpha \alpha'} \left( \frac{2\pi n}{T} \right) \sim \frac{T}{2\pi} \int_{-\pi/\Delta t}^{\pi/\Delta t} d\omega \log{\textrm{Det}} \widetilde{G}_{\alpha \alpha'}(\omega) \rightarrow \frac{T}{2\pi} \int_{-\infty}^{\infty} d\omega \log{\textrm{Det}} \widetilde{G}_{\alpha \alpha'}(\omega),
\end{equation}
up to terms which vanish faster than any polynomial in $T^{-1}$.
Thus $S_{\textrm{eff}}$ is proportional to $T$, and we only need to evaluate the proportionality constant.

Rather than calculate the integral directly, we follow Ref.~\cite{saad2019semiclassical} and evaluate the derivative $dS_{\textrm{eff}}/dT$ starting from Eq.~\eqref{eq:SFF_formal_action}.
It is convenient to rescale time as $t \rightarrow Tt$, so that $T$ becomes simply another parameter:
\begin{equation} \label{eq:SFF_formal_action_rescaled}
\begin{aligned}
S_{\textrm{eff}} &= \frac{T^2}{2} \int_0^1 dt dt' \sum_{\alpha \alpha'} (-1)^{\alpha + \alpha'} \left( \frac{J^2}{p} G_{\alpha \alpha'}(t, t')^p - F_{\alpha \alpha'}(t, t') G_{\alpha \alpha'}(t, t') \right) \\
&\qquad \qquad + \frac{1}{2} \log{\textrm{Det}} \Big[ i (-1)^{\alpha} \delta_{\alpha \alpha'} \big( \mu T^{-2} \partial_t^2 + z_{\alpha} \big) + (-1)^{\alpha + \alpha'} F_{\alpha \alpha'} \Big].
\end{aligned}
\end{equation}
Note that, since $S_{\textrm{eff}}$ is evaluated at a solution of the saddle point equations, we only need to differentiate the explicit factors of $T$:
\begin{equation} \label{eq:SFF_formal_action_total_time_derivative_v1}
\begin{aligned}
\frac{dS_{\textrm{eff}}}{dT} =& \; T \int_0^1 dt dt' \sum_{\alpha \alpha'} (-1)^{\alpha + \alpha'} \left( \frac{J^2}{p} G_{\alpha \alpha'}(t, t')^p - F_{\alpha \alpha'}(t, t') G_{\alpha \alpha'}(t, t') \right) \\
&\qquad \qquad - \frac{i\mu}{T^3} \int_0^1 dt \sum_{\alpha} (-1)^{\alpha} \partial_t^2 \Big[ i (-1)^{\alpha} \delta_{\alpha \alpha'} \big( \mu T^{-2} \partial_t^2 + z_{\alpha}(t) \big) + (-1)^{\alpha + \alpha'} F_{\alpha \alpha'} \Big]^{-1} \bigg|_{\alpha = \alpha', t = t'^+}.
\end{aligned}
\end{equation}
Returning to unscaled time and using Eq.~\eqref{eq:SFF_general_saddle_point_equations}, we have that
\begin{equation} \label{eq:SFF_formal_action_total_time_derivative_v2}
\frac{dS_{\textrm{eff}}}{dT} = -\frac{(p-1)J^2}{p} \int_0^T dt \sum_{\alpha \alpha'} (-1)^{\alpha + \alpha'} G_{\alpha \alpha'}(t)^p - \frac{i\mu}{T} \sum_{\alpha} (-1)^{\alpha} \partial_t^2 G_{\alpha \alpha}(0^+) = 0,
\end{equation}
again using Eqs.~\eqref{eq:Keldysh_component_symmetry_identity} and~\eqref{eq:thermal_energy_density_expression}.
Thus the proportionality constant is in fact zero, i.e., $S_{\textrm{eff}} = 0$.

\subsection{Evaluation of the SFF} \label{subsec:ergodic_SFF_evaluation}

To finally compute the SFF, we simply need to sum over all connected solutions, i.e., integrate over $\epsilon_{\textrm{aux}}$ and $\Delta$.
However, there are additional discrete symmetries which give further solutions: i) we can time-reverse the off-diagonal components, i.e., take $G_{ul}(t) = \mathcal{G}_{ul}(-t)$ and $G_{lu}(t) = \mathcal{G}_{lu}(-t)$; ii) if $p$ is even, we can take $G_{ul}(t) = -\mathcal{G}_{ul}(t)$ and $G_{lu}(t) = -\mathcal{G}_{lu}(t)$.
These must be summed over as well, giving an additional factor of $2(1 + \delta_{p \textrm{ even}})$, where $\delta_{p \textrm{ even}}$ is the indicator function on $p$ being even (1 if true, 0 if false).
Thus our final expression is
\begin{equation} \label{eq:SFF_connected_contribution_ergodic_phase}
\begin{aligned}
\mathbb{E} \Big[ \textrm{Tr} f(H) e^{-iHT} \textrm{Tr} f(H) e^{iHT} \Big] &\sim \left| \mathbb{E} \textrm{Tr} f(H) e^{-iHT} \right|^2 + \int \frac{d\epsilon_{\textrm{aux}}}{2\pi} f(\epsilon_{\textrm{aux}})^2 \int_0^T d\Delta \, 2 \big( 1 + \delta_{p \textrm{ even}} \big) e^0 \\
&= \left| \mathbb{E} \textrm{Tr} f(H) e^{-iHT} \right|^2 + 2 \big( 1 + \delta_{p \textrm{ even}} \big) T \int \frac{d\epsilon_{\textrm{aux}}}{2\pi} f(\epsilon_{\textrm{aux}})^2.
\end{aligned}
\end{equation}
The measure $1/2\pi$ can be derived using hydrodynamic methods~\cite{winerprx,saad2019semiclassical}, but its precise value is not essential for our purposes.
The key feature is simply that the linear-in-$T$ ramp has emerged.

However, keep in mind that Eq.~\eqref{eq:SFF_connected_contribution_ergodic_phase} is only valid if the filter function is such that all contributing values of $\epsilon_{\textrm{aux}}$ lie in the ergodic phase.
In the following section we modify this analysis to hold in the non-ergodic phase as well.
We shall see that it is necessary to incorporate the structure of multiple TAP states.

\section{The semiclassical ramp in the non-ergodic phase}
\label{sec:nonergodic_ramp}

As we have stressed repeatedly, the results of Sec.~\ref{sec:ergodic_ramp} rely heavily on having an equilibrium correlation function which decays to zero at late times.
Thus a new approach is needed to calculate the SFF in the non-ergodic phase, where $\mathcal{G}_{\alpha \alpha'}(t - t') \rightarrow q_{\textrm{EA}} \neq 0$ as $|t - t'| \rightarrow \infty$.
More specifically, we can no longer neglect the integral over the thermal branch in Eq.~\eqref{eq:Keldysh_EOM}, and $\mathcal{G}_{\alpha \alpha'}(t - t')$ no longer solves the SFF equations of motion (Eq.~\eqref{eq:SFF_general_saddle_point_equations}).

However, in the \textit{TAP} equations of motion, Eq.~\eqref{eq:Keldysh_TAP_EOM}, we \textit{can} neglect the thermal branch since $\mathcal G(t)-q_{\textrm{EA}}$ does decay to zero exponentially quickly.
This suggests that a viable strategy is to construct solutions for the SFF using the TAP correlation function.
Since TAP states are parametrized by the quantity $\mathcal{E}$ in Eq.~\eqref{eq:normalized_energy_definition}, it will be necessary to first modify the SFF path integral so as to involve $\mathcal{E}$.
We associate the magnetizations and overlap from the TAP approach with time-averaged functions of the spin configuration, namely
\begin{equation} \label{eq:spin_TAP_quantity_definitions}
m_i[\sigma] \equiv \frac{1}{T} \int_0^T dt \sigma_{iu}(t), \qquad q[\sigma] \equiv \frac{1}{T^2} \int_0^T dt dt' \frac{1}{N} \sum_i \sigma_{iu}(t) \sigma_{iu}(t').
\end{equation}
The choice to use only the upper contour in defining $m_i[\sigma]$ and $q[\sigma]$ will become convenient in Sec.~\ref{sec:HigherMoments}, but for now one could equally well use any other combination of branches, say the average of $\sigma_i(t)$ over the lower branch or over both branches symmetrically.
With these definitions, we introduce $\mathcal{E}$ via Eq.~\eqref{eq:normalized_energy_definition}.

\subsection{Effective action} \label{eq:nonergodic_effective_action}

To begin, insert an additional fat unity into the path integral:
\begin{equation} \label{eq:TAP_resolving_fat_unity}
\begin{aligned}
1 &= \int d\mathcal{E}_{\textrm{aux}} \delta \left[ N \mathcal{E}_{\textrm{aux}} - \frac{1}{J q[\sigma]^{p/2}} \sum_{(i_1 \cdots i_p)} J_{i_1 \cdots i_p} \left( \frac{1}{T} \int_0^T dt \sigma_{i_1 u}(t) \right) \cdots \left( \frac{1}{T} \int_0^T dt \sigma_{i_p u}(t) \right) \right].
\end{aligned}
\end{equation}
With this addition, the full path integral is
\begin{equation} \label{eq:TAP_resolved_original_path_integral}
\begin{aligned}
\textrm{SFF} &= \int \mathcal{D}P(J)\mathcal D\sigma^N d\mathcal{E}_{\textrm{aux}} d\lambda \exp \left[ \frac{i}{2} \sum_i \int_0^T dt \sum_{\alpha} (-1)^{\alpha} \Big[ \mu \big( \partial_t \sigma_{i \alpha}(t) \big)^2 - z_{\alpha}(t) \big( \sigma_{i \alpha}(t)^2 - 1 \big) \Big] \right] \\
&\qquad \cdot \exp \left[ -i \sum_{(i_1 \cdots i_p)} J_{i_1 \cdots i_p} \int_0^T dt \sum_{\alpha} (-1)^{\alpha} \sigma_{i_1 \alpha}(t) \cdots \sigma_{i_p \alpha}(t) \right] \\
&\qquad \qquad \cdot \exp \left[ iN\lambda \mathcal{E}_{\textrm{aux}} - \frac{i \lambda}{J q[\sigma]^{p/2}} \sum_{(i_1 \cdots i_p)} J_{i_1 \cdots i_p} \left( \frac{1}{T} \int_0^T dt \sigma_{i_1 u}(t) \right) \cdots \left( \frac{1}{T} \int_0^T dt \sigma_{i_p u}(t) \right) \right].
\end{aligned}
\end{equation}
Proceeding as usual --- averaging over disorder, introducing $G_{\alpha \alpha'}(t, t')$ and $F_{\alpha \alpha'}(t, t')$ as before, integrating out spins --- we arrive at
\begin{equation} \label{eq:TAP_resolved_SFF_path_integral}
\textrm{SFF}(T, f) = \int d\mathcal{E}_{\textrm{aux}} d\lambda \mathcal{D}G \mathcal{D}F \, f \big( \epsilon_u[\lambda, G] \big) f \big( \epsilon_l[\lambda, G] \big) e^{-NS_{\textrm{eff}}[\mathcal{E}_{\textrm{aux}}, \lambda, G, F]},
\end{equation}
\begin{equation} \label{eq:TAP_resolved_SFF_action}
\begin{aligned}
S_{\textrm{eff}}[\mathcal{E}_{\textrm{aux}}, \lambda, G, F] &= -i \lambda \mathcal{E}_{\textrm{aux}} - \frac{i}{2} \int_0^T dt \sum_{\alpha} (-1)^{\alpha} z_{\alpha}(t) \\
&\qquad + \frac{\lambda^2}{2p} + \frac{J \lambda}{p q[G]^{p/2}} \int_0^T dt \sum_{\alpha} (-1)^{\alpha} \left( \frac{1}{T} \int_0^T dt' G_{\alpha u}(t, t') \right)^p \\
&\qquad \qquad + \frac{1}{2} \int_0^T dt dt' \sum_{\alpha \alpha'} (-1)^{\alpha + \alpha'} \left( \frac{J^2}{p} G_{\alpha \alpha'}(t, t')^p - F_{\alpha \alpha'}(t, t') G_{\alpha \alpha'}(t, t') \right) \\
&\qquad \qquad \qquad + \frac{1}{2} \log{\textrm{Det}} \Big[ i (-1)^{\alpha} \delta_{\alpha \alpha'} \big( \mu \partial_t^2 + z_{\alpha} \big) + (-1)^{\alpha + \alpha'} F_{\alpha \alpha'} \Big],
\end{aligned}
\end{equation}
where we are denoting $q[G] \equiv T^{-2} \int_0^T dt dt' G_{uu}(t, t')$.
The argument of the filter function is modified as well; it is now
\begin{equation} \label{eq:TAP_resolved_energy_density_definition}
\epsilon_{\alpha}[\lambda, G] \equiv -\frac{\mu}{2} \partial_t^2 G_{\alpha \alpha}(0^+, 0) - \frac{iJ^2}{p} \int_0^T dt \sum_{\alpha'} (-1)^{\alpha'} G_{\alpha \alpha'}(t, 0)^p - \frac{iJ \lambda}{p q[G]^{p/2}} \left( \frac{1}{T} \int_0^T dt G_{\alpha u}(t, 0) \right)^p.
\end{equation}

Note that $\mathcal{E}_{\textrm{aux}}$ enters linearly into the action.
Thus if we were to integrate over $\mathcal{E}_{\textrm{aux}}$ at this point, we would obtain a $\delta$-function forcing $\lambda = 0$.
The action would then reduce to the ergodic-phase expression, Eq.~\eqref{eq:SFF_formal_action}.
While reassuring, this would not have accomplished anything, so we instead treat $\mathcal{E}_{\textrm{aux}}$ as a fixed parameter for now.
We obtain saddle point equations by differentiating Eq.~\eqref{eq:TAP_resolved_SFF_action} only with respect to $\lambda$, $z$, $G$, and $F$.

The saddle point equations, assuming time translation invariance from the outset, are
\begin{equation} \label{eq:TAP_resolved_EOM}
i \big( \mu \partial_t^2 + z_{\alpha} \big) G_{\alpha \alpha'}(t - t') + \int_0^T dt'' \sum_{\alpha''} (-1)^{\alpha''} F_{\alpha \alpha''}(t - t'') G_{\alpha'' \alpha'}(t'' - t') = (-1)^{\alpha} \delta_{\alpha \alpha'} \delta(t - t'),
\end{equation}
\begin{equation} \label{eq:TAP_resolved_self_energy_equation}
\begin{aligned}
F_{\alpha \alpha'}(t) = J^2 G_{\alpha \alpha'}(t)^{p-1} &+ \frac{J \lambda}{T q[G]^{p/2}} \big( \delta_{\alpha u} + \delta_{\alpha' u} \big) \left( \frac{\widetilde{G}_{\alpha \alpha'}(0)}{T} \right)^{p-1} \\
&- \frac{J \lambda}{T q[G]^{p/2 + 1}} \delta_{\alpha u} \delta_{\alpha' u} \sum_{\alpha''} (-1)^{\alpha''} \left( \frac{\widetilde{G}_{\alpha'' u}(0)}{T} \right)^p, 
\end{aligned}
\end{equation}
\begin{equation} \label{eq:TAP_resolved_potential_equation}
\mathcal{E}_{\textrm{aux}} = -\frac{iJT}{p q[G]^{p/2}} \sum_{\alpha} (-1)^{\alpha} \left( \frac{\widetilde{G}_{\alpha u}(0)}{T} \right)^p - \frac{i \lambda}{p},
\end{equation}
as well as the usual requirement $G_{\alpha \alpha}(0) = 1$.
Here $\widetilde{G}_{\alpha \alpha'}(\omega)$ is the Fourier transform of $G_{\alpha \alpha'}(t)$.

\subsection{Connected solutions} \label{subsec:TAP_resolved_connected_solutions}

With $\mathcal{E}_{\textrm{aux}}$ fixed, let $\mathcal{G}_{\alpha \alpha'}(t)$ be the solution to the TAP equation of motion (Eq.~\eqref{eq:Keldysh_TAP_EOM}) corresponding to inverse temperature $\beta_{\textrm{aux}}$.
Denote the Edwards-Anderson order parameter at $\mathcal{E}_{\textrm{aux}}$ and $\beta_{\textrm{aux}}$ by $q_{\textrm{EA}}$.
Also recall the various auxiliary quantities we defined in Sec.~\ref{subsec:QTAP}: the self-energy $\mathcal{F}_{\alpha \alpha'}(t) \equiv J^2 \mathcal{G}_{\alpha \alpha'}(t)^{p-1}$, the deviations $\Delta \mathcal{G}_{\alpha \alpha'}(t) \equiv \mathcal{G}_{\alpha \alpha'}(t) - q_{\textrm{EA}}$ and $\Delta \mathcal{F}_{\alpha \alpha'}(t) \equiv \mathcal{F}_{\alpha \alpha'}(t) - J^2 q_{\textrm{EA}}^{p-1}$, and the quantity $\Lambda \equiv \int_0^{\infty} dt \Delta \mathcal{G}^I(t)$.
We have that $\mathcal{G}_{\alpha \alpha'}(t)$ and $\mathcal{F}_{\alpha \alpha'}(t)$ obey Eq.~\eqref{eq:Keldysh_TAP_EOM}, which by taking $t$ and $t'$ to be far from the thermal branch can be written
\begin{equation} \label{eq:Keldysh_TAP_EOM_simplified}
i \big( \mu \partial_t^2 + z \big) \Delta \mathcal{G}_{\alpha \alpha'}(t - t') + \int_{-\infty}^{\infty} dt'' \sum_{\alpha''} (-1)^{\alpha''} \Delta \mathcal{F}_{\alpha \alpha''}(t - t'') \Delta \mathcal{G}_{\alpha'' \alpha'}(t'' - t') = (-1)^{\alpha} \delta_{\alpha \alpha'} \delta(t - t').
\end{equation}
We also have, as a result of Eq.~\eqref{eq:Keldysh_TAP_magnetization_alternate}, the relationship
\begin{equation} \label{eq:potential_GF_relationship}
\mathcal{E}_{\textrm{aux}} = \frac{2(p-1)J q_{\textrm{EA}}^{p/2-1} \Lambda}{p} + \frac{1}{2pJ q_{\textrm{EA}}^{p/2-1} \Lambda}.
\end{equation}
Finally, recall the expression for the complexity $\Sigma(\mathcal{E})$, the logarithm of the number of solutions to the TAP magnetization equations at $\mathcal{E}$:
\begin{equation} \label{eq:TAP_complexity_repeat}
\Sigma(\mathcal{E}) = \frac{1}{2} \left( 1 + 2\log{\frac{p}{2}} \right) - \frac{p \mathcal{E}^2}{2} + \frac{p^2}{8(p-1)} \left( \mathcal{E} + \sqrt{\mathcal{E}^2 - \mathcal{E}_{\textrm{th}}^2} \right)^2 + \log{\left( -\mathcal{E} + \sqrt{\mathcal{E}^2 - \mathcal{E}_{\textrm{th}}^2} \right)},
\end{equation}
where $\mathcal{E}_{\textrm{th}}^2 = 4(p-1)/p^2$.
Using Eq.~\eqref{eq:potential_GF_relationship}, we can express $\Sigma(\mathcal{E}_{\textrm{aux}})$ in terms of $\Lambda$ and $q_{\textrm{EA}}$:
\begin{equation} \label{eq:TAP_complexity_rewrite}
\Sigma(\mathcal{E}_{\textrm{aux}}) = -\frac{p-2}{2p} - \frac{1}{8pJ^2 q_{\textrm{EA}}^{p-2} \Lambda^2} + \frac{2(p-1)J^2 q_{\textrm{EA}}^{p-2} \Lambda^2}{p} - \frac{1}{2} \log{4J^2 q_{\textrm{EA}}^{p-2} \Lambda^2}.
\end{equation}

Our solution to the SFF saddle point equations, Eqs.~\eqref{eq:TAP_resolved_EOM} through~\eqref{eq:TAP_resolved_potential_equation}, is best written in the frequency domain (tildes denote Fourier transforms):
\begin{equation} \label{eq:TAP_resolved_G_solution_frequency}
\widetilde{G}_{\alpha \alpha'}(\omega) = Tq_{\textrm{EA}} \delta_{\omega 0} + \Delta \widetilde{\mathcal{G}}_{\alpha \alpha'}(\omega) + \frac{\widetilde{g}_{\alpha \alpha'}(\omega)}{T},
\end{equation}
\begin{equation} \label{eq:TAP_resolved_F_solution_frequency}
\widetilde{F}_{\alpha \alpha'}(\omega) = \left( TJ^2 q_{\textrm{EA}}^{p-1} + ipJ q_{\textrm{EA}}^{p/2-1} \big( \mathcal{E}_{\textrm{aux}} - 2J q_{\textrm{EA}}^{p/2-1} \Lambda \big) \big( \delta_{\alpha u} + \delta_{\alpha' u} \big) \right) \delta_{\omega 0} + \Delta \widetilde{\mathcal{F}}_{\alpha \alpha'}(\omega) + \frac{\widetilde{f}_{\alpha \alpha'}(\omega)}{T},
\end{equation}
\begin{equation} \label{eq:TAP_resolved_lambda_solution}
\lambda = ip \big( \mathcal{E}_{\textrm{aux}} - 2J q_{\textrm{EA}}^{p/2-1} \Lambda \big) + \frac{\delta}{T}.
\end{equation}
We again take $z_{\alpha}$ to be the equilibrium value corresponding to $\beta_{\textrm{aux}}$.
The precise form of the correction terms $\widetilde{g}_{\alpha \alpha'}(\omega)$, $\widetilde{f}_{\alpha \alpha'}(\omega)$, and $\delta$ is largely unimportant --- the essential feature is simply that they are $O(1)$ and the corrections are thus $O(T^{-1})$.
Note that in the time domain, this solution amounts to
\begin{equation} \label{eq:TAP_resolved_G_solution_time}
G_{\alpha \alpha'}(t) = q_{\textrm{EA}} + \sum_{n=-\infty}^{\infty} \Delta \mathcal{G}_{\alpha \alpha'}(t + nT) + \frac{g_{\alpha \alpha'}(t)}{T},
\end{equation}
\begin{equation} \label{eq:TAP_resolved_F_solution_time}
F_{\alpha \alpha'}(t) = J^2 q_{\textrm{EA}}^{p-1} + \sum_{n=-\infty}^{\infty} \Delta \mathcal{F}_{\alpha \alpha'}(t + nT) + \frac{ipJ q_{\textrm{EA}}^{p/2-1} \big( \mathcal{E}_{\textrm{aux}} - 2J q_{\textrm{EA}}^{p/2-1} \Lambda \big)}{T} \big( \delta_{\alpha u} + \delta_{\alpha' u} \big) + \frac{f_{\alpha \alpha'}(t)}{T}.
\end{equation}
The sums are convergent because $\Delta \mathcal{G}_{\alpha \alpha'}(t)$ and $\Delta \mathcal{F}_{\alpha \alpha'}(t)$ decay rapidly to zero as $|t| \rightarrow \infty$.

Although we have omitted it for notational simplicity, we can add a term $\delta_{\alpha \neq \alpha'} \Delta$ to the time arguments of $\Delta \mathcal{G}_{\alpha \alpha'}(t + nT)$ and $\Delta \mathcal{F}_{\alpha \alpha'}(t + nT)$ for any $\Delta \in [0, T)$, exactly as in Sec.~\ref{sec:ergodic_ramp}.
Due to the separate time translation symmetry on each branch of the SFF contour, all such solutions are equally valid and contribute the same action.
Thus we shall demonstrate the validity of Eqs.~\eqref{eq:TAP_resolved_G_solution_frequency} through~\eqref{eq:TAP_resolved_lambda_solution} and evaluate the action only for $\Delta = 0$, but then integrate over all $\Delta \in [0, T)$ in the final expression for the SFF.

Let us first confirm that our solution satisfies the saddle point equation for $\lambda$, Eq.~\eqref{eq:TAP_resolved_potential_equation}.
Referring to Eq.~\eqref{eq:Keldysh_zero_frequency_matrix}, we have that $\Delta \widetilde{\mathcal{G}}_{\alpha \alpha'}(0) = L + (-1)^{\alpha} 2i \Lambda \delta_{\alpha \alpha'}$.
Thus
\begin{equation} \label{eq:spin_TAP_overlap_expansion}
q[G] = q_{\textrm{EA}} + \frac{L + 2i \Lambda}{T} + O(T^{-2}),
\end{equation}
and Eq.~\eqref{eq:TAP_resolved_potential_equation} becomes
\begin{equation} \label{eq:TAP_resolved_potential_demonstration}
\mathcal{E}_{\textrm{aux}} = 2J q_{\textrm{EA}}^{p/2-1} \Lambda - \frac{i \lambda}{p} + O(T^{-1}).
\end{equation}
Solving for $\lambda$ indeed gives Eq.~\eqref{eq:TAP_resolved_lambda_solution}.
The $O(T^{-1})$ terms determine $\delta$ as a function of the other quantities.

Now turn to Eq.~\eqref{eq:TAP_resolved_self_energy_equation}.
In the frequency domain, the right-hand side evaluates to\footnote{Since $\widetilde{g}_{\alpha \alpha'}(\omega) = O(1)$ with respect to $T$, $g_{\alpha \alpha'}(t)$ decays to zero as $|t| \rightarrow \infty$ (at least to leading order).}
\begin{equation} \label{eq:TAP_resolved_self_energy_demonstration}
\begin{aligned}
&\int_0^T dt e^{i \omega t} J^2 \left( q_{\textrm{EA}} + \sum_{n=-\infty}^{\infty} \Delta \mathcal{G}_{\alpha \alpha'}(t + nT) \right)^{p-1} + ipJ q_{\textrm{EA}}^{p/2-1} \big( \mathcal{E}_{\textrm{aux}} - 2J q_{\textrm{EA}}^{p/2-1} \Lambda \big) \big( \delta_{\alpha u} + \delta_{\alpha' u} \big) \delta_{\omega 0} + O(T^{-1}) \\
&\qquad \qquad \stackrel{\textrm{set}}{=} \left( TJ^2 q_{\textrm{EA}}^{p-1} + ipJ q_{\textrm{EA}}^{p/2-1} \big( \mathcal{E}_{\textrm{aux}} - 2J q_{\textrm{EA}}^{p/2-1} \Lambda \big) \big( \delta_{\alpha u} + \delta_{\alpha' u} \big) \right) \delta_{\omega 0} + \Delta \widetilde{\mathcal{F}}_{\alpha \alpha'}(\omega) + \frac{\widetilde{f}_{\alpha \alpha'}(\omega)}{T}.
\end{aligned}
\end{equation}
Along the lines of Eq.~\eqref{eq:cross_terms_neglecting}, we have that
\begin{equation} \label{eq:non_ergodic_cross_terms_neglecting}
\begin{aligned}
J^2 \left( q_{\textrm{EA}} + \sum_{n=-\infty}^{\infty} \Delta \mathcal{G}_{\alpha \alpha'}(t + nT) \right)^{p-1} &= J^2 q_{\textrm{EA}}^{p-1} + J^2 \sum_{r=1}^{p-1} \binom{p-1}{r} q_{\textrm{EA}}^{p-1-r} \left( \sum_{n=-\infty}^{\infty} \Delta \mathcal{G}_{\alpha \alpha'}(t + nT) \right)^r \\
&\sim J^2 q_{\textrm{EA}}^{p-1} + J^2 \sum_{r=1}^{p-1} \binom{p-1}{r} q_{\textrm{EA}}^{p-1-r} \sum_{n=-\infty}^{\infty} \Delta \mathcal{G}_{\alpha \alpha'}(t + nT)^r \\
&= J^2 q_{\textrm{EA}}^{p-1} + \sum_{n=-\infty}^{\infty} \Delta \mathcal{F}_{\alpha \alpha'}(t + nT). 
\end{aligned}
\end{equation}
Thus, up to $O(1)$, both sides of Eq.~\eqref{eq:TAP_resolved_self_energy_demonstration} agree.

Finally, we confirm that Eq.~\eqref{eq:TAP_resolved_EOM} is satisfied.
At non-zero frequencies, we have
\begin{equation} \label{eq:TAP_resolved_EOM_non_zero_demonstration}
i \big( -\mu \omega^2 + z \big) \Delta \widetilde{\mathcal{G}}_{\alpha \alpha'}(\omega) + \sum_{\alpha''} (-1)^{\alpha''} \Delta \widetilde{\mathcal{F}}_{\alpha \alpha''}(\omega) \Delta \widetilde{\mathcal{G}}_{\alpha'' \alpha'}(\omega) + O(T^{-1}) \stackrel{\textrm{set}}{=} (-1)^{\alpha} \delta_{\alpha \alpha'},
\end{equation}
which agrees at $O(1)$ due to Eq.~\eqref{eq:Keldysh_TAP_EOM_simplified}.
At zero frequency, we instead have
\begin{equation} \label{eq:TAP_resolved_EOM_zero_demonstration_1}
\begin{aligned}
&iz \left( Tq_{\textrm{EA}} + \Delta \widetilde{\mathcal{G}}_{\alpha \alpha'}(0) + \frac{\widetilde{g}_{\alpha \alpha'}(0)}{T} \right) \\
&\quad + \sum_{\alpha''} (-1)^{\alpha''} \left( TJ^2 q_{\textrm{EA}}^{p-1} + ipJ q_{\textrm{EA}}^{p/2-1} \big( \mathcal{E}_{\textrm{aux}} - 2J q_{\textrm{EA}}^{p/2-1} \Lambda \big) \big( \delta_{\alpha u} + \delta_{\alpha'' u} \big) + \Delta \widetilde{\mathcal{F}}_{\alpha \alpha''}(0) + \frac{\widetilde{f}_{\alpha \alpha''}(0)}{T} \right) \\
&\qquad \qquad \qquad \qquad \qquad \qquad \qquad \qquad \qquad \qquad \qquad \cdot \left( Tq_{\textrm{EA}} + \Delta \widetilde{\mathcal{G}}_{\alpha'' \alpha'}(0) + \frac{\widetilde{g}_{\alpha'' \alpha'}(0)}{T} \right) \stackrel{\textrm{set}}{=} (-1)^{\alpha} \delta_{\alpha \alpha'}.
\end{aligned}
\end{equation}
The $O(T)$ terms come out to be
\begin{equation} \label{eq:TAP_resolved_EOM_zero_demonstration_2}
T \left( izq_{\textrm{EA}} + \sum_{\alpha''} (-1)^{\alpha''} \Big( J^2 q_{\textrm{EA}}^{p-1} \Delta \widetilde{\mathcal{G}}_{\alpha'' \alpha'}(0) + q_{\textrm{EA}} \Delta \widetilde{\mathcal{F}}_{\alpha \alpha''}(0) \Big) + ipJ q_{\textrm{EA}}^{p/2} \big( \mathcal{E}_{\textrm{aux}} - 2J q_{\textrm{EA}}^{p/2-1} \Lambda \big) \right) \stackrel{\textrm{set}}{=} 0.
\end{equation}
Yet from the TAP magnetization equations, Eq.~\eqref{eq:Keldysh_TAP_magnetization_equation}, it follows that
\begin{equation} \label{eq:Keldysh_TAP_magnetization_double_alternate}
\sum_{\alpha''} (-1)^{\alpha''} \Big( q_{\textrm{EA}} \Delta \widetilde{\mathcal{F}}_{\alpha \alpha''}(0) - (p-1)J^2 q_{\textrm{EA}}^{p-1} \Delta \widetilde{\mathcal{G}}_{\alpha \alpha''}(0) \Big) = -izq_{\textrm{EA}} - ipJ q_{\textrm{EA}}^{p/2} \mathcal{E}_{\textrm{aux}}.
\end{equation}
Thus Eq.~\eqref{eq:TAP_resolved_EOM_zero_demonstration_2} evaluates to
\begin{equation} \label{eq:TAP_resolved_EOM_zero_leading_term}
-ipJ q_{\textrm{EA}}^{p/2} \mathcal{E}_{\textrm{aux}} + pJ^2 q_{\textrm{EA}}^{p-1} \sum_{\alpha''} (-1)^{\alpha''} \Delta \widetilde{\mathcal{G}}_{\alpha \alpha''}(0) + ipJ q_{\textrm{EA}}^{p/2} \mathcal{E}_{\textrm{aux}} - 2ipJ^2 q_{\textrm{EA}}^{p-1} \Lambda = 0,
\end{equation}
using Eq.~\eqref{eq:Keldysh_zero_frequency_matrix}.
The $O(1)$ terms of Eq.~\eqref{eq:TAP_resolved_EOM_zero_demonstration_1} determine $\widetilde{g}_{\alpha \alpha'}(0)$ and $\widetilde{f}_{\alpha \alpha'}(0)$. We have therefore confirmed that all saddle point equations are solved by Eqs.~\eqref{eq:TAP_resolved_G_solution_frequency} through~\eqref{eq:TAP_resolved_lambda_solution}.

\subsection{Contribution of connected solutions} \label{subsec:nonergodic_contribution_connected_solutions}

It remains only to evaluate the action, Eq.~\eqref{eq:TAP_resolved_SFF_action}, at the above solution.
The action can be written as
\begin{equation} \label{eq:TAP_resolved_action_starting_point}
\begin{aligned}
S_{\textrm{eff}} &= -i \lambda \mathcal{E}_{\textrm{aux}} + \frac{\lambda^2}{2p} + \frac{JT \lambda}{p q[G]^{p/2}} \sum_{\alpha} (-1)^{\alpha} \left( \frac{\widetilde{G}_{\alpha u}(0)}{T} \right)^p \\
&\qquad + \frac{T}{2} \sum_{\alpha \alpha'} (-1)^{\alpha + \alpha'} \int_0^T dt \left( \frac{J^2}{p} G_{\alpha \alpha'}(t)^p - F_{\alpha \alpha'}(t) G_{\alpha \alpha'}(t) \right) \\
&\qquad \qquad + \frac{1}{2} \sum_{\omega} \log{\textrm{Det}} \Big[ i(-1)^{\alpha} \delta_{\alpha \alpha'} \big( -\mu \omega^2 + z \big) + (-1)^{\alpha + \alpha'} \widetilde{F}_{\alpha \alpha'}(\omega) \Big].
\end{aligned}
\end{equation}
Interestingly, we can determine $S_{\textrm{eff}}$ up to a single additive constant simply by noting that $dS_{\textrm{eff}}/d\mathcal{E}_{\textrm{aux}} = -i \lambda$ (recall that $S_{\textrm{eff}}$ is stationary with respect to variations in all quantities other than $\mathcal{E}_{\textrm{aux}}$).
With $\lambda$ given by Eq.~\eqref{eq:TAP_resolved_lambda_solution} and $q_{\textrm{EA}}^{p/2-1} \Lambda$ given by Eq.~\eqref{eq:Keldysh_TAP_overlap_closed_equation}, we can carry out the integral to obtain that
\begin{equation} \label{eq:nonergodic_action_almost_evaluation}
S_{\textrm{eff}}[\mathcal{E}_{\textrm{aux}}] = \frac{p \mathcal{E}_{\textrm{aux}}^2}{2} - \frac{p^2}{8(p-1)} \Big( \mathcal{E}_{\textrm{aux}} + \sqrt{\mathcal{E}_{\textrm{aux}}^2 - \mathcal{E}_{\textrm{th}}^2} \Big)^2 - \log{\Big( -\mathcal{E}_{\textrm{aux}} + \sqrt{\mathcal{E}_{\textrm{aux}}^2 - \mathcal{E}_{\textrm{th}}^2} \Big)} + C,
\end{equation}
for some unknown constant $C$.
Comparing to Eq.~\eqref{eq:TAP_complexity}, this is highly suggestive that $S_{\textrm{eff}} = -\Sigma(\mathcal{E}_{\textrm{aux}})$.
Of course, we do need to determine the remaining constant, and so we now turn to a more elaborate calculation.

Rather than substitute Eqs.~\eqref{eq:TAP_resolved_G_solution_frequency} and~\eqref{eq:TAP_resolved_F_solution_frequency} into Eq.~\eqref{eq:TAP_resolved_action_starting_point}, we instead use the simpler functions
\begin{equation} \label{eq:simplified_G_solution_frequency}
\widetilde{G}'_{\alpha \alpha'}(\omega) = Tq_{\textrm{EA}} \delta_{\omega 0} + \Delta \widetilde{\mathcal{G}}_{\alpha \alpha'}(\omega),
\end{equation}
\begin{equation} \label{eq:simplified_F_solution_frequency}
\widetilde{F}'_{\alpha \alpha'}(\omega) = \left( TJ^2 q_{\textrm{EA}}^{p-1} + ipJ q_{\textrm{EA}}^{p/2-1} \big( \mathcal{E}_{\textrm{aux}} - 2J q_{\textrm{EA}}^{p/2-1} \Lambda \big) \big( \delta_{\alpha u} + \delta_{\alpha' u} \big) \right) \delta_{\omega 0} + \Delta \widetilde{\mathcal{F}}_{\alpha \alpha'}(\omega) + \frac{\widetilde{f}_{\alpha \alpha'}(0)}{T} \delta_{\omega 0},
\end{equation}
and show that the error incurred in doing so vanishes at large $T$.

Let us demonstrate that the error is negligible first.
At any non-zero frequency, we have that
\begin{equation} \label{eq:simplified_solutions_nonzero_frequency}
\widetilde{G}_{\alpha \alpha'}(\omega) = \widetilde{G}'_{\alpha \alpha'}(\omega) + \frac{\widetilde{g}_{\alpha \alpha'}(\omega)}{T}, \qquad \widetilde{F}_{\alpha \alpha'}(\omega) = \widetilde{F}'_{\alpha \alpha'}(\omega) + \frac{\widetilde{f}_{\alpha \alpha'}(\omega)}{T}.
\end{equation}
The partial derivatives of $S_{\textrm{eff}}$ at nonzero $\omega$ are
\begin{equation} \label{eq:action_G_derivative_nonzero_frequency}
\frac{\partial S_{\textrm{eff}}}{\partial \widetilde{G}_{\alpha \alpha'}(\omega)} = \frac{1}{2} (-1)^{\alpha + \alpha'} \int_0^T dt e^{-i \omega t} \Big( J^2 G_{\alpha \alpha'}(t)^{p-1} - F_{\alpha \alpha'}(t) \Big),
\end{equation}
\begin{equation} \label{eq:action_F_derivative_nonzero_frequency}
\frac{\partial S_{\textrm{eff}}}{\partial \widetilde{F}_{\alpha \alpha'}(\omega)} = \frac{1}{2} (-1)^{\alpha + \alpha'} \bigg( \Big[ i(-1)^{\alpha} \delta_{\alpha \alpha'} \big( -\mu \omega^2 + z \big) + (-1)^{\alpha + \alpha'} \widetilde{F}_{\alpha' \alpha}(\omega) \Big]_{\alpha \alpha'}^{-1} - \int_0^T dt e^{-i \omega t} G_{\alpha \alpha'}(t) \bigg),
\end{equation}
which vanish when evaluated at $\widetilde{G}'_{\alpha \alpha'}(\omega) = \Delta \widetilde{\mathcal{G}}_{\alpha \alpha'}(\omega)$ and $\widetilde{F}'_{\alpha \alpha'}(\omega) = \Delta \widetilde{\mathcal{F}}_{\alpha \alpha'}(\omega)$.
Thus the $O(T^{-1})$ difference between $\widetilde{G}_{\alpha \alpha'}(\omega)$ and $\widetilde{G}'_{\alpha \alpha'}(\omega)$, as with $\widetilde{F}_{\alpha \alpha'}(\omega)$ and $\widetilde{F}'_{\alpha \alpha'}(\omega)$, translates only to an $O(T^{-2})$ difference in the action.
Even after summing over all $\omega \neq 0$, the total error\footnote{
Since the $G(t)^p$ term is not diagonal in the frequency domain, this argument requires a bit more care.
One can easily show that $\partial^2 S_{\textrm{eff}} / \partial \widetilde{G}(\omega) \partial \widetilde{G}(\omega')$ is $O(T^{-1})$ for $\omega \neq \pm \omega'$ and $O(1)$ for $\omega = \pm \omega'$.
Summing over all frequencies, the former case gives a total contribution $O(T^{-3}) O(T^2) = O(T^{-1})$ and the latter gives $O(T^{-2}) O(T) = O(T^{-1})$.
The total error is thus $O(T^{-1})$ as claimed.
} is only $O(T^{-1})$.

Neglecting non-zero frequencies, $\widetilde{F}'_{\alpha \alpha'}(\omega)$ is identical to $\widetilde{F}_{\alpha \alpha'}(\omega)$ and $\widetilde{G}'_{\alpha \alpha'}(\omega)$ differs only by $\widetilde{g}_{\alpha \alpha'}(0) \delta_{\omega 0}/T$.
In the time domain, the latter corresponds to
\begin{equation} \label{eq:semi_simplified_G_solution_time}
G_{\alpha \alpha'}(t) = G'_{\alpha \alpha'}(t) + \frac{\widetilde{g}_{\alpha \alpha'}(0)}{T^2} = q_{\textrm{EA}} + \sum_{n=-\infty}^{\infty} \Delta \mathcal{G}_{\alpha \alpha'}(t + nT) + \frac{\widetilde{g}_{\alpha \alpha'}(0)}{T^2}.
\end{equation}
Yet
\begin{equation} \label{eq:action_G_derivative_time}
\frac{\partial S_{\textrm{eff}}}{\partial G_{\alpha \alpha'}(t)} = \frac{T}{2} (-1)^{\alpha + \alpha'} \Big( J^2 G_{\alpha \alpha'}(t)^{p-1} - F_{\alpha \alpha'}(t) \Big) + O(1).
\end{equation}
When evaluated at $G'_{\alpha \alpha'}(t)$ and $F'_{\alpha \alpha'}(t)$, the $O(T)$ contribution vanishes (see Eq.~\eqref{eq:non_ergodic_cross_terms_neglecting}).
Thus $\partial S_{\textrm{eff}} / \partial G_{\alpha \alpha'}(t)$ is $O(1)$, and an $O(T^{-2})$ change to $G_{\alpha \alpha'}(t)$ leads only to an $O(T^{-1})$ change in the action even after integrating over $t$.

Since all errors are $O(T^{-1})$, we can safely evaluate $S_{\textrm{eff}}$ at Eqs.~\eqref{eq:simplified_G_solution_frequency} and~\eqref{eq:simplified_F_solution_frequency} rather than the full solution (we still use Eq.~\eqref{eq:TAP_resolved_lambda_solution} for $\lambda$).
The first line of Eq.~\eqref{eq:TAP_resolved_action_starting_point} can be computed straightforwardly.
It comes out to be
\begin{equation} \label{eq:nonergodic_action_evaluation_line_1}
\frac{p \big( \mathcal{E}_{\textrm{aux}} - 2J q_{\textrm{EA}}^{p/2-1} \Lambda \big)^2}{2} = \frac{1}{8pJ^2 q_{\textrm{EA}}^{p-2} \Lambda^2} - \frac{1}{p} + \frac{2J^2 q_{\textrm{EA}}^{p-2} \Lambda^2}{p},
\end{equation}
where we used Eq.~\eqref{eq:potential_GF_relationship} to obtain the right-hand side.

Next consider the bottom line.
Since $\widetilde{F}'_{\alpha \alpha'}(\omega) = \Delta \widetilde{\mathcal{F}}_{\alpha \alpha'}(\omega)$ for $\omega \neq 0$, while $\widetilde{F}'_{\alpha \alpha'}(0) = \widetilde{F}_{\alpha \alpha'}(0)$, we can write the determinant term as
\begin{equation} \label{eq:nonergodic_action_evaluation_determinant_1}
\begin{aligned}
&\frac{1}{2} \sum_{\omega} \log{\textrm{Det}} \Big[ i(-1)^{\alpha} \delta_{\alpha \alpha'} \big( -\mu \omega^2 + z \big) + (-1)^{\alpha + \alpha'} \Delta \widetilde{\mathcal{F}}_{\alpha \alpha'}(\omega) \Big] \\
&\qquad \qquad + \frac{1}{2} \log{\textrm{Det}} \Big[ iz(-1)^{\alpha} \delta_{\alpha \alpha'} + (-1)^{\alpha + \alpha'} \widetilde{F}_{\alpha \alpha'}(0) \Big] \\
&\qquad \qquad \qquad \qquad - \frac{1}{2} \log{\textrm{Det}} \Big[ iz(-1)^{\alpha} \delta_{\alpha \alpha'} + (-1)^{\alpha + \alpha'} \Delta \widetilde{\mathcal{F}}_{\alpha \alpha'}(0) \Big].
\end{aligned}
\end{equation}
The top line vanishes by exactly the same reasoning as in Sec.~\ref{subsec:ergodic_connected_action}: it is proportional to $T$ by the Euler-Maclaurin formula, and then must be zero since the derivative with respect to $T$ vanishes.
Given Eq.~\eqref{eq:Keldysh_zero_frequency_matrix}, the bottom line is simply
\begin{equation} \label{eq:nonergodic_action_evaluation_determinant_2}
\frac{1}{2} \log{\textrm{Det}} \Delta \widetilde{\mathcal{G}}_{\alpha \alpha'}(0) = \frac{1}{2} \log{4 \Lambda^2}.
\end{equation}
For the middle line we take an indirect approach.
We have that $iz(-1)^{\alpha} \delta_{\alpha \alpha'} + (-1)^{\alpha + \alpha'} \widetilde{F}_{\alpha \alpha'}(0)$ is the matrix inverse to $\widetilde{G}_{\alpha \alpha'}(0)$ (using the full solution for the latter, Eq.~\eqref{eq:TAP_resolved_G_solution_frequency}).
Written out,
\begin{equation} \label{eq:SFF_functions_matrix_inverse_z_basis}
\begin{pmatrix} iz + \widetilde{F}_{uu}(0) & -\widetilde{F}_{ul}(0) \\ -\widetilde{F}_{lu}(0) & -iz + \widetilde{F}_{ll}(0) \end{pmatrix} = \begin{pmatrix} \widetilde{G}_{uu}(0) & \widetilde{G}_{ul}(0) \\ \widetilde{G}_{lu}(0) & \widetilde{G}_{ll}(0) \end{pmatrix}^{-1} = \frac{1}{\textrm{Det} \widetilde{G}(0)} \begin{pmatrix} \widetilde{G}_{ll}(0) & -\widetilde{G}_{ul}(0) \\ -\widetilde{G}_{lu}(0) & \widetilde{G}_{uu}(0) \end{pmatrix}.
\end{equation}
Rather than this $(u, l)$ basis, express Eq.~\eqref{eq:SFF_functions_matrix_inverse_z_basis} in the $(u + l, u - l)$ basis (called ``classical''/``quantum'' in the Keldysh literature), denoted $(+, -)$:
\begin{equation} \label{eq:SFF_functions_matrix_inverse_x_basis}
\begin{pmatrix} \widetilde{F}_{--}(0) & iz + \widetilde{F}_{-+}(0) \\ iz + \widetilde{F}_{+-}(0) & \widetilde{F}_{++}(0) \end{pmatrix} = \frac{1}{\textrm{Det} \widetilde{G}(0)} \begin{pmatrix} \widetilde{G}_{--}(0) & -\widetilde{G}_{+-}(0) \\ -\widetilde{G}_{-+}(0) & \widetilde{G}_{++}(0) \end{pmatrix}.
\end{equation}
We can read off that $\textrm{Det} \widetilde{G}(0)^{-1} = \widetilde{F}_{++}(0) / \widetilde{G}_{++}(0)$.
Note that we only need $\widetilde{G}_{++}(0)$ and $\widetilde{F}_{++}(0)$ to $O(T)$ in order to calculate the determinant to $O(1)$.
Thus the middle line of Eq.~\eqref{eq:nonergodic_action_evaluation_determinant_1} evaluates to $(\log{J^2 q_{\textrm{EA}}^{p-2}})/2$, and the total contribution of the determinant term is
\begin{equation} \label{eq:nonergodic_action_evaluation_determinant_3}
\frac{1}{2} \log{4J^2 q_{\textrm{EA}}^{p-2} \Lambda^2}.
\end{equation}

Lastly consider the middle line of Eq.~\eqref{eq:TAP_resolved_action_starting_point}.
Since $G'_{\alpha \alpha'}(t) = \mathcal{G}_{\alpha \alpha'}(t)$ (up to exponentially small corrections), $\sum_{\alpha \alpha'} (-1)^{\alpha + \alpha'} G'_{\alpha \alpha'}(t)^p = 0$ by virtue of Eq.~\eqref{eq:Keldysh_component_symmetry_identity}.
We are left with
\begin{equation} \label{eq:nonergodic_action_evaluation_middle_line_1}
\begin{aligned}
&-\frac{T}{2} \sum_{\alpha \alpha'} (-1)^{\alpha + \alpha'} \int_0^T dt F'_{\alpha \alpha'}(t) G'_{\alpha \alpha'}(t) \\
&\qquad \sim -\frac{T}{2} \sum_{\alpha \alpha'} (-1)^{\alpha + \alpha'} \int_0^T dt \mathcal{F}_{\alpha \alpha'}(t) \mathcal{G}_{\alpha \alpha'}(t) \\
&\qquad \qquad - \frac{1}{2} \sum_{\alpha \alpha'} (-1)^{\alpha + \alpha'} \left( ipJ q_{\textrm{EA}}^{p/2-1} \big( \mathcal{E}_{\textrm{aux}} - 2J q_{\textrm{EA}}^{p/2-1} \Lambda \big) \big( \delta_{\alpha u} + \delta_{\alpha' u} \big) + \frac{\widetilde{f}_{\alpha \alpha'}(0)}{T} \right) \widetilde{G}'_{\alpha \alpha'}(0).
\end{aligned}
\end{equation}
The first term is again proportional to $\sum_{\alpha \alpha'} (-1)^{\alpha + \alpha'} G'_{\alpha \alpha'}(t)^p = 0$.
The second term would appear to be more problematic, since $\widetilde{f}_{\alpha \alpha'}(0)$ (for which we have not given an explicit expression) contributes at $O(1)$ due to $\widetilde{G}'_{\alpha \alpha'}(0)$ being $O(T)$.
However, we only need the component $\widetilde{f}_{--}(0)/T = \widetilde{F}_{--}(0)$, and from Eq.~\eqref{eq:SFF_functions_matrix_inverse_x_basis} we see that
\begin{equation} \label{eq:F_minus_component_evaluation}
\widetilde{F}_{--}(0) = \frac{1}{\textrm{Det} \widetilde{G}(0)} \widetilde{G}_{--}(0) = \frac{1}{\textrm{Det} \widetilde{G}(0)} \frac{\textrm{Det} \widetilde{G}(0) + \widetilde{G}_{+-}(0) \widetilde{G}_{-+}(0)}{\widetilde{G}_{++}(0)} = \frac{1 - 4J^2 q_{\textrm{EA}}^{p-2} \Lambda^2}{2Tq_{\textrm{EA}}} + O \left( \frac{1}{T^2} \right).
\end{equation}
Eq.~\eqref{eq:nonergodic_action_evaluation_middle_line_1} evaluates to
\begin{equation} \label{eq:nonergodic_action_evaluation_middle_line_2}
2pJ q_{\textrm{EA}}^{p/2-1} \Lambda \big( \mathcal{E}_{\textrm{aux}} - 2J q_{\textrm{EA}}^{p/2-1} \Lambda \big) - \frac{1}{2} + 2J^2 q_{\textrm{EA}}^{p-2} \Lambda^2 = \frac{1}{2} - 2J^2 q_{\textrm{EA}}^{p-2} \Lambda^2,
\end{equation}
again using Eq.~\eqref{eq:potential_GF_relationship}.

We finally have the large-$T$ limit of the action, given by the sum of Eqs.~\eqref{eq:nonergodic_action_evaluation_line_1},~\eqref{eq:nonergodic_action_evaluation_determinant_3}, and~\eqref{eq:nonergodic_action_evaluation_middle_line_2}:
\begin{equation} \label{eq:nonergodic_action_result}
S_{\textrm{eff}}[\mathcal{E}_{\textrm{aux}}] = \frac{p-2}{2p} + \frac{1}{8pJ^2 q_{\textrm{EA}}^{p-2} \Lambda^2} - \frac{2(p-1)J^2 q_{\textrm{EA}}^{p-2} \Lambda^2}{p} + \frac{1}{2} \log{4J^2 q_{\textrm{EA}}^{p-2} \Lambda^2}.
\end{equation}
Comparing to the complexity $\Sigma(\mathcal{E})$ given in Eq.~\eqref{eq:TAP_complexity_rewrite}, we see that $S_{\textrm{eff}}$ is precisely $-\Sigma(\mathcal{E}_{\textrm{aux}})$.

\subsection{Evaluation of the SFF} \label{subsec:final_SFF_evaluation}

We have shown that, at a given $\mathcal{E}_{\textrm{aux}}$ and for each value of inverse temperature $\beta_{\textrm{aux}}$, there is a solution to the SFF saddle point equations with $S_{\textrm{eff}} = -\Sigma(\mathcal{E}_{\textrm{aux}})$.
The full (connected) SFF is obtained by integrating over all $\mathcal{E}_{\textrm{aux}}$ and $\beta_{\textrm{aux}}$, as well as the symmetry-broken order parameter $\Delta$ (which contributes an overall factor of $T$) and an additional factor $2(1 + \delta_{p \textrm{ even}})$ from the discrete symmetries.
As in Sec.~\ref{sec:ergodic_ramp}, it is more convenient to integrate over the energy density $\epsilon(\mathcal{E}_{\textrm{aux}}, \beta_{\textrm{aux}})$.
We show in App.~\ref{sec:TAP_energy} that $\epsilon(\mathcal{E}, \beta)$ comes out to be precisely the argument of the filter function, Eq.~\eqref{eq:TAP_resolved_energy_density_definition}, when evaluated at the saddle point solution.

Our final result is that\footnote{The factor $\sqrt{pN/2\pi}$ comes from the integral over fluctuations in $\lambda$ --- the variance is $p/N$ (see Eq.~\eqref{eq:TAP_resolved_action_starting_point}), and the original fat unity introducing $\mathcal{E}_{\textrm{aux}}$ comes with a prefactor $N/2\pi$.}
\begin{equation} \label{eq:SFF_final_result}
\textrm{SFF}(T, f) = \big| \mathbb{E} \textrm{Tr} f(H) e^{-iHT} \big|^2 + 2 \big( 1 + \delta_{p \textrm{ even}} \big) T \sqrt{\frac{pN}{2\pi}} \int d\mathcal{E}_{\textrm{aux}} e^{N \Sigma(\mathcal{E}_{\textrm{aux}})} \int_{\epsilon_-(\mathcal{E}_{\textrm{aux}})}^{\epsilon_+(\mathcal{E}_{\textrm{aux}})} \frac{d\epsilon_{\textrm{aux}}}{2\pi} f(\epsilon_{\textrm{aux}})^2,
\end{equation}
where the inner integral runs only over the range $[\epsilon_-(\mathcal{E}_{\textrm{aux}}), \epsilon_+(\mathcal{E}_{\textrm{aux}})]$ in which solutions to the TAP equations exist.
Furthermore, one can easily generalize Eq.~\eqref{eq:SFF_final_result} by making the filter function $\mathcal{E}$-dependent, i.e., $f(\mathcal{E}, \epsilon_{\textrm{aux}})$.
The resulting quantity is the SFF for the projection of the system into certain TAP states.

Compare Eq.~\eqref{eq:SFF_final_result} for the non-ergodic phase to Eq.~\eqref{eq:SFF_connected_contribution_ergodic_phase} for the ergodic phase, and recall the discussion of block-diagonal Hamiltonians in Sec.~\ref{subsec:review_spectral_form_factor}.
Our result demonstrates that each metastable (i.e., TAP) state can be thought of its own quantum chaotic subspace, one which is independent of any others.
This is the central result of our paper.
While the qualitative idea has been proposed in previous work~\cite{Baldwin2017Clustering}, the present analysis both makes it precise and proves it.

\section{Higher moments of the evolution operator}
\label{sec:HigherMoments}

In this final section we consider higher moments of $\textrm{Tr} e^{-iHT}$, i.e., the quantities
\begin{equation} \label{eq:higher_SFF_definition}
\textrm{SFF}^{(n)}(T, f) \equiv \mathbb{E} \Big[ \Big( \textrm{Tr} f(H) e^{-iHT} \Big)^n \Big( \textrm{Tr} f(H) e^{iHT} \Big)^n \Big].
\end{equation}
The saddle points of these higher moments exhibit an interesting structure that will shed further light on the distribution of TAP states, although care must be taken in interpreting the results.
We first present the calculation and discuss afterwards.

\subsection{Effective action} \label{subsec:higher_moment_effective_action}

The effective action governing the $n$'th moment is derived in exactly the same manner as in Sec.~\ref{sec:nonergodic_ramp}.
The only major difference is that now spins have a ``replica'' index $a \in \{1, \cdots, n\}$ in addition to a contour index $\alpha \in \{u, l\}$.
We also include a separate fat unity defining $\mathcal{E}_{\textrm{aux}, a}$ for each replica.
The result is (compare to Eqs.~\eqref{eq:TAP_resolved_SFF_path_integral} and~\eqref{eq:TAP_resolved_SFF_action})
\begin{equation} \label{eq:higher_moment_SFF_path_integral}
\textrm{SFF}^{(n)}(T, f) = \int d\mathcal{E}_{\textrm{aux}} d\lambda \mathcal{D}G \mathcal{D}F \prod_{a=1}^n f \big( \epsilon_{au}[\lambda, G] \big) f \big( \epsilon_{al}[\lambda, G] \big) e^{-NS_{\textrm{eff}}[\mathcal{E}_{\textrm{aux}}, \lambda, G, F]},
\end{equation}
\begin{equation} \label{eq:higher_moment_SFF_action}
\begin{aligned}
S_{\textrm{eff}}[\mathcal{E}_{\textrm{aux}}, \lambda, G, F] &= -i \sum_a \lambda_a \mathcal{E}_{\textrm{aux}, a} - \frac{i}{2} \int_0^T dt \sum_{a \alpha} (-1)^{\alpha} z_{a \alpha}(t) \\
&\quad + \sum_{aa'} \frac{\lambda_a \lambda_{a'}}{2p q[G_{aa}]^{p/2} q[G_{a'a'}]^{p/2}} \left( \frac{1}{T^2} \int_0^T dt dt' G_{au, a'u}(t, t') \right)^p \\
&\quad \quad + \sum_{a'} \frac{J \lambda_{a'}}{pq[G_{a'a'}]^{p/2}} \int_0^T dt \sum_{a \alpha} (-1)^{\alpha} \left( \frac{1}{T} \int_0^T dt' G_{a \alpha, a' u}(t, t') \right)^p \\
&\quad \quad \quad + \frac{1}{2} \int_0^T dt dt' \sum_{aa'} \sum_{\alpha \alpha'} (-1)^{\alpha + \alpha'} \left( \frac{J^2}{p} G_{a \alpha, a' \alpha'}(t, t')^p - F_{a \alpha, a' \alpha'}(t, t') G_{a \alpha, a' \alpha'}(t, t') \right) \\
&\quad \quad \quad \quad + \frac{1}{2} \log{\textrm{Det}} \Big[ i (-1)^{\alpha} \delta_{aa'} \delta_{\alpha \alpha'} \big( \mu \partial_t^2 + z_{a \alpha} \big) + (-1)^{\alpha + \alpha'} F_{a \alpha, a' \alpha'} \Big],
\end{aligned}
\end{equation}
with energy densities
\begin{equation} \label{eq:higher_moment_energy_density_definition}
\begin{aligned}
\epsilon_{a \alpha}[\lambda, G] &= -\frac{\mu}{2} \partial_t^2 G_{a \alpha, a \alpha}(0^+, 0) \\
&\qquad \qquad - \frac{iJ^2}{p} \int_0^T dt \sum_{a' \alpha'} (-1)^{\alpha'} G_{a \alpha, a' \alpha'}(t, 0)^p - i \sum_{a'} \frac{J \lambda_{a'}}{pq[G_{a'a'}]^{p/2}} \left( \frac{1}{T} \int_0^T dt G_{a \alpha, a' u}(t, 0) \right)^p.
\end{aligned}
\end{equation}
The saddle point equations are therefore
\begin{equation} \label{eq:higher_moment_EOM}
\begin{aligned}
i \big( \mu \partial_t^2 + z_a \big) G_{a \alpha, a' \alpha'}(t - t') + \int_0^T dt'' \sum_{a'' \alpha''} (-1)^{\alpha''} F_{a \alpha, a'' \alpha''}(t - t'') &G_{a'' \alpha'', a' \alpha'}(t'' - t') \\
&= (-1)^{\alpha} \delta_{aa'} \delta_{\alpha \alpha'} \delta(t - t'),
\end{aligned}
\end{equation}
\begin{equation} \label{eq:higher_moment_self_energy_equation}
F_{a \alpha, a' \alpha'}(t) = J^2 G_{a \alpha, a' \alpha'}(t)^{p-1} + \frac{J}{T} \left( \frac{\lambda_a}{q[G_{aa}]^{p/2}} \delta_{\alpha u} + \frac{\lambda_{a'}}{q[G_{a'a'}]^{p/2}} \delta_{\alpha' u} \right) \left( \frac{\widetilde{G}_{a \alpha, a' \alpha'}(0)}{T} \right)^{p-1} + O(T^{-2}),
\end{equation}
\begin{equation} \label{eq:higher_moment_potential_equation}
\mathcal{E}_{\textrm{aux}, a} = -\frac{iJT}{pq[G_{aa}]^{p/2}} \sum_{a' \alpha'} (-1)^{\alpha'} \left( \frac{\widetilde{G}_{au, a' \alpha'}(0)}{T} \right)^p - \frac{i}{pq[G_{aa}]^{p/2}} \sum_{a'} \frac{\lambda_{a'}}{q[G_{a'a'}]^{p/2}} \left( \frac{\widetilde{G}_{au, a'u}(0)}{T} \right)^p.
\end{equation}

Note that Eqs.~\eqref{eq:higher_moment_EOM} through~\eqref{eq:higher_moment_potential_equation} have the following permutation symmetry with respect to replica indices.
Suppose that $G$, $F$, and $\lambda$ constitute a valid solution.
For any permutation $\pi$ of the set $\{1, \cdots, n\}$, define $\pi_{\alpha}(a)$ to be the permuted element $\pi(a)$ if $\alpha = l$ but simply the original element $a$ if $\alpha = u$.
Then the quantities $\overline{G}$, $\overline{F}$, and $\overline{\lambda}$ defined by
\begin{equation} \label{eq:higher_moment_permutation_symmetry}
\overline{G}_{a \alpha, a' \alpha'}(t, t') \equiv G_{\pi_{\alpha}(a) \alpha, \pi_{\alpha'}(a') \alpha'}(t, t'), \qquad \overline{F}_{a \alpha, a' \alpha'}(t, t') \equiv F_{\pi_{\alpha}(a) \alpha, \pi_{\alpha'}(a') \alpha'}(t, t'), \qquad \overline{\lambda}_a = \lambda_a,
\end{equation}
constitute an equally valid solution.
This symmetry has a nice graphical interpretation in terms of pairings between upper and lower contours, illustrated in Fig.~\ref{fig:multi_contour}: however contour $au$ is correlated with $a'l$ in a given solution, there is an alternate solution in which $au$ has the same correlation with $\pi(a')l$.

One trivial solution to the saddle point equations is to use the solution from Sec.~\ref{sec:nonergodic_ramp} for $a = a'$ while setting all cross-replica elements to zero.
The action then decomposes into a sum of single-replica actions, which we evaluated in Sec.~\ref{sec:nonergodic_ramp}.
In other words, this contribution to the $n$'th moment is simply $\textrm{SFF}(T, f)^n$.
However, by the permutation symmetry described above, we actually have $n!$ such contributions:
\begin{equation} \label{eq:higher_moment_disconnected_part}
\textrm{SFF}^{(n)}(T, f) = n! \cdot \textrm{SFF}(T, f)^n + \cdots,
\end{equation}
where the ellipses denote additional solutions.

\subsection{Connected solutions} \label{subsec:higher_moment_connected_solutions}

\begin{figure}
    \centering
    \includegraphics[width=.9\textwidth]{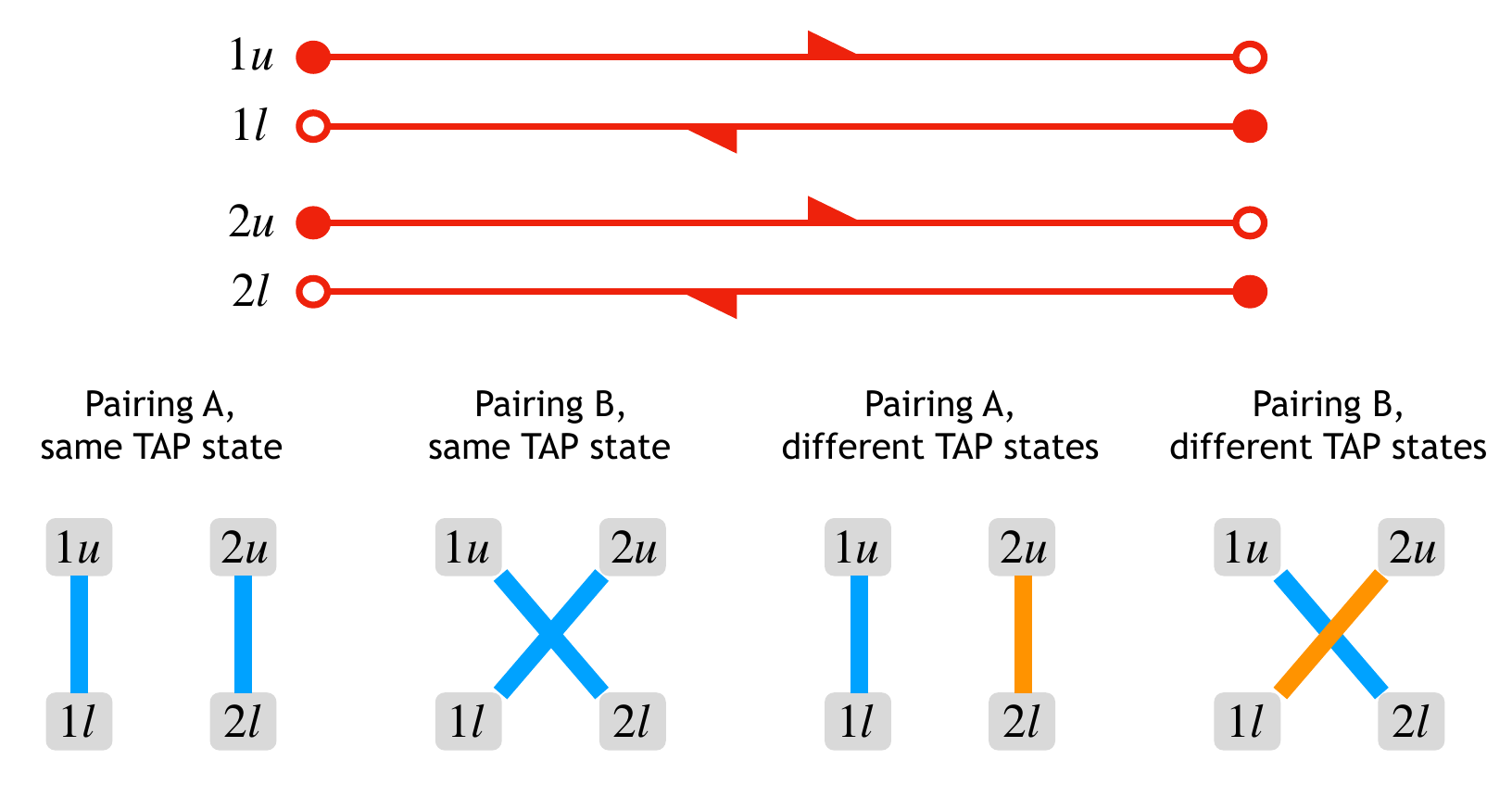}
    \caption{Graphical representation of the various saddle point solutions for the $n=2$ moment. The four contours --- $1u$, $1l$, $2u$, $2l$ --- are shown at the top. Below are the four varieties of solutions: each upper contour must be paired with a lower contour, but one is free to choose which replicas are paired, and there is further freedom in which TAP state each pair lies within (blue and orange lines indicate two different TAP states).}
    \label{fig:multi_contour}
\end{figure}

In general, for arbitrary values of $\mathcal{E}_{\textrm{aux}, a}$, we have been unable to find any further saddle points.
However, when some replicas have equal values of $\mathcal{E}_{\textrm{aux}}$, we can construct additional solutions.
Pick any set of inverse temperatures $\beta_a$ (not necessarily equal), and suppose that the replicas $\{1, \cdots, n\}$ partition into groups $A \equiv \{a_1, \cdots, a_{|A|}\}$, such that $\mathcal{E}_{\textrm{aux}, a}$ equals a common value $\mathcal{E}_{\textrm{aux}, A}$ for all $a \in A$.
We again take $G_{a \alpha, a \alpha'}(t - t')$ to be the solution from Sec.~\ref{sec:nonergodic_ramp}.
For $a$ and $a'$ in different groups, we still set $G_{a \alpha, a' \alpha'} = 0$.
For $a$ and $a'$ in the same group $A$, however, we now set
\begin{equation} \label{eq:higher_moment_cross_factor_G_ansatz}
G_{a \alpha, a' \alpha'}(t - t') = \big( q_{\textrm{EA}, a} q_{\textrm{EA}, a'} \big)^{1/2},
\end{equation}
where $q_{\textrm{EA}, a}$ is the Edwards-Anderson order parameter corresponding to $\mathcal{E}_{\textrm{aux}, A}$ and $\beta_a$.
This corresponds to the replicas lying within the same TAP state (see Fig.~\ref{fig:multi_contour}).
We can write this compactly as
\begin{equation} \label{eq:higher_moment_total_G_ansatz}
G_{a \alpha, a' \alpha'}(t) = \big( q_{\textrm{EA}, a} q_{\textrm{EA}, a'} \big)^{1/2} + \delta_{aa'} \Big( \Delta \mathcal{G}_{a, \alpha \alpha'}(t) + O(T^{-1}) \Big).
\end{equation}

Inserting into Eq.~\eqref{eq:higher_moment_potential_equation}, we have that $\lambda_a$ must obey
\begin{equation} \label{eq:higher_moment_lambda_determination_v1}
\sum_{a' \in A} \lambda_{a'} = ip \big( \mathcal{E}_{\textrm{aux}, A} - 2Jq_{\textrm{EA}, a}^{p/2-1} \Lambda_a \big) + O(T^{-1}).
\end{equation}
Note that, by virtue of Eq.~\eqref{eq:Keldysh_TAP_overlap_closed_equation}, $J q_{\textrm{EA}, a}^{p/2-1} \Lambda_a$ is a function solely of $\mathcal{E}_{\textrm{aux}, A}$.
Thus Eq.~\eqref{eq:higher_moment_lambda_determination_v1} is consistent among all $a \in A$.
The self-energy is then given by
\begin{equation} \label{eq:higher_moment_F_determination}
F_{a \alpha, a' \alpha'}(t) = J^2 \big( q_{\textrm{EA}, a} q_{\textrm{EA}, a'} \big)^{\frac{p-1}{2}} + \delta_{aa'} \Delta \mathcal{F}_{a, \alpha \alpha'}(t) + \frac{J}{T} \left( \sqrt{\frac{q_{\textrm{EA}, a'}^{p-1}}{q_{\textrm{EA}, a}}} \lambda_a \delta_{\alpha u} + \sqrt{\frac{q_{\textrm{EA}, a}^{p-1}}{q_{\textrm{EA}, a'}}} \lambda_{a'} \delta_{\alpha' u} \right) + O(T^{-1}).
\end{equation}
It remains only to check that Eq.~\eqref{eq:higher_moment_EOM} can be satisfied.
It is automatically solved at non-zero frequencies, since then $\widetilde{G}(\omega)$ and $\widetilde{F}(\omega)$ reduce to $\delta_{aa'} \Delta \widetilde{\mathcal{G}}(\omega)$ and $\delta_{aa'} \Delta \widetilde{\mathcal{F}}(\omega)$ respectively.
At zero frequency we confirm that the equation is solved to $O(T)$ (the $O(1)$ terms only determine subleading corrections).
Following the same steps as in Sec.~\ref{subsec:TAP_resolved_connected_solutions}, the left-hand side of Eq.~\eqref{eq:higher_moment_EOM} simplifies to
\begin{equation} \label{eq:higher_moment_G_determination}
JT \sqrt{q_{\textrm{EA}, a}^{p-1} q_{\textrm{EA}, a'}} \left( \sum_{a'' \in A} \lambda_{a''} + 2i(p-1) Jq_{\textrm{EA}, a}^{p/2-1} \Lambda_a + 2iJq_{\textrm{EA}, a'}^{p/2-1} \Lambda_{a'} - ip \mathcal{E}_{\textrm{aux}, A} \right) = 0,
\end{equation}
as desired.

Note that in this solution, only the sum $\sum_a \lambda_a$ is determined --- all orthogonal components of the vector $\lambda$ are free to take any values.
This does not imply that there are multiple such solutions, however.
Returning to the effective action in Eq.~\eqref{eq:higher_moment_SFF_action}, the fact that the saddle point equations determine only $G$, $F$, and $\sum_a \lambda_a$ means that, if we first integrate over them, the resulting $\lambda$-dependent action is of the form
\begin{equation} \label{eq:higher_moment_order_integrated_action}
S_{\textrm{eff}}[\mathcal{E}_{\textrm{aux}}, \lambda] = S \left[ \mathcal{E}_{\textrm{aux}}, \sum_a \lambda_a \right] - i \sum_{aa'} \sum_{b=2}^{|A|} \lambda_a u_{ab} u_{a'b} \mathcal{E}_{\textrm{aux}, a'},
\end{equation}
for some function $S$ of the single quantity $\sum_a \lambda_a$ (as well as all $\mathcal{E}_{\textrm{aux}}$) and for any choice of orthonormal basis vectors $u_{ab}$ orthogonal to the all-1 vector.
When we integrate over $\sum_a \lambda_a u_{ab}$, we thus get a $\delta$-function forcing $\sum_{a'} u_{a'b} \mathcal{E}_{\textrm{aux}, a'} = 0$.
Together, the $\delta$-functions force all $\mathcal{E}_{\textrm{aux}, a}$ to equal a common value $\mathcal{E}_{\textrm{aux}, A}$.
Not only is this consistent with our original assumption, it shows that our construction cannot work for any other values of $\mathcal{E}_{\textrm{aux}, a}$.

\subsection{Contribution of connected solutions} \label{subsec:higher_moment_connected_contribution}

To evaluate the action, note first of all that since the numbers $\beta_a$ define a continuous family of solutions, and since the action is by definition stationary at these solutions, all choices of $\beta_a$ must give the same value of the action.
We thus take all $\beta_a$ to equal a common value $\beta$ for simplicity.
The action evaluated at this solution still decomposes into a sum over groups, but now the contribution of a single group $A$ is
\begin{equation} \label{eq:higher_moment_action_evaluation_v1}
\begin{aligned}
S_{\textrm{eff}}[\mathcal{E}_{\textrm{aux}}] &= -i \mathcal{E}_{\textrm{aux}} \sum_{a \in A} \lambda_a + \frac{1}{2p} \left( \sum_{a \in A} \lambda_a \right)^2 + 2iJ q_{\textrm{EA}}^{p/2-1} \Lambda \sum_{a \in A} \lambda_a \\
&\qquad + \frac{T}{2} \sum_{aa'} \int_0^T dt \sum_{\alpha \alpha'} (-1)^{\alpha + \alpha'} \left( \frac{J^2}{p} G_{a \alpha, a \alpha'}(t)^p - F_{a \alpha, a \alpha'}(t) G_{a \alpha, a \alpha'}(t) \right) \\
&\qquad \qquad + \frac{1}{2} \log{\textrm{Det}} \Big[ i (-1)^{\alpha} \delta_{aa'} \delta_{\alpha \alpha'} \big( \mu \partial_t^2 + z \big) + (-1)^{\alpha + \alpha'} F_{a \alpha, a' \alpha'} \Big].
\end{aligned}
\end{equation}
Note that now $\mathcal{E}_{\textrm{aux}}$, $q_{\textrm{EA}}$, and $\Lambda$ are all independent of the replica $a$ (within a given group $A$).
We are also free to set all $\lambda_a = \lambda$, meaning that our saddle point solution simplifies to (in frequency space)
\begin{equation} \label{eq:higher_moment_simple_G_expression}
\widetilde{G}_{a \alpha, a' \alpha'}(\omega) = Tq_{\textrm{EA}} \delta_{\omega 0} + \delta_{aa'} \Big( \Delta \widetilde{\mathcal{G}}_{\alpha \alpha'}(\omega) + O(T^{-1}) \Big),
\end{equation}
\begin{equation} \label{eq:higher_moment_simple_F_expression}
\widetilde{F}_{a \alpha, a' \alpha'}(\omega) = \left( TJ^2 q_{\textrm{EA}}^{p-1} + \frac{ipJq_{\textrm{EA}}^{p/2-1} \big( \mathcal{E}_{\textrm{aux}} - 2Jq_{\textrm{EA}}^{p/2-1} \Lambda \big)}{|A|} \big( \delta_{\alpha u} + \delta_{\alpha' u} \big) \right) \delta_{\omega 0} + \delta_{aa'} \Delta \widetilde{\mathcal{F}}_{\alpha \alpha'}(\omega) + O(T^{-1}),
\end{equation}
\begin{equation} \label{eq:higher_moment_simple_lambda_expression}
\lambda = \frac{ip}{|A|} \big( \mathcal{E}_{\textrm{aux}} - 2Jq_{\textrm{EA}}^{p/2-1} \Lambda \big) + O(T^{-1}).
\end{equation}

Eq.~\eqref{eq:higher_moment_action_evaluation_v1} can be evaluated following the same procedure as in Sec.~\ref{subsec:nonergodic_contribution_connected_solutions}.
Directly substituting Eqs.~\eqref{eq:higher_moment_simple_G_expression} through~\eqref{eq:higher_moment_simple_lambda_expression} gives
\begin{equation} \label{eq:higher_moment_action_evaluation_v2}
S_{\textrm{eff}}[\mathcal{E}_{\textrm{aux}}] = \frac{p \mathcal{E}_{\textrm{aux}}^2}{2} - 2pJ^2 q_{\textrm{EA}}^{p-2} \Lambda^2 - \frac{Tq_{\textrm{EA}}}{2} \sum_{aa'} \sum_{\alpha \alpha'} (-1)^{\alpha + \alpha'} \widetilde{F}_{a \alpha, a' \alpha'}(0) - \frac{1}{2} \sum_{\omega} \log{\textrm{Det}} \widetilde{G}_{a \alpha, a' \alpha'}(\omega),
\end{equation}
and we again must determine certain components of $\widetilde{F}(0)$ and $\textrm{Det} \widetilde{G}(0)$.
As before, it is expedient to use the $(+,-)$ basis with respect to contour indices.
We also switch to the Fourier basis with respect to factor indices: from Eq.~\eqref{eq:higher_moment_simple_G_expression},
\begin{equation} \label{eq:higher_moment_G_Fourier_basis}
\begin{aligned}
\widetilde{G}_{b \alpha, b' \alpha'}(\omega) &\equiv \frac{1}{|A|} \sum_{aa'=1}^{|A|} e^{2\pi i(ab - a'b')/|A|} \widetilde{G}_{a \alpha, a' \alpha'}(\omega) \\
&= T|A|q_{\textrm{EA}} \delta_{b0} \delta_{b'0} \delta_{\omega 0} + \delta_{bb'} \Big( \Delta \widetilde{\mathcal{G}}_{\alpha \alpha'}(\omega) + O(T^{-1}) \Big).
\end{aligned}
\end{equation}
Thus $\textrm{Det} \widetilde{G}(\omega)$ factors with respect to $b$, and furthermore, $\sum_{\omega} \log{\textrm{Det} \widetilde{G}_b(\omega)} \sim 0$ for all $b \neq 0$ as in Secs.~\ref{subsec:ergodic_connected_action} and~\ref{subsec:nonergodic_contribution_connected_solutions}.
For $b = 0$, the determinant is calculated by comparing to the $b = 0$ block of $iz(-1)^{\alpha} + (-1)^{\alpha + \alpha'} \widetilde{F}(0)$, written in the $(+,-)$ basis (compare to Eq.~\eqref{eq:SFF_functions_matrix_inverse_x_basis}):
\begin{equation} \label{eq:higher_moment_matrix_inverse_x_basis}
\begin{pmatrix} \widetilde{F}_{0-, 0-}(0) & iz + \widetilde{F}_{0-, 0+}(0) \\ iz + \widetilde{F}_{0+, 0-}(0) & \widetilde{F}_{0+, 0+}(0) \end{pmatrix} = \frac{1}{\textrm{Det} \widetilde{G}_{0}(0)} \begin{pmatrix} \widetilde{G}_{0-, 0-}(0) & -\widetilde{G}_{0+, 0-}(0) \\ -\widetilde{G}_{0-, 0+}(0) & \widetilde{G}_{0+, 0+}(0) \end{pmatrix}.
\end{equation}
We see that $\textrm{Det} \widetilde{G}_0(0) = \widetilde{G}_{0+,0+}(0)/\widetilde{F}_{0+,0+}(0) \sim 1/J^2 q_{\textrm{EA}}^{p-2}$, and $\widetilde{F}_{0-,0-}(0)$ (which is in fact the only element of $\widetilde{F}(0)$ needed in Eq.~\eqref{eq:higher_moment_action_evaluation_v2}) is given by $\widetilde{G}_{0-,0-}(0)/\textrm{Det} \widetilde{G}_0(0) \sim (1 - 4J^2 q_{\textrm{EA}}^{p-2} \Lambda^2)/2T|A|q_{\textrm{EA}}$.
The action evaluates to
\begin{equation} \label{eq:higher_moment_action_evaluation_v3}
S_{\textrm{eff}}[\mathcal{E}_{\textrm{aux}}] = \frac{p-2}{2p} + \frac{1}{8p J^2 q_{\textrm{EA}}^{p-2} \Lambda^2} - \frac{2(p-1) J^2 q_{\textrm{EA}}^{p-2} \Lambda^2}{p} + \frac{1}{2} \log{4J^2 q_{\textrm{EA}}^{p-2} \Lambda^2},
\end{equation}
which is again precisely $-\Sigma(\mathcal{E}_{\textrm{aux}})$.

\subsection{Evaluation of the Higher SFF} \label{subsec:higher_moment_evaluation_SFF}

In the above calculation, note that we get a single contribution of complexity for the \textit{entire} group $A$.
However, there is still a factor $(2T)^{|A|} (1 + \delta_{p \textrm{ even}})^{2|A|-1}$ due to the separate time translation, time reversal, and reflection symmetries of each replica\footnote{The contribution $(1 + \delta_{p \textrm{ even}})^{2|A|-1}$, rather than $(1 + \delta_{p \textrm{ even}})^{2|A|}$, is because reflecting all spin configurations does not change the values of any overlaps.}.
Finally, the sum over all connected solutions amounts to a sum over the possible ways of partitioning $n$ elements, in addition to the $n!$ ways of pairing upper and lower contours.
Using $P \equiv \{A_1, \cdots, A_{|P|}\}$ to denote a partition, we have that
\begin{equation} \label{eq:higher_moment_final_expression}
\begin{aligned}
\textrm{SFF}^{(n)}(T, f) = n! \sum_P &\prod_{A \in P} 2^{|A|} \big( 1 + \delta_{p \textrm{ even}} \big)^{2|A| - 1} T^{|A|} \sqrt{\frac{pN}{2\pi}} \int d\mathcal{E}_{\textrm{aux}, A} e^{N \Sigma(\mathcal{E}_{\textrm{aux}, A})} \\
&\qquad \qquad \cdot \prod_{a \in A} \int_{\epsilon_-(\mathcal{E}_{\textrm{aux}, A})}^{\epsilon_+(\mathcal{E}_{\textrm{aux}, A})} \frac{\textrm{d}\epsilon_{\textrm{aux}, a}}{2\pi} f(\epsilon_{\textrm{aux}, a})^2.
\end{aligned}
\end{equation}
In particular, suppose the filter function is chosen so as to have a small width $\Delta \mathcal{E} \ll 1/N$ around a certain value $\mathcal{E}$ (as in Sec.~\ref{subsec:final_SFF_evaluation}, the above calculation can easily be modified to allow for $\mathcal{E}$-dependent filter functions).
Then the $n$'th moment simplifies to
\begin{equation} \label{eq:higher_moment_special_final_expression}
\textrm{SFF}^{(n)}(T, f) = n! \sum_P \left( \big( 1 + \delta_{p \textrm{ even}} \big)^{-1} \sqrt{\frac{pN}{2\pi}} e^{N \Sigma(\mathcal{E})} \Delta \mathcal{E} \right)^{|P|} \left( 2 \big( 1 + \delta_{p \textrm{ even}} \big)^2 T \int_{\epsilon_-(\mathcal{E})}^{\epsilon_+(\mathcal{E})} \frac{\textrm{d}\epsilon_{\textrm{aux}}}{2\pi} f(\epsilon_{\textrm{aux}})^2 \right)^n.
\end{equation}

Eq.~\eqref{eq:higher_moment_special_final_expression} has a nice interpretation as the $n$'th moment of a sum over a Poisson-distributed number of Gaussians.
To be precise, suppose we have an infinite sequence of i.i.d.\ complex Gaussians, $\{Z_i\}_{i=1}^{\infty}$, each with $\mathbb{E} Z_i = 0$ and $\mathbb{E} Z_i Z_i^* = \sigma^2$.
Consider the sum $S \equiv \sum_{i=1}^M Z_i$, where $M$ is itself a Poisson-distributed random variable with mean $\mu$.
The $n$'th moment of $SS^*$, averaging over both Gaussians and $M$, can be written
\begin{equation} \label{eq:Poisson_distributed_sum_moment_v1}
\mathbb{E} \big[ S^n S^{*n} \big] = \sum_{m=0}^{\infty} p_{\mu}(m) \mathbb{E} \left[ \left( \sum_{i=1}^m Z_i \right)^n \left( \sum_{i=1}^m Z_i^* \right)^n \right] = n! \sum_{m=0}^{\infty} p_{\mu}(m) \big( m \sigma^2 \big)^n
\end{equation}
where $p_{\mu}(m)$ denotes the Poisson distribution of mean $\mu$, and Wick's theorem is used for the latter equality.
It is known that the $n$'th moment of a Poisson distribution is $\sum_P \mu^{|P|}$, where the sum is again over all partitions of $n$ elements.
Thus
\begin{equation} \label{eq:Poisson_distributed_sum_moment_v2}
\mathbb{E} \big[ S^n S^{*n} \big] = n! \sum_P \mu^{|P|} \sigma^{2n}.
\end{equation}
If we associate $\sigma^2$ with the SFF of a single TAP state at $\mathcal{E}$,
\begin{equation} \label{eq:single_TAP_state_SFF}
\sigma^2 = 2 \big( 1 + \delta_{p \textrm{ even}} \big)^2 T \int_{\epsilon_-(\mathcal{E})}^{\epsilon_+(\mathcal{E})} \frac{\textrm{d}\epsilon_{\textrm{aux}}}{2\pi} f(\epsilon_{\textrm{aux}})^2,
\end{equation}
and associate $\mu$ with the number of TAP states,
\begin{equation} \label{eq:number_TAP_states}
\mu = \big( 1 + \delta_{p \textrm{ even}} \big)^{-1} \sqrt{\frac{pN}{2\pi}} e^{N \Sigma(\mathcal{E})} \Delta \mathcal{E},
\end{equation}
then Eqs.~\eqref{eq:higher_moment_special_final_expression} and~\eqref{eq:Poisson_distributed_sum_moment_v2} are identical.

It is quite tempting to interpret this as saying that the number of TAP states at $\mathcal{E}$ is Poisson-distributed with mean given by Eq.~\eqref{eq:number_TAP_states}, and that each TAP state has a Gaussian-distributed value of $\textrm{Tr} e^{-iHT}$ with variance (i.e., SFF) given by Eq.~\eqref{eq:single_TAP_state_SFF}.
We are not aware of any results in the literature which would contradict such a claim.
However, keep in mind that $\textrm{SFF}^{(n)}$ has perturbative corrections around each saddle point which are suppressed by powers of $N$, whereas every connected partition in Eq.~\eqref{eq:higher_moment_special_final_expression} is suppressed \textit{exponentially} relative to the fully disconnected one, whose contribution is given by Eq.~\eqref{eq:higher_moment_disconnected_part}\footnote{
At sufficiently low energies, where the function $\Sigma(\mathcal{E})$ is negative, the situation is reversed and the fully connected partition dominates.
The issue remains, however, that we do not calculate perturbative corrections around the dominant saddle point.
Furthermore, the relevance of these moment calculations to \textit{individual} realizations of the PSM is much more suspect when $\Sigma(\mathcal{E}) < 0$.
}.
Thus we cannot claim to have rigorously computed the $n$'th moment to any level of accuracy beyond the disconnected piece.
Nonetheless, the structure of saddle points which we have identified is highly suggestive and warrants further investigation.

\section{Concluding Remarks} \label{sec:conclusion}

The focus of this paper was the derivation of Eq.~\eqref{eq:SFF_final_result}, which gives the connected SFF of the quantum PSM in the non-ergodic phase.
Our result demonstrates that each metastable state (i.e., TAP state) can be considered as an independent chaotic phase with an independent RMT-like Hamiltonian, at least as far as level statistics are concerned.

It is interesting to compare our result for the PSM with Ref.~\cite{saad2019semiclassical}, which performs a similar calculation for the SYK model (the latter is structurally quite similar but with fermionic degrees of freedom).
In the ergodic phase, we find an essentially identical result using extremely similar methods.
Below the dynamical transition, however, the PSM displays an enhanced ramp quite different from that of the SYK model, which has no analogous phase.

Since we only calculate the SFF up to times polynomial in system size, our results are consistent with but do not test the distinction between localized and delocalized phases shown in Fig.~\ref{fig:spin_glass_cartoons}, which is only relevant beyond the exponentially long timescale corresponding to tunneling between TAP states.
We leave it for future work to incorporate such instanton effects into the path integral, expecting that they will reduce the SFF to the random matrix result precisely in the non-ergodic delocalized phase (and even then only beyond the exponential tunneling timescale).
At the same time, consideration of exponential scales can allow one to identify a plethora of additional dynamical phases (see Ref.~\cite{Zhao2014Three} for an example), and so the structure of instantons in these spin glass models may be quite rich.

In addition to the SFF, we have also considered higher moments of the evolution operator.
We identified an important family of saddle points and evaluated its contribution to these higher SFFs (Eq.~\eqref{eq:higher_moment_special_final_expression}).
The results suggest that: i) the number of TAP states at a given energy is Poisson-distributed, and ii) the numbers of TAP states at different energies are independent.
However, since we have not evaluated the perturbative corrections around each saddle point, which at finite complexity would generically dominate over any subleading saddle points, we cannot claim to have an accurate calculation.
It is another direction for future work to study the distribution of TAP states more systematically.

Our results can be further understood by comparing to Refs.~\cite{PhysRevE.102.060202} and~\cite{winerprx} on one hand and Refs.~\cite{2001} and~\cite{Facoetti_2019} on the other.
The first set of papers argues that for a system which separates into weakly coupled sectors, the SFF enhancement is the sum of return probabilities over all configurations.
If the time evolution can be considered as an effective Markov process with transfer rates between sectors given by some matrix $M$, then the SFF enhancement factor is $\textrm{Tr} e^{MT}$.
The second set of papers argues that for a \textit{classical} spin glass undergoing Markovian stochastic dynamics with generator $M$, the number of TAP states can be calculated --- and perhaps even defined --- as $\textrm{Tr} e^{MT}$.
In this sense, the present paper can be considered as a ``missing link'' that extends the results of Refs.~\cite{2001} and~\cite{Facoetti_2019} to quantum systems.

The fact that SFF enhancement is related to return probabilities suggests that the spectral statistics of spin glasses may contain information on aging dynamics as well.
Another open question is whether the \textit{equilibrium} replica-symmetry-breaking transition has any consequences for spectral statistics.
These, as well as those already mentioned, are all promising directions for future work.

Finally, let us briefly comment on the case $p=2$, which --- being a Gaussian model but for the global spherical constraint (and still integrable in any case~\cite{Cugliandolo_2018}) --- exhibits very different behavior than the $p > 2$ models considered here.
The spectral statistics of the analogous SYK model with two-point interactions among fermions have been studied in the mean-field limit \cite{winerprl,Liao_2020}, and found to have ramps growing exponentially in time.
One explanation for these ramps is the spontaneous breaking of a hidden $SU(2)^k$ symmetry in the saddle point equations, since for $p=2$ the matrix $G$ has a separate $SU(2)$ conjugation symmetry at each frequency.
Inspections show that the $p=2$ spherical model has the same symmetry at tree level, although it is broken by higher loop effects.
A full analysis of this system would be yet another excellent topic for further research.

\section*{Acknowledgements}

This work was supported by the following: the U.S. Department of Energy, Office of Science, Basic Energy Sciences under award number DE-SC0001911 (V.G.); the Joint Quantum Institute (M.W.); the Air Force Office of Scientific Research under award numbers FA9550-17-1-0180 (M.W.) and FA9550-19-1-0360 (B.S.); the U.S. Department of Energy, Office of Science, Office of Advanced Scientific Computing Research, Accelerated Research for Quantum Computing program ``FAR-QC'' (R.B.); the DoE ASCR Quantum Testbed Pathfinder program under award number DE-SC0019040 (C.L.B.); the DoE ASCR Accelerated Research in Quantum Computing program under award number DE-SC0020312 (C.L.B.); the DoE QSA, AFOSR, AFOSR MURI, NSF PFCQC program, NSF QLCI under award number OMA-2120757 (C.L.B.); DoE award number DE-SC0019449 (C.L.B.), ARO MURI, and DARPA SAVaNT ADVENT (C.L.B.).
 This material is based upon work supported by the National Science Foundation Graduate Research Fellowship Program under Grant No. DGE 1840340 (R.B.), and by the National Science Foundation NRC postdoctoral fellowship program (C.L.B.).

\renewcommand{\thechapter}{6}

\chapter{Discussion}
\label{chapter:disc}
In this thesis, we studied the connected spectral form factor in a variety of contexts, including systems with conserved or nearly-conserved quantities, hydrodynamic systems, systems with a wide variety of spontaneous symmetry breaking patterns, systems with oscillating sound modes, and glasses.

In chapter \ref{chapter:hydro}, we studied how slow dynamics can affect the SFF. We paid special interest to the case where the slow dynamics were hydrodynamic modes coming from a local conservation law or symmetry. In particular, we studied diffusive and subdiffusive hydrodynamics with and without interactions.

Chapter \ref{chapter:ssb} took a look at what happens when the symmetries are spontaneously broken. We find that, counterintuitively, spontaneously broken symmetries actually enchance the SFF even more than unbroken ones.

Chapter \ref{chapter:soundPole} took the next step, analyzing the spectral form factors of systems with sound poles, a more realistic model of hydrodynamics than the diffusive cases studied in chapter \ref{chapter:hydro}.

Finally, chapter \ref{chapter:glass} is a case study on the spectral statistics of a glassy system. This is a very different sort of slow dynamics than the kind studied in earlier chapters. Nonetheless, we find the same agreement between long-time dynamics and long-time spectral statistics.

Most of this thesis was, either directly or indirectly, a study of the total return probability or TRP (the sole exception being chapter \ref{chapter:ssb}, where the assumptions relating TRP to the spectral form factor break down somewhat). 

While the relation between the TRP and the spectral statistics is interesting and useful, it is important to remember that each of them is worthy of study in their own right, even in contexts where they are not closely related.

While this paper focuses almost entirely on the spectral form factor, there are other quantities in spectral statistics. For instance, the level-spacing ratio studied in \cite{PhysRevB.75.155111,Atas_2013,KARAMPAGIA2022122453} characterizes the ratio between two consecutive gaps in the energy spectrum. This quantity is well-understood in the context of random matrix ensembles, but how physical processes like hydrodynamic behavior affect this ratio is still a mystery.

On the other hand, there are generalizations of the SFF that study time contours with more than two closed loops. If the SFF is the variance of $Z(iT)$, one can ask about higher order statistics, and in particular higher-order cumulants. Going back to the diagonal approximation discussed in section \ref{sec:specQuantumChaos}, we note that the SFF is the subject of a remarkable coincidence: the number of closed paths grows exponentially as $e^{\lambda T}$ while the amplitudes shrink as $e^{-\lambda T/2}$, preventing any exponential behavior in the SFF, which is proportional to the number of paths times the amplitude squared. Higher-order moments won't have this cancellation, and thus might provide a route to measuring the Lyapunov exponent directly from spectra statistics. This would be an important step in characterizing the many notions of chaos discussed in sections \ref{sec:classChaos},\ref{sec:quanChaos}, such as thermalization versus spectral statistics versus Lyapunov growth.

The TRP is not an inherently quantum quantity, and related notions have appeared in the study of dynamical systems for decades \cite{artinMazur,Ruelle2002DynamicalZF}. One paper that connects this dynamical quantity to classical glasses is \cite{Facoetti_2019}. However, there are still huge areas to be explored with this paradigm. We will discuss three of them here.

One of the most important open problems in computer science is the study of descent in a complex landscape \cite{pml1Book,kingma2017adam}. For many of these landscapes, using analytic techniques from physics is completely intractable, but there are a number of situations where the tools of dynamical mean field theory \cite{Zou_2024,Mignacco_2021} allow us to track systems across high-dimensional landscapes. The TRP is an excellent diagnostic of our system's exploration of the loss landscape. It can compare different optimizers (stochastic gradient descent versus momentum \cite{kingma2017adam,hao2021adaptive} versus learned optimizers \cite{harrison2022closer}).

At the same time, there still countless questions in physics where the approach to equilibrium is confusing and the TRP may represent a potential resolution. One particular class of problems is the study of extended objects such as strings \cite{Kibble_1976,hirth1992theory}, planar defects \cite{mermin1979topological}, or phase boundaries. The TRPs of such situations could be studied by taking a probabilistic model of their dynamics, such as a dynamical mean string field theory \cite{Iqbal_2022} or the KPZ equation \cite{KPZ,corwin2011kardarparisizhang,Quastel_2015} and putting them on periodic time.

A final extension of the total return probability is the extension to periodically driven systems. The TRP can count the number of non-equilibrium steady states. One can imagine systems analogous to glasses, with a large number of attractive basins in phase space and an exponentially large TRP even after many cycles.
\titleformat{\chapter}{\normalfont\large}{Appendix \thechapter:}{1em}{}
\renewcommand{\chaptername}{Appendix}

\appendix
\renewcommand{\thechapter}{A1}

\chapter{Appendices for Chapter 2}
\label{chapter:app1}
\section{Appendix: Review of the SSS Wormhole Solution}
\label{app:SSS}
In \cite{Stanford2018} the authors (SSS) evaluate the ramp contribution exactly for the SYK model. They start by noting that the SFF is a partition function on two contours. In particular, if we denote the collective fields of a single copy of the system by $\Psi$, the SFF is given by a path integral on two copies of the system with periodic time coordinates,
\begin{equation}
    \sff(T,f(E)=1)=\int \mathcal D\Psi_1D\Psi_2 \exp\left (i \int d^d x\int_0^T dt \{ L[\Psi_1]-L[\Psi_2]\}\right)
\end{equation}
The essential insight in their paper is that for systems with many degrees of freedom, like the SYK model at large $N$, this path integral can be evaluated by saddle-point methods and that a non-trivial family of saddle points give the ramp. These are thermofield double solutions at inverse temperature $\beta_{\text{aux}}$, suitably adjusted using images to account for the different boundary conditions, that correlate the two contours. At large $T$ such solutions always approximately solve the SFF two contour equations of motion because they solve the equations of motion on a $e^{-\beta_{\text{aux}}H}e^{iHT}e^{-iHT}$ contour (where the time evolutions trivially cancel) and the contours are identical in the `bulk' of the forward and backward legs. Implicitly, we are appealing to the forgetfulness of chaotic systems, which here means that the solutions are exponentially insensitive to the boundary conditions.


\section{Appendix: CTP Formulation of Hydrodynamics}
\label{app:CTP}

The Closed Time Path (CTP) formalism \cite{glorioso2018lectures,crossley2017effective} is an effective theory of hydrodynamics on the Schwinger-Keldysh contour. In \cite{Chen_Lin_2019} a simplified version describing just energy diffusion is used to derive long-time-tails for two-point functions in hydrodynamics. We will largely follow their conventions. 
One starts with a partition function
\begin{equation}
    Z_{CTP}[A_1^\mu(t,x),A_2^\mu(t,x)]=\frac {1}{\tr e^{-\beta H}} \tr \mathcal P e^{-\beta H}\exp(-i H T+\int dx dt A_1^\mu J_{1\mu})\exp(i H T-\int dx dt A_2^\mu J_{2\mu}),
\end{equation}
where $A$ can be thought of as an external gauge field coupling to the conserved currents.

We express $Z_{CTP}$ as $e^{I[A_1,A_2]}$, where $I$ is a nonlocal action. The main assumption is that after `integrating in' slow modes the action will become local. In a standard hydrodynamic system, the only slow modes correspond to conservation laws and these modes are brought in to enforce those laws. We have
\begin{equation}
    \exp(I[A^\mu_1,A^\mu_2])=\int \mathcal D \phi_1\mathcal D \phi_2\exp(i\int dtdx L[B^\mu_1=A^\mu_1+\partial^\mu \phi_1,B^\mu_2=A^\mu_2+\partial^\mu \phi_2)
    \label{eq:CTPAction}
\end{equation}
Where $L$ is a (generally complex) action functional. 

As equation \eqref{eq:CTPAction} makes manifest, the action doesn't depend on the $\phi$s except through the modified gauge fields $A_\mu+\partial_\mu \phi$. If we change variables to
\begin{equation}
\begin{split}
    \phi_r=\frac{\phi_1+\phi_2}{2}\\
    \phi_a=\phi_1-\phi_2
\end{split}
\end{equation}
we can derive additional identities. For instance unitary implies that 
\begin{equation}\left\langle \prod_i \phi_i(t_i)\phi_a(0)\right\rangle =0\end{equation}
will always be zero whenever $\forall_i t_i<0$. In other words, if the chronologically latest insertion is $a$-type, the expectation value is zero. This theorem is called the last time theorem, and is explored in detail in \cite{gao2018ghostbusters}. This, in turn, can be used to derive the fact that all terms in $L$ have at least one factor of an $a$ field. It is also worth explicitly noting the implications of the last time theorem for two point functions. We also always have
\begin{equation}
\begin{split}
    G_{aa}(t)=0,\\
    G_{ar}(t)=\theta_+(t) f(t).
\end{split}
\end{equation}
Additional constraints on CTP Lagrangians derived in other works include the fact all factors of the $r$ variables come with at least one time derivative and the KMS condition that $S[\phi_1(t),\phi_2(t)]=S[\phi_2(i\beta-t),\phi_1(-t)]$.

It is also worth noting that we can formulate in terms of slightly different variables, replacing $\partial_t \phi_r$ with the energy density $\epsilon$. For the theory in \cite{Chen_Lin_2019} used in the main text, at the Gaussian level $\epsilon$ is equal to $c \beta^{-1} \partial_t\phi_r$ plus higher derivative corrections. The general formula for $\epsilon$ is obtained by differentiating the action with respect to $A^0_a$, since it couples to $\epsilon$. For one conserved quantity at leading order in derivatives, the most general quadratic Lagrangian consistent with the requirements of derivatives and KMS symmetry is
\begin{equation}
  L=-\phi_a\left(\partial_t\epsilon-D\nabla^2\epsilon\right)+i\beta^{-2}\kappa(\nabla \phi_a)^2.
\end{equation}

To make the physics more transparent, it is useful to introduce an auxiliary variable $F(x,t)$. We can rewrite our action as
\begin{equation}
    L=-\phi_a\left(\partial_t\epsilon-D\nabla^2\epsilon - F\right)+i\frac{1}{4\beta^{-2}\kappa} F \nabla^{-2} F.
\end{equation}
The action is now linear in $\phi_a$, meaning that $\phi_a$ serves as a Lagrange multiplier enforcing the stochastic partial differential equation
\begin{equation}
    \partial_t\epsilon-D\partial_x^2\epsilon+F=0,
\end{equation}
where $F$ is now interpreted as a fluctuating force. This then leads to the probability distribution of energy modes discussed in the main text.

\section{Appendix: Direct Evaluation of Hydro Integral}
\label{app:direct}

The spectral form factor of any system is given by a path integral over two copies of the system:
\begin{equation}
\sff=\overline{\int \mathcal D \psi_1\mathcal D \psi_2 \exp(iS[\psi_1]-iS[\psi_2])},
\end{equation}
where the overline denotes some kind of disorder average, for example, over some random couplings or smearing over a time window. The path integral has periodic boundary conditions with period $T$ and we use $\psi_i$ as a stand-in for all the fundamental fields on contour $i$. When considering the thermal SFF, the individual contours involved can include thermal segments, or more generally, some portion that implements a filter function.

In the spirit of hydrodynamics, and in the context of the simple theory of diffusion in \eqref{eq:CTPLag}, we will replace these fields with a field $\epsilon$ representing energy density and field $\phi_a$ enforcing the hydrodynamic equations. The price for this is introducing cross-terms between the contours. The integral is now
\begin{equation}
\begin{split}
\sff=\int \mathcal D \epsilon\mathcal D \phi_a \exp(iS_{\text{hydro}}),\\
S_{\text{hydro}}=\int dVdt\bigg(-\phi_a(\partial_t-D\Delta)\epsilon+iD\kappa\phi_a \Delta\phi_a\bigg).
\end{split}
\label{eq:hydroSFFIntegral}
\end{equation}
If we break the time integration into $ T/\Delta t$ segments, then the measure is 
\begin{equation}
  \mathcal{D} \epsilon \mathcal D \phi_a = \prod_{x} \prod_{\ell=0}^{T/\Delta t-1} \frac{d \epsilon(x, t = \ell \Delta t) d\phi_a(x,t = \ell \Delta t)}{2\pi}.
\end{equation}
The $2\pi$s are to enforce proper normalization of delta functions imposing the hydro equations. 

$S_{\text{hydro}}$ is a translation-invariant Gaussian function, so we can break path integral \eqref{eq:hydroSFFIntegral} into a product over spatial modes $k$ with $D\Delta=\lambda_k$, and then over temporal frequencies. These dimensionless frequencies, the eigenvalues of the $ dt \partial_t$ matrix, are the $T/\Delta t$ complex numbers $i\omega$ obeying $(i\omega+1)^{T/\Delta t}=1$. Going to the basis of these modes, we have
\begin{equation}
\begin{split}
\sff=\prod_k \sff_k , \\
\sff_k=\prod_\omega \frac{1}{i\omega-\lambda_k \Delta t}, 
\end{split}
\label{eq:hydroProducts}
\end{equation}
Note that $\sff_k$ is a product over roots of unity of the form $\prod \left[ i\omega +1 - (1+ \lambda_k \Delta t)\right]$. For odd $T/\Delta t$, the result is
\begin{equation}
\sff_k=\left(1-(1+\lambda_k \Delta t)^{T/\Delta t}\right)^{-1} =_{\Delta t \rightarrow 0} \left(1-e^{\lambda_k T}\right)^{-1}.
\end{equation}
After multiplying together the contribution from different momentum modes, we precisely recover the return probability formula derived in the main text. 
The one mode which can't be evaluated this way is the hydrodynamic zero mode corresponding to the total energy, since it has a vanishing action. Instead, one is forced to do the full integral $\int \frac{dE d\phi_a}{2\pi}$ with $\int d\phi_a = T$. For systems with time-reversal symmetry there is an additional solution where contour 2 is reversed, so the $2\pi$ in the denominator becomes $\pi$. Thus, the quadratic hydro theory with periodic temporal boundary conditions correctly recovers \eqref{eq:bigAnswer}.

\renewcommand{\thechapter}{A2}

\chapter{Appendices for Chapter 3}
\label{chapter:app2}
\section{Appendix: SFF For Unbroken Non-Abelian Symmetries}
\label{app:nonAb}
For systems with non-Abelian symmetry group $G$, the Hilbert space can be decomposed into a sum of irreducible representations $R$ of the Hamiltonian with dimension $|R|$. These are the non-Abelian analogs of the charge sectors in systems with $U(1)$ symmetries. Just as no term in an $U(1)$-invariant Hamiltonian can mix states with different charges, no term in a $G$-invariant Hamiltonian can mix states in different irreducible representation. Written in the $|R|$ subspace, the Hamiltonian should decompose into a tensor product of a chaotic random matrix and an $|R|$ by $|R|$ identity. In equations we have 
\begin{equation}
    H=\bigoplus_R I_{|R|}\otimes H_R,
\end{equation}
where each $H_R$ is an independent matrix with independent eigenvalues. If we write $Z_R(T,f)=\tr e^{-iH_RT}f(H_R)$, then $Z(T,f)=\sum_R |R| Z_R(T,f)$. Squaring this, and setting the cross-terms to 0, we recover equation \ref{eq:nonabelianBase}. For systems with spontaneous symmetry breaking, the $Z_R$s are no longer independent and it no longer makes sense to set the cross-terms to $0$. Resolving this complication is the key goal of this paper.
\section{Appendix: Internal Charge}
\label{app:IntCharge}

For many realistic systems, there is charge/momentum contained within the state $\psi$,  not just encoded in the pattern of $\phi$s.  A nice example of this is a superconductor. There is some charge in the condensate,  but plenty of other charge in the system,  including in uncondensed electrons and atomic nuclei. Likewise,  if we look at a mechanical example of rotational SSB like a planet with a cloud of moons orbiting a star, there is some orbital angular momentum of the overall system,  but the system of planets and moons has its own intrinsic angular momentum.  What sort of Hamiltonian captures this situation?

Let's say the states are characterized by an order parameter $\phi$, an internal angular momentum $L$, and an internal state $\ket \psi$. 
The simplest $G$-invariant Hamiltonian we can write for this is given by equation \eqref{eq:HRandom} for with internal-$G$-invariant $H_0$ and $H_i$s. This candidate $H$ actually has two copies of the symmetry, an internal one and an external one. 

To make the model more realistic, we need to have some coupling which breaks us down to just one copy of the symmetry. First let's do it for $Z_n$. We can parameterize the sectors by order parameter $0\leq \phi <n$ and $q$. Then we have a coupling connecting the $\phi,q$ and $\phi,q'$ sectors proportional to $e^{2\pi i /n \phi(q-q')}$.  This obviously doesn't commute with bland translation or bland conservation of $q$.  But it does retain an overall $Z_n$ symmetry.
Written in block matrix form, such a Hamiltonian for a $Z_4$ symmetry might take the form
\begin{equation}
H=\begin{pmatrix}
H_0^0&I&0&I\\
I&H_0^1&I&0\\
0&I&H_0^2&I\\
I&0&I&H_0^3
\end{pmatrix}
\end{equation}
Where $H_0^\phi$ is, itself, a block matrix of the form
\begin{equation}
H_0^\phi=\begin{pmatrix}
Q_{00}&e^{\frac{2\pi i}4\phi(0-1)}Q_{01}&e^{\frac{2\pi i}4\phi(0-2)}Q_{02}&e^{\frac{2\pi i}4\phi(0-3)}Q_{03}\\
e^{\frac{2\pi i}4\phi(1-0)}Q_{10}&Q_{11}&e^{\frac{2\pi i}4\phi(1-2)}Q_{12}&e^{\frac{2\pi i}4\phi(1-3)}Q_{13}\\
e^{\frac{2\pi i}4\phi(2-0)}Q_{20}&e^{\frac{2\pi i}4\phi(2-1)}Q_{21}&Q_{22}&e^{\frac{2\pi i}4\phi(2-3)}Q_{23}\\
e^{\frac{2\pi i}4\phi(3-0)}Q_{30}&e^{\frac{2\pi i}4\phi(3-1)}Q_{31}&e^{\frac{2\pi i}4\phi(3-2)}Q_{32}&Q_{33}
\end{pmatrix}
\end{equation}
While the $Q_{ii}$ matrices all have to be square, Hermitian matrices and $Q_{ij}=Q_{ji}^\dagger$, in general the internal-charge-i subspace and the internal-charge-j subspace can be of totally difference sizes.

If we have a more general group,  the elements are parameterized by order parameter $\phi$,  irrep $R$ and vector $k$ within that irrep.  We connect $\phi,R$ and $\phi,R'$ by $R_{G(\phi)}MR'_{G^{-1}(\phi)}$ where $G(\phi)$ is a group element that gets us to $\phi$, and $M$ is invariant under the unbroken part of the symmetry.

\section{Appendix: Long-Time Behavior with Non-Abelian Discrete Groups}
\label{app:longTime}

The enhancement factor for non-Abelian discrete symmetry groups is $\tr[e^{-\text{Trans}(E)T}]$ with transfer matrix
\begin{equation}
\begin{split}
 \text{Trans} =\frac 12 \sum r_i (M_i\otimes I-I\otimes M_i^T)(M_i^T\otimes I-I\otimes M_i)+\\
 (M_i^T\otimes I-I\otimes M_i)(M_i\otimes I-I\otimes M_i^T).
\end{split}
\end{equation}
In this section, we count the zero modes of the transfer matrix. Such a zero mode must be annihilated by $M_i \otimes I-I\otimes M_i^T$ for all $M$s, which is a heavy constraint. 

For our analysis, we will need to decompose a vector space $C^{\Phi}$ into irreps. This space is a representation of $G$, with the $G$ matrices forming permutation matrices which permute the elements of $\Phi$ according to the group action. The matrices $M_i$ are matrices acting on this space which commute with every element of $G$. We can decompose the representation $C^\Phi$ into irreducible representations of $R$ as
\begin{equation}
    C^{\Phi}=\bigoplus_R R^{\oplus K_R},
\end{equation}
where each representation $R$ appears $K_R$ times. A vector $v$ in $C^\Phi$ can be written as $v_{R,k,\mu}$, where $R$ denotes the irreducible representation it transforms in, $0\leq k<K_R$ indicates which copy of $R$, and $\mu$ is the index within the representation $R$. In this case, the requirement that the $M_i$s commute with elements of $G$ mean that they can act only on the $k$ index, in a way not depending on the $\mu$ index. If we sum over a large enough collection of $M$s (in particular, enough $M$s so that their action on the $k$ indices don't all commute) then the only vectors on $C^{\Phi\otimes \Phi}$ which are annihilated by the transfer matrix are ones in which the left and right $k$ indices are maximally entangled. We are free to choose the representation $R$ and the indices $\mu_1$ and $\mu_2$ for the right and left replicas. So we have
$\sum_R |R|^2$ zero modes. This is also the random matrix theory prediction for the long-time ramp enhancement.

\section{Appendix: Deriving Hydro from the Discrete Case}
\label{app:HydroDerive}

Let's talk about how a term like equation \eqref{eq:transfer2} gives rise to a Goldstone-like theory.  It can be thought of as decoupled diffusion on both the right and left copies of the system,  with positive and negative imaginary diffusivities.  Of course, this is just QFT with a canonical kinetic term
\begin{equation}
L=\frac 12\phi_1\partial_t^2\phi_1-\frac 12\phi_2\partial_t^2\phi_2=\phi_a\partial_t^2\phi_r.
\label{eq:Lra}
\end{equation}
This is already similar to the Goldstone theory, but it needs an $aa$ term. This comes from the contribution of the matrix in equation \eqref{eq:transferEq}. This matrix can be thought of as generating an un-intuitive sort of random walk on $\Phi^2$.  The most un-intuitive part is that the transfer probabilities aren't all positive. We can see from the formula that there is a morass of positive and negative signs. In order to build intuition, let's consider the case of $Z_n$ symmetry acting on $n$ elements. We know that the matrices $M_i$ are indexed by $\Phi^2/G$, which in this case is just the collection of jump sizes,  ranging from the trivial jump to jumping $n-1$ to the right. For simplicity, let's just look at $(M_1)_{ij}=\delta_{i,j+1}$,  the nearest-neighbor jump. 

What sort of transfer matrix does this give rise to? Remembering that the transfer matrix is a linear map from the vector space $R^{\Phi^2}$ to itself and thus has a total of 4 $\Phi$ indices,  it is unilluminating to try to write the whole thing.  But we can write that it is 
\begin{equation}T_{ij,i'j'}\propto 2\delta_{i,i'}\delta_{j,j'}-\delta_{i,i'+1}\delta_{j,j'+1}-\delta_{i,i'-1}\delta_{j,j'-1}
\end{equation}
If we go from a discrete $Z_n$ to a continuous $U(1)$,  this transfer matrix corresponds to $\phi_r$ undergoing diffusion while $\phi_a$ doesn't change at all.  When we combine this with the transfer matrix in equation \eqref{eq:transfer2} we add the covariances/correlators. Equation \eqref{eq:Lra} only has $ar$ correlators. Adding in an $rr$ correlator results in a Lagrangian 
\begin{equation}
L=\phi_a\partial_t^2\phi_r+iC\phi_a\partial_t^2\phi_a.
\end{equation}
This is a generic hydrodynamic action for a superfluid, obtained entirely through taking the continuous limit of the transfer matrices.

\section{Appendix: Time-Reversal Symmetry}
\label{app:TR}

Let's consider spontaneous time-reversal symmetry breaking. First, what is a good RMT-like toy model of the phenomenon? Consider a block Hamiltonian of the form 
\begin{equation}
    H=\begin{pmatrix}H_0&H_1\\H_1&H_0^*\end{pmatrix},
    \label{eq:trmodel}
\end{equation} 
where $H_0$ is an $N\times N$ GUE matrix and whose real and imaginary parts have variance $J^2/N$, and $H_1$ is a randomly selected GOE matrix whose elements are all independent and have variance $J_1^2/N$. This matrix has an anti-unitary time reversal symmetry which conjugates the elements and switches the two blocks (which we will label the $+$ and $-$ blocks). But within each block, there is no time-reversal symmetry. Since the system will choose a block and only slowly tunnel back and forth, we say that the system spontaneously breaks the time-reversal symmetry.

We can study this using a similar instanton-like approach as with $Z_2$ SSB. We start with a doubled system.  Since one copy of the system has two sectors, a pair of copies has $2\times 2=4$ sectors.  If the system is in the $++$ or $--$ sectors, it has exactly the same transfer matrix as a GUE matrix with two sectors.  The processes contributing are the $++\to--/--\to++$ process with amplitude given by Fermi's golden rule, and the negative-amplitude $++\to++/--\to--$ processes. This gives an enhancement of $\tr e^{MT}=1+e^{-r T}$, (where $r$ is the transition rate) over the GUE result. However, we can also time-reverse the left contour with respect to the right.  Now we have a system starting in either the $+-$ or $-+$ states,  and the left replica is performing a time-reversed version of the right-replica's evolution. This gives another $1+e^{-rT}$ contribution. The overall enhancement factor is thus $2(1+e^{-rT})$ times the GUE ramp, which is of course $1+e^{-rT}$ times the GOE ramp. 

One thing which bears discussion and which is not fully understood is what happens when the Thouless time becomes comparable to the Heisenberg time for individual subsystems, in large part because we don't have a path integral-like picture for how the ramp gives way to the plateau. For GOE-like systems, plateau-like behavior sets in gradually even before the Heisenberg time. One guess then to assume that the ramp behavior for time reversal SSB is thus $(1+e^{-rT})\text{SFF}_{\text{GOE}}(T)$ even when $t_{\text{Thouless}}\lesssim t_{\text{Heis}}$.
\begin{figure}
\centering
    \includegraphics[width=.5\textwidth]{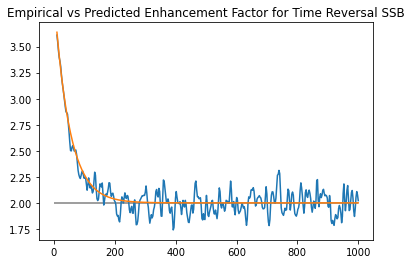}
    \caption{Predicted (orange) and observed (blue) enhancement relative to GOE expectation for the model in equation \eqref{eq:trmodel}.}
    \label{fig:timeBreak}
\end{figure}
\renewcommand{\thechapter}{A3}

\chapter{Appendices for Chapter 4}
\label{chapter:app3}
\section{More On The Total Return Probability}
\label{app:TRP}
\begin{figure}
    \centering
    \includegraphics{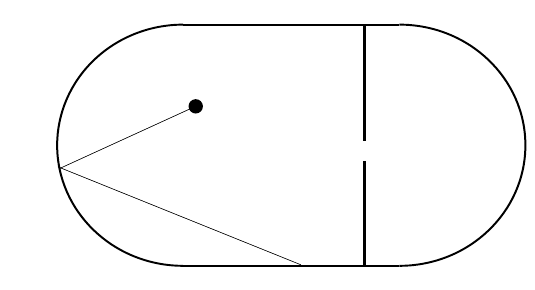}
    \caption{A simple example of chaotic dynamics with two slowly thermalizing sectors. The ball will quickly spread out within the Hilbert Space of a given chamber, but only much more slowly will it transfer between the two sectors. During this time the connected SFF will be larger than the RMT prediction.}
    \label{fig:bunSectors}
\end{figure}
\begin{figure}
    \centering
    \includegraphics[scale=0.5]{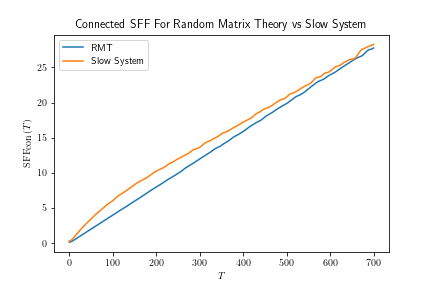}
    \includegraphics[scale=0.5]{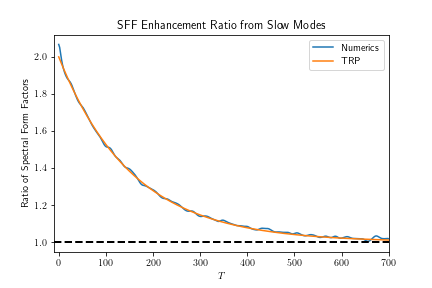}
    \caption{(Left) the RMT vs nearly block diagonal connected SFFs during the ramp regime. (Right) A comparison of the numerical ratio (blue) versus the TRP in equation \ref{eq:TwoBlockTRP} (orange).}
    \label{fig:twoSFFs}
\end{figure}

The Total Return Probability (TRP) is a measure of how much a system has thermalized after time $T$. Remarkably, it has also been shown \cite{winerprx} to be the enhancement factor for the connected SFF at time $T$. One simple example of a system with slow thermalization is a single quantum particle bouncing around a partitioned Bunimovich stadium in figure \ref{fig:bunSectors} (not to be confused with the much more complicated system discussed in section \ref{sec:freeStadium}, a many-body system filling a Bunimovich stadium). The two chambers of this stadium are our two sectors. This particle will bounce around and quickly (after the Ehrenfest time) be spread out within a given sector. At this point $p_{1 \to 1}$ and $p_{2 \to 2}$ will both be 1 and the TRP will be 2. As more time passes, both $p_{1 \to 1}$ and $p_{2 \to 2}$ will decrease towards the equilibrium probabilities of being in chamber 1 or 2 respectively. Eventually the TRP will be one.
More RMT-focused readers might prefer the example where $H$ is a matrix with
\begin{equation}
    H=\begin{pmatrix}
    H_1&H_{\textrm{small}}\\
    H_{\textrm{small}}^\dagger&H_2
    \end{pmatrix}
\end{equation}
where some weak coupling $H_{\textrm{small}}$ induces jumping between two RMT-like sectors. In this case the TRP can be calculated using Fermi's golden rule. We get
\begin{equation}
    \trp(T)=1+e^{-\lambda T}
    \label{eq:TwoBlockTRP}
\end{equation}
for a specific value of $\lambda$ characterizing the strength of the off-block-diagonal matrix elements. An example plot is seen in figure \ref{fig:twoSFFs}.

A more complicated case would be a single particle diffusing through a complicated medium. In this case the sectors would be regions of space and $p_{i \to i}$ would be calculated by solving the diffusion equation. This procedure would agree with previous results obtained through sigma model methods\cite{Kravtsov1994LevelSI,andreev1995spectral}. An even more complicated case would be a many-body system with a conserved charge, where the sectors would be coarse-grained charge distributions $\rho_i(x)$. $p_{i \to i}$ is calculated using fluctating hydrodynamics as discussed in subsection \ref{subsec:DPT} and in more detail in \cite{winerprx}. The results agree with a large-$q$ Floquet calculation done in \cite{Friedman_2019}.
\section{SYK2 At Finite Temperature}
\label{app:SYK2}

In \cite{PhysRevLett.125.250601,winerprl} the authors investigated the many-body SFF in a specific system of free-fermions with random two-body interactions. By specifically studying the many-body SFF of these
They find an exponential ramp in the infinite-temperature SFF, in particular with a ramp of the form $SFF\propto N^{\#T}$, where the number depended on the specific model under consideration. This exponential ramp behavior can be argued to come from the same cumulant expansion as equation \eqref{eq:varTerm}.

In the case of the SYK2 model with no charge/number conservation, we have
\begin{equation}
    H=\frac 12 \psi_iM^{ij}\psi_j,
\end{equation}
where $M$ is an anti-symmetric real matrix. One can show that the eigenvalues of $M$ follow a semicircle law, and obey GUE level repulsion (the reason for this is that even though the matrix is real, the eigenvalues come in pairs $\pm \lambda$, and there is repulsion for both the positive and negative eigenvalues). We can write the full-system SFF at inverse temperature $\beta$ as
\begin{equation}
    \log SFF=\sum_{\lambda>0}2\textrm{Re}\log(1+\exp(-\beta \lambda+iT \lambda))=\sum_{j\neq 0}\frac{(-1)^j}j\sum_{\lambda>0}\exp(-\beta j \lambda+ijT \lambda))=\sum_{j\neq 0}\frac{(-1)^j}{j}Z(jT,j\beta)  
\end{equation}
Note the similarity between this form and equation \eqref{eq:stadium}. And while at infinite temperature all terms of the cumulant expansion are needed, we will show that at finite temperature only the first two are important. 

For most random matrix ensembles, we can write the probability density as
\begin{equation}
    dP(M)=\exp(-N\tr V(M))dM.
\end{equation}
In terms of the eigenvalues density $\rho(\lambda)=\frac 1N\sum_i \delta(\lambda-\lambda_1)$ and the eigenvalues distribution density as 
\begin{equation}
    dP(\rho)\propto \exp(-\int N^2 V(\lambda)\rho(\lambda) d\lambda+\int N^2\rho(\lambda_1)\rho(\lambda_2)\log|\lambda_1-\lambda_2|d\lambda_1d\lambda_2+\int N\rho(\lambda) \log \rho(\lambda)d\lambda)\mathcal D \rho
    \label{eq:measure}
\end{equation}
At large $N$, this is the quadratic functional
\begin{equation}
    dP(\rho)=\exp(-\int N^2 V(\lambda)\rho(\lambda) d\lambda+\int N^2 \rho(\lambda_1)\rho(\lambda_2)\log|\lambda_1-\lambda_2|d\lambda_1d\lambda_2)\mathcal D \rho.
    \label{eq:measureQuad}
\end{equation}
This means that we can analyze quantities related to $\rho$ by linear transformations (including $Z(T,f)$, or the resolvent $R$) using only the first two cumulants. This breaks down only when $\rho$ gets extremely close to zero. Let us start with $\rho=\bar \rho$, the saddle point, and see what temperatures cause trouble.

For instance, the SFF of the SYK2 model can be written
\begin{equation}
    SFF=\int dP(\rho) \exp(N\int \rho \log(1+\exp(iT\lambda-\beta \lambda))d\lambda+cc)
\end{equation}
We can make use of
\begin{equation}
    \log(1+e^{ix})=\sum_n \log \frac{x-(2n+1)\pi}{(2n+1)\pi}
\end{equation}
This means we can write the SFF as 
\begin{equation}
    SFF=\int dP(\rho) \exp(N\int \rho \sum_n \log \frac{x-(2n+1)\pi/(T+i\beta)}{(2n+1)\pi/(T+i\beta)}+\log \frac{x-(2n+1)\pi/(T-i\beta)}{(2n+1)\pi/(T-i\beta)}d\lambda)
\end{equation}
Combining with the quadratic large $N$ measure in equation \eqref{eq:measureQuad} we can solve for the new saddle-point value of $\rho$. We have
\begin{equation}
\begin{split}
    \rho=\rho_0+\delta \rho\\
    \delta \rho=\frac 1 N \sum_n (2n+1)\frac {\beta}{T^2+\beta^2} \frac{1}{(x-\frac{(2n+1)\pi T}{T^2+\beta^2})^2+(2n+1)^2\pi^2\frac {\beta^2}{(T^2+\beta^2)^2}}
\end{split}
\end{equation}
In the limit of small $\beta$, this is just a train of delta functions of mass $\frac 1 N$. The deepest well will be of depth $\frac {T^2+\beta^2}{N\pi^2 \beta}$. If this becomes $O(1)$, comparable to $\rho_0$, the cumulant expansion breaks down. For any fixed nonzero $\beta$ this isn't a concern, except at very long times. But if $\beta \sim N^{-1}$, we get the more complicated $N^{\#T}$ behavior.

In the regime where the cumulant expansion works, and where $T\gg \beta $, we can evaluate the first two terms $c_1,c_2$ in the cumulant expansion. The first is non-universal and dependent on the density of states:
\begin{equation}
    c_1=\textrm E\sum_{j\neq 0}\frac {(-1)^j}j Z(ijT-j\beta)=\sum_j{(-1)^j}j\int d\lambda \rho(\lambda) \exp(ijT\lambda-j\beta\lambda)
\end{equation}
In the case of the semicircle law, this becomes a sum of Bessel functions.

\begin{figure}
    \centering
    \includegraphics[scale=0.4]{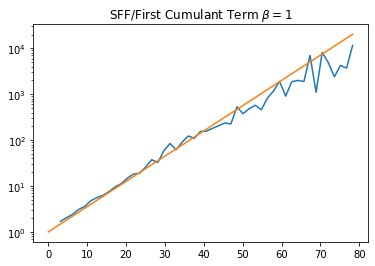}
    \includegraphics[scale=0.4]{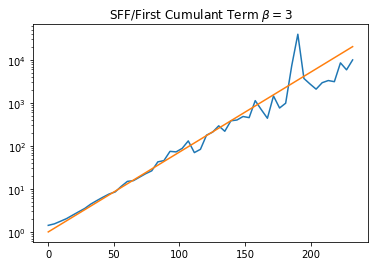}
    \includegraphics[scale=0.4]{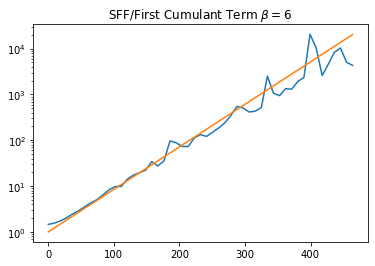}\\
    \includegraphics[scale=0.4]{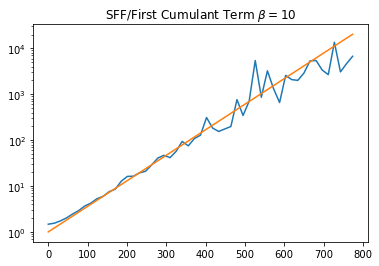}
    \includegraphics[scale=0.4]{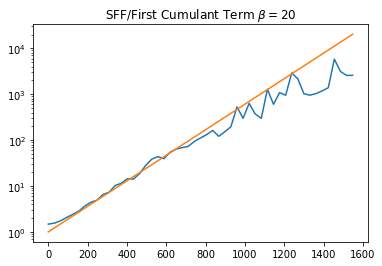}
    \includegraphics[scale=0.4]{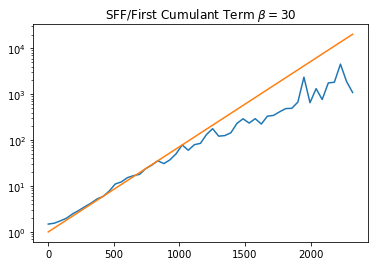}
    \caption{The quantity in blue is a numerical estimate for the SFF divided by the theoretical value from just the first cumulant. The value in orange is the second cumulant term, which seems to explain all the remaining discrepancy. At later times the two curves seem to diverge, suggesting that at sufficiently long times later terms in the cumulant expansion might be necessary.}
    \label{fig:SYK2Graphs}
\end{figure}

The next term is 
\begin{equation}
    c_2=\textrm {var}\sum_{j\neq 0}{(-1)^j}jZ(ijT-j\beta)=\frac{1}{\bbeta\pi}\sum_{j\neq 0}\frac 1{j^2}\int d\lambda \exp(-2j\beta \lambda)jT
\end{equation}
The sum can be evaluated for any values of the parameters. In the case where $\beta E_{\text{max}}\gg1$, that sum becomes simple:
\begin{equation}
    \textrm {var}\sum_{j\neq 0}{(-1)^j}jZ(ijT-j\beta)=\frac{1}{\bbeta \pi}\sum_{j\neq 0} \frac{T}{2\beta j^2}=\frac{\pi T}{12\beta}
\end{equation}
We can graph $\log \left(\text{SFF}(T,\beta)\right)-c_1$ and $c_2$ as functions of time. We see good alignment in figure \ref{fig:SYK2Graphs}.

These findings can be interestingly compared with the so-called Sigma model, one of the state-of-the-art tools in spectral statistics. This technique consists of studying a many-body version of a one-body chaotic system \cite{Altland_2021,barney2023spectral,Altland_2021Operator}. By calculating the partition function of a Sigma Model, one can extract the spectral determinant, and thus the SFF, of the system of interest. The work in this paper as well as \cite{PhysRevLett.125.250601,winerprl} shows that not just the partition function but also the SFF of the many-body system contains signatures of the single-body SFF.

\section{Nonlinear Hydrodynamics and Interacting Sound Waves}
\label{app:Interaction}

In this appendix, we sketch the inclusion of hydrodynamic interactions using a diagrammatic perturbation theory along the lines discussed in~\cite{winerprx}. We consider the basic diagramatic setup and, as an example, evaluate a single diagram. We do not give a comprehensive analysis of interaction effects and we discuss some key open issues at the end of the section.

Let's modify our action in equation \eqref{eq:lhydro1} to include interactions among the various modes. We now have a Lagrangian that includes a huge number of possible cubic and higher-order interactions. This leads to perturbative corrections to $\log \Zs$. For concreteness, we will choose the Lagrangian
\begin{equation}
        L_{\text{Kel}}=\frac 12 \phi_a\left(\partial_t^2+\frac{2\Gamma}{c^2} \partial_t^3-c^2\partial_\mu^2\right)\phi_r+\frac{2i\Gamma}{\beta c^2}\phi_a\partial_t^2\phi_a+\lambda(\partial_\mu \phi_r)^2 \partial_t \phi_r \partial_t \phi_a,
\end{equation}
though there are other quadratic terms with comparable effects, and the action we wrote isn't even KMS invariant.

We evaluate corrections to $\Zs$ diagrammatically, according to the rules in \cite{winerprx}. The propagators in the CTP formalism are
\begin{equation}
    \begin{split}
        G^{CTP}_{ra}(t,k)=\textrm{Re} \frac {e^{-\Gamma k^2t}}{ic|k|-\Gamma k^2} \exp\left(ic|k|t\right)\theta(t),\\
        G^{CTP}_{rr}(t,k)=\textrm{Re} \frac {e^{-\Gamma k^2|t|}}{ic|k|-\Gamma k^2} \exp\left(ic|k||t|\right).
    \end{split}
\end{equation}
In the DPT formalism, propagators are wrapped around the time circle according to
\begin{equation}
    G^{DPT}(t,k)=\sum_{n=-\infty}^\infty G^{CTP}(t+nT,k).
    \label{eq:GDPT}
\end{equation}
Note that the IR divergence for a sound-pole system is much less pronounced than in the case of diffusive hydrodynamics.

\begin{figure}
    \centering
    \includegraphics[scale=0.3]{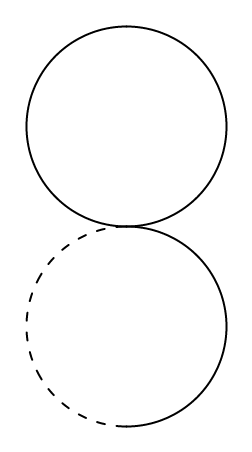}
    \caption{The dashed line represents $\phi_a$, the solid line represents $\phi_r$.}
    \label{fig:figureEight}
\end{figure}

With our propagators in hand, it is time to evaluate Feynman diagrams. Since the SFF is a partition function on an unorthodox manifold, the correction to the log of the SFF is given by a sum of bubble diagrams. The leading diagram is given in figure \ref{fig:figureEight}.
This can be factored to
\begin{equation}
    \Delta L=-\lambda\left(\int d^dk_1 k_1^2 G^{DPT}_{rr}(t=0,k_1)\right)\left(\int d^dk_2 \partial_t^2G^{DPT}_{ra}(t=0,k_2)\right) 
    \label{eq:2ints}
\end{equation}
The first of the integrals is UV divergent. We will impose a hard cutoff at $k=\Lambda$. In keeping with realistic hydrodynamics, we will chose $\Lambda\ll \frac c\Gamma$. With this cutoff, the first integral works out to $\frac{S_{d-1}}{d+2}\frac{\Gamma}{c^2}\Lambda^{d+2}$.
The second integral can be evaluated as
\begin{equation}
\begin{split}
\int d^dk_2 \partial_t^2G^{DPT}_{ra}(t=0,k)=\sum_{j=1}^\infty \int d^dk \partial_t^2G^{CTP}_{ra}(jT,k)=
\textrm{Re}\sum_{j=1}^\infty \int d^dk (ic|k|-\Gamma k^2){e^{-j\Gamma k^2T}}\exp\left(ijc|k|T\right)=\\
S_{d-1}\textrm{Re}\sum_{j=1}^\infty \int_0^\infty dk (ick-\Gamma k^2)|k|^{d-1}{e^{-j\Gamma k^2T}}\exp\left(ijckT\right)=\\
\frac{S_{d-1}}{2}\sum_{j=1}^\infty \int_{-\infty}^\infty dk (ick-\Gamma k^2)|k|^{d-1}{e^{-j\Gamma k^2T}}\exp\left(ijckT\right)
\end{split}
\label{eq:GraSum}
\end{equation}
Much like equation \eqref{eq:dip}, the integral in equation \eqref{eq:GraSum} has long-time behavior if $d$ is even, and a quick decay if $d$ is odd. The integral works out to
\begin{equation}
    \int_{-\infty}^\infty dk (ick-\Gamma k^2)|k|^{d-1}{e^{-j\Gamma k^2T}}\exp\left(ijckT\right)= 
\begin{cases}
  \sqrt{\frac{2\pi }{j\Gamma T}}\left(c\frac 1{(jT)^d}\partial_c^d-\Gamma\frac 1{(jT)^{d+1}}\partial_c^{d+1}\right)\exp(-j\frac{c^2}{4\Gamma}T)&\text{if }d \textrm{ odd}\\
  c\frac{(d)!}{(ijcT)^{d+1}}+O(T^{-d-2})          &\text{if }d \textrm{ even}
\end{cases}
\label{eq:cases}
\end{equation}
For odd $d$, the sum in equation \eqref{eq:GraSum} is dominated by the $j=1$ term, while for even $d$ we extract a value of $c\zeta(d+1)\frac{(d)!}{(icT)^{d+1}}+O(T^{-d-2})$.

Multiplying everything together, we get
\begin{equation}
    \Delta \text{(action)}/(VT)\approx\begin{cases}
        -\lambda \frac{S_{d-1}^2}{2(d+2)}\frac{\Gamma}{c^2}\Lambda^{d+2}\sqrt{\frac{2\pi }{\Gamma T}}\left(c\frac 1{(T)^d}\partial_c^d-\Gamma\frac 1{(T)^{d+1}}\partial_c^{d+1}\right)\exp(-\frac{c^2}{4\Gamma}T)&\text{if }d \textrm{ odd},\\
        -\lambda \frac{S_{d-1}^2}{2(d+2)}\frac{\Gamma}{c}\Lambda^{d+2}\zeta(d+1)\frac{(d)!}{(icT)^{d+1}}&\text{if }d \textrm{ even}.\\
    \end{cases}
\end{equation}
This can be interpreted as a renormalization of $c$ in equation \eqref{eq:dip}. More complicated diagrams have less straightforward interpretations, but any individual diagram can be evaluated using these Feynman rules. 

There are also more complicated effects with no obvious diagrammatic interpretation. All of the effects in this section are renormalizations to the first cumulant in the exponent. But presumably higher cumulants of the single-particle density of states can also be renormalized by interactions. We do not know how to systematically investigate these effects. For example, in writing equation \eqref{eq:2ints} we are assuming a density of states where the number of Laplacian eigenvalues between $k^2$ and $(k+\delta k)^2$ is $\frac{V}{(2\pi)^d}S_{d-1}k^{d-1}\delta k$. More realistically, this density would be a fluctuating variable. Would the overlaps between different eigenstates depend on the relative energies and on the density of states? We don't know. The higher-cumulant analogues of the calculation in this appendix are an open question.
\renewcommand{\thechapter}{A4}

\chapter{Appendices for Chapter 5}
\label{chapter:app4}

\section{Derivation of Schwinger-Keldysh TAP equations} \label{sec:TAP_derivation}

Here we derive the TAP equations on the Schwinger-Keldysh contour, given by Eqs.~\eqref{eq:Keldysh_TAP_EOM} and~\eqref{eq:Keldysh_TAP_magnetization_equation} of the main text.
Our derivation is a straightforward generalization of that in Ref.~\cite{Biroli_2001} for the thermal circle, but since we are not aware of it appearing in the literature, we aim to make this section as self-contained as possible.

We begin with the expression for the TAP action, given by Eq.~\eqref{eq:Keldysh_TAP_action} and reproduced here:
\begin{equation} \label{eq:Keldysh_TAP_action_reproduction}
\begin{aligned}
iNS_{\textrm{TAP}}[m, \mathcal{G}, \eta] &\equiv \log{\int \mathcal{D}\sigma^N \exp \left[ i \sum_i S_i^0 - i \eta \int_{\mathcal{C}} dt \sum_{(i_1 \cdots i_p)} J_{i_1 \cdots i_p} \sigma_{i_1}(t) \cdots \sigma_{i_p}(t) \right]} \\
&\qquad \qquad + \frac{iN}{2} \int_{\mathcal{C}} dt z(t) - i \int_{\mathcal{C}} dt \sum_i h_i(t) m_i(t) + \frac{iN}{2} \int_{\mathcal{C}} dt dt' \Lambda(t, t') \mathcal{G}(t, t'),
\end{aligned}
\end{equation}
where $\mathcal{C}$ denotes the Schwinger-Keldysh contour and
\begin{equation} \label{eq:Keldysh_TAP_noninteracting_action_reproduction}
S_i^0 \equiv \int_{\mathcal{C}} dt \left( \frac{\mu}{2} \big( \partial_t \sigma_i(t) \big)^2 - \frac{z(t)}{2} \sigma_i(t)^2 + h_i(t) \sigma_i(t) \right) - \frac{1}{2} \int_{\mathcal{C}} dt dt' \Lambda(t, t') \sigma_i(t) \sigma_i(t').
\end{equation}
Note that we have included an additional parameter $\eta$ governing the strength of interactions.
We calculate $S_{\textrm{TAP}}$ by expanding in $\eta$, ultimately setting $\eta = 1$.
One can show that all terms beyond second order vanish in the thermodynamic limit~\cite{Plefka1982Convergence,Biroli_2001}, so we need only calculate $S_{\textrm{TAP}}$ and its first two derivatives at $\eta = 0$.
Then
\begin{equation} \label{eq:TAP_action_reconstruction}
S_{\textrm{TAP}}[m, \mathcal{G}, 1] \sim S_{\textrm{TAP}}[m, \mathcal{G}, 0] + \frac{\partial S_{\textrm{TAP}}[m, \mathcal{G}, 0]}{\partial \eta} + \frac{1}{2} \frac{\partial^2 S_{\textrm{TAP}}[m, \mathcal{G}, 0]}{\partial \eta^2}.
\end{equation}
We proceed term by term.

\subsubsection*{Zeroth order}

At zeroth order in $\eta$, we have that
\begin{equation} \label{eq:TAP_action_zeroth_order_start}
\begin{aligned}
iNS_{\textrm{TAP}}[m, \mathcal{G}, 0] &= \sum_i \log{\int \mathcal{D}\sigma_i e^{iS_i^0}} \\
&\quad + \frac{iN}{2} \int_{\mathcal{C}} dt z(t) - i \int_{\mathcal{C}} dt \sum_i h_i(t) m_i(t) + \frac{iN}{2} \int_{\mathcal{C}} dt dt' \Lambda(t, t') \mathcal{G}(t, t').
\end{aligned}
\end{equation}
The first term evaluates to
\begin{equation} \label{eq:noninteracting_partition_evaluation}
\begin{aligned}
&\int \mathcal{D}\sigma_i \exp \left[ -\frac{i}{2} \int_{\mathcal{C}} dt dt' \sigma_i(t) \left( \delta(t - t') \big( \mu \partial_t^2 + z(t) \big) + \Lambda(t, t') \right) \sigma_i(t') + i \int_{\mathcal{C}} dt h_i(t) \sigma_i(t) \right] \\
&\qquad \qquad \qquad = \textrm{Det} \Big[ i (\mu \partial_t^2 + z) + i \Lambda \Big]^{-\frac{1}{2}} \exp \left[ -\frac{1}{2} \int_{\mathcal{C}} dt dt' h_i(t) \Big( i (\mu \partial_t^2 + z) + i \Lambda \Big)^{-1}(t, t') h_i(t') \right].
\end{aligned}
\end{equation}
We further have that
\begin{equation} \label{eq:TAP_noninteracting_correlations}
\big< \sigma_i(t) \sigma_i(t') \big> - \big< \sigma_i(t) \big> \big< \sigma_i(t') \big> = \Big( i (\mu \partial_t^2 + z) + i \Lambda \Big)^{-1}(t, t'),
\end{equation}
\begin{equation} \label{eq:TAP_noninteracting_magnetizations}
\big< \sigma_i(t) \big> = i \int_{\mathcal{C}} dt' \Big( i (\mu \partial_t^2 + z) + i \Lambda \Big)^{-1}(t, t') h_i(t').
\end{equation}
Note that the left-hand sides are constrained to be $\mathcal{G}(t, t') - Q(t, t')$ and $m_i(t)$ (summing over $i$ for the former), where $Q(t, t') \equiv N^{-1} \sum_i m_i(t) m_i(t')$.
Thus we can use Eqs.~\eqref{eq:TAP_noninteracting_correlations} and~\eqref{eq:TAP_noninteracting_magnetizations} to express the action in terms of them.
The first term of Eq.~\eqref{eq:Keldysh_TAP_noninteracting_action_reproduction} becomes
\begin{equation} \label{eq:TAP_noninteracting_evaluation_1}
\frac{N}{2} \log{\textrm{Det}} \Big[ \mathcal{G} - Q \Big] + \frac{N}{2} \int_{\mathcal{C}} dt dt' \Big( \mathcal{G} - Q \Big)^{-1}(t, t') Q(t, t'),
\end{equation}
and the second line can be written
\begin{equation} \label{eq:TAP_noninteracting_evaluation_2}
-N \int_{\mathcal{C}} dt dt' \Big( \mathcal{G} - Q \Big)^{-1}(t, t') Q(t, t') - \frac{iN \mu}{2} \int_{\mathcal{C}} dt \partial_t^2 \mathcal{G}(t, t') \Big|_{t' = t} + \frac{N}{2} \int_{\mathcal{C}} dt dt' \Big( \mathcal{G} - Q \Big)^{-1}(t, t') \mathcal{G}(t, t').
\end{equation}
Combining the two, we have that (up to an unimportant constant)
\begin{equation} \label{eq:TAP_noninteracting_action_result}
iS_{\textrm{TAP}}[m, \mathcal{G}, 0] = \frac{1}{2} \log{\textrm{Det}} \Big[ \mathcal{G} - Q \Big] - \frac{i \mu}{2} \int_{\mathcal{C}} dt \partial_t^2 \mathcal{G}(t, t') \Big|_{t' = t}.
\end{equation}

\subsubsection*{First order}

Strictly speaking, since $h_i(t)$ and $\Lambda(t, t')$ are chosen so that $\langle \sigma_i(t) \rangle = m_i(t)$ and $\sum_i \langle \sigma_i(t) \sigma_i(t') \rangle = N\mathcal{G}(t, t')$ at every value of $\eta$, they themselves are functions of $\eta$ (as is $z(t)$).
However, the Legendre-transform structure of the TAP action ensures that the total derivative with respect to $\eta$ equals the partial derivative holding fields fixed (for example, the contribution to $\partial S_{\textrm{TAP}} / \partial \eta$ from $\partial h_i(t) / \partial \eta$ comes with a factor $\langle \sigma_i(t) \rangle - m_i(t) = 0$).
Thus we have simply that
\begin{equation} \label{eq:TAP_first_order_evaluation_general}
\frac{\partial S_{\textrm{TAP}}[m, \mathcal{G}, \eta]}{\partial \eta} = -\frac{1}{N} \int_{\mathcal{C}} dt \sum_{(i_1 \cdots i_p)} J_{i_1 \cdots i_p} \big< \sigma_{i_1}(t) \cdots \sigma_{i_p}(t) \big>.
\end{equation}
At $\eta = 0$, the spins are non-interacting and the expectation value factors:
\begin{equation} \label{eq:TAP_first_order_evaluation_noninteracting}
\frac{\partial S_{\textrm{TAP}}[m, \mathcal{G}, 0]}{\partial \eta} = -\frac{1}{N} \int_{\mathcal{C}} dt \sum_{(i_1 \cdots i_p)} J_{i_1 \cdots i_p} m_{i_1}(t) \cdots m_{i_p}(t).
\end{equation}

\subsubsection*{Second order}

The second derivative requires a bit more work.
From Eq.~\eqref{eq:TAP_first_order_evaluation_general}, we have that
\begin{equation} \label{eq:TAP_second_derivative_starting_point}
\begin{aligned}
\frac{\partial^2 S_{\textrm{TAP}}[m, \mathcal{G}, \eta]}{\partial \eta^2} &= \frac{i}{N} \int_{\mathcal{C}} dt dt' \sum_{II'} J_I J_{I'} \Big( \big< \sigma_I(t) \sigma_{I'}(t') \big> - \big< \sigma_I(t) \big> \big< \sigma_{I'}(t') \big> \Big) \\
&\quad - \frac{i}{N} \int_{\mathcal{C}} dt dt' \sum_{Ii'} J_I \frac{\partial h_{i'}(t')}{\partial \eta} \Big( \big< \sigma_I(t) \sigma_{i'}(t') \big> - \big< \sigma_I(t) \big> \big< \sigma_{i'}(t') \big> \Big) \\
&\quad + \frac{i}{2N} \int_{\mathcal{C}} dt dt' \sum_{Ii'} J_I \frac{\partial z(t')}{\partial \eta} \Big( \big< \sigma_I(t) \sigma_{i'}(t')^2 \big> - \big< \sigma_I(t) \big> \big< \sigma_{i'}(t')^2 \big> \Big) \\
&\quad + \frac{i}{2N} \int_{\mathcal{C}} dt dt' dt'' \sum_{Ii'} J_I \frac{\partial \Lambda(t', t'')}{\partial \eta} \Big( \big< \sigma_I(t) \sigma_{i'}(t') \sigma_{i'}(t'') \big> - \big< \sigma_I(t) \big> \big< \sigma_{i'}(t') \sigma_{i'}(t'') \big> \Big),
\end{aligned}
\end{equation}
where for brevity, we use $I$ to denote $(i_1 \cdots i_p)$ and $\sigma_I(t)$ to denote $\sigma_{i_1}(t) \cdots \sigma_{i_p}(t)$.
To evaluate the partial derivatives on the right-hand side, which we only need at $\eta = 0$, note that
\begin{equation} \label{eq:TAP_Legendre_relations}
N \frac{\partial S_{\textrm{TAP}}[m, \mathcal{G}, \eta]}{\partial m_i(t)} = -h_i(t), \qquad \frac{\partial S_{\textrm{TAP}}[m, \mathcal{G}, \eta]}{\partial \rho(t)} = \frac{1}{2} z(t), \qquad \frac{\partial S_{\textrm{TAP}}[m, \mathcal{G}, \eta]}{\partial \mathcal{G}(t, t')} = \frac{1}{2} \Lambda(t, t'),
\end{equation}
where we set $\sum_i \langle \sigma_i(t)^2 \rangle = N \rho(t)$ (even though ultimately $\rho(t) = 1$).
Thus, using Eq.~\eqref{eq:TAP_first_order_evaluation_noninteracting},
\begin{equation} \label{eq:TAP_field_derivatives_evaluation}
\begin{gathered}
\frac{\partial h_i(t)}{\partial \eta} = -N \frac{\partial^2 S_{\textrm{TAP}}[m, \mathcal{G}, 0]}{\partial m_i(t) \partial \eta} = \sum_{(i_1 \cdots i_p)} J_{i_1 \cdots i_p} \frac{\partial \big( m_{i_1}(t) \cdots m_{i_p}(t) \big)}{\partial m_i(t)}, \\
\frac{\partial z(t)}{\partial \eta} = 2 \frac{\partial^2 S_{\textrm{TAP}}[m, \mathcal{G}, 0]}{\partial \rho(t) \partial \eta} = 0, \qquad \frac{\partial \Lambda(t, t')}{\partial \eta} = 2 \frac{\partial^2 S_{\textrm{TAP}}[m, \mathcal{G}, 0]}{\partial \mathcal{G}(t, t') \partial \eta} = 0.
\end{gathered}
\end{equation}
The second derivative simplifies to
\begin{equation} \label{eq:TAP_second_derivative_simplified}
\begin{aligned}
\frac{\partial^2 S_{\textrm{TAP}}[m, \mathcal{G}, \eta]}{\partial \eta^2} &= \frac{i}{N} \int_{\mathcal{C}} dt dt' \sum_{II'} J_I J_{I'} \Big( \big< \sigma_I(t) \sigma_{I'}(t') \big> - \big< \sigma_I(t) \big> \big< \sigma_{I'}(t') \big> \Big) \\
&\quad - \frac{i}{N} \int_{\mathcal{C}} dt dt' \sum_{II'i'} J_I J_{I'} \Big( \big< \sigma_I(t) \sigma_{i'}(t') \big> - \big< \sigma_I(t) \big> \big< \sigma_{i'}(t') \big> \Big) \frac{\partial \big< \sigma_{I'}(t') \big>}{\partial m_{i'}(t')}.
\end{aligned}
\end{equation}
Just as one can show that higher-order $\eta$ derivatives vanish in the thermodynamic limit, one can also show that it is safe to replace $J_I J_{I'}$ by its average in Eq.~\eqref{eq:TAP_second_derivative_simplified}~\cite{Plefka1982Convergence,Biroli_2001}.
Again using that expectation values factor at $\eta = 0$, we therefore have that
\begin{equation} \label{eq:TAP_second_order_evaluation}
\frac{\partial^2 S_{\textrm{TAP}}[m, \mathcal{G}, 0]}{\partial \eta^2} = \frac{iJ^2}{p} \int_{\mathcal{C}} dt dt' \bigg[ \mathcal{G}(t, t')^p - Q(t, t')^p - p \Big( \mathcal{G}(t, t') - Q(t, t') \Big) Q(t, t')^{p-1} \bigg].
\end{equation}

\subsubsection*{Full TAP action}

Inserting Eqs.~\eqref{eq:TAP_noninteracting_action_result},~\eqref{eq:TAP_first_order_evaluation_noninteracting}, and~\eqref{eq:TAP_second_order_evaluation} into Eq.~\eqref{eq:TAP_action_reconstruction}, we find that
\begin{equation} \label{eq:TAP_action_full_result}
\begin{aligned}
iS_{\textrm{TAP}} &= \frac{1}{2} \log{\textrm{Det}} \Big[ \mathcal{G} - Q \Big] - \frac{i \mu}{2} \int_{\mathcal{C}} dt \partial_t^2 \mathcal{G}(t, t') \Big|_{t' = t} - \frac{i}{N} \int_{\mathcal{C}} dt \sum_{(i_1 \cdots i_p)} J_{i_1 \cdots i_p} m_{i_1}(t) \cdots m_{i_p}(t) \\
&\qquad \qquad - \frac{J^2}{2p} \int_{\mathcal{C}} dt dt' \bigg[ \mathcal{G}(t, t')^p + (p-1) Q(t, t')^p - p \mathcal{G}(t, t') Q(t, t')^{p-1} \bigg].
\end{aligned}
\end{equation}
As discussed in the main text, the TAP equations come from setting to zero the derivatives with respect to $m_i(t)$ and $\mathcal{G}(t, t')$.
However, since $\mathcal{G}(t, t)$ is not free to vary (it must equal 1 by the spherical constraint), we include a Lagrange multiplier that we again denote by $z(t)$.
The TAP equations are thus
\begin{equation} \label{eq:original_TAP_correlation_equation}
\Big( \mathcal{G} - Q \Big)^{-1}(t, t') - i\mu \partial_t^2 \delta(t - t') - J^2 \big( \mathcal{G}(t, t')^{p-1} - Q(t, t')^{p-1} \big) = iz(t) \delta(t - t'),
\end{equation}
\begin{equation} \label{eq:original_TAP_magnetization_equation}
\begin{aligned}
\int_{\mathcal{C}} dt' \Big( \mathcal{G} - Q \Big)^{-1}(t, t') m_i(t') &+ i \sum_{(i_1 \cdots i_p)} J_{i_1 \cdots i_p} \frac{\partial \big( m_{i_1}(t) \cdots m_{i_p}(t) \big)}{\partial m_i(t)} \\
&- (p-1)J^2 \int_{\mathcal{C}} dt' \Big( \mathcal{G}(t, t') - Q(t, t') \Big) Q(t, t')^{p-2} m_i(t') = 0.
\end{aligned}
\end{equation}
Time translation invariance allows us to further simplify by setting $m_i(t) = m_i$ and $z(t) = z$.
Note that then $Q(t, t') = N^{-1} \sum_i m_i^2 \equiv q_{\textrm{EA}}$.
After some rearrangement, the TAP equations can be written
\begin{equation} \label{eq:final_TAP_correlation_equation}
i \big( \mu \partial_t^2 + z \big) \Big( \mathcal{G}(t, t') - q_{\textrm{EA}} \Big) + J^2 \int_{\mathcal{C}} dt'' \Big( \mathcal{G}(t, t')^{p-1} - q_{\textrm{EA}}^{p-1} \Big) \Big( \mathcal{G}(t'', t') - q_{\textrm{EA}} \Big) = \delta(t - t'),
\end{equation}
\begin{equation} \label{eq:final_TAP_magnetization_equation}
J^2 \int_{\mathcal{C}} dt' \Big( \mathcal{G}(t, t')^{p-1} - (p-1) q_{\textrm{EA}}^{p-2} \mathcal{G}(t, t') + (p-2) q_{\textrm{EA}}^{p-1} \Big) m_i = -izm_i - i \sum_{(i_1 \cdots i_p)} J_{i_1 \cdots i_p} \frac{\partial (m_{i_1} \cdots m_{i_p})}{\partial m_i}.
\end{equation}
These are precisely Eqs.~\eqref{eq:Keldysh_TAP_EOM} and~\eqref{eq:Keldysh_TAP_magnetization_equation} from the main text.

\section{Energy of a TAP state} \label{sec:TAP_energy}

Here we derive an expression for the total energy density $\epsilon$ in terms of a solution $\mathcal{G}(t, t')$ to the TAP equations, Eqs.~\eqref{eq:final_TAP_correlation_equation} and~\eqref{eq:final_TAP_magnetization_equation}.

Note that, despite the complicated derivation, $\mathcal{G}(t, t')$ is simply the contour-ordered expectation value of $N^{-1} \sum_i \sigma_i(t) \sigma_i(t')$.
Thus, as we discuss below in App.~\ref{sec:filter_functions}, the kinetic energy per spin is given by
\begin{equation} \label{eq:TAP_kinetic_energy_expression}
\frac{\mu}{2N \Delta t^2} \sum_i \Big< \big[ \sigma_i(t + 2\Delta t) - \sigma_i(t + \Delta t) \big] \big[ \sigma_i(t + \Delta t) - \sigma_i(t) \big] \Big> = -\frac{\mu}{2} \partial_t^2 \mathcal{G}(t^+, t),
\end{equation}
where $t^+$ denotes $t + \Delta t$.

To evaluate the potential energy, we replace $J_{i_1 \cdots i_p}$ by $(1 + \zeta(t)) J_{i_1 \cdots i_p}$ in the original path integral.
On the one hand,
\begin{equation} \label{eq:TAP_potential_energy_formal_expression}
\frac{\partial}{\partial \zeta(t)} S_{\textrm{TAP}} \bigg|_{\zeta = 0} = -\frac{1}{N} \sum_{(i_1 \cdots i_p)} J_{i_1 \cdots i_p} \big< \sigma_{i_1}(t) \cdots \sigma_{i_p}(t) \big>.
\end{equation}
On the other hand, it is easy to see how the explicit calculation of $S_{\textrm{TAP}}$ is modified by $\zeta(t)$:
\begin{equation} \label{eq:TAP_action_zeta_modification}
\begin{aligned}
S_{\textrm{TAP}} = S_{\textrm{TAP}} \bigg|_{\zeta = 0} &- \frac{1}{N} \int_{\mathcal{C}} dt \zeta(t) \sum_{(i_1 \cdots i_p)} J_{i_1 \cdots i_p} m_{i_1}(t) \cdots m_{i_p}(t) \\
&+ \frac{iJ^2}{2p} \int_{\mathcal{C}} dt dt' \big( \zeta(t) + \zeta(t') \big) \bigg[ \mathcal{G}(t, t')^p + (p-1) Q(t, t')^p - p \mathcal{G}(t, t') Q(t, t')^{p-1} \bigg] + O(\zeta^2).
\end{aligned}
\end{equation}
Thus
\begin{equation} \label{eq:TAP_potential_energy_actual_expression}
\begin{aligned}
\frac{1}{N} \sum_{(i_1 \cdots i_p)} J_{i_1 \cdots i_p} \big< \sigma_{i_1}(t) \cdots \sigma_{i_p}(t) \big> &= \frac{1}{N} \sum_{(i_1 \cdots i_p)} J_{i_1 \cdots i_p} m_{i_1}(t) \cdots m_{i_p}(t) \\
&\quad - \frac{iJ^2}{p} \int_{\mathcal{C}} dt' \bigg[ \mathcal{G}(t, t')^p + (p-1) Q(t, t')^p - p \mathcal{G}(t, t') Q(t, t')^{p-1} \bigg],
\end{aligned}
\end{equation}
and the total energy per spin is given by
\begin{equation} \label{eq:TAP_full_energy_expression}
\begin{aligned}
\epsilon = -\frac{\mu}{2} \partial_t^2 \mathcal{G}(t^+, t) &+ \frac{1}{N} \sum_{(i_1 \cdots i_p)} J_{i_1 \cdots i_p} m_{i_1}(t) \cdots m_{i_p}(t) \\
&- \frac{iJ^2}{p} \int_{\mathcal{C}} dt' \bigg[ \mathcal{G}(t, t')^p + (p-1) Q(t, t')^p - p \mathcal{G}(t, t') Q(t, t')^{p-1} \bigg].
\end{aligned}
\end{equation}

Writing out the various components of $\mathcal{G}$ explicitly and using time translation invariance (as well as taking $t$ to be far from the thermal branch), we have that
\begin{equation} \label{eq:TAP_full_energy_Keldysh_expression}
\epsilon = -\frac{\mu}{2} \partial_t^2 \mathcal{G}_{\alpha \alpha}(0^+) + Jq_{\textrm{EA}}^{p/2} \mathcal{E} - \frac{iJ^2}{p} \int dt' \sum_{\alpha'} (-1)^{\alpha'} \bigg[ \mathcal{G}_{\alpha \alpha'}(t - t')^p - p q_{\textrm{EA}}^{p-1} \mathcal{G}_{\alpha \alpha'}(t - t') + (p-1) q_{\textrm{EA}}^p \bigg],
\end{equation}
where $\mathcal{E}$ is defined as in the main text (Eq.~\eqref{eq:normalized_energy_definition}).
Despite the appearance, this expression is independent of both $t$ and $\alpha$.
Compare it to the argument of the filter function from Sec.~\ref{sec:nonergodic_ramp} (Eq.~\eqref{eq:TAP_resolved_energy_density_definition}), reproduced here:
\begin{equation} \label{eq:TAP_resolved_energy_density_reproduction}
\epsilon_{\alpha}[\lambda, G] = -\frac{\mu}{2} \partial_t^2 G_{\alpha \alpha}(0^+) - \frac{iJ^2}{p} \int_0^T dt' \sum_{\alpha'} (-1)^{\alpha'} G_{\alpha \alpha'}(t')^p - \frac{iJ \lambda}{pq[G]^{p/2}} \left( \frac{1}{T} \int_0^T dt G_{\alpha u}(t, 0) \right)^p.
\end{equation}
Evaluated at the saddle point given in Sec.~\ref{subsec:TAP_resolved_connected_solutions}, Eq.~\eqref{eq:TAP_resolved_energy_density_reproduction} comes out to be (up to corrections small in $T^{-1}$)
\begin{equation} \label{eq:TAP_resolved_energy_density_equivalence}
\epsilon_{\alpha}[\lambda, G] = -\frac{\mu}{2} \partial_t^2 \mathcal{G}_{\alpha \alpha}(0^+) + Jq_{\textrm{EA}}^{p/2} \mathcal{E} - 2J^2 q_{\textrm{EA}}^{p-1} \Lambda - \frac{iJ^2}{p} \int dt' \sum_{\alpha'} (-1)^{\alpha'} \Big( \mathcal{G}_{\alpha \alpha'}(t')^p - q_{\textrm{EA}}^p \Big),
\end{equation}
which, since $2i \Lambda \equiv \int dt' \sum_{\alpha'} (-1)^{\alpha'} \mathcal{G}_{\alpha \alpha'}(t)$, is precisely the energy density of the TAP state, Eq.~\eqref{eq:TAP_full_energy_Keldysh_expression}.

\section{Accounting for filter functions}
\label{sec:filter_functions}

The purpose of this section is to derive Eqs.~\eqref{eq:SFF_formal_path_integral} through~\eqref{eq:energy_density_definition} of the main text.
In the original definition of the SFF,
\begin{equation} \label{eq:generic_SFF_definition_restated}
\textrm{SFF}(T, f) = \overline{\textrm{Tr} f(H) e^{-iHT} \, \textrm{Tr} f(H) e^{iHT}},
\end{equation}
we need to be careful in how we treat the filter functions, since $f(H)$ does depend on the specific disorder realization.
It is convenient to assume that $f(H)$ is an analytic function of energy \textit{density}, i.e., it can be expanded as
\begin{equation} \label{eq:filter_function_expansion_definition}
f(H) = \sum_{n=0}^{\infty} \frac{c_n}{n!} \left( \frac{H}{N} \right)^n,
\end{equation}
where the coefficients $c_n$ do not depend on $N$.
This is not a significant restriction, since we can still choose $f(H)$ to be as tightly concentrated around any given energy density $\epsilon_0$ as we wish.

Inserting Eq.~\eqref{eq:filter_function_expansion_definition} into Eq.~\eqref{eq:generic_SFF_definition_restated} gives
\begin{equation} \label{eq:SFF_expanded_expression}
\textrm{SFF}(T, f) = \sum_{n_u n_l} \frac{c_{n_u} c_{n_l}}{n_u! n_l!} \frac{1}{N^{n_u + n_l}} \overline{\textrm{Tr} H^{n_u} e^{-iHT} \, \textrm{Tr} H^{n_l} e^{iHT}}.
\end{equation}
We construct the SFF path integral as usual: write $H = H_{\textrm{q}} + H_{\textrm{cl}}$ with $H_{\textrm{q}} \equiv \sum_i \pi_i^2 / 2\mu$ the kinetic energy and $H_{\textrm{cl}}$ the potential energy (including a Lagrange multiplier for the spherical constraint), approximate
\begin{equation} \label{eq:generic_Trotterization}
e^{i H T} \sim (e^{i H_{\textrm{q}} \Delta t} e^{i H_{\textrm{cl}} \Delta t})^\frac{T}{\Delta t},
\end{equation}
insert resolutions of the identity alternating between the $|\sigma \rangle$ and the $|\pi \rangle$ basis, perform the Gaussian integrals over momenta, and finally take $\Delta t \rightarrow 0$.
The only subtlety is in where and how we insert the factors of $H$ during this process.
While any arrangement has to ultimately give the same answer, we can simplify things by making judicious choices.

First write each $H$ as $\Delta t \sum_t H/T$ and insert one term at every time slice, i.e.,
\begin{equation} \label{eq:H_insertion_distributing}
\begin{aligned}
\textrm{Tr} H^{n_u} e^{-iHT} \textrm{Tr} H^{n_l} e^{iHT} = \left( \frac{\Delta t}{T} \right)^{n_u + n_l} &\sum_{t_1 \cdots t_{n_u}} \textrm{Tr} e^{-iH(T - t_{n_u})} H e^{-iH(t_{n_u} - t_{n_u-1})} \cdots e^{-iH(t_2 - t_1)} H e^{-iHt_1} \\
\cdot &\sum_{t_1 \cdots t_{n_l}} \textrm{Tr} e^{iH(T - t_{n_l})} H e^{iH(t_{n_l} - t_{n_l-1})} \cdots e^{iH(t_2 - t_1)} H e^{iHt_1}.
\end{aligned}
\end{equation}
Note that, to leading order in $\Delta t$, no two factors of $H$ coincide at the same time.
Furthermore, each factor of $H$ is itself two terms, whose operators we arrange within the Trotterized exponential (Eq.~\eqref{eq:generic_Trotterization}) as indicated:
\begin{equation} \label{eq:exponential_insertion_placement}
\cdots e^{i H_{\textrm{q}} \Delta t} e^{i H_{\textrm{cl}} \Delta t} \left( \frac{1}{2\mu} \sum_i \underbracket{\pi_i} e^{i H_{\textrm{q}} \Delta t} e^{i H_{\textrm{cl}} \Delta t} \underbracket{\pi_i} \, + \, e^{i H_{\textrm{q}} \Delta t} e^{i H_{\textrm{cl}} \Delta t} \underbracket{H_{\textrm{cl}}} \right) e^{i H_{\textrm{q}} \Delta t} e^{i H_{\textrm{cl}} \Delta t} \cdots.
\end{equation}
With this arrangement, when we carry out the rest of the path integral procedure, we end up with the integrand
\begin{equation} \label{eq:H_insertion_resulting_integrand}
\frac{\mu}{2 \Delta t^2} \sum_i \big[ \sigma_i(t + 2 \Delta t) - \sigma_i(t + \Delta t) \big] \big[ \sigma_i(t + \Delta t) - \sigma_i(t) \big] + \sum_{(i_1 \cdots i_p)} J_{i_1 \cdots i_p} \sigma_{i_1}(t) \cdots \sigma_{i_p}(t).
\end{equation}
While these expressions are for the lower contour, those for the upper contour are exactly analogous.
Thus the $n_u n_l$ term of the SFF becomes
\begin{equation} \label{eq:SFF_single_expansion_term_path_integral}
\textrm{Tr} H^{n_u} e^{-iHT} \textrm{Tr} H^{n_l} e^{iHT} = \int \mathcal{D}\sigma^N e^{iS} \left( \frac{\Delta t}{T} \right)^{n_u + n_l} \sum_{t_1 \cdots t_{n_u}} H_u(t_1) \cdots H_u(t_{n_u}) \sum_{t_1 \cdots t_{n_l}} H_l(t_1) \cdots H_l(t_{n_l}),
\end{equation}
where the action $S$ is, in continuum notation,
\begin{equation} \label{eq:SFF_original_spin_action}
S = \int_0^T dt \sum_{\alpha} (-1)^{\alpha} \left[ \sum_i \left( \frac{\mu}{2} \big( \partial_t \sigma_i(t) \big)^2 - \frac{z(t)}{2} \big( \sigma_{i \alpha}(t)^2 - 1 \big) \right) - \sum_{(i_1 \cdots i_p)} J_{i_1 \cdots i_p} \sigma_{i_1 \alpha}(t) \cdots \sigma_{i_p \alpha}(t) \right],
\end{equation}
and we've defined (in a slight abuse of notation) the random function
\begin{equation} \label{eq:path_integral_energy_function}
H_{\alpha}(t) \equiv \frac{\mu}{2 \Delta t^2} \sum_i \big[ \sigma_{i \alpha}(t + 2 \Delta t) - \sigma_{i \alpha}(t + \Delta t) \big] \big[ \sigma_{i \alpha}(t + \Delta t) - \sigma_{i \alpha}(t) \big] + \sum_{(i_1 \cdots i_p)} J_{i_1 \cdots i_p} \sigma_{i_1 \alpha}(t) \cdots \sigma_{i_p \alpha}(t).
\end{equation}

Eq.~\eqref{eq:SFF_single_expansion_term_path_integral} must now be averaged over the random couplings.
Due to the term of the action linear in $J_{i_1 \cdots i_p}$, each coupling acquires an expectation value:
\begin{equation} \label{eq:annealed_coupling_expectation_value}
\overline{J_{i_1 \cdots i_p} e^{iS}} = -\frac{iJ^2 (p-1)!}{C_{i_1 \cdots i_p} N^{p-1}} \int_0^T dt' \sum_{\alpha'} (-1)^{\alpha'} \sigma_{i_1 \alpha'}(t') \cdots \sigma_{i_p \alpha'}(t').
\end{equation}
Thus in the average of Eq.~\eqref{eq:SFF_single_expansion_term_path_integral}, one contribution is the fully disconnected term in which each insertion of $H_{\alpha}(t)$ gets replaced by its mean value, denoted $N \epsilon_{\alpha}(t)$:
\begin{equation} \label{eq:SFF_energy_density_expression}
N \epsilon_{\alpha}(t) \equiv \frac{\mu}{2} \sum_i \big[ \partial_t \sigma_{i \alpha}(t^+) \big] \big[ \partial_t \sigma_{i \alpha}(t) \big] - \frac{iNJ^2}{p} \int_0^T dt' \sum_{\alpha'} (-1)^{\alpha'} \left( \frac{1}{N} \sum_i \sigma_{i \alpha}(t) \sigma_{i \alpha'}(t') \right)^p.
\end{equation}
In fact, all other contributions, which come from contractions of pairs of couplings about their means, are necessarily subleading in $1/N$: whereas the disconnected term contributes one factor of $N$ for every insertion of $H$, each contraction accounts for two insertions but contributes only one factor of $N$ between the two.
Thus, to leading order,
\begin{equation} \label{eq:SFF_single_expansion_term_average}
\overline{\textrm{Tr} H^{n_u} e^{-iHT} \textrm{Tr} H^{n_l} e^{iHT}} \sim \int \mathcal{D}\sigma^N \big( N \epsilon'_u \big)^{n_u} \big( N \epsilon'_l \big)^{n_l} \overline{e^{iS}},
\end{equation}
where we've defined $\epsilon'_{\alpha} \equiv T^{-1} \int dt \epsilon_{\alpha}(t)$.

Returning to the SFF, we see that we can simply resum the series for the filter functions, but now with the \textit{deterministic} function $\epsilon'$ in place of the random operator $H$:
\begin{equation} \label{eq:SFF_expression_filter_function_treated}
\textrm{SFF}(T, f) = \int \mathcal{D}\sigma^N f \big( \epsilon'_u \big) f \big( \epsilon'_l \big) \overline{e^{iS}}.
\end{equation}
The average of $e^{iS}$ is treated exactly as in Subsec.~\ref{subsec:Schwinger_Keldysh_path_integral}, albeit with the extra index $\alpha$.
It results in the effective action given by Eq.~\eqref{eq:SFF_formal_action} in the main text.
Note that, when we introduce the order parameter $G_{\alpha \alpha'}(t, t')$ (see Eq.~\eqref{eq:partition_function_fat_unity}), we can also substitute it into the expression for $\epsilon'$, giving the function
\begin{equation} \label{eq:energy_density_first_definition}
\epsilon_{\alpha}[G] \equiv -\frac{\mu}{2T} \int dt \partial_t^2 G_{\alpha \alpha}(t^+, t) - \frac{iJ^2}{pT} \int_0^T dt dt' \sum_{\alpha'} (-1)^{\alpha'} G_{\alpha \alpha'}(t, t')^p.
\end{equation}
The only difference from Eq.~\eqref{eq:energy_density_definition} is that the latter assumes $G_{\alpha \alpha'}(t, t')$ is time-translation invariant.
This is true of the saddle-point value, and the replacement is justified here because $f(\epsilon[G])$ is independent of $N$ and is thus evaluated solely at the saddle point in the large-$N$ limit.
We have therefore arrived at Eqs.~\eqref{eq:SFF_formal_path_integral} through~\eqref{eq:energy_density_definition} of the main text.

\renewcommand{\baselinestretch}{1}
\small\normalsize

\bibliographystyle{unsrt-phys}
\bibliography{reference1,reference2,reference3,reference4,reference5,reference6}

\end{document}